\documentclass[authoryear]{elsarticle}

\usepackage[english]{babel}

\usepackage[fig,chgbar]{fmocdmac}

\AtEndPreamble
  {

  }

\newtheorem{definition}{Definition}
\newtheorem{construction}{Construction}
\newtheorem{proposition}{Proposition}
\newtheorem{lemma}{Lemma}

\newtheorem{theorem}{Theorem}
\newtheorem{corollary}{Corollary}

\newtheorem{example}{Example}
\newtheorem{openproblem}{Open Problem}

% ...

% ...

\cmdtxtoparname{Meta}[Meta-ADIF]

\cmdmthargfun{hsp}

% ...

\usrmth{sup}{}{argfun}

\usrmth{psf}{}{argfun}

\usrmth{Cnt}{}{sym}[\odot]

% ...

\cmdmthsetext{TeamAsg}[TAsg]
\usrmth{Team}{Asg}{set}[X]

\newcommand{\EmpTeamAsg}{\emptyset}
\newcommand{\TrvTeamAsg}{\{\emptyfun\}}

\cmdmthsetext{HypAsg}[HAsg]
\usrmth{Hyp}{Asg}{str}[X]

\newcommand{\EmpHypAsg}{\emptyset}
\newcommand{\NulHypAsg}{\{\emptyset\}}
\newcommand{\TrvHypAsg}{\{\!\{\emptyfun\}\!\}}

\cmdmthsetext{FunAsg}[FAsg]
\usrmth{Fun}{Asg}{str}[F]

\usrmth{eql}{}{lrel}[=]
\usrmth{noteql}{}{lrel}[\neq]
\usrmth{eqv}{}{lrel}[\equiv]
\usrmth{noteqv}{}{lrel}[\not\equiv]
\usrmth{inc}{}{lrel}[\sqsubseteq]
\usrmth{notinc}{}{lrel}[\not\sqsubseteq]
\usrmth{unt}{}{lrel}[\parallel]
\usrmth{notunt}{}{lrel}[\nparallel]

\cmdmthargset{Chc}
\usrmth{chc}{Fun}{str}[\Gamma]

\cmdmthargset{Fnc}

\cmdmthargfun{par}
\cmdmthargfun{ext}
\cmdmthargfun{cyl}

\cmdmthsym{PhsI}[\shortdownarrow]
\cmdmthsym{PhsII}[\circlearrowright]

\cmdmthargfun{cht}

\cmdmthargfun{Rel}
\cmdmthargfun{Team}

% ...

\newcommand{\XRed}[1][]{{\color{red!50!black}\XSet[1#1]}}
\newcommand{\XBlue}[1][]{{\color{blue!50!black}\XSet[2#1]}}
\newcommand{\XGreen}[1][]{{\color{green!25!black}\XSet[3#1]}}
\newcommand{\asgRed}[1][]{{\color{red!50!black}\asgElm[#1]}}
\newcommand{\asgBlue}[1][]{{\color{blue!50!black}\asgElm[#1]}}
\newcommand{\asgGreen}[1][]{{\color{green!25!black}\asgElm[#1]}}

\newcommand{\iffExpl}[1]{\mathrel{\stackrel{\text{\tiny #1}}{\Leftrightarrow}}}

\newcommand{\Team}[1][]{{\XSet[#1]}}

\newcommand{\TeamInDual}[1][]{\Team[#1]'}
\newcommand{\TeamInDDual}{\mathring{\Team}}
\newcommand{\TeamHat}[1][]{\hat{\Team}_{#1}}

\newcommand{\domainValue}[1][]{{\aElm[#1]}}

\newcommand{\IFmodels}{\cmodels[\DIF][\forall]}

\newcommand{\EAmodels}{{\cmodels[][\QEA]}}
\newcommand{\AEmodels}{{\cmodels[][\QAE]}}

\newcommand{\asgChc}[1][i]{\asgElm((\FFun_j)_{j\leq #1})}
\newcommand{\VarX}[1][]{{X_{#1}}}
\newcommand{\Buckets}{\mathcal{B}}
\newcommand{\ChoiceFunsBuckets}{\text{ChcB}}

\newcommand{\choiceFunBucket}{\mu}
\newcommand{\SkolemizedSOF}{\text{Sk}}
\newcommand{\choiceFunBucketWinning}{\hat{\choiceFunBucket}}

\AfterEndPreamble{

\tikzstyle{thm}=[draw, rectangle]
\tikzstyle{lmm}=[draw, rectangle]
\tikzstyle{cor}=[draw, rectangle]
\tikzstyle{prp}=[draw, rectangle]
\tikzstyle{edgedep} = [->,>=latex,rounded corners=5pt]

\newcommand{\figHypAsgOrd}{
\begin{tikzpicture}[>=stealth', scale = 0.75, every node/.style={scale=.75}]

  \draw[very thick] (0, 0) rectangle (5, 5.75);

  \draw (-0.25, 6) node [below right, fill=white] {$\HypAsg[1]$};

  \draw[thick, color = red!50!black] (0.5, 4) rectangle (4.5, 5.25);
  \draw[thick, color = red!50!black] (0.5, 2.25) rectangle (4.5, 3.5);
  \draw[thick, color = red!50!black] (0.5, 0.5) rectangle (4.5, 1.75);

  \draw (0.25, 5.5) node [below right, fill=white] {$\XRed[1]$};
  \draw (0.25, 3.75)  node [below right, fill=white] {$\XRed[2]$};
  \draw (0.25, 2) node [below right, fill=white] {$\XRed[3]$};

  \draw (1.75, 4.625) node {$\asgElm[1]\;\asgElm[3]\;\asgElm[5]$};
  \draw (1.75, 2.875) node {$\asgElm[6]\;\asgElm[7]\;\asgElm[8]$};
  \draw (1.75, 1.125) node {$\asgElm[7]\;\asgElm[10]$};

  \draw[color = blue!50!black, dashed] (3, 4.25) rectangle (4.25, 5);
  \draw[color = blue!50!black, dashed] (3, 2.5) rectangle (4.25, 3.25);
  \draw[color = blue!50!black, dashed] (3, 0.75) rectangle (4.25, 1.5);

  \draw (3.625, 4.625) node {\color{blue!50!black}$\asgElm[2]\;\asgElm[4]$};
  \draw (3.625, 2.875) node {\color{blue!50!black}$\asgElm[2]\;\asgElm[4]$};
  \draw (3.625, 1.125) node {\color{blue!50!black}$\asgElm[6]\;\asgElm[9]$};

  \draw (6, 2.875) node {\LARGE$\inc$};

  \draw[very thick] (7, 0) rectangle (12, 5.75);

  \draw (6.75, 6) node [below right, fill=white] {$\HypAsg[2]$};

  \draw[thick, color = blue!50!black] (7.5, 4) rectangle (11.5, 5.25);
  \draw[thick, color = blue!50!black] (7.5, 2.25) rectangle (11.5, 3.5);
  \draw[thick] (7.5, 0.5) rectangle (11.5, 1.75);

  \draw (7.25, 5.5) node [below right, fill=white] {$\XBlue[1]$};
  \draw (7.25, 3.75) node [below right, fill=white] {$\XBlue[2]$};
  \draw (7.25, 2) node [below right, fill=white] {$\XSet[23]$};

  \draw (9.5, 4.625) node {\color{blue!50!black}$\asgElm[2]\;\asgElm[4]$};

  \draw (9.5, 2.875) node {\color{blue!50!black}$\asgElm[6]\;\asgElm[9]$};

  \draw (9.5, 1.125) node {$\asgElm[6]\;\asgElm[9]\;\asgElm[11]$};

  \draw[thick, ->] (4.30, 4.625) -- (7.45, 4.625);
  \draw[thick, ->] (4.30, 2.875) -- (7.45, 4.625);
  \draw[thick, ->] (4.30, 1.125) -- (7.45, 2.875);

\end{tikzpicture}
}

\newcommand{\figGraphThmHypRefMeta}{

\begin{figure}[ht]
	\begin{center}
		\begin{tikzpicture}
			\node[thm] at (0, 0) (thm13) {Theorem~\ref{thm:hyprefMeta}};
			\node[prp] at (-6, 0) (lmm10) {Lemma~\ref{lmm:mon}};
			\node[lmm] at (-3, -0.7) (lmm13) {Lemma~\ref{lmm:monMeta}};

			\draw[edgedep] (lmm10) -- (thm13);
			\draw[edgedep] (lmm13) -| (thm13);
			\draw[edgedep] (lmm10) -| (lmm13);
		\end{tikzpicture}
	\end{center}
	\caption{Dependency graph of Theorem~\ref{thm:hyprefMeta}.}
\end{figure}
}

\newcommand{\figGraphThmDblDltMeta}{
\begin{figure}[ht]
	\begin{center}
		\begin{tikzpicture}
			\node[thm] at (0, 0) (thm14) {Theorem~\ref{thm:dbldltMeta}};
			\node[thm] at (-6, 0.7) (thm13) {Theorem~\ref{thm:hyprefMeta}};
			\node[prp] at (-6, 0) (prp1) {Proposition~\ref{prp:empnultrv}};
			\node[lmm] at (-3, -0.7) (lmm2) {Lemma~\ref{lmm:dltii}};
			\node[lmm] at (-6, -1.4) (lmm1) {Lemma~\ref{lmm:dlti}};
			\node[] at (-9, 0.7) (dots) {$\ldots$};

			\draw[edgedep] (prp1) -- (thm14);
			\draw[edgedep] (lmm2) -| (thm14);
			\draw[edgedep] (lmm1) -| (thm14);
			\draw[edgedep] (thm13) -| (thm14);
			\draw[edgedep] (prp1) -| (lmm2);
			\draw[edgedep] (lmm1) -| (lmm2);
			\draw[edgedep] (dots) -- (thm13);
		\end{tikzpicture}
	\end{center}
	\caption{Dependency graph of Theorem~\ref{thm:dbldltMeta}.}
\end{figure}
}

\newcommand{\figGraphThmBnlLawI}{
\begin{figure}[ht]
	\begin{center}
		\begin{tikzpicture}
			\node[thm] at (0, 0) (thm3) {Theorem~\ref{thm:blnlawi}};
			\node[lmm] at (-3, 0) (lmm3) {Lemma~\ref{lmm:empnulhyp}};

			\draw[edgedep] (lmm3) -- (thm3);
		\end{tikzpicture}
	\end{center}
	\caption{Dependency graph of Theorem~\ref{thm:blnlawi}.}
\end{figure}
}

\newcommand{\figGraphThmHypRef}{
\begin{figure}[ht]
	\begin{center}
		\begin{tikzpicture}
			\node[thm] at (0, 0) (thm3) {Theorem~\ref{thm:hypref}};
			\node[lmm] at (-3, 0) (lmm3) {Theorem~\ref{thm:hyprefMeta}};

			\draw[edgedep] (lmm3) -- (thm3);
		\end{tikzpicture}
	\end{center}
	\caption{Dependency graph of Theorem~\ref{thm:hypref}.}
\end{figure}
}

\newcommand{\figGraphThmDblDlt}{
\begin{figure}[ht]
	\begin{center}
		\begin{tikzpicture}
			\node[thm] at (0, 0) (thm3) {Theorem~\ref{thm:dbldlt}};
			\node[lmm] at (-3, 0) (lmm3) {Theorem~\ref{thm:dbldltMeta}};

			\draw[edgedep] (lmm3) -- (thm3);
		\end{tikzpicture}
	\end{center}
	\caption{Dependency graph of Theorem~\ref{thm:dbldlt}.}
\end{figure}
}

\newcommand{\figGraphThmPrfExt}{
\begin{figure}[ht]
	\begin{center}
		\begin{tikzpicture}
			\node[thm] at (0, 0) (thm3) {Theorem~\ref{thm:prfext}};
			\node[lmm] at (-3, 0) (lmm3) {Theorem~\ref{thm:prfextMeta}};

			\draw[edgedep] (lmm3) -- (thm3);
		\end{tikzpicture}
	\end{center}
	\caption{Dependency graph of Theorem~\ref{thm:prfext}.}
\end{figure}
}

\newcommand{\figGraphThmPrfExtMeta}{
	\begin{figure}[ht]
		\begin{center}
			\begin{tikzpicture}
				\node[thm] at (0, 0) (thm4) {Theorem~\ref{thm:prfextMeta}};
				\node[thm] at (-3, 0) (thm14) {Theorem~\ref{thm:dbldltMeta}};
				\node[] at (-6, 0) (dots) {$\ldots$};

				\draw[edgedep] (thm14) -- (thm4);
				\draw[edgedep] (dots) -- (thm14);
			\end{tikzpicture}
		\end{center}
		\caption{Dependency graph of Theorem~\ref{thm:prfextMeta}.}
	\end{figure}
}

\newcommand{\figGraphThmFolSemAdq}{
	\begin{figure}[ht]
		\begin{center}
			\begin{tikzpicture}
				\node[thm] at (0, 0) (thm5) {Theorem~\ref{thm:folsemadq}};
				\node[lmm] at (-6, 1.4) (lmm1) {Lemma~\ref{lmm:dlti}};
				\node[lmm] at (-6, 0.7) (lmm2) {Lemma~\ref{lmm:dltii}};
				\node[lmm] at (-3, 0.7) (lmm4) {Lemma~\ref{lmm:foldlt}};
				\node[lmm] at (-3, 0) (lmm5) {Lemma~\ref{lmm:folqnt}};
				\node[lmm] at (-3, -0.7) (lmm6) {Lemma~\ref{lmm:folblncon}};
				\node[] at (-9, 0.7) (dots) {$\ldots$};

				\draw[edgedep] (lmm4) -| (thm5);
				\draw[edgedep] (lmm5) -- (thm5);
				\draw[edgedep] (lmm6) -| (thm5);
				\draw[edgedep] (lmm1) -| (lmm4);
				\draw[edgedep] (lmm2) -- (lmm4);
				\draw[edgedep] (dots) -- (lmm2);
			\end{tikzpicture}
		\end{center}
		\caption{Dependency graph of Theorem~\ref{thm:folsemadq}.}
	\end{figure}
}

\newcommand{\figGraphThmDifSemAdq}{
	\begin{figure}[ht]
		\begin{center}
			\begin{tikzpicture}
				\node[thm] at (0, 0) (thm6) {Theorem~\ref{thm:difsemadq}};
				\node[thm] at (-3, -0.7) (thm2) {Theorem~\ref{thm:dbldlt}};
				\node[] at (-6, -0.7) (dots) {$\ldots$};
				\node[lmm] at (-6, 0) (lmm1) {Lemma~\ref{lmm:dlti}};
				\node[lmm] at (-3, 0) (lmm8) {Lemma~\ref{lmm:teapar}};
				\node[lmm] at (-3, 0.7) (lmm7) {Lemma~\ref{lmm:cylext}};

				\draw[edgedep] (lmm7) -| (thm6);
				\draw[edgedep] (lmm8) -- (thm6);
				\draw[edgedep] (lmm1) -- (lmm8);
				\draw[edgedep] (thm2) -| (thm6);
				\draw[edgedep] (dots) -- (thm2);
			\end{tikzpicture}
		\end{center}
		\caption{Dependency graph of Theorem~\ref{thm:difsemadq}.}
	\end{figure}
}

\newcommand{\figGraphThmQntInt}{
	\begin{figure}[ht]
		\begin{center}
			\begin{tikzpicture}
				\node[thm] at (0, 0) (thm7) {Theorem~\ref{thm:qntint}};
				\node[thm] at (-3, -0.7) (thm5) {Theorem~\ref{thm:folsemadq}};
				\node[thm] at (-3, 0.7) (thm4) {Theorem~\ref{thm:prfextMeta}};
				\node[lmm] at (-3, 0) (lmm9) {Lemma~\ref{lmm:extint}};
				\node[lmm] at (-6, 0) (lmm2) {Lemma~\ref{lmm:dltii}};
				\node[] at (-6, 0.7) (dots1) {$\ldots$};
				\node[] at (-9, 0) (dots2) {$\ldots$};
				\node[] at (-6, -0.7) (dots3) {$\ldots$};

				\draw[edgedep] (thm4) -| (thm7);
				\draw[edgedep] (lmm9) -- (thm7);
				\draw[edgedep] (thm5) -| (thm7);
				\draw[edgedep] (lmm2) -- (lmm9);
				\draw[edgedep] (dots1) -- (thm4);
				\draw[edgedep] (dots2) -- (lmm2);
				\draw[edgedep] (dots3) -- (thm5);
			\end{tikzpicture}
		\end{center}
		\caption{Dependency graph of Theorem~\ref{thm:qntint}.}
	\end{figure}
}

\newcommand{\figGraphThmHST}{
	\begin{figure}[ht]
		\begin{center}
			\begin{tikzpicture}
				\node[thm] at (0, 0) (thm8) {Theorem~\ref{thm:hst}};
				\node[thm] at (-3, 0.7) (thm4) {Theorem~\ref{thm:prfextMeta}};
				\node[thm] at (-3, 0) (thm7) {Theorem~\ref{thm:qntint}};
				\node[] at (-6, 0.7) (dots1) {$\ldots$};
				\node[] at (-6, 0) (dots2) {$\ldots$};

				\draw[edgedep] (thm4) -| (thm8);
				\draw[edgedep] (thm7) -- (thm8);
				\draw[edgedep] (dots1) -- (thm4);
				\draw[edgedep] (dots2) -- (thm7);
			\end{tikzpicture}
		\end{center}
		\caption{Dependency graph of Theorem~\ref{thm:hst}.}
	\end{figure}
}

\newcommand{\figGraphThmAdifSolInt}{
	\begin{figure}[ht]
		\begin{center}
			\begin{tikzpicture}
				\node[thm] at (0, 0) (thm10) {Theorem~\ref{thm:adifsolint}};
				\node[thm] at (-3, 0.7) (thm5) {Theorem~\ref{thm:folsemadq}};
				\node[thm] at (-3, 0) (thm8) {Theorem~\ref{thm:hst}};
				\node[] at (-6, 0.7) (dots1) {$\ldots$};
				\node[] at (-6, 0) (dots2) {$\ldots$};

				\draw[edgedep] (thm5) -| (thm10);
				\draw[edgedep] (thm8) -- (thm10);
				\draw[edgedep] (dots1) -- (thm5);
				\draw[edgedep] (dots2) -- (thm8);
			\end{tikzpicture}
		\end{center}
		\caption{Dependency graph of Theorem~\ref{thm:adifsolint}.}
	\end{figure}
}

\newcommand{\figGraphThmSolAdfInt}{
	\begin{figure}[ht]
		\begin{center}
			\begin{tikzpicture}
				\node[thm] at (0, 0) (thm11) {Theorem~\ref{thm:soladfint}};
				\node[thm] at (-3, 0) (thm8) {Theorem~\ref{thm:hst}};
				\node[] at (-6, 0) (dots1) {$\ldots$};

				\draw[edgedep] (thm8) -- (thm11);
				\draw[edgedep] (dots1) -- (thm8);
			\end{tikzpicture}
		\end{center}
		\caption{Dependency graph of Theorem~\ref{thm:soladfint}.}
	\end{figure}
}

\newcommand{\figGraphThmGamThrSem}{
	\begin{figure}[ht]
		\begin{center}
			\begin{tikzpicture}
				\node[thm] at (0, 0) (thm12) {Theorem~\ref{thm:gamthrsem}};
				\node[lmm] at (-3, 0.7) (lmm13) {Lemma~\ref{lmm:MetaSkol}};
				\node[lmm] at (-3, 0) (lmm14) {Lemma~\ref{lmm:BucketSoundness}};
				\node[thm] at (-3, -0.7) (thm8) {Theorem~\ref{thm:hst}};
				\node[] at (-6, -0.7) (dots1) {$\ldots$};

				\draw[edgedep] (lmm13) -| (thm12);
				\draw[edgedep] (lmm14) -- (thm12);
				\draw[edgedep] (thm8) -| (thm12);
				\draw[edgedep] (dots1) -- (thm8);
			\end{tikzpicture}
		\end{center}
		\caption{Dependency graph of Theorem~\ref{thm:gamthrsem}.}
	\end{figure}
}

}

\hypersetup
  {
  pdftitle  = {Alternating (In)Dependence-Friendly Logic},
  pdfauthor = {D. Bellier, M. Benerecetti, D. Della Monica, F. Mogavero}
  }

\journal{}

\begin{document}

  \title{Alternating (In)Dependence-Friendly Logic}
  \author{Dylan Bellier, Massimo Benerecetti, Dario Della Monica, and
    Fabio Mogavero}



\begin{abstract}

Hintikka and Sandu originally proposed \emph{Independence Friendly Logic} (\IF)
as a first-order logic of \emph{imperfect information} to describe
\emph{game-theoretic phenomena} underlying the semantics of natural language.
The logic allows for expressing independence constraints among quantified
variables, in a similar vein to Henkin quantifiers, and has a nice
\emph{game-theoretic semantics} in terms of \emph{imperfect information games}.
However, the \IF semantics exhibits some limitations.
It treats the players asymmetrically, considering only one of the two players as
having imperfect information when evaluating truth, \resp, falsity, of a
sentence.
In addition, the truth and falsity of sentences coincide with the existence of a
uniform winning strategy for one of the two players in the semantic imperfect
information game.
As a consequence, \IF does admit undetermined sentences, which are neither true
nor false, thus failing the law of excluded middle.
In this paper, we investigate an extension of \IF, called \emph{Alternating
Dependence/Independence Friendly Logic} (\ADIF), tailored to overcome these
limitations.
To this end, we introduce a novel \emph{compositional semantics}, generalising
the one based on trumps proposed by Hodges for \IF.
The new semantics
\begin{inparaenum}[(i)]
  \item
    allows for meaningfully restricting both players at the same time,
  \item
    enjoys the property of game-theoretic determinacy,
  \item
    recovers the law of excluded middle for sentences, and
  \item
    grants \ADIF the full descriptive power of \emph{Second Order Logic}.
\end{inparaenum}
We also provide an equivalent \emph{Herbrand-Skolem semantics} and a
\emph{game-theoretic semantics} for the prenex fragment of \ADIF, the latter
being defined in terms of a determined infinite-duration game that precisely
captures the other two semantics on finite structures.

\end{abstract}



  \maketitle



\section{Introduction}

\emph{Informational independence} is a phenomenon that emerges quite naturally
in \emph{game theory}, as players in a game make moves based on what they know
about the state of the current play~\citep{NM44}.
In games such as Chess or Go, both players have \emph{perfect information} about
the current state of the play and the moves they and their adversary have
previously made.
For other games, like the card-games Poker and Bridge, the players have to make
decisions based only on \emph{partial} (\ie, \emph{imperfect})
\emph{information} on the state of the play.
In other words, in these latter games, players have to make decisions
\emph{informationally independent} of some of the choices made by the other
players.
Given the tight connection between games and logics, think for instance at
\emph{game-theoretic semantics}~\citep{Lor61,Lor68,Hin73}, a number of proposals
have been put forward to reason with or about informational independence, most
notably, \emph{Independence-Friendly Logic}~\citep{HS89}, \emph{Dependence
Logic}~\citep{Vaa07}, and logics derived
thereof~\citep{Gal12,GV13a,Kuu13,CGW13,Kuu15}.

Independence-Friendly Logic (\IF) was originally introduced by~\cite{HS89}, and
later extensively studied, \eg, in~\cite{MSS11}, as an extension of
\emph{First-Order Logic} (\FOL)~\citep{HA28} with informational independence as
first-class notion, and with applications in semantics of \emph{natural
language} in mind.
Unlike in \FOL, where quantified variables always functionally depend on all the
previously quantified ones, the values for quantified variables in \IF can be
chosen independently of the values of some specific variables quantified before
in the formula.
This is syntactically represented by means of the so called \emph{slashed
operator} notation, where, for instance, $(\LExs \varElm/\WSet) \varphiFrm$ is
intended to mean that variable $\varElm$ must be chosen independently (\ie,
without knowledge) of the values of the variables contained in the set $\WSet$.
The logic has a nice game-theoretic semantics~\citep{HS97}, given in terms of
games of imperfect information, where a sentence is true if the verifier player,
usually called Eloise, has a \emph{strategy} to win the semantic game.
If the falsifier player, Abelard, has a \emph{winning strategy}, then the
sentence is declared false.
Since games with imperfect information are considered here, neither situation
may occur, as the specific game may be \emph{undetermined}.
In this case, the corresponding sentence is neither true nor false, therefore
establishing a failure of the \emph{law of excluded middle}.
\cite{Hod97a} later developed a compositional semantics for \IF, by defining
satisfaction \wrt a set of assignment, called \emph{trump} (\aka \emph{teams},
in later iterations of the idea), instead of a single assignment as in classic
Tarskian semantics~\citep{Tar36,Tar44} of \FOL.
The high level intuition here is that a trump encodes the informational
uncertainty about what is the actual current assignment.

Dependence Logic~\citep{Vaa07} (\DL) takes a slightly different approach to the
problem, by separating quantifiers from dependence specification.
This is achieved by adding to \FOL the so called \emph{dependence atoms} of the
form $=\!(\xVec, \yvarElm)$, with the intended meaning that the value of
variable $\yvarElm$ is completely determined by, hence functionally dependent
on, the value of variables in the vector $\xVec$.
The separation of dependence constraints and quantifiers can express very
naturally dependencies on both quantified and non quantified variables, and
allows for a quite flexible approach to reasoning about dependence and
independence.
\DL has also been extended with other types of atoms like, \eg,
\emph{independence atoms}~\citep{GV13a} and \emph{inclusion/exclusion
atoms}~\citep{Gal12}.
The logic is expressively equivalent to both \IF and the existential fragment of
\emph{Second Order Logic} (\SOL)~\citep{HA38,Chu56,Sha91}.
As such, \DL still allows for undetermined sentences and is not closed under
classical negation.
To recover closure under negation and, consequently, the law of excluded middle,
\cite{Vaa07} introduced \emph{Team Logic} (\TL), an extension of \DL with the so
called \emph{contradictory negation} $\LNot$, an idea already investigated by
\cite{Hin96} in the context of \IF, where it was allowed only in front of a
sentence.
\TL is substantially more expressive than \DL, reaching the full descriptive
power of \SOL, covering, thus, the entire polynomial hierarchy~\citep{Sto76}.
However, in order to recover the nice properties of \FOL, such as the duality
between Boolean connectives and quantifiers, \TL requires two different versions
of the propositional connectives, $\LNeg$ and $\LNot$ for negation, $\LCon$ and
$\oplus$ for conjunction, $\LDis$ and $\otimes$ for disjunction, as well as an
additional pseudo quantifier $!\varElm$ called \emph{shriek}.
This approach also bears significant consequences.
In particular, \TL lacks any meaningful direct game-theoretic interpretation, as
also pointed out by~\cite{Vaa07}, which \DL still retains, mainly thanks to its
equivalence with existential \SOL.

There is a well-known connection between logics to reason with or about
informational independence and the extension of first-order logic with the
\emph{partially ordered} (\aka \emph{branching} or \emph{Henkin})
\emph{quantifiers}, originally proposed by~\cite{Hen61} to overcome the linear
dependence intrinsic in classic quantifier prefixes (see also~\cite{KM95} for a
comprehensive survey on the topic).  For instance, the sentence {\footnotesize
  $\left({\!\!\!
\begin{array}{c@{}c}
  {\forall \varElm[1]} & {\exists \yvarElm[1]}
\\
  {\forall \varElm[2]} & {\exists \yvarElm[2]}
\end{array}
}\!\!\!\right)$}$\varphi$
states that for all $\varElm[1]$ and $\varElm[2]$, there exists a value for
$\yvarElm[1]$, that only depends on $\varElm[1]$, and a value for $\yvarElm[2]$,
that only depends on $\varElm[2]$, such that $\varphi$ is true.
Sentences like this can easily be expressed in \IF by means of suitable variable
independence schemata.
For the sentence in the example, $\forall \varElm[1] \forall \varElm[2] \exists
(\yvarElm[1] / \{ \varElm[2] \}) \exists (\yvarElm[2] / \{ \varElm[1] \})
\varphi$ is an equivalent \IF sentence.
Similarly to \IF, the prenex fragment of the logic with Henkin quantifiers,
where a Henkin quantifier prefix is followed by a \FOL sentence~\citep{Wal70},
is known to be expressively equivalent to $\Sigma_{1}^{1}$, the existential
fragment of \SOL, while the full (non-prenex) logic was proved to be able to
express $\Delta_{2}^{1}$-properties by~\cite{End70}.

As observed by~\cite{BG86}, logics with Henkin quantifiers exhibit an asymmetric
nature from a game-theoretical viewpoint, in that they typically consider only
whether the existential player, Eloise, has a winning strategy that proves a
formula true.
This is, instead, solved in \IF, at the cost of indeterminacy of the logic, by
introducing two satisfaction relations, one for truth and one for falsity, and
by defining them in terms of uniform strategies for the players~\citep{MSS11}.
More specifically, a strategy for a player, either Eloise or Abelard, is said to
be uniform if for every variable $\varElm$, which is controlled by that player
and is required to be independent of a set of variables $\WSet$, the strategy
always chooses the same value in all the states of the game that differ only for
the values of the variables in $\WSet$.
To win the game and prove the sentence true, Eloise is required to have a
uniform strategy that wins every play induced by her strategy.
These compatible plays need not be compatible with any uniform strategy of the
adversary, meaning that when evaluating truth of a sentence, no restrictions to
the universal quantifiers controlled by Abelard actually apply.
A similar situation happens when evaluating falsity of a sentence.
In this case, Abelard, needs to have a uniform strategy that wins all the
compatible plays.
Here, the constraints on the existential variables are ignored.
The imperfect information nature of these games manifests itself in the
uniformity requirements that leads to indeterminacy of the logic.
This, in turn, implies that some sentences are neither true nor false.

The situation described above is also reflected in Hodges' separate use of
trumps and co-trumps in the compositional semantics he proposed for \IF.
His idea of using sets of assignments allows for mimicking the uniformity
constraints on the strategies in a compositional way.
Essentially, a trump records all the states, represented here as assignments,
the game could be in, depending on the possible choices made by Abelard and the
corresponding responses by Eloise.
These assignments correspond, intuitively, to the (partial) plays compatible
with the strategy followed by Eloise when evaluating the formula.
A trump can, then, encode the uncertainty that Eloise has about the actual
current state of the play, in that assignments that only differ for the
variables in $\WSet$ are indistinguishable to Eloise when she has to choose the
value of a variable $\varElm$ that is independent of the variables in $\WSet$.
This allows Eloise to make her choice in each such state in a uniform way and
adhere to the constraints on her variables when trying to prove the truth of the
formula.
Analogously, a co-trump encodes the states induced by the possible choices of
Eloise and allows Abelard to behave uniformly when he wants to falsify the
formula.

In this work we investigate a conservative extension of \IF, called
\emph{Alternating Dependence/Independence Friendly Logic} (\ADIF), tailored to
take the restrictions of both players into account at the same time, namely both
when evaluating truth and when evaluation falsity, and to overcome the
indeterminacy of the logic.
To this end, we generalise trumps/teams in such a way that the choices of both
players are recorded in the semantic structure \wrt which formulae are
evaluated, enabling both of them to make their choices in accordance with the
uniformity constraints required by the independence restrictions specified in
the quantifiers.
This approach leads to the notion of \emph{hyperteam}, defined as a set of
teams, which provides a two-level structure, where each level is intuitively
associated with one of the two players and encodes the uncertainty that the
opponent has about the actual choices up to that stage of the play.
From another perspective, the structure can be viewed as encoding all the
possible plays in the underlying game, comprising the choices of one player as
well as the possible responses of the opponent.
With all this information at hand, then, we can easily obtain the plays of the
dual game, namely the one in which the two players exchange their roles.
The change of roles between the players, in turn, precisely corresponds to the
game-theoretic interpretation of negation.
This allows us to include negation to the logic in a very natural way and, at
the same time, recover the law of the excluded middle, which is lost in \IF,
by avoiding undetermined sentences, and have a fully symmetric treatment of
the independence constraints on the universal and existential quantifiers.
We show that the resulting logic is at least as powerful as full \SOL and \TL.
We also provide a novel game-theoretic semantics for the prenex fragment of the
logic, by means of a determined infinite-duration first-order game with a
parity-like winning condition, that we call \emph{independence game}.
For any \ADIF-sentence $\varphiSnt$ and finite relational structure $\AStr$, we
can build an independence game $\GameName[\varphiSnt][\AStr]$ such that Eloise
has a winning strategy \iff $\varphi$ is true in $\AStr$.
As a byproduct, given that there exists a translation of \TL into \ADIF,
independence games indirectly provide a game-theoretic interpretation for \TL.

\section{Alternating Dependence/Independence-Friendly Logic}
\label{sec:adif}

\emph{Alternating Dependence/In\-dependence-Friendly Logic} (\ADIF, for short)
is defined as an extension of \FOL.
Therefore, throughout the work we shall assume, as it is customary, a countable
set of variables $\VarSet$ and a generic signature $\LSig \defeq \tuple {\RName}
{\artFun}$ comprised of a set $\RName$ of relation symbols, including the
interpreted relation `$=$' for equality, and a function $\artFun \colon \RName
\rightarrow \SetN$ providing the arity of each relation in $\RName$.
We also fix, if not stated otherwise, an $\LSig$-structure $\AStr \defeq \tuple
{\ASet} {\{ \RRel[][\AStr] \}_{\RRel \in \RName}}$, with domain of the discourse
$\ASet$, interpretation $\RRel[][\AStr] \subseteq \ASet[][\artFun(\RRel)]$ of
each relation $\RRel \in \RName$, and size $\card{\AStr} \defeq \card{\ASet}$.


\subsection{Syntax}
\label{sec:adif;sub:syn}

In the same vein as \IF, \ADIF augments \FOL with \emph{restricted quantifiers},
where restrictions specify possible \emph{dependence/independence constraints}.
Basically, these restrictions allow for formulae of the form $\LExs[][_{+\WSet}]
\varElm \ldotp \varphiFrm$ and $\LExs[][_{-\WSet}] \varElm \ldotp \varphiFrm$,
whose intuitive reading is best understood in game-theoretic terms, where the
existential quantifiers are controlled by player Eloise (the \emph{verifier}),
while universal quantifiers are controlled by player Abelard (the
\emph{falsifier}).
Then, the intended meaning of $\LExs[][_{+\WSet}] \varElm \ldotp \varphiFrm$
(\resp, $\LExs[][_{-\WSet}] \varElm \ldotp \varphiFrm$) is that Eloise has to
choose a value for $\varElm$, solely depending on (\resp, independently of) the
values of the variables in $\WSet$, that makes $\varphiFrm$ true.
Similarly, $\LAll[][_{+\WSet}] \varElm \ldotp \varphiFrm$ (\resp,
$\LAll[][_{-\WSet}] \varElm \ldotp \varphiFrm$) means that Abelard has to be
able to choose, solely depending on (\resp, independently of) the variables in
$\WSet$, a value for $\varElm$ that allows him to prove the argument
$\varphiFrm$ false.
In other words, the decoration $\pm\WSet$ specifies what information is
available to the player associated with the logic quantifier when she or he has
to make the choice.
In case of $+\WSet$, only the variables in that set are available, when $-\WSet$
is present, instead, only the variables outside the set, \ie, in the complement
$\VarSet \setminus \WSet$, are visible.

\begin{definition}[\ADIF Syntax]
  \label{def:syn(adif)}
  The \emph{Alternating Dependence/Independence-Friendly Logic} (\ADIF, for
  short) is the set of formulae built according to the following grammar, where
  $\RRel \in \RName$, $\xVec \in \VarSet[][\art{\RRel}]$, $\varElm \in
  \VarSet$, and $\WSet \subseteq \VarSet$ with $\card{\WSet} \!<\! \omega$:
  \[
    \varphiFrm \seteq \Ff \mid \Tt \mid \RRel(\xVec) \mid \LNeg \varphiFrm \mid
    \varphiFrm \LCon \varphiFrm \mid \varphiFrm \LDis \varphiFrm \mid
    \LExs[][_{\pm\WSet}] \varElm \ldotp \varphiFrm \mid \LAll[][_{\pm\WSet}]
    \varElm \ldotp \varphiFrm.
  \]
  \ADF (\resp, \AIF) denotes the fragment were only \emph{dependence} ($+\WSet$)
  (\resp, \emph{independence} (-\WSet)) \emph{constructs} are permitted.
\end{definition}

Predicative logics usually rely on a notion of \emph{free placeholder} to
correctly define the meaning of a formula and \ADIF is no exception.
In \ADIF, however, we distinguish between \emph{support} and \emph{free
variables}.
Specifically, support variables are the ones occurring in some atom
$\RRel(\xVec)$ that need to be assigned a value in order to evaluate the truth
of the formula. The free variables, instead, also include those occurring in
some dependence/independence constraint.
By $\sup{} \colon \ADIF \to \pow{\VarSet}$ we denote the function collecting all
support variables $\sup{\varphiFrm}$ of a formula $\varphiFrm$, defined as
follows:
\begin{itemize}[$\bullet$]
\item
  \begin{inparaitem}
  \item[]
    $\sup{\bot}, \sup{\top} \defeq \emptyset$;
  \item[\hspace{0.5em}$\bullet$]
    $\sup{\RRel(\xVec)} \defeq \xVec$;
  \item[\hspace{0.5em}$\bullet$]
    $\sup{\LNeg \varphiFrm} \defeq \sup{\varphiFrm}$;
  \end{inparaitem}
\item
  $\sup{\varphiFrm[1] \Cnt[][]\, \varphiFrm[2]} \defeq \sup{\varphiFrm[1]} \cup
  \sup{\varphiFrm[2]}$, for all connective symbols $\Cnt \in \{ \LCon, \LDis
  \}$;
\item
  $\sup{\Qnt[][_{\pm\WSet}] \varElm \ldotp \varphiFrm} \defeq \sup{\varphiFrm}
  \setminus \{ \varElm \}$, for all quantifier symbols $\Qnt \in \{ \LExs,
  \LAll \}$.
\end{itemize}
The free-variable function $\free{} \colon \ADIF \to \pow{\VarSet}$ is defined
similarly, except for the quantifier case, which is reported in the following:
\begin{itemize}[$\bullet$]
\item
  $\free{\Qnt[][_{\pm\WSet}] \varElm \ldotp \varphiFrm} \defeq
  (\free{\varphiFrm} \setminus \{ \varElm \}) \cup \denot{\pm\WSet}$, if
  $\varElm \in \free{\varphiFrm}$, and $\free{\Qnt[][_{\pm\WSet}] \varElm
  \ldotp \varphiFrm} \defeq \free{\varphiFrm}$, otherwise, for all quantifier
  symbols $\Qnt \in \{ \LExs, \LAll \}$, with $\denot{\pm\WSet}$ denoting the
  set $\WSet$, for the symbol `$+$', and its complement $\VarSet \setminus
  \WSet$, for the symbol `$-$'.
\end{itemize}
Obviously, it holds that $\sup{\varphiFrm} \subseteq \free{\varphiFrm}$.
A \emph{sentence} $\varphiSnt$ is a formula with $\free{\varphiSnt} =
\emptyset$.
If $\sup{\varphiSnt} = \emptyset$, but $\free{\varphiSnt} \neq \emptyset$, then
$\varphiSnt$ is just a \emph{pseudo sentence}.
As an example, $\varphiSnt = \LAll[][+\emptyset] \varElm \ldotp
\LExs[][+\emptyset] \yvarElm \ldotp (\varElm = \yvarElm)$ is a sentence, while
$\varphiSnt' = \LAll[][+\emptyset] \varElm \ldotp \LExs[][+\zvarElm] \yvarElm
\ldotp (\varElm = \yvarElm)$ is a pseudo sentence, since $\sup{\varphiSnt'} =
\emptyset$, but $\free{\varphiSnt'} = \{ \zvarElm \}$.
We also define $\LExs \varElm \ldotp \varphiFrm \defeq \LExs[][_{+\WSet}]
\varElm \ldotp \varphiFrm$ and $\LAll \varElm \ldotp \varphiFrm \defeq
\LAll[][_{+\WSet}] \varElm \ldotp \varphiFrm$, where $\WSet \defeq
\sup{\varphiFrm} \setminus \{ \varElm \}$.
From now on, by \FOL we mean the syntactic fragment of \ADIF composed of
formulae that only use the last two quantifiers.
For such formulae, it holds that $\sup{\varphiFrm} = \free{\varphiFrm}$.
As we shall show in Section~\ref{sec:adq}, this fragment semantically
corresponds to classic \FOL as defined by~\cite{Tar36}.
Similarly, we shall later identify a richer fragment of \ADIF that semantically
corresponds to \IF as formalised by~\cite{Hod97a}.

Before giving the formal definition of the compositional semantics, it is worth
providing just few examples of properties expressible in \ADIF.
In discussing these examples, then, we shall rely on the informal game-theoretic
interpretation of the quantifiers given above.

Let us picture a two-turn game where Player~$1$, who chooses first, controls the
variable $\varElm$ and Player~$2$, who chooses second, controls $\yvarElm$.
Let $\psiFrm(\varElm, \yvarElm)$ be the goal of Player~$2$ and consider the
following two \ADF sentences:
\begin{align*}
  {\varphiSnt[1] \seteq \LAll \varElm \ldotp \LExs[][+\varElm] \yvarElm \ldotp
  \psiFrm(\varElm, \yvarElm)};
& \hspace{2.5em}
  {\varphiSnt[2] \seteq \LExs \varElm \ldotp \LAll[][+\varElm] \yvarElm \ldotp
  \LNeg \psiFrm(\varElm, \yvarElm).}
\end{align*}
Sentence $\varphiFrm[1]$, whenever true, requires Player~$2$, in this case
Eloise, be able to respond to every choice for $\varElm$ made by Player~$1$, in
this case Abelard, so that goal $\psiFrm(\varElm, \yvarElm)$ is always
satisfied.
This corresponds to the existence of a winning strategy for Eloise, namely a
strategy that wins every induced play in the game, for the objective
$\psiFrm(\varElm, \yvarElm)$.
On the contrary, with inverted roles, the truth of $\varphiFrm[2]$ ensures that
there is a choice of Eloise such that, no matter what Abelard chooses,
$\psiFrm(\varElm, \yvarElm)$ cannot be achieved.
This means that Abelard cannot have a winning strategy for $\psiFrm(\varElm,
\yvarElm)$.
If $\varphiFrm[2]$ is false, instead, it is Abelard who has a winning strategy
for $\psiFrm(\varElm, \yvarElm)$, while the falsity of $\varphiFrm[1]$ ensures
the existence of a choice of Abelard such that, no matter what Eloise chooses,
$\psiFrm(\varElm, \yvarElm)$ cannot be achieved.
Note that both sentences belong to the \FOL fragment introduced above and their
semantics also corresponds to the Tarskian one.
However, the \ADF sentences
\begin{align*}
  {\varphiSnt[3] \seteq \LAll \varElm \ldotp \LExs[][+\emptyset] \yvarElm \ldotp
  \psiFrm(\varElm, \yvarElm)};
& \hspace{2.5em}
  {\varphiSnt[4] \seteq \LExs \varElm \ldotp \LAll[][+\emptyset] \yvarElm \ldotp
  \LNeg \psiFrm(\varElm, \yvarElm)}
\end{align*}
add imperfect information to the picture and have no \FOL analogue.
Sentence $\varphiFrm[3]$ still postulates the existence of a winning strategy for
Eloise, but this time also requires that, when making the choice for $\yvarElm$,
the player has no access to any information and, in particular, to the value
chosen for $\varElm$ by the opponent.
We call such a strategy $\emptyset$-uniform.
Similarly, $\varphiFrm[4]$, when true, witnesses the non-existence of such a
$\emptyset$-uniform winning strategy for Abelard.
The \ADIF pseudo sentences
\begin{align*}
  {\varphiSnt[5] \seteq \LAll \varElm \ldotp \LExs[][-\varElm] \yvarElm \ldotp
  \psiFrm(\varElm, \yvarElm)};
& \hspace{2.5em}
  {\varphiSnt[6] \seteq \LExs \varElm \ldotp \LAll[][-\varElm] \yvarElm \ldotp
  \LNeg \psiFrm(\varElm, \yvarElm)}
\end{align*}
have a very similar  meaning to $\varphiFrm[3]$ and $\varphiFrm[4]$,
respectively, with the exception that $\yvarElm$ can depend on any variable
different from $\varElm$, since $\free{\varphiSnt[5]} = \free{\varphiSnt[6]} =
\VarSet \setminus \varElm$.

Consider now a three-turn game, extending the previous one, where, after the
move of Player~$2$, Player~$1$ chooses the value for another variable under its
control, let us call this  $\zvarElm$.
The \ADF sentence
\[
  \varphiSnt[7] \seteq \LExs \varElm \ldotp \LAll[][+\emptyset] \yvarElm \ldotp
  \LExs[][+\varElm] \zvarElm \ldotp \left( \psiFrm[1](\varElm, \yvarElm) \LCon
  \psiFrm[2](\yvarElm, \zvarElm) \right)
\]
is a bit more involved.
First of all, it states that Player~$2$, \ie, Abelard, cannot see the choice
made for $\varElm$.
In addition, while Player~$1$, \ie, Eloise, is not aware of $\yvarElm$, she has
access to the value previously chosen for $\xvarElm$ by herself.
The sentence, whenever true, ensure the existence of a choice by Eloise which
ensures that Abelard cannot prevent $\psiFrm[1](\varElm, \yvarElm)$ from
happening, no matter what he chooses.
Moreover, Eloise can respond to any of these latter choices for $\yvarElm$ and
win objective $\psiFrm[2](\yvarElm, \zvarElm)$ by only looking at the value of
$\varElm$.
This means that Abelard is not able to prevent $\psiFrm[1](\varElm, \yvarElm)$
and, a the same time, Eloise has a $\xvarElm$-uniform strategy to win
$\psiFrm[2](\yvarElm, \zvarElm)$ at the same time.

\subsection{Semantics}
\label{sec:adif;sub:sem}

The semantics we define for \ADIF follows an approach similar to~\citep{Hod97a},
where a compositional semantics for \IF was first proposed.
Hodges' idea was to expand an assignment for the free variables to a set of
assignments, a trump in his terminology (\aka team~\citep{Vaa07}), with the
intuition of capturing Eloise's uncertainty on the actual state of the semantic
game underlying the logic~\citep{HS89}.
This essentially corresponds to recording the possible choices made by the
opponent, \ie, Abelard, for its own variables.
Hodges' semantics, though able to correctly capture \IF, is, however, not
adequate for our purposes.
Indeed, by design, it is intrinsically asymmetric, treating the two players
differently.
More specifically, a single set of assignments only provides complete
information about the choices of one of the two players (\ie, Abelard in trumps
and Eloise in co-trumps) and only allows to restrict the choices of the adversary.
This is also connected with the lack of classic properties of negation,
specifically the law of excluded middle.

In order to overcome these limitations, incorporate negation into \ADIF in a
natural way and obtain a fully determined logic, we propose here a
generalisation of Hodges' approach.
To give semantics to an \ADIF formula $\varphiFrm$, we then proceed as follows.
Similarly to Hodges, the idea is that the interpretations of the free variables
correspond to the choices that the two players could make up to the current
stage of the game, \ie, the stage where the formula $\varphiFrm$ has to be
evaluated.
These possible choices are organised in a two-level structure, \ie, a set of
sets of assignments, each level summarising the information about the choices a
player may have made in previous turns.
In order to evaluate the formula $\varphiFrm$, then, a player chooses a set of
assignments, while its opponent chooses one assignment in that set where
$\varphiFrm$ must hold.
We shall use a flag $\alpha \in \{ \QEA, \QAE \}$, called \emph{alternation
flag}, to keep track of which player is assigned to which level of choice.
If $\alpha = \QEA$, Eloise chooses the set of assignments, while Abelard chooses
one of those assignments; if $\alpha = \QAE$, the dual reasoning applies.
In a sense, the level associated with a given player, say Eloise, encodes the
uncertainty that the opponent Abelard has about her actual choices up to that
stage.

Given a flag $\alpha \in \{ \QEA, \QAE \}$, we denote by $\dual{\alpha}$ the
dual flag, \ie, $\dual{\alpha} \in \allowbreak \{ \QEA, \QAE \}$ with
$\dual{\alpha} \neq \alpha$.
Let $\AsgSet \defeq \VarSet \pto \ASet$ be the set of (partial)
\emph{assignments} over $\VarSet$, namely partial functions from variables to
values in the structure domain $\ASet$.
Given a set of variables $\VSet \subseteq \VarSet$, we denote by $\AsgSet(\VSet)
\defeq \set{ \asgElm \in \AsgSet }{ \dom{\asgElm} = \VSet }$ the assignments
defined on $\VSet$ and by $\AsgSet[\subseteq](\VSet) \defeq \set{ \asgElm \in
\AsgSet }{ \VSet \subseteq \dom{\asgElm} }$ the set of assignments defined on
some superset of $\VSet$.
A \emph{team (of assignments)} is a set of total assignments all defined on the
same set of variables.
Formally, $\TeamAsgSet \defeq \set{ \TeamAsg \subseteq \AsgSet(\VSet) }{ \VSet
\subseteq \VarSet }$ collects all possible teams over some subset $\VSet$ of
$\VarSet$, $\TeamAsgSet(\VSet) \defeq \set{ \TeamAsg \in \TeamAsgSet }{ \TeamAsg
\subseteq \AsgSet(\VSet) }$ contains those over $\VSet$ and
$\TeamAsgSet[\subseteq](\VSet) \defeq \set{ \TeamAsg \in \TeamAsgSet }{ \TeamAsg
\subseteq \AsgSet[\subseteq](\VSet) }$ the teams defined on supersets of
$\VSet$.
The idea described above is, then, captured by the notion of \emph{hyperteam (of
assignments)}, namely a set of teams defined over some arbitrary set $\VSet
\subseteq \VarSet$:
\[
  \HypAsgSet \defeq \set{ \HypAsg \subseteq \TeamAsgSet(\VSet) }{ \VSet
  \subseteq \VarSet }.
\]
By $\HypAsgSet(\VSet) \defeq \set{ \HypAsg \in \HypAsgSet }{ \HypAsg \subseteq
\pow{\AsgSet(\VSet)} }$ we denote the set of hyperteams over $\VSet$, while
$\HypAsgSet[\subseteq](\VSet) \defeq \set{ \HypAsg \in \HypAsgSet }{ \HypAsg
\subseteq \pow{\AsgSet[\subseteq](\VSet)} }$ contains the hyperteams defined on
supersets of $\VSet$.
All the assignments inside a team $\TeamAsg \in \TeamAsgSet$ or hyperteam
$\HypAsg \in \HypAsgSet$ are defined on the same variables, whose sets are
indicated by $\var{\TeamAsg}$ and $\var{\HypAsg}$, respectively.
We shall call the empty set of teams $\EmpHypAsg$ the \emph{empty hyperteam},
every set containing the empty team, \eg, $\NulHypAsg$, a \emph{null hyperteam},
and the set $\TrvHypAsg$ containing a single team comprised only of the empty
assignment the \emph{trivial hyperteam}.
Essentially, the trivial hyperteam encodes the situation in which none of the
players has made any choice yet and, hence, contains the minimal ``consistent''
state of a game.
In this sense, then, null and empty hyperteams do not convey any meaningful
information about the possible state of a game and are included here mainly for
technical reasons, as they allow for a cleaner formal definition of the
semantics.
For this reason, we shall refer to every hyperteam which is neither the empty
hyperteam nor a null hyperteam with the term \emph{proper hyperteam}.

For any pair of hyperteams $\HypAsg[1], \HypAsg[2] \in \HypAsgSet$, we write
$\HypAsg[1] \inc \HypAsg[2]$ to state that, for all teams $\TeamAsg[1] \in
\HypAsg[1]$, there exists a team $\TeamAsg[2] \in \HypAsg[2]$ such that
$\TeamAsg[2] \subseteq \TeamAsg[1]$.
As usual, $\HypAsg[1] \eqv \HypAsg[2]$ denotes the fact that both $\HypAsg[1]
\inc \HypAsg[2]$ and $\HypAsg[2] \inc \HypAsg[1]$ hold true.
Obviously, $\HypAsg[1] \subseteq \HypAsg[2]$ implies $\HypAsg[1] \inc
\HypAsg[2]$, which, in turn, implies $\var{\HypAsg[1]} = \var{\HypAsg[2]}$.
It is clear that the relation $\inc$ is both reflexive and transitive, hence it
is a preorder; as an immediate consequence, $\eqv$ is an equivalence relation.
In particular, we shall show (see Corollary~\ref{cor:hypeqv} later in this
section) that $\eqv$ captures the intuitive notion of equivalence between
hyperteams, in the sense that two equivalent hyperteams \wrt $\eqv$ do satisfy
the same \ADIF formulae.
Figure~\ref{fig:hypasgord} provides a graphical representation of the preorder
relation $\inc$.

\begin{figure}[h]
  \centering{\figHypAsgOrd}
  \caption{\label{fig:hypasgord} Two hyperteams with $\HypAsg[1] \inc
    \HypAsg[2]$, but $\HypAsg[2] \not\inc \HypAsg[1]$.}
\end{figure}

\begin{example}
  \label{exm:hypasgordi}
  In Figure~\ref{fig:hypasgord}, the hyperteam $\HypAsg[1]$ is $\inc$-included
  in the hyperteam $\HypAsg[2]$, since, for each team $\XSet$ in $\HypAsg[1]$,
  there is a team in $\HypAsg[2]$ that is set-included in $\XSet$.
  For instance, the team $\XRed[1]$ of $\HypAsg[1]$ contains the assignments
  $\asgElm[1]$, $\asgElm[2]$, $\asgElm[3]$, $\asgElm[4]$ and $\asgElm[5]$, so,
  it includes the team $\XBlue[1]$ of $\HypAsg[2]$ composed of $\asgElm[2]$ and
  $\asgElm[4]$.
  Note that not all teams in $\HypAsg[2]$ are included in a team in $\HypAsg[1]$
  and different teams of $\HypAsg[1]$ can choose the same team of $\HypAsg[2]$
  to include.
\end{example}

Since we are dealing with imperfect information, we need a way to define a
notion of indistinguishability relative to dependence constraints, intuitively,
those specified in quantifiers.
Given a hyperteam $\HypAsg \in \HypAsgSet$ and a set of variables $\WSet
\subseteq \VarSet$, we define $\HypAsg \!\rst[\WSet] \defeq \allowbreak \set{
\TeamAsg \!\rst[\WSet] }{ \TeamAsg \in \HypAsg }$ and $\TeamAsg \!\rst[\WSet]
\defeq \allowbreak \set{ \asgElm \!\rst[\WSet] }{ \asgElm \in \TeamAsg }$, where
$\asgElm \!\rst[\WSet]$ is the restriction of the assignment $\asgElm$ to the
domain $\dom{\asgElm} \cap \WSet$.
We can, then, compare hyperteams relative to $\WSet$ by writing $\HypAsg[1]
\eql[\WSet] \HypAsg[2]$ for $\HypAsg[1] \!\rst[\WSet] \eql \HypAsg[2]
\!\rst[\WSet]$, meaning that the two hyperteams are indistinguishable when only
variables in $\WSet$ are considered.
Similarly, $\HypAsg[1] \eqv[\WSet] \HypAsg[2]$ stands for $\HypAsg[1]
\!\rst[\WSet] \eqv \HypAsg[2] \!\rst[\WSet]$ and means that they are equivalent
on $\WSet$, while $\HypAsg[1] \inc[\WSet] \HypAsg[2]$ abbreviates $\HypAsg[1]
\!\rst[\WSet] \inc \HypAsg[2] \!\rst[\WSet]$ and relativises the ordering to a
dependence constraint.
Obviously, $\HypAsg[1] \eql[\WSet] \HypAsg[2]$, $\HypAsg[1] \eqv[\WSet]
\HypAsg[2]$, and $\HypAsg[1] \inc[\WSet] \HypAsg[2]$ imply $\HypAsg[1]
\eql[\WSet'] \HypAsg[2]$, $\HypAsg[1] \eqv[\WSet'] \HypAsg[2]$, and $\HypAsg[1]
\inc[\WSet'] \HypAsg[2]$, respectively, for all $\WSet' \subseteq \WSet$.

\begin{example}
  \label{exm:hypasgordii}
  In Figure~\ref{fig:hypasgord}, $\HypAsg[2]$ is not $\inc$-included in
  $\HypAsg[1]$, as none of the teams of $\HypAsg[2]$ includes a team of
  $\HypAsg[1]$.
  Now, assume the existence of a set of variables $\WSet$ that makes $\{
  \asgElm[1], \asgElm[3], \asgElm[4], \asgElm[5], \asgElm[6], \asgElm[7],
  \asgElm[10] \} \!\rst[\WSet]$ collapse to $\{ \asgElm[1] \} \!\rst[\WSet]$.
  Then, we have:
  \[
    \begin{array}{l|l}
      \multicolumn{1}{c}{{\HypAsg[2] \!\rst[\WSet]}}
    &
      \multicolumn{1}{c}{{\HypAsg[1] \!\rst[\WSet]}}
    \\
      {\XSet[21] \!\rst[\WSet] = \{ \asgElm[1] \!\rst[\WSet], \asgElm[2]
      \!\rst[\WSet] \}}
    &
      {\XSet[11] \!\rst[\WSet] = \{ \asgElm[1] \!\rst[\WSet], \asgElm[2]
      \!\rst[\WSet] \}}
    \\
      {\XSet[22] \!\rst[\WSet] = \{ \asgElm[1] \!\rst[\WSet], \asgElm[9]
      \!\rst[\WSet] \}}
    &
      {\XSet[12] \!\rst[\WSet] = \{ \asgElm[1] \!\rst[\WSet], \asgElm[2]
      \!\rst[\WSet], \asgElm[8] \!\rst[\WSet] \}}
    \\
      {\XSet[23] \!\rst[\WSet] = \{ \asgElm[1] \!\rst[\WSet], \asgElm[9]
      \!\rst[\WSet], \asgElm[11] \!\rst[\WSet] \}}
    &
      {\XSet[13] \!\rst[\WSet] = \{ \asgElm[1] \!\rst[\WSet], \asgElm[9]
      \!\rst[\WSet] \}}
  \end{array}
  \]
  Now, team $\XSet[11] \!\rst[\WSet]$ is included in $\XSet[21] \!\rst[\WSet]$
  and team $\XSet[13] \!\rst[\WSet]$ is included in both $\XSet[22]
  \!\rst[\WSet]$ and $\XSet[23] \!\rst[\WSet]$.
  Therefore, $\HypAsg[2] \inc[\WSet] \HypAsg[1]$ and, so, $\HypAsg[1]
  \eqv[\WSet] \HypAsg[2]$, since $\HypAsg[1] \inc \HypAsg[2]$.
\end{example}

The alternating semantics is given by means of a satisfaction relation between a
hyperteams $\HypAsg$ and a formula $\varphiFrm$, \wrt a given interpretation of
the  players in $\HypAsg$, that is \wrt an alternation flag $\alpha\in \{ \QEA,
\QAE \}$.
As a consequence, we shall introduce two satisfaction relations,
$\cmodels[][\QEA]$ and $\cmodels[][\QAE]$, one for each interpretation of
players in the hyperteam.
The intuition is that, when the alternation flag $\alpha$ is $\QEA$, then a team
is chosen existentially by Eloise and all its assignments, chosen universally by
Abelard, must satisfy $\varphi$.
Conversely, when $\alpha$ is $\QAE$, then all teams, chosen universally by
Abelard, must contain at least one assignments, chosen existentially by Eloise,
that satisfies $\varphi$.

The definition of the semantics relies on three basic operations on hyperteams:
the \emph{dualisation} swaps the role of the two players in a hyperteam,
allowing for connecting the two satisfaction relations and a symmetric treatment
of quantifiers later on; the \emph{extension} directly handles quantifications;
finally, the \emph{partition} deals with disjunction and conjunction.

Let us consider the \emph{dualisation operator} first.
Given a hyperteam $\HypAsg$, the dual hyperteam $\dual{\HypAsg}$ exchanges the
role of the two players \wrt $\HypAsg$.
This means that, if Eloise is the player choosing the team in $\HypAsg$ and
Abelard the one choosing the assignment in the team, it will be Abelard who
chooses the team in $\dual{\HypAsg}$ and Eloise the one who chooses the
assignment.
To ensure that the semantics of the underlying game is not altered when
exchanging the order of choice for the two players, we need to reshuffle the
assignments in $\HypAsg$ so as to simulate the original dependencies between the
choices.
To this end, for a hyperteam $\HypAsg$, we introduce the set
\[
  \ChcSet{\HypAsg} \defeq \set{ \chcFun \colon \HypAsg \to \AsgSet }{ \forall
  \TeamAsg \in \HypAsg \ldotp \chcFun(\TeamAsg) \in \TeamAsg }
\]
of \emph{choice functions}, whose definition implicitly assumes the axiom of
choice, whenever the structure domain $\ASet$ is infinite.
$\ChcSet{\HypAsg}$ contains all the functions $\chcFun$ that, for every team
$\XSet$ in $\HypAsg$, pick a specific assignment $\chcFun(\XSet)$ in that set.
Each such function simulates a possible choice of the second player of $\HypAsg$
depending on the choice of (the team chosen by) the first player.
The dual hyperteam $\dual{\HypAsg}$, then, collects the images of the choice
functions in $\ChcSet{\HypAsg}$.
We, thus, obtain a hyperteam in which the choice order of the two players is
inverted:
\[
  \dual{\HypAsg} \defeq \set{ \img{\chcFun} }{ \chcFun \in \ChcSet{\HypAsg} }.
\]

It is immediate to check that the only hyperteams equivalent to the empty or
null ones are themselves and they are also dual of one another.
Therefore, the class of proper hyperteams is closed under dualisation.
In addition, the trivial hyperteam is self-dual.

\begin{proposition}[name =, restate = prpempnultrv]
  \label{prp:empnultrv}
  \begin{inparaenum}[1)]
    \item
      $\HypAsg \eqv \EmpHypAsg$ \iff $\HypAsg = \EmpHypAsg$ \iff $\dual{\HypAsg}
      = \NulHypAsg$;
    \item
      $\HypAsg \eqv \NulHypAsg$ \iff $\EmpTeamAsg \!\in\! \HypAsg$ \iff
      $\dual{\HypAsg} = \EmpHypAsg$.
      Moreover,
    \item
      $\dual{\TrvHypAsg} = \TrvHypAsg$.
      Finally,
    \item
      $\HypAsg$ is proper \iff $\dual{\HypAsg}$ is proper as well.
  \end{inparaenum}
\end{proposition}

\begin{example}
  \label{exm:hypdlt}
  Consider the following two dual hyperteams
  \[
    \HypAsg=
    \begin{Bmatrix}
      \,\XRed = \{ \asgRed[11], \asgRed[12] \},\, \\
      \,\XBlue = \{ \asgBlue[21], \asgBlue[22] \},\, \\
      \,\XGreen = \{ \asgGreen[3] \}\,
    \end{Bmatrix}
    \text{\ \ and\ \ \ }
    \dual{\HypAsg} =
    \begin{Bmatrix}
      {\img{\chcFun[1]}} = \{ \asgRed[11], \asgBlue[21], \asgGreen[3] \}, \\
      {\img{\chcFun[2]}} = \{ \asgRed[11], \asgBlue[22], \asgGreen[3] \}, \\
      {\img{\chcFun[3]}} = \{ \asgRed[12], \asgBlue[21], \asgGreen[3] \}, \\
      {\img{\chcFun[4]}} = \{ \asgRed[12], \asgBlue[22], \asgGreen[3] \}
      \phantom{,}
    \end{Bmatrix},
  \]
  where the teams are $\XRed = \{ \asgRed[11], \asgRed[12] \}$, $\XBlue = \{
  \asgBlue[21], \asgBlue[22] \}$, and $\XGreen = \{ \asgGreen[3] \}$.
  Every team in $\dual{\HypAsg}$ is obtained as the image of one of the four
  choice functions $\chcFun[i] \in \ChcSet{\HypAsg}$, each choosing exactly one
  assignment from $\XRed$, one from $\XBlue$, and the unique one from $\XGreen$.
  Intuitively, in $\HypAsg$ the strategy of the first player, say Eloise, can
  only choose the colour of the final assignments (either red for $\XRed$, blue
  for $\XBlue$, or green for $\XGreen$), while the one for Abelard decides which
  assignment of each colour will be picked.
  After dualisation, the two players exchange the order in which they choose.
  Therefore, Abelard, starting first in $\dual{\HypAsg}$, will select one of
  the four choice functions, which picks an assignment for each colour.
  Eloise, choosing second, by using her strategy that selects the colour will
  give the final assignment.
  In other words, the original strategies of the players encoded in the
  hyperteam, as well as their dependencies, are preserved, regardless of the
  swap of their role in the dual hyperteam.
  The example also shows that, as we shall prove shortly (see
  Theorem~\ref{thm:dbldlt} later in this section), if we dualise a hyperteam
  $\HypAsg$ and, at the same time, swap the original interpretation $\alpha\in
  \{ \QEA, \QAE \}$ of the player to $\dual{\alpha}$, we obtain that the pair
  $(\dual{\HypAsg}, \dual{\alpha})$ gives an equivalent representation of the
  information contained in the original pair $(\HypAsg, \alpha)$.
\end{example}

Dualisation enjoys an \emph{involution property} similar to the classic Boolean
negation: by applying the dualisation twice, we obtain a hyperteam equivalent to
the original one.
This confirms that the operation preserves the entire information encoded in the
hyperteams.

\begin{lemma}[name = Dualisation I, restate = lmmdlti]
  \label{lmm:dlti}
  For all hyperteams $\HypAsg \in \HypAsgSet$, it holds that $\HypAsg
  \eqv[\WSet] \dual{\dual{\HypAsg}}$, for all $\WSet \subseteq \VarSet$.
  In addition, $\HypAsg \subseteq \dual{\dual{\HypAsg}}$, if $\HypAsg$ is
  proper.
\end{lemma}

The proof of this lemma, together with those of all the non-trivial results in
the main paper, can be found in appendix.

Observe the clear analogy between the structure of hyperteams with alternation
flag $\QEA$ (\resp, $\QAE$) and the structure of DNF (\resp, CNF) Boolean
formulae, where the dualisation swaps between two equivalent forms.
The following lemma formally states that this operation swaps the role of the
two players, while still preserving the original dependencies among their
choices.

\begin{lemma}[name = Dualisation II, restate = lmmdltii]
  \label{lmm:dltii}
  The following equivalences hold true, for all hyperteams $\HypAsg \in
  \HypAsgSet$ and property $\PsiSet \subseteq \AsgSet$.
  \begin{enumerate}[1)]
    \item\label{lmm:dltii(ea)}
      Statements~\ref{lmm:dltii(ea:org)} and~\ref{lmm:dltii(ea:dlt)} are
      equivalent:
      \begin{enumerate}[a)]
        \item\label{lmm:dltii(ea:org)}
          there exists a team $\TeamAsg \in \HypAsg$ (\resp, $\TeamAsg \in
          \dual{\HypAsg}$) such that $\TeamAsg \subseteq \PsiSet$;
        \item\label{lmm:dltii(ea:dlt)}
          for all teams $\TeamAsg' \in \dual{\HypAsg}$ (\resp, $\TeamAsg' \in
          \HypAsg$), it holds that $\TeamAsg' \cap \PsiSet \neq \EmpTeamAsg$.
      \end{enumerate}
    \item\label{lmm:dltii(ee)}
      Statements~\ref{lmm:dltii(ee:org)} and~\ref{lmm:dltii(ee:dlt)} are
      equivalent:
      \begin{enumerate}[a)]
        \item\label{lmm:dltii(ee:org)}
          there exists a team $\TeamAsg \in \HypAsg$ such that $\TeamAsg \cap
          \PsiSet \neq \EmpTeamAsg$;
        \item\label{lmm:dltii(ee:dlt)}
          there exists a team $\TeamAsg' \in \dual{\HypAsg}$ such that
          $\TeamAsg' \cap \PsiSet \neq \EmpTeamAsg$.
      \end{enumerate}
    \item\label{lmm:dltii(aa)}
      Statements~\ref{lmm:dltii(aa:org)} and~\ref{lmm:dltii(aa:dlt)} are
      equivalent:
      \begin{enumerate}[a)]
        \item\label{lmm:dltii(aa:org)}
          for all teams $\TeamAsg \in \HypAsg$, it holds that $\TeamAsg
          \subseteq \PsiSet$;
        \item\label{lmm:dltii(aa:dlt)}
          for all teams $\TeamAsg' \in \dual{\HypAsg}$, it holds that $\TeamAsg'
          \subseteq \PsiSet$.
      \end{enumerate}
  \end{enumerate}
\end{lemma}

Item~\ref{lmm:dltii(ea)} provides the semantic meaning of the operation, stating
that if there exists a team in $\HypAsg$ all of whose assignments satisfy some
property $\PsiSet$, then each team in $\dual{\HypAsg}$ has an assignment
satisfying the property, and \viceversa.
This directly connects the two interpretations of hyperteams, $\QAE$ and $\QEA$.
Item~\ref{lmm:dltii(ee)} establishes that no assignment is lost from the
original teams in $\HypAsg$, while Item~\ref{lmm:dltii(aa)} asserts that no new
assignments are added to $\dual{\HypAsg}$.
It could be proved that any two operators that satisfies the three conditions in
the lemma will produce equivalent hyperteams, in the sense of $\eqv[\WSet]$,
when applied to the same hyperteam.

Quantifications are taken care of by the \emph{extension operator}.
Let $\FncSet{} \defeq \AsgSet \to \ASet$ be the set of functions that map
assignments to a value in the domain $\ASet$ of the structure
$\AStr$.
Essentially, these objects play the role of the Skolem functions in Skolem
semantics or, equivalently, of the strategies in game-theoretical semantics.
To account for possible imperfect information, we need to ensure that these
functions choose values uniformly on indistinguishable assignments.
This constraint is captured by restricting the functions so that they must
choose the same value for assignments that are indistinguishable \wrt some given
set of variables $\WSet$.
Formally:
\[
  \FncSet[\WSet]{} \defeq \set{ \FFun \in \FncSet{} }{ \forall \asgElm \in
  \AsgSet \ldotp \FFun(\asgElm) = \FFun(\asgElm \!\rst[\WSet]) }.
\]
Clearly, $\FncSet{} = \FncSet[\denot{+\VarSet}]{} =
\FncSet[\denot{-\emptyset}]{}$.
The \emph{extension of an assignment} $\asgElm \in \AsgSet$ by a function $\FFun
\in \FncSet{}$ for a variable $\xElm \in \VarSet$ is defined as
$\extFun{\asgElm, \FFun, \xElm} \defeq {\asgElm}[\xElm \mapsto \FFun(\asgElm)]$,
which extends $\asgElm$ with $\xElm$ by assigning to it the value
$\FFun(\asgElm)$ prescribed by the function $\FFun$.
The \emph{extension operation} can then be lifted to teams $\TeamAsg \in
\TeamAsgSet$ in the obvious way, \ie, by setting $\extFun{\XSet, \FFun,
\xElm} \defeq \set{ \extFun{\asgElm, \FFun, \xElm} }{ \asgElm \in \XSet }$.
This operation embeds into $\XSet$ the entire player strategy encoded by
$\FFun$.
Finally, the \emph{extension of a hyperteam} $\HypAsg \in \HypAsgSet$
with $\xElm$ is simply the set of extensions with $\xElm$ of all its teams by
all possible functions:
\[
  \extFun[\WSet]{\HypAsg, \varElm} \defeq \set{ \extFun{\TeamAsg, \FFun,
  \varElm} }{ \TeamAsg \in \HypAsg, \FFun \in \FncSet[\WSet]{} }.
\]
The extension operation essentially embeds into $\HypAsg$ all possible
($\WSet$-uniform) strategies for choosing the value of $\xElm$, each one encoded
by a function $\FFun$ in $\FncSet[\WSet]{}$.

\begin{example}
  \label{exm:ext}
  Let $\HypAsg = \{ \TeamAsg[1] \!=\! \{ \asgElm[1], \asgElm[2] \}, \TeamAsg[2]
  \!=\! \{  \asgElm[1], \asgElm[3] \} \}$ be a hyperteam.
  To extend $\emptyset$-uniformly $\HypAsg$ with variable $\varElm$ over the
  structure domain $\ASet = \{ 0, 1 \}$, one needs to extend each team in
  $\HypAsg$ with the two $\emptyset$-uniform (\ie, constant) functions
  $\FFun[0](\asgElm) = 0$ and $\FFun[1](\asgElm) = 1$:
  \[
    \extFun[\emptyset]{\HypAsg, \varElm} =
    \left\{
    \begin{array}{cl}
      {\extFun{\TeamAsg[1], \FFun[0], \varElm}} =
    &\!\!\!\!\!
      {\{ {\asgElm[1]}[\varElm \mapsto 0], {\asgElm[2]}[\varElm \mapsto 0] \}}
    \\
      {\extFun{\TeamAsg[1], \FFun[1], \varElm}} =
    &\!\!\!\!\!
      {\{ {\asgElm[1]}[\varElm \mapsto 1], {\asgElm[2]}[\varElm \mapsto 1] \}}
    \\
      {\extFun{\TeamAsg[2], \FFun[0], \varElm}} =
    &\!\!\!\!\!
      {\{ {\asgElm[1]}[\varElm \mapsto 0], {\asgElm[3]}[\varElm \mapsto 0] \}}
    \\
      {\extFun{\TeamAsg[2], \FFun[1], \varElm}} =
    &\!\!\!\!\!
      {\{ {\asgElm[1]}[\varElm \mapsto 1], {\asgElm[3]}[\varElm \mapsto 1] \}}
    \end{array}
    \right\}\!.
  \]
\end{example}

Conjunctions and disjunctions are dealt with by means of the \emph{partition
operator}.
We provide here the intuition for disjunction, the dual reasoning applies to
conjunction.
Assume that the two players of $\HypAsg$, defined over the variables $\{
\varElm, \yvarElm \}$, are interpreted according to the alternation flag $\QAE$:
Abelard chooses the team and Eloise chooses the assignment in the team.
In our setting, then, in order to satisfy, \eg, $(\varElm = 0) \LDis (\varElm =
1)$, Eloise has to show that, for each team $\TeamAsg \in \HypAsg$ chosen by
Abelard, she has a way to select one of the disjuncts $\varElm = i$, with $i \in
\{ 0, 1 \}$, so that the given team has an assignment satisfying the disjunct.
To capture Eloise's choice on which disjunct to choose based on the team given
by Abelard, we define, for a hyperteam $\HypAsg$, the following set
\[
  \parFun[]{\HypAsg} \defeq \set{ (\HypAsg[1], \HypAsg[2]) \in \pow{\HypAsg}
  \times \pow{\HypAsg} }{ \HypAsg[1] \cap \HypAsg[2] = \emptyset \land
  \HypAsg[1] \cup \HypAsg[2] = \HypAsg }\!,
\]
which collects all the possible bipartitions of $\HypAsg$.
Intuitively, the hyperteam $\HypAsg[1]$ will be used to satisfy $\varElm = 0$,
while $\HypAsg[2]$ will be used for $\varElm = 1$.
Basically, $\parFun{\HypAsg}$ contains all the possible strategies by means of
which Eloise can try to satisfy the two disjuncts.
Then, we say that Eloise satisfies the disjunction if there is a pair
$(\HypAsg[1]', \HypAsg[2]')$ (hence, a hyperteam-partition strategy) in that set
such that $\HypAsg[1]'$ satisfies the left disjunct and $\HypAsg[2]'$ satisfies
the right one.

The compositional semantics of \ADIF can be, then, defined as follows, where
$\xVec[][\asgElm]$ denotes the tuple of elements of the underlying structure
$\AStr$ obtained by applying the assignment $\asgElm$ to the tuple of variables
$\xVec$ component-wise.

\begin{definition}[\ADIF Semantics]
  \label{def:sem(adif)}
  The \emph{Hodges' alternating semantic relation} ${\AStr, \HypAsg
  \cmodels[][\alpha] \varphiFrm}$ for \ADIF is inductively defined as follows,
  for all \ADIF formulae $\varphiFrm$, hyperteams $\HypAsg \in
  \HypAsgSet[\subseteq](\sup{\varphiFrm})$, and alternation flags $\alpha \in \{
  \QEA, \QAE \}$:
  \begin{enumerate}[1)]
    \item\label{def:sem(adif:fbv)}
      \begin{inparaenum}[a)]
        \item\label{def:sem(adif:fbv:ea)}
          $\AStr, \HypAsg \cmodels[][\QEA] \Ff$ if $\EmpTeamAsg \in \HypAsg$;
        \hfill
        \item\label{def:sem(adif:fbv:ae)}
          $\AStr, \HypAsg \cmodels[][\QAE] \Ff$ if $\HypAsg = \EmpHypAsg$;
      \end{inparaenum}
    \item\label{def:sem(adif:tbv)}
      \begin{inparaenum}[a)]
        \item\label{def:sem(adif:tbv:ae)}
          $\AStr, \HypAsg \cmodels[][\QAE] \Tt$ if $\EmpTeamAsg \not\in
          \HypAsg$;
        \hfill
        \item\label{def:sem(adif:tbv:ea)}
          $\AStr, \HypAsg \cmodels[][\QEA] \Tt$ if $\HypAsg \neq \EmpHypAsg$;
      \end{inparaenum}
    \item\label{def:sem(adif:rel)}
      \begin{enumerate}[a)]
        \item\label{def:sem(adif:rel:ea)}
          $\AStr, \HypAsg \cmodels[][\QEA] \RRel(\xVec)$ if there exists a team
          $\TeamAsg \in \HypAsg$ such that, for all assignments $\asgElm \in
          \TeamAsg$, it holds that $\xVec[][\asgElm] \in \RRel[][\AStr]$;
        \item\label{def:sem(adif:rel:ae)}
          $\AStr, \HypAsg \cmodels[][\QAE] \RRel(\xVec)$ if, for all teams
          $\TeamAsg \in \HypAsg$, there exists an assignment $\asgElm \in
          \TeamAsg$ such that $\xVec[][\asgElm] \in \RRel[][\AStr]$;
      \end{enumerate}
    \item\label{def:sem(adif:neg)}
      $\AStr, \HypAsg \cmodels[][\alpha] \neg \phiFrm$ if $\AStr, \HypAsg
      \notcmodels[][\dual{\alpha}] \phiFrm$;
    \item\label{def:sem(adif:con)}
      \begin{enumerate}[a)]
        \item\label{def:sem(adif:con:ea)}
          $\AStr, \HypAsg \cmodels[][\QEA] \phiFrm[1] \LCon \phiFrm[2]$ if, for
          all bipartitions $(\HypAsg[1], \HypAsg[2]) \in \parFun{\HypAsg}$, it
          holds that $\AStr, \HypAsg[1] \cmodels[][\QEA] \phiFrm[1]$ or $\AStr,
          \HypAsg[2] \cmodels[][\QEA] \phiFrm[2]$;
        \item\label{def:sem(adif:con:ae)}
          $\AStr, \HypAsg \cmodels[][\QAE] \phiFrm[1] \LCon \phiFrm[2]$ if
          $\AStr, \dual{\HypAsg} \cmodels[][\QEA] \phiFrm[1] \LCon \phiFrm[2]$;
      \end{enumerate}
    \item\label{def:sem(adif:dis)}
      \begin{enumerate}[a)]
        \item\label{def:sem(adif:dis:ea)}
          $\AStr, \HypAsg \cmodels[][\QEA] \phiFrm[1] \LDis \phiFrm[2]$ if
          $\AStr, \dual{\HypAsg} \cmodels[][\QAE] \phiFrm[1] \LDis \phiFrm[2]$;
        \item\label{def:sem(adif:dis:ae)}
          $\AStr, \HypAsg \cmodels[][\QAE] \phiFrm[1] \LDis \phiFrm[2]$ if there
          exists a bipartition $(\HypAsg[1], \HypAsg[2]) \in
          \parFun{\HypAsg}$ such that $\AStr, \HypAsg[1] \cmodels[][\QAE]
          \phiFrm[1]$ and $\AStr, \HypAsg[2] \cmodels[][\QAE] \phiFrm[2]$;
      \end{enumerate}
    \item\label{def:sem(adif:exs)}
      \begin{enumerate}[a)]
        \item\label{def:sem(adif:exs:ea)}
          $\AStr, \HypAsg \cmodels[][\QEA] \LExs[][_{\pm\WSet}] \varElm \ldotp
          \phiFrm$ if $\AStr, \extFun[\denot{\pm\WSet}]{\HypAsg, \varElm}
          \cmodels[][\QEA] \phiFrm$;
        \item\label{def:sem(adif:exs:ae)}
          $\AStr, \HypAsg \cmodels[][\QAE] \LExs[][_{\pm\WSet}] \varElm \ldotp
          \phiFrm$ if $\AStr, \dual{\HypAsg} \cmodels[][\QEA]
          \LExs[][_{\pm\WSet}] \varElm \ldotp \phiFrm$;
      \end{enumerate}
    \item\label{def:sem(adif:all)}
      \begin{enumerate}[a)]
        \item\label{def:sem(adif:all:ea)}
          $\AStr, \HypAsg \cmodels[][\QEA] \LAll[][_{\pm\WSet}] \varElm \ldotp
          \phiFrm$ if $\AStr, \dual{\HypAsg} \cmodels[][\QAE]
          \LAll[][_{\pm\WSet}] \varElm \ldotp \phiFrm$;
        \item\label{def:sem(adif:all:ae)}
          $\AStr, \HypAsg \cmodels[][\QAE] \LAll[][_{\pm\WSet}] \varElm \ldotp
          \phiFrm$ if $\AStr, \extFun[\denot{\pm\WSet}]{\HypAsg, \varElm}
          \cmodels[][\QAE] \phiFrm$.
      \end{enumerate}
  \end{enumerate}
\end{definition}

Items~\ref{def:sem(adif:fbv)} and~\ref{def:sem(adif:tbv)} take care of the
Boolean constants, requiring, \eg, $\Tt$ to be satisfied by all hyperteams,
except for the empty one, under the $\QEA$ interpretation, and the null one,
under $\QAE$.
A dual reasoning applies to $\Ff$.
The other base case for atomic formulae, Item~\ref{def:sem(adif:rel)}, is
trivial and follows the interpretation of the alternation flag.
Negation, in accordance with the classic game-theoretic interpretation, is dealt
with by Item~\ref{def:sem(adif:neg)} by exchanging the interpretation of the
players of the hyperteam.
The semantics of the remaining Boolean connectives
(Items~\ref{def:sem(adif:con)} and~\ref{def:sem(adif:dis)}) and quantifiers
(Items~\ref{def:sem(adif:exs)} and~\ref{def:sem(adif:all)}) is a direct
application of the partition and extension operators previously defined.
Observe that swapping between $\cmodels[][\QEA]$ and $\cmodels[][\QAE]$
(Items~\ref{def:sem(adif:con:ae)}, \ref{def:sem(adif:dis:ea)},
\ref{def:sem(adif:exs:ae)} and~\ref{def:sem(adif:all:ea)}) is done according to
Lemma~\ref{lmm:dltii} and represents the fundamental point where our approach
departs from Hodges' semantics~\citep{Hod97a,Hod97b}.

For every \ADIF formula $\varphi$ and alternation flag $\alpha \in \{ \QEA, \QAE
\}$, we say that $\varphi$ is \emph{$\alpha$-satisfiable on $\AStr$}, in symbols
$\AStr \cmodels[][\alpha] \varphi$, if there exists a proper hyperteam $\HypAsg
\in \HypAsgSet(\sup{\varphi})$ such that $\AStr, \HypAsg \cmodels[][\alpha]
\varphi$.
As already mentioned before, here we are not considering the empty and null
hyperteams as potential hyperteams, since these do not convey meaningful information.
We simply say that $\varphi$ is \emph{$\alpha$-satisfiable} \iff it is
$\alpha$-satisfiable on some structure $\AStr$.
Also, $\varphi$ \emph{$\alpha$-implies} (\resp, is \emph{$\alpha$-equivalent}
to) an \ADIF formula $\phi$ \emph{on $\AStr$}, in symbols $\varphi
\implies[\AStr][\alpha] \phi$ (\resp, $\varphi \cequiv[\AStr][\alpha] \phi$),
whenever $\AStr, \HypAsg \cmodels[][\alpha] \varphi$ implies $\AStr, \HypAsg
\cmodels[][\alpha] \phi$ (\resp, $\AStr, \HypAsg \cmodels[][\alpha] \varphi$
\iff $\AStr, \HypAsg \cmodels[][\alpha] \phi$), for all $\HypAsg \in
\HypAsgSet[\subseteq](\sup{\varphi} \cup \sup{\phi})$.
%
If the implication (\resp, equivalence) holds for all structures $\AStr$, we
just state that $\varphi$ \emph{$\alpha$-implies} (\resp, is
\emph{$\alpha$-equivalent} to) $\phi$, in symbols $\varphi \implies[][\alpha]
\phi$ (\resp, $\varphi \cequiv[][\alpha] \phi$).
Finally, we say that $\varphi$ is \emph{satisfiable} if it is both $\QEA$- and
$\QAE$-satisfiable, and $\varphi$ \emph{implies} (\resp, is \emph{equivalent}
to) $\phi$, in symbols $\varphi \implies \phi$ (\resp, $\varphi \eqv \phi$), if
both $\varphi \implies[][\QEA] \phi$ and $\varphi \implies[][\QAE] \phi$ (\resp,
$\varphi \cequiv[][\QEA] \phi$ and $\varphi \cequiv[][\QAE] \phi$) hold true.




\subsection{Examples}
\label{sec:adif;sub:exm}

To familiarise with the proposed compositional semantics of \ADIF, we now
present few examples of evaluation of formulae via a step by step unravelling of
all the semantic rules involved.

\begin{example}
  \label{exm:snt} Consider the sentence $\varphiSnt[4] = \LExs \varElm \ldotp
  \LAll[][+\emptyset] \yvarElm \ldotp \LNeg \psiFrm(\varElm, \yvarElm)$ from
  above, where we instantiate $\psiFrm(\varElm, \yvarElm)$ as $(\varElm =
  \yvarElm)$.
  We evaluate $\varphiSnt[4]$ in the binary structure $\AStr = \tuple {\{ 0, 1
    \}} {=^{\AStr}}$ against the trivial hyperteam $\TrvHypAsg$.
  The alternation flag is of no consequence, since $\TrvHypAsg$ is self-dual
  (see Proposition~\ref{prp:empnultrv}), hence, we can choose $\alpha = \QEA$,
  without loss of generality.
  We want to check whether $\AStr, \TrvHypAsg \cmodels[][\QEA] \varphiSnt[4]$.
  The semantic rule for the existential quantifier $\LExs \varElm$ requires to
  compute the extension $\extFun[\emptyset]{\TrvHypAsg, \varElm}$ of
  $\TrvHypAsg$.
  This results in
  \[
    \AStr, \TrvHypAsg \cmodels[][\QEA] \LExs \varElm \ldotp \LAll[][+\emptyset]
    \yvarElm \ldotp \LNeg (\varElm = \yvarElm)
  \ \ \iff\ \
    \AStr, \HypAsg \cmodels[][\QEA] \LAll[][+\emptyset] \yvarElm \ldotp \LNeg
    (\varElm = \yvarElm),
  \]
  where $\HypAsg = \{\!\{ \varElm\!:\!0 \}, \{ \varElm\!:\!1 \}\!\}$.
  The rule for the universal quantifier $\LAll[][+\emptyset] \yvarElm$ requires
  to dualise the hyperteam and switch the flag to $\QAE$.
  Since every team of $\HypAsg$ is a singleton, there is only one possible
  choice function, thus, the result is
  \[
    \AStr, \HypAsg \cmodels[][\QEA] \LAll[][+\emptyset] \yvarElm \ldotp \LNeg
    (\varElm = \yvarElm)
  \ \ \iff\ \
    \AStr, \dual{\HypAsg} \cmodels[][\QAE] \LAll[][+\emptyset] \yvarElm \ldotp
    \LNeg (\varElm = \yvarElm),
  \]
  where $\dual{\HypAsg} = \{\!\{ \varElm\!:\!0, \varElm\!:\!1 \}\!\}$.
  Now the quantifier $\LAll[][+\emptyset] \yvarElm$ and the alternation flag
  $\QAE$ are coherent, and we extend the hyperteam to obtain
  $\extFun[\emptyset]{\dual{\HypAsg}, \yvarElm}$, where only constant functions
  can be used for the extensions, since $\yvarElm$ cannot depend on $\varElm$.
  The result is, then,
  \[
    \AStr, \dual{\HypAsg} \cmodels[][\QAE] \LAll[][+\emptyset] \yvarElm \ldotp
    \LNeg (\varElm = \yvarElm)
  \ \ \iff\ \
    \AStr, \extFun[\emptyset]{\dual{\HypAsg}, \yvarElm} \cmodels[][\QAE]
    \LNeg (\varElm = \yvarElm),
  \]
  where
  $\extFun[\emptyset]{\dual{\HypAsg}, \yvarElm} =
  \left\{\hspace{-0.35em}\left\{\!\!\!
  \begin{array}{cc}
    {\varElm}\!:\!0 \\
    {\yvarElm}\!:\!0
  \end{array}
  \!\!,\!\!\!
  \begin{array}{cc}
    {\varElm}\!:\!1 \\
    {\yvarElm}\!:\!0
  \end{array}
  \!\!\!\right\}\!,\!\left\{\!\!\!
  \begin{array}{c}
    {\varElm}\!:\!0\\
    {\yvarElm}\!:\!1
  \end{array}
  \!\!,\!\!\!
  \begin{array}{c}
    {\varElm}\!:\!1\\
    {\yvarElm}\!:\!1
  \end{array}
  \!\!\!\right\}\hspace{-0.35em}\right\}$.
  The rule for the negation operation $\LNeg$ dualises the flag and, in
  addition, requires the hyperteam $\extFun[\emptyset]{\dual{\HypAsg},
  \yvarElm}$ not to satisfy the atom $(\varElm = \yvarElm)$ under $\QEA$.
  This means that every team in $\extFun[\emptyset]{\dual{\HypAsg}, \yvarElm}$
  must contain an assignment that falsifies the atom.
  But this is indeed the case, since every team has an assignment $\asgElm$ such
  that $\asgElm(\varElm) \neq \asgElm(\yvarElm)$.
  Hence, $\varphiSnt[4]$ evaluates to true in $\AStr$ against $\TrvHypAsg$.
  Observe that, on the contrary, the sentence $\varphiSnt[3] = \LAll \varElm
  \ldotp \LExs[][+\emptyset] \yvarElm \ldotp \psiFrm(\varElm, \yvarElm)$ from
  above evaluates to false in $\AStr$ against $\TrvHypAsg$, being
  equivalent to the negation of $\varphiSnt[4]$.
  Indeed, in this case, following the semantic rules for the quantifiers, we
  would still end up with the same hyperteam
  $\extFun[\emptyset]{\dual{\HypAsg}, \yvarElm}$ against which we need to
  evaluate the matrix $\varElm = \yvarElm$.
  However, this time the alternation flag would be $\QEA$ and, as we already
  noted above, every team in $\extFun[\emptyset]{\dual{\HypAsg}, \yvarElm}$
  contains one assignment falsifying $\varElm = \yvarElm$.
\end{example}

\begin{example}
  \label{exm:psdsnt}
  Consider the pseudo sentence $\varphiSnt[6] = \LExs \varElm \ldotp
  \LAll[][-\varElm]  \yvarElm \ldotp \LNeg \psiFrm(\varElm, \yvarElm)$ from
  above, where again we instantiate $\psiFrm(\varElm, \yvarElm)$
  as $(\varElm = \yvarElm)$.
  The exact same reasoning followed in Example~\ref{exm:snt} shows that
  $\varphiSnt[6]$ is true in $\AStr$ against the trivial hyperteam $\TrvHypAsg$.
  Consequently, the pseudo sentence $\varphiSnt[5] = \LAll \varElm \ldotp
  \LExs[][-\varElm] \yvarElm \ldotp \psiFrm(\varElm, \yvarElm)$ is false in
  $\AStr$ against $\TrvHypAsg$, being equivalent to the negation of
  $\varphiSnt[6]$.
  These two pseudo sentences, however, are not equivalent to the sentences
  $\varphiSnt[4]$ and $\varphiSnt[3]$, respectively.
  To see this, let us evaluate $\varphiSnt[5]$ in $\AStr$ against the hyperteam
  $\HypAsg = \{\!\{ \zvarElm\!:\!0, \zvarElm\!:\!1 \}\!\}$ \wrt the alternation
  flag $\alpha = \QAE$.
  Note that $\zvarElm \in \free{\varphiSnt[5]} = \denot{-\varElm} = \VarSet
  \setminus \{ \varElm \}$.
  The semantic rule for $\LAll \varElm$ requires to compute the extension
  $\extFun[\emptyset]{\HypAsg, \varElm}$ of $\HypAsg$.
  This results in
  \[
    \AStr, \HypAsg \cmodels[][\QAE] \LAll \varElm \ldotp \LExs[][-\varElm]
    \yvarElm \ldotp (\varElm = \yvarElm)
  \ \ \iff\ \
    \AStr, \extFun[\emptyset]{\HypAsg, \varElm} \cmodels[][\QAE]
    \LExs[][-\varElm] \yvarElm \ldotp (\varElm = \yvarElm),
  \]
  where
  $\extFun[\emptyset]{\HypAsg, \varElm} =
  \left\{\hspace{-0.35em}\left\{\!\!\!
  \begin{array}{cc}
    {\zvarElm}\!:\!0 \\
    {\varElm}\!:\!0
  \end{array}
  \!\!,\!\!\!
  \begin{array}{cc}
    {\zvarElm}\!:\!1 \\
    {\varElm}\!:\!0
  \end{array}
  \!\!\!\right\}\!,\!\left\{\!\!\!
  \begin{array}{c}
    {\zvarElm}\!:\!0\\
    {\varElm}\!:\!1
  \end{array}
  \!\!,\!\!\!
  \begin{array}{c}
    {\zvarElm}\!:\!1\\
    {\varElm}\!:\!1
  \end{array}
  \!\!\!\right\}\hspace{-0.35em}\right\}$.
  The rule for $\LExs[][-\varElm] \yvarElm$ requires to dualise the hyperteam
  and switch the flag to $\QEA$.
  Since both teams in $\extFun[\emptyset]{\HypAsg, \varElm}$ contains two
  assignments, there are four choice functions in total, leading to
  \[
    \AStr, \extFun[\emptyset]{\HypAsg, \varElm} \cmodels[][\QAE]
    \LExs[][-\varElm] \yvarElm \ldotp (\varElm = \yvarElm)
  \ \ \iff\ \
    \AStr, \dual{\extFun[\emptyset]{\HypAsg, \varElm}} \cmodels[][\QEA]
    \LExs[][-\varElm] \yvarElm \ldotp (\varElm = \yvarElm),
  \text{where}
  \]
  \[
  \dual{\extFun[\emptyset]{\HypAsg, \varElm}} =
  \left\{\hspace{-0.35em}
  \begin{array}{c}
    {\TeamAsg[1]} = \\
    \left\{\hspace{-0.35em}
    \begin{array}{cc}
      {\zvarElm}\!:\!0 \\
      {\varElm}\!:\!0
    \end{array}
    \!\!,\!\!\!
    \begin{array}{c}
      {\zvarElm}\!:\!0\\
      {\varElm}\!:\!1
    \end{array}
    \hspace{-0.35em}\right\}
  \end{array}
  \!\!,\!\!\!
  \begin{array}{c}
    {\TeamAsg[2]} = \\
    \left\{\hspace{-0.35em}
    \begin{array}{cc}
      {\zvarElm}\!:\!0 \\
      {\varElm}\!:\! 0
    \end{array}
    \!\!,\!\!\!
    \begin{array}{c}
      {\zvarElm}\!:\!1\\
      {\varElm}\!:\!1
    \end{array}
    \hspace{-0.35em}\right\}
  \end{array}\!\!,\!\!\!
  \begin{array}{c}
    {\TeamAsg[3]} = \\
    \left\{\hspace{-0.35em}
    \begin{array}{cc}
      {\zvarElm}\!:\!1 \\
      {\varElm}\!:\!0
    \end{array}
    \!\!,\!\!\!
    \begin{array}{c}
      {\zvarElm}\!:\!0\\
      {\varElm}\!:\!1
    \end{array}
    \hspace{-0.35em}\right\}
  \end{array}
  \!\!,\!\!\!
  \begin{array}{c}
    {\TeamAsg[4]} = \\
    \left\{\hspace{-0.35em}
    \begin{array}{cc}
      {\zvarElm}\!:\!1 \\
      {\varElm}\!:\!0
    \end{array}
    \!\!,\!\!\!
    \begin{array}{c}
      {\zvarElm}\!:\!1\\
      {\varElm}\!:\!1
    \end{array}
    \hspace{-0.35em}\right\}
  \end{array}
  \!\!\right\}\!.
  \]
  The extension $\der{\HypAsg} \defeq \extFun[\VarSet \setminus \varElm]{
  \dual{\extFun[\emptyset]{\HypAsg, \varElm}}, \yvarElm } = \extFun[\{ \zvarElm
  \}]{ \dual{\extFun[\emptyset]{\HypAsg, \varElm}}, \yvarElm }$ with the four
  functions that can only depend on $\zvarElm$, happens to contains $12$ teams
  and cannot be displayed here.
  However, it should be easy to check that among these teams one can find $\XSet
  \defeq \extFun{\XSet[2], \FFun, \yvarElm} = \{ \asgElm[1], \asgElm[2] \}$,
  where $\asgElm[1](\zvarElm) = \asgElm[1](\varElm) = \asgElm[1](\yvarElm) = 0$,
  $\asgElm[2](\zvarElm) = \asgElm[2](\varElm) = \asgElm[2](\yvarElm) = 1$, and
  $\FFun(\asgElm) = \asgElm(\zvarElm)$.
  Now, the final step requires checking whether $\AStr, \der{\HypAsg}
  \cmodels[][\QEA] (\varElm = \yvarElm)$.
  Since every assignment in $\XSet$ satisfies $(\varElm = \yvarElm)$, the pseudo
  sentence is proved true in $\AStr$ against $\HypAsg$.
  As an immediate consequence, $\varphiSnt[6]$ evaluates to false in $\AStr$
  against $\HypAsg$.
  Instead, it is possible to show that the evaluations of $\varphiSnt[3]$ and
  $\varphiSnt[4]$ remain unchanged on $\HypAsg$, \ie, they are again false and
  true, respectively, due to the fact that they are sentences (this is a direct
  consequence of Corollary~\ref{cor:hypeqv}, proved later on).
\end{example}

The above example should clarify the reasoning behind the choice of the name
\emph{pseudo sentences}, for those formulae $\varphiFrm$ with $\sup{\varphiFrm}
= \emptyset$, but $\free{\varphiFrm} \neq \emptyset$.
As for sentences, a pseudo sentence can be verified against an arbitrary
hyperteam; however, similarly to formulae, its truth may depend on the specific
hyperteam.

\begin{example}
  \label{exm:runexm}
  Consider the sentence $\varphiSnt[7] = \LExs \varElm \ldotp
  \LAll[][+\emptyset] \yvarElm \ldotp \LExs[][+\varElm] \zvarElm \ldotp (
  \psiFrm[1](\varElm, \yvarElm) \allowbreak \LCon \psiFrm[2](\yvarElm,
  \zvarElm) )$ from above, where we instantiate
  $\psiFrm[1](\varElm, \yvarElm)$ as $(\varElm = \yvarElm)$ and
  $\psiFrm[2](\yvarElm, \zvarElm)$ as $(\yvarElm = \zvarElm)$.
  We evaluate this sentence against the same structure $\AStr$ of the previous
  examples and the trivial hyperteam $\TrvHypAsg$.
  Observe also that $\varphiFrm[7]$ shares most of the quantifier prefix of
  sentence $\varphiSnt[4]$ in Example~\ref{exm:snt}.
  As a consequence, by applying the same steps as before, we end up with the
  following equivalence:
  \[
    \AStr, \TrvHypAsg \cmodels[][\QEA] \varphiSnt[7]
  \ \ \iff\ \
    \AStr, \extFun[\emptyset]{\dual{\HypAsg}, \yvarElm} \cmodels[][\QAE]
    \LExs[][+\varElm] \zvarElm \ldotp (\varElm = \yvarElm) \LCon (\yvarElm =
    \zvarElm),
  \]
  where
  $\extFun[\emptyset]{\dual{\HypAsg}, \yvarElm} =
  \left\{\hspace{-0.35em}\left\{\!\!\!
  \begin{array}{cc}
    {\varElm}\!:\!0 \\
    {\yvarElm}\!:\!0
  \end{array}
  \!\!,\!\!\!
  \begin{array}{cc}
    {\varElm}\!:\!1 \\
    {\yvarElm}\!:\!0
  \end{array}
  \!\!\!\right\}\!,\!\left\{\!\!\!
  \begin{array}{c}
    {\varElm}\!:\!0\\
    {\yvarElm}\!:\!1
  \end{array}
  \!\!,\!\!\!
  \begin{array}{c}
    {\varElm}\!:\!1\\
    {\yvarElm}\!:\!1
  \end{array}
  \!\!\!\right\}\hspace{-0.35em}\right\}$.
  Applying the rule for $\LExs[][+\varElm] \zvarElm$ requires dualisation first,
  leading to
  \[
    \AStr, \TrvHypAsg \cmodels[][\QEA] \varphiSnt[7]
  \ \ \iff\ \
    \AStr, \dual{\extFun[\emptyset]{\dual{\HypAsg}, \yvarElm}} \cmodels[][\QEA]
    \LExs[][+\varElm] \zvarElm \ldotp (\varElm = \yvarElm) \LCon (\yvarElm =
    \zvarElm),
  \text{where}
  \]
  \[
  \dual{\extFun[\emptyset]{\dual{\HypAsg}, \yvarElm}} =
  \left\{\hspace{-0.35em}
  \begin{array}{c}
    {\TeamAsg[1]} = \\
    \left\{\hspace{-0.35em}
    \begin{array}{cc}
      {\varElm}\!:\!0 \\
      {\yvarElm}\!:\!0
    \end{array}
    \!\!,\!\!\!
    \begin{array}{c}
      {\varElm}\!:\!0\\
      {\yvarElm}\!:\!1
    \end{array}
    \hspace{-0.35em}\right\}
  \end{array}
  \!\!,\!\!\!
  \begin{array}{c}
    {\TeamAsg[2]} = \\
    \left\{\hspace{-0.35em}
    \begin{array}{cc}
      {\varElm}\!:\!0 \\
      {\yvarElm}\!:\! 0
    \end{array}
    \!\!,\!\!\!
    \begin{array}{c}
      {\varElm}\!:\!1\\
      {\yvarElm}\!:\!1
    \end{array}
    \hspace{-0.35em}\right\}
  \end{array}\!\!,\!\!\!
  \begin{array}{c}
    {\TeamAsg[3]} = \\
    \left\{\hspace{-0.35em}
    \begin{array}{cc}
      {\varElm}\!:\!1 \\
      {\yvarElm}\!:\!0
    \end{array}
    \!\!,\!\!\!
    \begin{array}{c}
      {\varElm}\!:\!0\\
      {\yvarElm}\!:\!1
    \end{array}
    \hspace{-0.35em}\right\}
  \end{array}
  \!\!,\!\!\!
  \begin{array}{c}
    {\TeamAsg[4]} = \\
    \left\{\hspace{-0.35em}
    \begin{array}{cc}
      {\varElm}\!:\!1 \\
      {\yvarElm}\!:\!0
    \end{array}
    \!\!,\!\!\!
    \begin{array}{c}
      {\varElm}\!:\!1\\
      {\yvarElm}\!:\!1
    \end{array}
    \hspace{-0.35em}\right\}
  \end{array}
  \!\!\right\}\!.
  \]
  The extension $\der{\HypAsg} \defeq \extFun[\{ \varElm \}]{
  \dual{\extFun[\emptyset]{\dual{\HypAsg}, \yvarElm}}, \zvarElm}$ can only use
  functions that depend on $\varElm$ alone and there are four of them.
  Similarly to the previous example, the hyperteam $\der{\HypAsg}$ ends up
  containing $12$ teams.
  Among these teams one can find $\XSet \defeq \extFun{\XSet[2], \FFun,
  \zvarElm} = \{ \asgElm[1], \asgElm[2] \}$, where $\asgElm[1](\varElm) =
  \asgElm[1](\yvarElm) = \asgElm[1](\zvarElm) = 0$, $\asgElm[2](\varElm) =
  \asgElm[2](\yvarElm) = \asgElm[2](\zvarElm) = 1$, and $\FFun(\asgElm) =
  \asgElm(\varElm)$.
  Now, the final step requires checking whether $\AStr, \der{\HypAsg}
  \cmodels[][\QEA] (\varElm = \yvarElm) \LCon (\yvarElm = \zvarElm)$.
  By the rule for the conjunction connective, this is true if $\AStr,
  \der{\HypAsg}[1] \cmodels[][\QEA] (\varElm = \yvarElm)$ or $\AStr,
  \der{\HypAsg}[2] \cmodels[][\QEA] (\yvarElm = \zvarElm)$, for all
  bipartitions $(\der{\HypAsg}[1], \der{\HypAsg}[2]) \in
  \parFun{\hat{\HypAsg}}$.
  Obviously, any such partition would contain $\XSet$ either in
  $\der{\HypAsg}[1]$ or in $\der{\HypAsg}[2]$.
  Since every assignment in $\XSet$ satisfies both $(\varElm = \yvarElm)$ and
  $(\yvarElm = \zvarElm)$, the sentence is proved true in $\AStr$ against
  $\TrvHypAsg$.
\end{example}




\subsection{Fundamentals}
\label{sec:adif;sub:fun}

\ADIF enjoys several classic properties, such as \emph{Boolean laws} and the
canonical representation for formulae in \emph{negation normal form} (\nnf, for
short), that are usually expected to hold for a logic closed under negation.

We start with the following very basic result, characterising the truth of
formulae over the null and empty hyperteams.

\begin{lemma}[name = Empty \& Null Hyperteams, restate = lmmempnulhyp]
  \label{lmm:empnulhyp}
  The following hold true for every \ADIF formula $\varphiFrm$ and hyperteam
  $\HypAsg \in \HypAsgSet[\subseteq](\sup{\varphiFrm})$:
  \begin{enumerate}[1)]
    \item\label{lmm:empnulhyp(ea)}
      \begin{inparaenum}[a)]
        \item\label{lmm:empnulhyp(ea:emp)}
          $\AStr, \EmpHypAsg \notcmodels[][\QEA] \varphiFrm$;
        \hfill
        \item\label{lmm:empnulhyp(ea:nul)}
          $\AStr, \HypAsg \cmodels[][\QEA] \varphiFrm$, where $\EmpTeamAsg \in
          \HypAsg$;
      \end{inparaenum}
    \item\label{lmm:empnulhyp(ae)}
      \begin{inparaenum}[a)]
        \item\label{lmm:empnulhyp(ae:emp)}
          $\AStr, \EmpHypAsg \cmodels[][\QAE] \varphiFrm$;
        \hfill
        \item\label{lmm:empnulhyp(ae:nul)}
          $\AStr, \HypAsg \notcmodels[][\QAE] \varphiFrm$, where $\EmpTeamAsg
          \in \HypAsg$.
      \end{inparaenum}
  \end{enumerate}
\end{lemma}

The preorder $\inc$ on hyperteams introduced above captures
the intuitive notion of satisfaction strength \wrt \ADIF formulae.
Basically, if $\HypAsg[1] \inc \HypAsg[2]$, the hyperteam $\HypAsg[1]$
satisfies, \wrt the $\QEA$ (\resp, $\QAE$) semantic relation, less (\resp, more)
formulae than the hyperteam $\HypAsg[2]$.
Actually, a stronger version of this property holds, when the $\inc$-preorder is
restricted to the set of free variables of the formula.
This property is trivial for atomic formulae and can easily be proved by
structural induction for the non-atomic ones.

\begin{theorem}[name = Hyperteam Refinement, restate = thmhypref]
  \label{thm:hypref}
  Let $\varphiFrm$ be an \ADIF formula and $\HypAsg, \HypAsg' \in
  \HypAsgSet[\subseteq](\sup{\varphiFrm})$ two hyperteams with ${\HypAsg
  \inc[\free{\varphiFrm}] \HypAsg'}$.
  Then:
  \begin{enumerate}[1)]
    \item\label{thm:hypref(ea)}
      if $\AStr, \HypAsg \cmodels[][\QEA] \varphiFrm$ then $\AStr, \HypAsg'
      \cmodels[][\QEA] \varphiFrm$;
    \item\label{thm:hypref(ae)}
      if $\AStr, \HypAsg' \cmodels[][\QAE] \varphiFrm$ then $\AStr, \HypAsg
      \cmodels[][\QAE] \varphiFrm$.
  \end{enumerate}
\end{theorem}

As an immediate consequence, we obtain the following result.

\begin{corollary}[name = Hyperteam Equivalence, restate = corhypeqv]
  \label{cor:hypeqv}
  Let $\varphiFrm$ be an \ADIF formula and $\HypAsg, \HypAsg' \in
  \HypAsgSet[\subseteq](\sup{\varphiFrm})$ two hyperteams with ${\HypAsg
  \eqv[\free{\varphiFrm}] \HypAsg'}$.
  Then:
  \[
    \AStr, \HypAsg \cmodels[][\alpha] \varphiFrm
  \text{\ \ \iff\ \ }
    \AStr, \HypAsg' \cmodels[][\alpha] \varphiFrm.
  \]
\end{corollary}

Since, by definition, an \ADIF sentence $\varphiSnt$ satisfies
$\free{\varphiSnt} = \emptyset$, we can test its truth by just looking at its
satisfaction \wrt the trivial hyperteam $\TrvHypAsg$, as every proper hyperteam
is equivalent to $\TrvHypAsg$ on the empty set of variables.
Recall that this property is, instead, not necessarily enjoyed by a pseudo
sentence, as already observed in Example~\ref{exm:psdsnt}.

\begin{corollary}[name = Sentence Satisfiability, restate = corsensat]
  \label{cor:sensat}
  Let $\varphiSnt$ be an \ADIF sentence.
  Then, $\varphiSnt$ is $\alpha$-satisfiable \iff $\AStr, \TrvHypAsg
  \cmodels[][\alpha] \varphiFrm$, for some $\LSig$-structure $\AStr$.
\end{corollary}

As mentioned in Example~\ref{exm:hypdlt}, swapping the players of a hyperteam
$\HypAsg$, \ie, switching the alternation flag, and swapping the choices of the
players, \ie, dualising $\HypAsg$, have the same effect as far as satisfaction
is concerned.
Recall in addition that, by Lemma~\ref{lmm:dlti}, the dualisation enjoys the
involution property.
Consequently, dualising both the alternation flag $\alpha$ and the hyperteam
$\HypAsg$ preserves truth of formulae.
These observations are formalised by the following result.

\begin{theorem}[name = Double Dualisation, restate = thmdbldlt]
  \label{thm:dbldlt}
  For every \ADIF formula $\varphiFrm$ and hyperteam $\HypAsg \in
  \HypAsgSet[\subseteq](\sup{\varphiFrm})$, it holds that $\AStr, \HypAsg
  \cmodels[][\alpha] \varphiFrm$ \iff $\AStr, \dual{\dual{\HypAsg}}
  \cmodels[][\alpha] \varphiFrm$ \iff $\AStr, \dual{\HypAsg}
  \cmodels[][\dual{\alpha}] \varphiFrm$.
\end{theorem}

The above property also grants that formulae satisfaction, implication, and
equivalence do not depend on the specific interpretation $\alpha$ of hyperteams:
a positive answer for $\alpha$ implies the same for $\dual{\alpha}$.
This \emph{invariance} corresponds to the intuition that Eloise and Abelard both
agree on the true and false formulae, as well on the concept of logical
consequence and equivalence.

\begin{corollary}[name = Interpretation Invariance, restate = corintinv]
  \label{cor:intinv}
  Let $\varphiFrm$ and $\phiFrm$ be \ADIF formulae.
  Then, $\varphiFrm$ is $\QEA$-satisfiable \iff $\varphiFrm$ is
  $\QAE$-satisfiable.
  Also, $\varphiFrm \implies[][\QEA] \phiFrm$ \iff $\varphiFrm \implies[][\QAE]
  \phiFrm$ and $\varphiFrm \cequiv[][\QEA] \phiFrm$ \iff $\varphiFrm
  \cequiv[][\QAE] \phiFrm$.
\end{corollary}

Given the game-theoretic nature of hyperteams and negation, \ADIF does not enjoy
\emph{logical determinacy}, \ie, the property stating that a model either
satisfies a formula or its negation, \wrt the same semantic relation.
However, it satisfies the \emph{game-theoretic determinacy} stated below, which
corresponds to the following intuition: if a player cannot prove the truth of a
formula, then the other player can prove the truth of its negation.

\begin{corollary}[name = Game-Theoretic Determinacy, restate = cordet]
  \label{cor:det}
  Let $\varphiFrm$ be an \ADIF formula and $\HypAsg \in
  \HypAsgSet[\subseteq](\sup{\varphiFrm})$ a hyperteam.
  Then:
  \begin{enumerate}[1)]
    \item\label{cor:det(flg)}
      either $\AStr, \HypAsg \cmodels[][\alpha] \varphiFrm$ or $\AStr, \HypAsg
      \cmodels[][\dual{\alpha}] \neg \varphiFrm$;
    \item\label{cor:det(hyp)}
      either $\AStr, \HypAsg \cmodels[][\alpha] \varphiFrm$ or $\AStr,
      \dual{\HypAsg} \cmodels[][\alpha] \neg \varphiFrm$.
  \end{enumerate}
\end{corollary}

Since, as observed before, the truth of sentences can be tested on the trivial
hyperteam $\TrvHypAsg$, regardless of the specific alternation flag $\alpha$,
the classic \emph{law of excluded middle} does hold for sentences.
In the following, by $\AStr \models \varphiSnt$ we denote the fact that
$\varphiSnt$ is both $\QEA$-satisfiable and $\QAE$-satisfiable on $\AStr$.

\begin{corollary}[name = Law of Excluded Middle , restate = corthrexcprn]
  \label{cor:lawexcmid}
  Let $\varphiSnt$ be an \ADIF sentence.
  Then, either $\AStr \models \varphiSnt$ or $\AStr \models \neg
  \varphiSnt$.
\end{corollary}

Thanks to the above properties, we can establish the following elementary
Boolean laws, which, in turn, allow for a canonical representation of formulae
in \nnf, as stated in Corollary~\ref{cor:nnf}.

\begin{theorem}[name = Boolean Laws, restate = thmblnlawi]
  \label{thm:blnlawi}
  Let $\varphiFrm[1]$, $\varphiFrm[2]$ and $\varphiFrm$ be \ADIF formulae.
  Then:
  \begin{enumerate}[1)]
    \item\label{thm:blnlawi(neg)}
      \begin{inparaenum}[a)]
        \item\label{thm:blnlawi(neg:fbv)}
          $\neg \Ff \equiv \Tt$;
        \hfill
        \item\label{thm:blnlawi(neg:tbv)}
          $\neg \Tt \equiv \Ff$;
        \hfill
        \item\label{thm:blnlawi(neg:inv)}
          $\varphiFrm \equiv \neg \neg \varphiFrm$;
      \end{inparaenum}
    \item\label{thm:blnlawi(conbln)}
      \begin{inparaenum}[a)]
        \item\label{thm:blnlawi(conbln:ann)}
          $\varphiFrm \LCon \Ff \equiv \Ff \LCon \varphiFrm \equiv \Ff$;
        \hfill
        \item\label{thm:blnlawi(conbln:idn)}
          $\varphiFrm \LCon \Tt \equiv \Tt \LCon \varphiFrm \equiv \varphiFrm$;
      \end{inparaenum}
    \item\label{thm:blnlawi(disbln)}
      \begin{inparaenum}[a)]
        \item\label{thm:blnlawi(disbln:ann)}
          $\varphiFrm \LDis \Tt \equiv \Tt \LDis \varphiFrm \equiv \Tt$;
        \hfill
        \item\label{thm:blnlawi(disbln:idn)}
          $\varphiFrm \LDis \Ff \equiv \Ff \LDis \varphiFrm \equiv \varphiFrm$;
      \end{inparaenum}
    \item\label{thm:blnlawi(com)}
      \begin{inparaenum}[a)]
        \item\label{thm:blnlawi(com:con)}
          $\varphiFrm[1] \LCon \varphiFrm[2] \equiv \varphiFrm[2] \LCon
          \varphiFrm[1]$;
        \hfill
        \item\label{thm:blnlawi(com:dis)}
          $\varphiFrm[1] \LDis \varphiFrm[2] \equiv \varphiFrm[2] \LDis
          \varphiFrm[1]$;
      \end{inparaenum}
    \item\label{thm:blnlawi(con)}
      \begin{inparaenum}[a)]
        \item\label{thm:blnlawi(con:elm)}
          $\varphiFrm[1] \LCon \varphiFrm[2] \implies \varphiFrm[1]$;
        \hfill
        \item\label{thm:blnlawi(con:ass)}
          $\varphiFrm[1] \LCon (\varphiFrm \LCon \varphiFrm[2]) \equiv
          (\varphiFrm[1] \LCon \varphiFrm) \LCon \varphiFrm[2]$;
      \end{inparaenum}
    \item\label{thm:blnlawi(dis)}
      \begin{inparaenum}[a)]
        \item\label{thm:blnlawi(dis:int)}
          $\varphiFrm[1] \implies \varphiFrm[1] \LDis \varphiFrm[2]$;
        \hfill
        \item\label{thm:blnlawi(dis:ass)}
          $\varphiFrm[1] \LDis (\varphiFrm \LDis \varphiFrm[2]) \equiv
          (\varphiFrm[1] \LDis \varphiFrm) \LDis \varphiFrm[2]$;
      \end{inparaenum}
    \item\label{thm:blnlawi(dml)}
      \begin{inparaenum}[a)]
        \item\label{thm:blnlawi(dml:con)}
          $\varphiFrm[1] \LCon \varphiFrm[2] \equiv \neg (\neg \varphiFrm[1]
          \LDis \neg \varphiFrm[2])$;
        \hfill
        \item\label{thm:blnlawi(dml:dis)}
          $\varphiFrm[1] \LDis \varphiFrm[2] \equiv \neg (\neg \varphiFrm[1]
          \LCon \neg \varphiFrm[2])$;
      \end{inparaenum}
    \item\label{thm:blnlawi(qnt)}
      \begin{inparaenum}[a)]
        \item\label{thm:blnlawi(qnt:exs)}
          $\LExs[][_{\pm\WSet}] \varElm \ldotp \varphiFrm \equiv \neg
          (\LAll[][_{\pm\WSet}] \varElm \ldotp \neg \varphiFrm)$;
        \hfill
        \item\label{thm:blnlawi(qnt:all)}
          $\LAll[][_{\pm\WSet}] \varElm \ldotp \varphiFrm \equiv \neg
          (\LExs[][_{\pm\WSet}] \varElm \ldotp \neg \varphiFrm)$.
      \end{inparaenum}
  \end{enumerate}
\end{theorem}

\begin{corollary}[name = Negation Normal Form, restate = cornnf]
  \label{cor:nnf}
  Every \ADIF formula is equivalent to an \ADIF formula in \nnf.
\end{corollary}

Currently, we do not know whether \ADIF does enjoy a \emph{prenex normal form}
(\pnf, for short).
For this reason, in Sections~\ref{sec:metthr} and~\ref{sec:gamthrsem}, we shall
mainly consider formulae that are already in \pnf.

\begin{openproblem}[name = \ADIF Prenex Normal Form]
  \label{opn:adifpnf}
  Is every \ADIF formula equivalent to an \ADIF formula in \pnf?
\end{openproblem}

For technical convenience, we shall now generalise the extension operator to
(finite) quantifier prefixes $\qntElm$, whose set is denoted by $\QntSet$.
Notice that, \wlogx, we only consider prefixes where each variable $\varElm$
\label{pag:Qassumption}
\begin{inparaenum}[(i)]
\item
  is quantified at most once,
\item
  does not occur in the dependence/independence constraint set
  $\denot{\pm\WSet}$ of its quantifier $\Qnt[][\pm\WSet] \varElm$, and
\item
  cannot be quantified in the scope of a quantifier $\Qnt[][\pm\WSet] \yvarElm$
  whose dependence/independence constraint set $\denot{\pm\WSet}$ includes
  $\varElm$ itself.
\end{inparaenum}
By $\var{\qntElm}$ and $\dep{\qntElm}$ we denote the set of variables quantified
in $\qntElm$ and the union of all dependence/independence constraint sets
occurring in $\qntElm$, respectively.
Given a hyperteam $\HypAsg$ and an alternation flag $\alpha$, the operator
$\extFun[\alpha]{\HypAsg, \qntElm}$ corresponds to iteratively applying the
extension operator to $\HypAsg$, for all quantifiers occurring in $\qntElm$, in
that specific order.
To this end, we first introduce the notion of \emph{coherence} of a quantifier
symbol $\Qnt \in \{ \exists, \forall \}$ with an alternation flag $\alpha \in \{
\QEA, \QAE \}$ as follows: $\Qnt$ is \emph{$\alpha$-coherent} if either $\alpha
= \QEA$ and $\Qnt = \exists$ or $\alpha = \QAE$ and $\Qnt = \forall$.
Now, the application of a quantifier $ \Qnt[][_{\pm\WSet}] \varElm$ to
$\HypAsg$, denoted by $\extFun[\alpha]{\HypAsg, \Qnt[][_{\pm\WSet}] \varElm}$,
follows the semantics of quantifiers, as defined in
Items~\ref{def:sem(adif:exs)} and~\ref{def:sem(adif:all)} of
Definition~\ref{def:sem(adif)}.
More precisely, it just corresponds to the extension of $\HypAsg$ with
$\varElm$, when $\Qnt$ is $\alpha$-coherent.
Conversely, when $\Qnt$ is $\dual{\alpha}$-coherent, we need to dualise the
extension with $\varElm$ of the dual of $\HypAsg$.
Formally:
\[
  \extFun[\alpha]{\HypAsg, \Qnt[][_{\pm\WSet}] \varElm} \defeq
  \begin{cases}
    {\extFun[\denot{\pm\WSet}]{\HypAsg, \varElm}},
      & \text{ if } {\Qnt} \text{ is }  \alpha\text{-coherent}; \\\\[-0.95em]
    {\dual{\extFun[\denot{\pm\WSet}]{\dual{\HypAsg}, \varElm}}},
      & \text{ otherwise}.
  \end{cases}
\]
The operator naturally lifts to arbitrary quantification prefixes $\qntElm$:
\begin{inparaenum}[1)]
  \item
    $\extFun[\alpha]{\HypAsg, \epsilon} \defeq \HypAsg$;
  \item
    $\extFun[\alpha]{\HypAsg, \Qnt[][_{\pm\WSet}] \varElm \ldotp \qntElm} \defeq
    \extFun[\alpha]{\extFun[\alpha]{\HypAsg, \Qnt[][_{\pm\WSet}] \varElm},
    \qntElm}$.
\end{inparaenum}
We also define $\extFun[\alpha]{\qntElm} \defeq \extFun[\alpha]{\TrvHypAsg,
\qntElm}$.
A simple structural induction on a quantifier prefix $\qntElm \in \QntSet$,
shows that a hyperteam $\HypAsg$ $\alpha$-satisfies a formula $\qntElm \phi$
\iff its $\alpha$-extension \wrt $\qntElm$ $\alpha$-satisfies its matrix $\phi$.

\begin{theorem}[name = Prefix Extension, restate = thmprfext]
  \label{thm:prfext}
  Let $\qntElm \phi$ be an \ADIF formula, where $\qntElm \in \QntSet$ is a
  quantifier prefix and $\phi$ is an arbitrary \ADIF formula.
  Then, $\AStr, \HypAsg \cmodels[][\alpha] \qntElm \phi$ \iff $\AStr,
  \extFun[\alpha]{\HypAsg, \qntElm} \cmodels[][\alpha] \phi$, for all
  hyperteams $\HypAsg \in \HypAsgSet[\subseteq](\sup{\qntElm \phi})$.
\end{theorem}







\section{Adequacy}
\label{sec:adq}

In this section, we show that Hodges' alternating semantics based on hyperteams
is \emph{adequate}, \ie, it is a \emph{conservative extension}, precisely
capturing both Tarski's satisfaction for \FOL and Hodges' semantics of \IF (see
Definitions~\ref{def:syn(dif)} and~\ref{def:sem(dif)} for an equivalent
syntactic variant of \IF), when restricted to the corresponding fragments, as
formally stated in Theorems~\ref{thm:folsemadq} and~\ref{thm:difsemadq} below.


\subsection{First-Order Logic}
\label{sec:adq;sub:fol}

We can now prove that, when focusing on the \FOL fragment of \ADIF, as defined
in Section~\ref{sec:adif;sub:syn}, the satisfaction relation of
Definition~\ref{def:sem(adif)} corresponds to the classic Tarskian satisfaction.
This \emph{\FOL adequacy} property holds trivially for atomic formulae and, in
order to extend it to the remaining \FOL components, we make use of the
following three lemmata, which take care of dualisation, quantifiers, and binary
Boolean connectives, respectively.

As extensively discussed before, the dualisation swaps the role of the two
players, while still preserving the original dependencies among their choices.
Indeed, if a \FOL property is satisfied by a hyperteam \wrt a given alternation
flag, it is satisfied by its dual version \wrt the dual flag, as formally
stated in the lemma below.

\begin{lemma}[name = \FOL Dualisation, restate = lmmfoldlt]
  \label{lmm:foldlt}
  The following equivalences hold, for all \FOL formulae $\varphiFrm$ and
  hyperteams $\HypAsg \in \HypAsgSet[\subseteq](\sup{\varphiFrm})$.
  \begin{enumerate}[1)]
    \item\label{lmm:foldlt(ea)}
      Statements~\ref{lmm:foldlt(ea:org)} and~\ref{lmm:foldlt(ea:dlt)} are
      equivalent:
      \begin{enumerate}[a)]
        \item\label{lmm:foldlt(ea:org)}
          there exists a team $\TeamAsg \!\in\! \HypAsg$ such that $\AStr,
          \asgElm \cmodels[\FOL] \varphiFrm$, for all assignments $\asgElm
          \!\in\! \TeamAsg$;
        \item\label{lmm:foldlt(ea:dlt)}
          for all teams $\TeamAsg \!\in\! \dual{\HypAsg}$, there exists an
          assignment $\asgElm \!\in\! \TeamAsg$ such that $\AStr, \asgElm
          \cmodels[\FOL] \varphiFrm$.
      \end{enumerate}
    \item\label{lmm:foldlt(ae)}
      Statements~\ref{lmm:foldlt(ae:org)} and~\ref{lmm:foldlt(ae:dlt)} are
      equivalent:
      \begin{enumerate}[a)]
        \item\label{lmm:foldlt(ae:org)}
          for all teams $\TeamAsg \!\in\! \HypAsg$, there exists an assignment
          $\asgElm \!\in\! \TeamAsg$ such that $\AStr, \asgElm \cmodels[\FOL]
          \varphiFrm$;\!
        \item\label{lmm:foldlt(ae:dlt)}
          there exists a team $\TeamAsg \!\in\! \dual{\HypAsg}$ such that
          $\AStr, \asgElm \cmodels[\FOL] \varphiFrm$, for all assignments
          $\asgElm \!\in\! \TeamAsg$.
      \end{enumerate}
  \end{enumerate}
\end{lemma}

The following lemma states that the extension operator provides an adequate
semantics for classic \FOL quantifications, when applied to all support
variables.
Statement~\ref{lmm:folqnt(ea)} considers Eloise's choices, when the
interpretation of the hyperteam is $\QEA$, while Statement~\ref{lmm:folqnt(ae)}
takes care of Abelard's choices, when the interpretation is the dual $\QAE$.

\begin{lemma}[name = \FOL Quantifiers, restate = lmmfolqnt]
  \label{lmm:folqnt}
  The following equivalences hold, for all \FOL formulae $\varphiFrm$, variables
  $\varElm \!\in\! \VarSet$, and hyperteams $\HypAsg \!\in\!
  \HypAsgSet[\subseteq](\VSet)$ with $\VSet \defeq \sup{\varphiFrm} \setminus \{
  \varElm \}$.
  \begin{enumerate}[1)]
    \item\label{lmm:folqnt(ea)}
      Statements~\ref{lmm:folqnt(ea:org)} and~\ref{lmm:folqnt(ea:ext)} are
      equivalent:
      \begin{enumerate}[a)]
        \item\label{lmm:folqnt(ea:org)}
          there exists a team $\TeamAsg \in \HypAsg$ such that $\AStr, \asgElm
          \cmodels[\FOL] \LExs \varElm \ldotp \varphiFrm$, for all $\asgElm \in
          \TeamAsg$;
        \item\label{lmm:folqnt(ea:ext)}
          there exists a team $\TeamAsg \in \extFun[\VSet]{\HypAsg, \varElm}$
          such that $\AStr, \asgElm \cmodels[\FOL] \varphiFrm$, for all $\asgElm
          \in \TeamAsg$.
      \end{enumerate}
    \item\label{lmm:folqnt(ae)}
      Statements~\ref{lmm:folqnt(ae:org)} and~\ref{lmm:folqnt(ae:ext)} are
      equivalent:
      \begin{enumerate}[a)]
        \item\label{lmm:folqnt(ae:org)}
          for all teams $\TeamAsg \in \HypAsg$, there exists $\asgElm \in
          \TeamAsg$ such that $\AStr, \asgElm \cmodels[\FOL] \LAll \varElm
          \ldotp \varphiFrm$;
        \item\label{lmm:folqnt(ae:ext)}
          for all teams $\TeamAsg \in \extFun[\VSet]{\HypAsg, \varElm}$, there
          exists $\asgElm \in \TeamAsg$ such that $\AStr, \asgElm \cmodels[\FOL]
          \varphiFrm$.
      \end{enumerate}
  \end{enumerate}
\end{lemma}

Finally, the partition operator precisely mimics the semantics of the binary
Boolean connectives when the corrected interpretation of the underlying
hyperteam is considered.

\begin{lemma}[name = \FOL Boolean Connectives, restate = lmmfolblncon]
  \label{lmm:folblncon}
  The following equivalences hold, for all \FOL formulae $\varphiFrm[1]$ and
  $\varphiFrm[2]$ and hyperteams $\HypAsg \!\in\! \HypAsgSet[\subseteq](\VSet)$
  with $\VSet \defeq \sup{\varphiFrm[1]} \cup \sup{\varphiFrm[2]}$.
  \begin{enumerate}[1)]
    \item\label{lmm:folblncon(ea)}
      Statements~\ref{lmm:folblncon(ea:org)} and~\ref{lmm:folblncon(ea:par)} are
      equivalent:
      \begin{enumerate}[a)]
        \item\label{lmm:folblncon(ea:org)}
          there exists a team $\TeamAsg \in \HypAsg$ such that $\AStr, \asgElm
          \cmodels[\FOL] \varphiFrm[1] \wedge \varphiFrm[2]$, for all $\asgElm
          \in \TeamAsg$;
        \item\label{lmm:folblncon(ea:par)}
          for each bipartition $(\HypAsg[1], \HypAsg[2]) \in
          \parFun[]{\HypAsg}$, there exist an index $i \in \{ 1, 2 \}$ and a
          team $\TeamAsg \in \HypAsg[i]$ such that $\AStr, \asgElm
          \cmodels[\FOL] \varphiFrm[i]$, for all $\asgElm \in \TeamAsg$.
      \end{enumerate}
    \item\label{lmm:folblncon(ae)}
      Statements~\ref{lmm:folblncon(ae:org)} and~\ref{lmm:folblncon(ae:par)} are
      equivalent:
      \begin{enumerate}[a)]
        \item\label{lmm:folblncon(ae:org)}
          for all teams $\TeamAsg \in \HypAsg$, there exists $\asgElm \in
          \TeamAsg$ such that $\AStr, \asgElm \cmodels[\FOL] \varphiFrm[1] \vee
          \varphiFrm[2]$;
        \item\label{lmm:folblncon(ae:par)}
          there exists a bipartition $(\HypAsg[1], \HypAsg[2]) \in
          \parFun[]{\HypAsg}$ such that, for all indexes $i \in \{ 1, 2 \}$ and
          teams $\TeamAsg \in \HypAsg[i]$, it holds that $\AStr, \asgElm
          \cmodels[\FOL] \varphiFrm[i]$, for some $\asgElm \in \TeamAsg$.
      \end{enumerate}
  \end{enumerate}
\end{lemma}

We can now state the \FOL adequacy property for \ADIF.

\begin{theorem}[name = \FOL Adequacy, restate = thmfolsemadq]
  \label{thm:folsemadq}
  For all \FOL formulae $\varphiFrm$ and hyperteams $\HypAsg \in
  \HypAsgSet[\subseteq](\sup{\varphiFrm})$, it holds that:
  \begin{enumerate}[1)]
    \item\label{thm:folsemadq(ea)}
      $\AStr, \HypAsg \cmodels[][\QEA] \varphiFrm$ \iff there exists a team
      $\TeamAsg \in \HypAsg$ such that, for all assignments $\asgElm \in
      \TeamAsg$, it holds that $\AStr, \asgElm \cmodels[\FOL] \varphiFrm$;
    \item\label{thm:folsemadq(ae)}
      $\AStr, \HypAsg \cmodels[][\QAE] \varphiFrm$ \iff, for all teams $\TeamAsg
      \in \HypAsg$, there exists an assignment $\asgElm \in \TeamAsg$ such that
      $\AStr, \asgElm \cmodels[\FOL] \varphiFrm$.
  \end{enumerate}
\end{theorem}




\subsection{Dependence/Independence-Friendly Logic}
\label{sec:adq;sub:dif}

\emph{Dependence/Independence-Friendly Logic}~\citep{Vaa07,HS89} can be viewed
as a (syntactic variant of a) fragment of \ADIF, where
\begin{inparaenum}[i)]
  \item
    negation can only occur in front of atoms and
  \item
    just one kind of quantifier can be restricted, depending on a flag $\beta
    \in \{ \forall, \exists\}$.
\end{inparaenum}

\begin{definition}[\DIF Syntax]
  \label{def:syn(dif)}
  The $\exists/\forall$-\emph{Dependence/Independence-Friendly Logic}
  ($\exists/\forall$-\DIF, for short) is the set of formulae built according to
  the following grammar, where $\RRel \in \RName$, $\xVec \in
  \VarSet[][\art{\RRel}]$, $\varElm \in \VarSet$, and $\WSet \subseteq \VarSet$
  with $\card{\WSet} < \omega$:
  \begin{itemize}
  \item[$\exists$-\DIF]
    $\varphiFrm \seteq \RRel(\xVec) \mid \neg \RRel(\xVec) \mid \varphiFrm
    \LCon[][] \varphiFrm \mid \varphiFrm \LDis[][] \varphiFrm \mid
    \LExs[][_{\pm\WSet}] \varElm \ldotp \varphiFrm \mid \LAll[][-\emptyset]
    \varElm \ldotp \varphiFrm$.
  \item[$\forall$-\DIF]
    $\varphiFrm \seteq \RRel(\xVec) \mid \neg \RRel(\xVec) \mid \varphiFrm
    \LCon[][] \varphiFrm \mid \varphiFrm \LDis[][] \varphiFrm \mid
    \LExs[][-\emptyset] \varElm \ldotp \varphiFrm \mid \LAll[][_{\pm\WSet}]
    \varElm \ldotp \varphiFrm$.
  \end{itemize}
\end{definition}

Hodges' semantics of \DIF formulae is defined on teams.
There are two types of semantics rules, one for each flag $\beta \in \{ \LExs,
\LAll \}$, which are dual of one another.
The $\forall$-semantics is the classic one reported in~\cite{Hod97a}, also
denoted as `$+$'-semantics in~\cite{MSS11}, while the $\exists$-semantics
corresponds to the negation of the `$-$'-semantics.
Before recalling the definitions of these two semantics, we need to provide two
additional operators.
For the Boolean connectives, we define a partition operation for teams as
follows: $\parFun{\TeamAsg} \defeq \set{ (\TeamAsg[1], \TeamAsg[2]) \in
\pow{\TeamAsg} \times \pow{\TeamAsg} }{ \TeamAsg[1] \cap \TeamAsg[2] = \emptyset
\land \TeamAsg[1] \cup \TeamAsg[2] = \TeamAsg }$.
The rule for quantifier uses the extension operator $\extFun{}$, when the
quantifier is not coherent with the flag $\beta$.
When the quantifier is coherent, instead, the semantics requires a
\emph{cylindrification} operator on teams.
Intuitively, the cylindrification of a team $\TeamAsg$ \wrt some variable
$\varElm$ extends each of its assignments with every possible value for
$\varElm$.
Formally, $\cylFun{\TeamAsg, \varElm} \defeq \set{ {\asgElm}[\varElm \mapsto
\aElm] }{ \asgElm \in \TeamAsg, \aElm \in \ASet }$.

\begin{definition}[\DIF Semantics]
  \label{def:sem(dif)}
  The \emph{Hodges' semantic relation} $\AStr, \TeamAsg \cmodels[_{\DIF}][\beta]
  \varphiFrm$ for $\dual{\beta}$-\DIF is inductively defined as follows, for
  all $\dual{\beta}$-\DIF formulae $\varphiFrm$ and teams $\TeamAsg \subseteq
  \AsgSet[\subseteq]( \free{\varphiFrm})$, with $\beta, \dual{\beta} \in \{
  \exists, \forall \}$ and $\beta \neq \dual{\beta}$:
  \begin{enumerate}[1)]
  \item\label{def:sem(dif:a)}
    \begin{enumerate}[a)]
    \item\label{def:sem(dif:rel:a)}
      $\AStr, \TeamAsg \cmodels[_{\DIF}][\forall] \RRel(\xVec)$ if, for all
      $\asgElm \in \TeamAsg$, it holds that $\xVec[][\asgElm] \in
      \RRel[][\AStr]$;
    \item\label{def:sem(dif:neg:a)}
      $\AStr, \TeamAsg \cmodels[_{\DIF}][\forall] \neg \RRel(\xVec)$ if, for
      all $\asgElm \in \TeamAsg$, it holds that $\xVec[][\asgElm]
      \not\in \RRel[][\AStr]$;
    \item\label{def:sem(dif:con:a)}
      $\AStr, \TeamAsg \cmodels[_{\DIF}][\forall] \varphiFrm[1] \LCon
      \varphiFrm[2]$ if $\AStr, \TeamAsg \cmodels[_{\DIF}][\forall]
      \varphiFrm[1]$ and $\AStr, \TeamAsg \cmodels[_{\DIF}][\forall]
      \varphiFrm[2]$;
    \item\label{def:sem(dif:dis:a)}
      $\AStr, \TeamAsg \cmodels[_{\DIF}][\forall] \varphiFrm[1] \LDis[][]
      \varphiFrm[2]$ if $\AStr, \TeamAsg[1] \cmodels[_{\DIF}][\forall]
      \varphiFrm[1]$ and $\AStr, \TeamAsg[2] \cmodels[_{\DIF}][\forall]
      \varphiFrm[2]$, for some bipartition $(\TeamAsg[1], \TeamAsg[2]) \in
      \parFun[]{\TeamAsg}$;
    \item\label{def:sem(dif:exs:a)}
      $\AStr, \TeamAsg \cmodels[_{\DIF}][\forall] \LExs[][_{\pm\WSet}] \varElm
      \ldotp \varphiFrm[]$ if $\AStr, \extFun{\TeamAsg, \FFun, \varElm}
      \cmodels[_{\DIF}][\forall] \varphiFrm[]$, for some $\FFun \in
      \FncSet[\denot{\pm\WSet}]{}$;
    \item\label{def:sem(dif:all:a)}
      $\AStr, \TeamAsg \cmodels[_{\DIF}][\forall] \LAll[][-\emptyset] \varElm
      \ldotp \varphiFrm[]$ if $\AStr, \cylFun{\TeamAsg, \varElm}
      \cmodels[_{\DIF}][\forall] \varphiFrm[]$;
    \end{enumerate}
  \item\label{def:sem(dif:e)}
    \begin{enumerate}[a)]
    \item\label{def:sem(dif:rel:e)}
      $\AStr, \TeamAsg \cmodels[_{\DIF}][\exists] \RRel(\xVec)$ if there
      exists $\asgElm \in \TeamAsg$ such that $\xVec[][\asgElm] \in
      \RRel[][\AStr]$;
    \item\label{def:sem(dif:neg:e)}
      $\AStr, \TeamAsg \cmodels[_{\DIF}][\exists] \neg \RRel(\xVec)$ if there
      exists $\asgElm \in \TeamAsg$ such that $\xVec[][\asgElm] \notin
      \RRel[][\AStr]$;
    \item\label{def:sem(dif:con:e)}
      $\AStr, \TeamAsg \cmodels[_{\DIF}][\exists] \varphiFrm[1] \LCon[][]
      \varphiFrm[2]$ if $\AStr, \TeamAsg[1] \cmodels[_{\DIF}][\exists]
      \varphiFrm[1]$ or $\AStr, \TeamAsg[2] \cmodels[_{\DIF}][\exists]
      \varphiFrm[2]$, for all bipartitions $(\TeamAsg[1], \TeamAsg[2]) \in
      \parFun[]{\TeamAsg}$;
    \item\label{def:sem(dif:dis:e)}
      $\AStr, \TeamAsg \cmodels[_{\DIF}][\exists] \varphiFrm[1] \LDis[][]
      \varphiFrm[2]$ if $\AStr, \TeamAsg \cmodels[_{\DIF}][\exists]
      \varphiFrm[1]$ or $\AStr, \TeamAsg \cmodels[_{\DIF}][\exists]
      \varphiFrm[2]$;
    \item\label{def:sem(dif:exs:e)}
      $\AStr, \TeamAsg \cmodels[_{\DIF}][\exists] \LExs[][-\emptyset] \varElm
      \ldotp \varphiFrm[]$ if $\AStr, \cylFun{\TeamAsg, \varElm}
      \cmodels[_{\DIF}][\exists] \varphiFrm[]$;
    \item\label{def:sem(dif:all:e)}
      $\AStr, \TeamAsg \cmodels[_{\DIF}][\exists] \LAll[][_{\pm\WSet}]
      \varElm \ldotp \varphiFrm[]$ if $\AStr, \extFun[]{\TeamAsg, \FFun,
      \varElm} \cmodels[_{\DIF}][\exists] \varphiFrm[]$, for all $\FFun \in
      \FncSet[\denot{\pm\WSet}]{}$.
    \end{enumerate}
  \end{enumerate}
\end{definition}

In order to show that \ADIF is indeed a conservative extension of \DIF, we need
to be able to simulate the semantics on teams with the one on hyperteams.
As a first step, we lift the cylindrification operator to hyperteams in the
obvious way, by defining $\cylFun{\HypAsg, \varElm} \defeq \set{
\cylFun{\TeamAsg, \varElm} }{ \TeamAsg \in \HypAsg }$.
While the semantics of \ADIF does not provide a primitive operator for
cylindrification, this operation is easily simulated by first dualising the
hyperteam, then by applying the extension for $\xElm$ uniformly over all the
variables in the domain of $\HypAsg$, and, finally dualising the result again.
The following lemma establishes the equivalence of these two different
operations.

\begin{lemma}[name = Cylindrical Extension, restate = lmmcylext]
  \label{lmm:cylext}
  Let $\HypAsg \in \HypAsgSet$ be a hyperteam.
  Then, $\cylFun{\HypAsg, \varElm} \eqv \dual{\extFun[\WSet]{\dual{\HypAsg},
  \varElm}}$, for all variables $\varElm \in \VarSet$ and sets of variables
  $\WSet$, with $\var{\HypAsg} \subseteq \WSet \subseteq \VarSet$.
\end{lemma}

A similar problem arises with the team partitioning operator that is not present
in the semantics of \ADIF.
Once again, the dualisation operator, together with the hyperteam partitioning
operator, allows for simulating it.
More specifically, we first apply the dualisation to the hyperteam $\HypAsg$,
then the partitioning to obtain $(\HypAsg[1], \HypAsg[2])\in
\parFun[]{\dual{\HypAsg}}$, and finally dualise again the two resulting
hyperteam and obtain $\dual{\HypAsg[1]}$ and $\dual{\HypAsg[2]}$, each of which
happens to contain teams that would result from the team partitioning operator
applied to teams in $\HypAsg$.

\begin{lemma}[name = Team Partitioning, restate = lmmteapar]
  \label{lmm:teapar}
  Let $\HypAsg \in \HypAsgSet$ be a hyperteam.
  Then:
  \begin{enumerate}[1)]
  \item\label{lmm:teapar(hyp)}
    for all hyperteam bipartitions $(\HypAsg[1], \HypAsg[2]) \in
    \parFun[]{\dual{\HypAsg}}$ and teams $\YSet[1] \in \dual{\HypAsg[1]}$ and
    $\YSet[2] \in \dual{\HypAsg[2]}$, there exists a team $\TeamAsg \in
    \HypAsg$ such that $\TeamAsg \subseteq \YSet[1] \cup \YSet[2]$;
  \item\label{lmm:teapar(tea)}
    for all teams $\TeamAsg \in \HypAsg$ and team bipartitions $(\TeamAsg[1],
    \TeamAsg[2]) \in \parFun[]{\TeamAsg}$, there exist a hyperteam bipartition
    $(\HypAsg[1], \HypAsg[2]) \in \parFun[]{\dual{\HypAsg}}$ and two teams
    $\YSet[1] \in \dual{\HypAsg[1]}$ and $\YSet[2] \in \dual{\HypAsg[2]}$ such
    that $\YSet[1] \subseteq \TeamAsg[1]$ and $\YSet[2] \subseteq \TeamAsg[2]$.
  \end{enumerate}
\end{lemma}

Based on these two lemmata, one can prove the following theorem, which
establishes the required adequacy result.

\begin{theorem}[name = \DIF Adequacy, restate = thmdifsemadq]
  \label{thm:difsemadq}
  For all $\exists/\forall$-\DIF formulae
  $\varphiFrm[][_{\exists}]/\varphiFrm[][_{\forall}]$ and hyperteams $\HypAsg
  \in \HypAsgSet[\subseteq](\sup{}(
  \varphiFrm[][_{\exists}]/\varphiFrm[][_{\forall}]\, ))$, it holds that:
  \begin{enumerate}[1)]
    \item\label{thm:difsemadq(ea)}
      $\AStr, \HypAsg \cmodels[][\QEA]\! \varphiFrm[][_{\exists}]$ \iff there
      exists a team $\TeamAsg \in \HypAsg$ such that $\AStr, \TeamAsg
      \cmodels[_{\DIF}][\forall] \varphiFrm[][_{\exists}]$;
    \item\label{thm:difsemadq(ae)}
      $\AStr, \HypAsg \cmodels[][\QAE]\! \varphiFrm[][_{\forall}]$ \iff, for all
      teams $\TeamAsg \in \HypAsg$, it holds that $\AStr, \TeamAsg
      \cmodels[_{\DIF}][\exists] \varphiFrm[][_{\forall}]$.
  \end{enumerate}
\end{theorem}

\begin{example}
  \label{exm:nondet}
  In Example~\ref{exm:snt}, it has been observed that the two \ADIF sentences
  $\varphiSnt[3] = \LAll \varElm \ldotp \LExs[][+\emptyset] \yvarElm \ldotp
  (\varElm = \yvarElm)$ and $\varphiSnt[4] = \LExs \varElm \ldotp
  \LAll[][+\emptyset] \yvarElm \ldotp \LNeg (\varElm = \yvarElm)$ evaluate
  to false and true, respectively, in the binary structure $\AStr = \tuple {\{
  0, 1 \}} {=^{\AStr}}$ against the trivial hyperteam $\TrvHypAsg$.
  We also claimed that they are the semantic negation of each other, something
  that now can be easily proved thanks to Corollary~\ref{cor:lawexcmid} and
  Theorem~\ref{thm:blnlawi}.
  Notice now that all these properties hold true for the two $\exists$-\DIF and
  $\forall$-\DIF sentences $\varphiSnt[3]' = \LAll[][-\emptyset] \varElm \ldotp
  \LExs[][+\emptyset] \yvarElm \ldotp (\varElm = \yvarElm)$ and $\varphiSnt[4]'
  = \LExs[][-\emptyset] \varElm \ldotp \LAll[][+\emptyset] \yvarElm \ldotp
  \LNeg (\varElm = \yvarElm)$.
  At this point, we can show that the truth and falsity of $\varphiSnt[3]'$ and
  $\varphiSnt[4]'$ convey different meanings when evaluated in \IF
  (equivalently, \DIF).
  First, recall that,~\cite{Hod97a} (see also~\cite{MSS11}) defines an \IF
  sentence $\varphiSnt$ to be \emph{true} in $\AStr$, if $\AStr, \TrvTeamAsg
  \cmodels[][+] \varphiSnt$, and \emph{false} on $\AStr$, if $\AStr, \TrvTeamAsg
  \cmodels[][-] \varphiSnt$.
  As observed above, this means that $\varphiSnt$ is \emph{true} in $\AStr$, if
  $\AStr, \TrvTeamAsg \cmodels[_{\DIF}][\forall] \varphiSnt[\exists]$, and
  \emph{false} in $\AStr$, if $\AStr, \TrvTeamAsg \notcmodels[_{\DIF}][\exists]
  \varphiSnt[\forall]$, where $\varphiSnt[\exists]$ and $\varphiSnt[\forall]$
  are the $\exists$-\DIF and $\forall$-\DIF sentences obtained from
  $\varphiSnt$ by substituting $-\emptyset$ for the restrictions of the
  universal and existential quantifiers, respectively.
  Now, both $\varphiSnt[3]'$ and $\varphiSnt[4]'$ are \IF sentences.
  Moreover, as stated above, $\varphiSnt[3]'$ is an $\exists$-\DIF sentence,
  while $\varphiSnt[4]'$ is a $\forall$-\DIF sentence.
  Thus, due to Theorem~\ref{thm:difsemadq}, we immediately obtain $\AStr,
  \TrvTeamAsg \notcmodels[_{\DIF}][\forall] \varphiSnt[3]'$ and $\AStr,
  \TrvTeamAsg \cmodels[_{\DIF}][\exists] \varphiSnt[4]'$, since $\AStr,
  \TrvHypAsg \notcmodels \varphiSnt[3]'$ and $\AStr,
  \TrvHypAsg \cmodels \varphiSnt[4]'$.
  As a consequence, when evaluated in \IF, $\varphiSnt[3]'$ is not true and
  $\varphiSnt[4]'$ is not false.
  However, $\varphiSnt[3]'$ is not false and $\varphiSnt[4]'$ is not true
  either, which implies that the two sentences are undetermined.
  Indeed, $\AStr, \TrvTeamAsg \cmodels[_{\DIF}][\exists] \varphiSnt[3\forall]'$
  and $\AStr, \TrvTeamAsg \notcmodels[_{\DIF}][\forall] \varphiSnt[4\exists]'$,
  where $\varphiSnt[3\forall]' = \LAll[][-\emptyset] \varElm \ldotp
  \LExs[][-\emptyset] \yvarElm \ldotp (\varElm = \yvarElm)$ and
  $\varphiSnt[4\exists]' = \LExs[][-\emptyset] \varElm \ldotp
  \LAll[][-\emptyset] \yvarElm \ldotp \LNeg (\varElm = \yvarElm)$, since
  $\AStr, \TrvHypAsg \cmodels \varphiSnt[3\forall]'$ and $\AStr, \TrvHypAsg
  \notcmodels \varphiSnt[4\exists]'$.
\end{example}







\section{Meta Theory}
\label{sec:metthr}

We now introduce a \emph{meta-level interpretation} of the quantifiers by means
of a Herbrand-Skolem semantics extending the compositional one based on
hyperteams, which results to be essential for
\begin{inparaenum}[1)]
  \item
    the solution of the \emph{model-checking problem},
  \item
    the proof that \ADIF covers the entire \emph{polynomial hierarchy}, by means
    of an encoding of \emph{Second-Order Logic} (\SOL, for
    short)~\citep{HA38,Chu56,Sha91} and \emph{Team Logic} (\TL, for
    short)~\citep{Vaa07}, and
  \item
    the adequacy of the \emph{game-theoretic semantics} presented in
    Section~\ref{sec:gamthrsem}.
\end{inparaenum}


\subsection{Meta Extension}
\label{sec:metthr;sub:metext}

The game-theoretic interpretation of the quantifiers
$\LExs[][_{\pm\WSet}]\varElm$ and $\LAll[][_{\pm\WSet}]\varElm$ implicitly
identifies strategies for Eloise and Abelard satisfying the
$\denot{\pm\WSet}$-uniformity constraint.
The \emph{meta extension} of \ADIF we propose here makes these strategies
explicit, by augmenting the logic with the two quantifiers,
$\LEExs[][_{\pm\WSet}] \varElm$ and $\LAAll[][_{\pm\WSet}] \varElm$, ranging
over $\denot{\pm\WSet}$-uniform Herbrand/Skolem functions~\citep{Bus98a}.
Intuitively, $\LEExs[][_{\pm\WSet}] \varElm \ldotp \varphiFrm$ ensures the
existence of a $\denot{\pm\WSet}$-uniform Skolem function assigning values to
$\varElm$ that satisfies $\varphiFrm$, while $\LAAll[][_{\pm\WSet}] \varElm
\ldotp \varphiFrm$ verifies $\varphiFrm$, for all values assigned to $\varElm$
by some $\denot{\pm\WSet}$-uniform Herbrand function.

\begin{definition}[\Meta Syntax]
  \label{def:syn(met)}
  \!\!The \emph{\ADIF Meta Extension} (\Meta, for short) is the set of formulae
  built according to Definition~\ref{def:syn(adif)} extended as follows, where
  $\varElm \in \VarSet$ and $\WSet \subseteq \VarSet$ with $\card{\WSet} <
  \omega$:
  \[
    \varphiFrm \seteq \ADIF \mid \LEExs[][_{\pm\WSet}] \varElm \ldotp \varphiFrm
    \mid \LAAll[][_{\pm\WSet}] \varElm \ldotp \varphiFrm.
  \]
\end{definition}

The set of \emph{support variables} $\sup{\varphiFrm}$ of a \Meta formula
$\varphiFrm$ is defined as in ADIF, with the additional two simple cases
$\sup{\Qnt[][_{\pm\WSet}] \varElm \ldotp \varphiFrm} \defeq \sup{\varphiFrm}
\setminus \{ \varElm \}$, for $\Qnt \in \{ \LEExs, \LAAll \}$.
The definition of free variables is, instead, quite more intricate and requires
the introduction of the following supplemental functions of \emph{free variables
under (meta) dependency context} $\free{} \colon \Meta \times (\VarSet \pto
\pow{\VarSet}) \to \pow{\VarSet}$ and \emph{dependence variables under (meta)
dependency context} $\dep{} \colon \Meta \times (\VarSet \pto \pow{\VarSet}) \to
\pow{\VarSet}$, where by \emph{dependency context} we mean any partial function
$\iotaFun \in \VarSet \pto \pow{\VarSet}$.
The \emph{transitive closure} of $\iotaFun$ is a dependency context
$\iotaFun[][*] \in \dom{\iotaFun} \to \pow{\VarSet}$ such that, for each
variable $\varElm \in \dom{\iotaFun}$ in its domain, $\iotaFun[][*](\varElm)$ is
the smallest set of variables such that
\begin{inparaenum}[(a)]
\item
  $\iotaFun(\varElm) \subseteq \iotaFun[][*](\varElm)$ and
\item
  $\iotaFun(\yvarElm) \subseteq \iotaFun[][*](\varElm)$, for all variables
  $\yvarElm \in \iotaFun[][*](\varElm) \cap \dom{\iotaFun}$.
\end{inparaenum}
Finally, $\iotaFun$ is \emph{acyclic} if $\varElm \not\in
\iotaFun[][*](\varElm)$, for all variables $\varElm \in \dom{\iotaFun}$.
The functions $\free{}$ and $\dep{}$ can be defined in a mutual recursive
fashion as follows.
\begin{itemize}[$\bullet$]
\item
  $\free{\bot, \iotaFun}, \free{\top, \iotaFun} \defeq \emptyset$;
\item
  $\free{\RRel(\xVec), \iotaFun} \defeq \xVec \cup \bigcup \set{
  \iotaFun[][*](\varElm) }{ \varElm \in \xVec \cap \dom{\iota} }$;
\item
  $\free{\LNeg \varphiFrm, \iotaFun} \defeq \free{\varphiFrm, \iotaFun}$;
\item
  $\free{\varphiFrm[1] \Cnt[][]\, \varphiFrm[2], \iotaFun} \defeq
  \free{\varphiFrm[1], \iotaFun} \cup \free{\varphiFrm[2], \iotaFun}$, for $\Cnt
  \in \{ \LCon, \LDis \}$;
\item
  $\free{\Qnt[][_{\pm\WSet}] \varElm \ldotp \varphiFrm, \iotaFun} \defeq
  (\free{\varphiFrm, \iotaFun'} \setminus \{ \varElm \}) \cup \denot{\pm\WSet}$,
  if $\varElm \in \free{\varphiFrm, \iotaFun'}$, and  $\free{\Qnt[][_{\pm\WSet}]
  \varElm \ldotp \varphiFrm, \iotaFun} \allowbreak  \defeq \free{\varphiFrm,
  \iotaFun'}$, otherwise, where $\iotaFun' \defeq  \iotaFun \setminus \{
  \varElm \}$, for $\Qnt \in \{ \LExs, \LAll \}$;
\item
  $\free{\Qnt[][_{\pm\WSet}] \varElm \ldotp \varphiFrm, \iotaFun} \defeq
  \free{\varphiFrm, \iotaFun'}$, if $\varElm \in \dep{\varphiFrm, \iotaFun'}$,
  and $\free{\Qnt[][_{\pm\WSet}] \varElm \ldotp \varphiFrm, \iotaFun} \defeq
  \free{\varphiFrm, \iotaFun'} \setminus \{ \varElm \}$, otherwise, where
  $\iotaFun' \defeq {\iotaFun}[\varElm \mapsto \denot{\pm\WSet}]$, for $\Qnt \in
  \{ \LEExs, \LAAll \}$.
\end{itemize}
Intuitively, a variable $\yvarElm$ can be free in a \Meta formula $\varphiFrm$
under a dependency context $\iotaFun$ for one (or more) of the following three
reasons (always outside the scope of a quantifier for $\yvarElm$ that masks it):
\begin{inparaenum}[(i)]
  \item
    it is explicitly used in some relational symbol;
  \item
    it occurs in the (transitive) dependency set $\iotaFun[][*](\varElm)$ of
    some meta quantified variable $\varElm$ used in a relational symbol;
  \item
    it appears in the dependence/independence constraint set $\denot{\pm\WSet}$
    of some first-order quantifier $\Qnt[][_{\pm\WSet}] \varElm$ of a free
    variable $\varElm$.
\end{inparaenum}
Notice that a meta quantifier of a variable $\varElm$ masks such a variable
only if it does not appear in the set of (first-order) dependence variables of
its matrix.
\begin{itemize}[$\bullet$]
\item
  $\dep{\bot, \iotaFun}, \dep{\top, \iotaFun}, \dep{\RRel(\xVec), \iotaFun}
  \defeq \emptyset$;
\item
  $\dep{\LNeg \varphiFrm, \iotaFun} \defeq \dep{\varphiFrm, \iotaFun}$;
\item
  $\dep{\varphiFrm[1] \Cnt[][]\, \varphiFrm[2], \iotaFun} \defeq
  \dep{\varphiFrm[1], \iotaFun} \cup \dep{\varphiFrm[2], \iotaFun}$, for $\Cnt
  \in \{ \LCon, \LDis \}$;
\item
  $\dep{\Qnt[][_{\pm\WSet}] \varElm \ldotp \varphiFrm, \iotaFun} \defeq
  (\dep{\varphiFrm, \iotaFun'} \setminus \{ \varElm \}) \cup \denot{\pm\WSet}$,
  if $\varElm \in \free{\varphiFrm, \iotaFun'}$, and $\dep{\Qnt[][_{\pm\WSet}]
  \varElm \ldotp \varphiFrm, \iotaFun} \allowbreak \defeq \dep{\varphiFrm,
  \iotaFun'}$, otherwise, where $\iotaFun' \defeq  \iotaFun \setminus \{
  \varElm \}$, for $\Qnt \in \{ \LExs, \LAll \}$;
\item
  $\dep{\Qnt[][_{\pm\WSet}] \varElm \ldotp \varphiFrm, \iotaFun} \defeq
  \dep{\varphiFrm, \iotaFun'}$, where $\iotaFun' \defeq {\iotaFun}[\varElm
  \mapsto \denot{\pm\WSet}]$, for $\Qnt \in \{ \LEExs, \LAAll \}$.
\end{itemize}
Intuitively, a variable $\yvarElm$ belongs to the set $\dep{\varphiFrm,
\iotaFun}$ if it appears in the dependence/independence constraint set
$\denot{\pm\WSet}$ of some first-order quantifier $\Qnt[][_{\pm\WSet}] \varElm$
of a free variable $\varElm$ and, at the same time, is not removed, \ie, is not
under the scope of another first-order quantifier for $\yvarElm$ itself.
Notice that the dependencies of the variables quantified by a meta-quantifier,
which are maintained by the dependency context $\iotaFun$, are not taken into
account here, as they are only used to determine which variables are free.
At this point, the set of \emph{free variables} $\free{\varphiFrm}$ of a \Meta
formula $\varphiFrm$ is defined as $\free{\varphiFrm, \emptyfun}$.

To keep track of the Herbrand/Skolem functions already quantified, we use a
\emph{function assignment} $\FunAsg \in \FunAsgSet \defeq \VarSet \pto
\FncSet{}$ mapping each variable $\varElm \in \VSet \defeq \dom{\FunAsg}$ to a
function $\FunAsg(\varElm) \in \FncSet{}$.
To extend a hyperteam $\HypAsg \in \HypAsgSet(\USet)$ with $\FunAsg$, we make
use of the \emph{extension operator} $\extFun{\HypAsg, \FunAsg} \defeq \set{
\extFun{\TeamAsg, \FunAsg} }{ \TeamAsg \in \HypAsg }$, where
\begin{inparaenum}[(i)]
\item
  $\extFun{\TeamAsg, \FunAsg} \defeq \set{ \asgElm \in \cylFun{\TeamAsg, \VSet}
  }{ \forall \varElm \in \VSet \setminus \USet \ldotp \asgElm(\varElm) =
    \FunAsg(\varElm)(\asgElm) }$ is the extension of the team $\TeamAsg$ over
  the variables in $\VSet$, so that the value $\asgElm(\varElm)$ given by an
  assignment $\asgElm$ to each (not yet assigned) variable $\varElm \in \VSet
  \setminus \USet$ is coherent with the one prescribed by $\FunAsg(\varElm)$ and
\item
  $\cylFun{\TeamAsg, \VSet} \defeq \set{ \asgElm \in \AsgSet(\USet \cup \VSet)
  }{ \asgElm \rst[\USet] \in \TeamAsg }$ is the cylindrification of a team
  $\TeamAsg \in \TeamAsgSet(\USet)$ \wrt the set of variables $\VSet \setminus
  \USet$.
\end{inparaenum}
Finally, a function assignment $\FunAsg \in \FunAsgSet$ is \emph{acyclic} if,
for all variables $\varElm \in \dom{\FunAsg}$, it holds that $\FunAsg(\varElm)
\in \FncSet[\iotaFun(\varElm)]{}$, for some acyclic dependency context $\iotaFun
\in \VarSet \pto \pow{\VarSet}$, with $\dom{\FunAsg} \subseteq \dom{\iotaFun}$.

\begin{definition}[\Meta Semantics]
  \label{def:sem(met)}
  The \emph{Hodges' alternating semantic relation} $\AStr, \FunAsg, \HypAsg
  \cmodels[][\alpha] \varphiFrm$ for \Meta is inductively defined as follows,
  for all \Meta formulae $\varphiFrm$, function assignments $\FunAsg \in
  \FunAsgSet$, hyperteams $\HypAsg \in \HypAsgSet[\subseteq](\sup{\varphiFrm}
  \setminus \dom{\FunAsg})$, and alternation flags $\alpha \in \{ \QEA, \QAE
  \}$:
  \begin{enumerate}[1)]
  \item[\ref*{def:sem(adif:fbv)},\ref*{def:sem(adif:tbv)},%
    \ref*{def:sem(adif:neg)}-\ref*{def:sem(adif:all)})]\label{def:sem(met:adif)}
    All \ADIF cases, but those ones of the atomic relations, are defined by
    lifting, in the obvious way, the corresponding items of
    Definition~\ref{def:sem(adif)} to function assignments, \ie, the latter play
    no role;
  \setcounter{enumi}{2}
  \item\label{def:sem(met:rel)}
    \begin{enumerate}[a)]
    \item\label{def:sem(met:rel:ea)}
      $\AStr, \FunAsg, \HypAsg \cmodels[][\QEA] \RRel(\xVec)$ if there exists a
      team $\TeamAsg \in \extFun{\HypAsg, \FunAsg}$ such that, for all
      assignments $\asgElm \in \TeamAsg$, it holds that $\xVec[][\asgElm] \in
      \RRel[][\AStr]$;
    \item\label{def:sem(met:rel:ae)}
      $\AStr, \FunAsg, \HypAsg \cmodels[][\QAE] \RRel(\xVec)$ if, for all teams
      $\TeamAsg \in \extFun{\HypAsg, \FunAsg}$, there exists an assignment
      $\asgElm \in \TeamAsg$ such that $\xVec[][\asgElm] \in \RRel[][\AStr]$;
    \end{enumerate}
  \setcounter{enumi}{8}
  \item\label{def:sem(met:eexs)}
    $\AStr, \FunAsg, \HypAsg \cmodels[][\alpha] \LEExs[][_{\pm\WSet}] \varElm
    \ldotp \phiFrm$ if $\AStr, {\FunAsg}[\varElm \mapsto \FFun], \HypAsg
    \cmodels[][\alpha] \phiFrm$, for some function $\FFun \in
    \FncSet[\denot{\pm\WSet}]{}$;
  \item\label{def:sem(met:aall)}
    $\AStr, \FunAsg, \HypAsg \cmodels[][\alpha] \LAAll[][_{\pm\WSet}] \varElm
    \ldotp \phiFrm$ if $\AStr, {\FunAsg}[\varElm \mapsto \FFun], \HypAsg
    \cmodels[][\alpha] \phiFrm$, for all functions $\FFun \in
    \FncSet[\denot{\pm\WSet}]{}$.
  \end{enumerate}
\end{definition}

Essentially, to evaluate an atomic formula $\RRel(\xVec)$, we extend $\HypAsg$
with the functions dictated by $\FunAsg$ and then we check the assignments
following the alternation given by the flag $\alpha \in \{\QAE,\QEA\}$ as in
plain \ADIF.
Indeed, Item~\ref{def:sem(met:rel)} above can be re-stated in the following
equivalent form, which allows for a unified treatment of the alternation flags:
\[
  \AStr, \FunAsg, \HypAsg \cmodels[][\alpha] \RRel(\xVec),
\text{\ if }
  \AStr, \extFun{\HypAsg, \FunAsg} \cmodels[][\alpha] \RRel(\xVec),
\]
where the second occurrence of the satisfaction relation $\cmodels[][\alpha]$
refers to the Hodges' alternating semantic relation for \ADIF, as per
Item~\ref{def:sem(adif:rel)} of Definition~\ref{def:sem(adif)}.
The semantics of the \emph{meta quantifiers} $\LEExs[][_{\pm\WSet}] \varElm$ and
$\LAAll[][_{\pm\WSet}] \varElm$ is the classic second-order one, where the
functions chosen at the meta level are stored in the assignment $\FunAsg$.

The notions of satisfaction, implication, and equivalence, given at the end of
Section~\ref{sec:adif;sub:sem} immediately lift to \Meta.
In addition, all relevant results proved for \ADIF in
Section~\ref{sec:adif;sub:fun} clearly lift to the \Meta semantics of \ADIF
formulae.
These results are, indeed, proved in this generalised form in~\ref{app:metthr}.
In particular, satisfaction in \ADIF and in \Meta coincide.

\begin{proposition}[name =, restate = prpadifmeta]
  \label{prp:adifmeta}
  $\AStr, \HypAsg \cmodels[][\alpha] \varphiFrm$ \iff $\AStr, \emptyfun, \HypAsg
  \cmodels[][\alpha] \varphiFrm$, for every \ADIF formula $\varphiFrm$ and
  hyperteam $\HypAsg \in \HypAsgSet[\subseteq](\sup{\varphiFrm})$.
\end{proposition}

At first glance, the semantic rule for the meta quantifiers might seem to mimic
the corresponding quantifier rule of \DIF and \TL, as in both cases a choice of
a Skolem/Herbrand function is involved.
However, unlike in \DIF and \TL, the application of the functions to the
hyperteam is delayed until the evaluation of an atomic formula.
This makes the behaviour of quantifications in the two semantics diverge
significantly.
Such a difference is also mirrored in the more complex definition of free
variables given above.

The following lemma characterises the connection between the compositional
semantics of first-order quantifications $\LExs[][_{\pm\WSet}] \varElm$ and
$\LAll[][_{\pm\WSet}] \varElm$ and the corresponding choice of a Skolem/Herbrand
function.

\begin{lemma}[name = Extension Interpretation, restate = lmmextint]
  \label{lmm:extint}
  The following four equivalences hold true, for all hyperteams $\HypAsg \in
  \HypAsgSet(\VSet)$ over $\VSet \subseteq \VarSet$, properties $\PsiSet
  \subseteq \AsgSet(\VSet \cup \{ \varElm \})$ over $\VSet \cup \{ \varElm \}$
  with $\varElm \in \VarSet \setminus \VSet$, sets of variables $\WSet
  \subseteq \VarSet$, and quantifier symbols $\Qnt \in \{ \LExs, \LAll \}$.
  \begin{enumerate}[1)]
    \item\label{lmm:extint(inccoh)}
      Statements~\ref{lmm:extint(inccoh:org)} and~\ref{lmm:extint(inccoh:int)}
      are equivalent, whenever $\Qnt$ is $\alpha$-coherent:
      \begin{enumerate}[a)]
        \item\label{lmm:extint(inccoh:org)}
          there exists $\TeamAsg' \in \extFun[\alpha]{\HypAsg,
          \Qnt[][_{\pm\WSet}] \varElm}$ such that $\TeamAsg' \subseteq \PsiSet$;
        \item\label{lmm:extint(inccoh:int)}
          there exist $\FFun \in \FncSet[\denot{\pm\WSet}]{}$ and $\TeamAsg \in \HypAsg$
          such that $\extFun{\TeamAsg, \FFun, \varElm} \subseteq \PsiSet$.
      \end{enumerate}
    \item\label{lmm:extint(intcoh)}
      Statements~\ref{lmm:extint(intcoh:org)} and~\ref{lmm:extint(intcoh:int)}
      are equivalent, whenever $\Qnt$ is $\alpha$-coherent:
      \begin{enumerate}[a)]
        \item\label{lmm:extint(intcoh:org)}
          for all $\TeamAsg' \in \extFun[\alpha]{\HypAsg, \Qnt[][_{\pm\WSet}]
          \varElm}$, it holds that $\TeamAsg' \cap \PsiSet \neq \EmpTeamAsg$;
        \item\label{lmm:extint(intcoh:int)}
          for all $\FFun \in \FncSet[\denot{\pm\WSet}]{}$ and $\TeamAsg \in \HypAsg$, it
          holds that $\extFun{\TeamAsg, \FFun, \varElm} \cap \PsiSet \neq
          \EmpTeamAsg$.
      \end{enumerate}
    \item\label{lmm:extint(incnch)}
      Statements~\ref{lmm:extint(incnch:org)} and~\ref{lmm:extint(incnch:int)}
      are equivalent, whenever $\Qnt$ is $\dual{\alpha}$-coherent:
      \begin{enumerate}[a)]
        \item\label{lmm:extint(incnch:org)}
          there exists $\TeamAsg' \in \extFun[\alpha]{\HypAsg,
          \Qnt[][_{\pm\WSet}] \varElm}$ such that $\TeamAsg' \subseteq \PsiSet$;
        \item\label{lmm:extint(incnch:int)}
          for all $\FFun \in \FncSet[\denot{\pm\WSet}]{}$, it holds that
          $\extFun{\TeamAsg, \FFun, \varElm} \subseteq \PsiSet$, for some
          $\TeamAsg \in \HypAsg$.
      \end{enumerate}
    \item\label{lmm:extint(intnch)}
      Statements~\ref{lmm:extint(intnch:org)} and~\ref{lmm:extint(intnch:int)}
      are equivalent, whenever $\Qnt$ is $\dual{\alpha}$-coherent:
      \begin{enumerate}[a)]
        \item\label{lmm:extint(intnch:org)}
          for all $\TeamAsg' \in \extFun[\alpha]{\HypAsg, \Qnt[][_{\pm\WSet}]
          \varElm}$, it holds that $\TeamAsg' \cap \PsiSet \neq \EmpTeamAsg$;
        \item\label{lmm:extint(intnch:int)}
          there is $\FFun \in \FncSet[\denot{\pm\WSet}]{}$ such that $\extFun{\TeamAsg,
          \FFun, \varElm} \cap \PsiSet \neq \EmpTeamAsg$, for all $\TeamAsg \in
          \HypAsg$.
      \end{enumerate}
  \end{enumerate}
\end{lemma}

Equivalences~\ref{lmm:extint(inccoh)} and~\ref{lmm:extint(intnch)}, when $\Qnt =
\LExs$, implicitly state that an existential quantification can always be
simulated by an existential choice of a suitable Skolem function, independently
of the alternation flag $\alpha$ for the hyperteam.
Dually, Equivalences~\ref{lmm:extint(intcoh)} and~\ref{lmm:extint(incnch)}, when
$\Qnt = \LAll$, state that a universal quantification can be simulated by a
universal choice of a suitable Herbrand function, again regardless of  $\alpha$.
These observations can be formulated in \Meta as follows.

\begin{theorem}[name = Quantifier Interpretation, restate = thmqntint]
  \label{thm:qntint}
  The following equivalences hold true, for all \FOL formulae $\phiFrm$,
  variables $\varElm \in \VarSet$, sets of variables $\WSet \subseteq \VarSet$
  with $\varElm \not\in \denot{\pm\WSet}$, acyclic function assignments
  $\FunAsg \in \FunAsgSet$ with $\dom{\FunAsg} \cap \denot{\pm\WSet} =
  \emptyset$, and hyperteams $\HypAsg \in \HypAsgSet[\subseteq]((\sup{\phiFrm}
  \setminus \{ \varElm \}) \setminus \dom{\FunAsg})$ with $\varElm \notin
  \var{\HypAsg}$:
  \begin{enumerate}[1)]
    \item\label{thm:qntint(exs)}
      $\AStr, \FunAsg, \HypAsg \cmodels[][\alpha] \LExs[][_{\pm\WSet}] \varElm
      \ldotp \phiFrm$ \iff $\AStr, \FunAsg, \HypAsg \cmodels[][\alpha]
      \LEExs[][_{\pm\WSet}] \varElm \ldotp \phiFrm$;
    \item\label{thm:qntint(all)}
      $\AStr, \FunAsg, \HypAsg \cmodels[][\alpha] \LAll[][_{\pm\WSet}] \varElm
      \ldotp \phiFrm$ \iff $\AStr, \FunAsg, \HypAsg \cmodels[][\alpha]
      \LAAll[][_{\pm\WSet}] \varElm \ldotp \phiFrm$.
  \end{enumerate}
\end{theorem}

Given an \ADIF formula $\qntElm \phiFrm$ with quantifier prefix $\qntElm \in
\QntSet$ and \FOL matrix $\phiFrm$, we can convert each quantification in
$\qntElm$, from inside out, into the corresponding meta quantification, by
suitably iterating the result reported above.
The meta quantifiers in the obtained prefix are in reverse order with respect to
the order of corresponding standard quantifiers in the original prefix.
To formalise this idea, we introduce the \emph{Herbrand-Skolem prefix} function
$\hspFun{}$ as follows:
\begin{enumerate}[a)]
  \item
    $\hspFun{\varepsilon} \defeq \varepsilon$;
  \item
    $\hspFun{\qntElm \ldotp \LExs[][_{\pm\WSet}] \varElm} \defeq
    \LEExs[][_{\pm\WSet}] \varElm \ldotp \hspFun{\qntElm}$;
  \item
    $\hspFun{\qntElm \ldotp \LAll[][_{\pm\WSet}] \varElm} \defeq
    \LAAll[][_{\pm\WSet}] \varElm \ldotp \hspFun{\qntElm}$.
\end{enumerate}
We can show that $\qntElm \phiFrm \equiv \hspFun{\qntElm} \phiFrm$, by
exploiting Theorem~\ref{thm:prfext}.
This conversion resembles a merging of the standard Skolem/Herbrand-isation
procedures~\citep{Hei67,Bus98a} that convert a \FOL sentence either into an
equi-satisfiable/equi-valid \FOL sentence without existential/universal
quantifiers, or into an equivalent \SOL formula.
Note that the herbrandisation approach here is connected with the notion of
Kreisel counterexample~\citep{Kre51,Kre52} applied to \DIF~\citep{MSS11}.

\begin{theorem}[name = Herbrand-Skolem Theorem, restate = thmhst]
  \label{thm:hst}
  Let $\qntElm \phiFrm$ be an \ADIF formula in \pnf with quantifier prefix
  $\qntElm \!\in\! \QntSet$ and \FOL matrix $\phiFrm$.
  Then, $\AStr, \FunAsg, \HypAsg \cmodels[][\alpha] \qntElm \phiFrm$ \iff
  $\AStr, \FunAsg, \HypAsg \cmodels[][\alpha] \hspFun{\qntElm} \phiFrm$, for
  all acyclic function assignments $\FunAsg \in \FunAsgSet$ with $\dom{\FunAsg}
  \cap \dep{\qntElm} = \emptyset$ and hyperteams $\HypAsg \in
  \HypAsgSet[\subseteq](\sup{\qntElm \phiFrm} \setminus \dom{\FunAsg})$ with
  $\var{\HypAsg} \cap \var{\qntElm} = \emptyset$.
\end{theorem}

\begin{example}
  \label{exm:runexmmet}
  Let us consider again the sentence from Example~\ref{exm:runexm}, \ie,
  $\varphiSnt[7] = \LExs \varElm \ldotp \LAll[][+\emptyset] \yvarElm \ldotp
  \LExs[][+\varElm] \zvarElm \ldotp (\xvarElm = \yvarElm) \LCon (\yvarElm =
  \zvarElm)$.
  We already saw that the sentence is true in the original binary structure
  $\AStr$ of the same example.
  If we convert $\varphiSnt[7]$ into \Meta via the function $\hspFun{}$, we
  obtain $\LEExs[][+\varElm] \zvarElm \ldotp \LAAll[][+\emptyset] \yvarElm
  \ldotp \LEExs[][+\emptyset] \varElm \ldotp (\xvarElm = \yvarElm) \LCon
  (\yvarElm = \zvarElm)$.
  To show this sentence true in $\AStr$, it suffices to assign to $\zvarElm$ the
  identity function that copies the value assigned to $\varElm$.
  Then, whatever value is chosen for $\yvarElm$, the same value can be assigned
  to $\varElm$.
  By the semantics of \Meta, the result immediately follows.
\end{example}

Thanks to this Herbrand/Skolem-isation procedure, we can transform a \ADIF
sentence in \pnf into a \Meta sentence in \pnf, where only the meta-quantifiers
$\LEExs[][_{\pm\WSet}] \varElm$ and $\LAll[][_{\pm\WSet}] \varElm$ occur.
Since one needs only polynomial space in the size of the underlying structure to
represent the quantified functions, the same approach used for \FOL model
checking is applicable here too.

\begin{theorem}[name = Model-Checking Problem, restate = thmmcprb]
  \label{thm:mcprb}
  Let $\AStr$ be a finite structure and $\varphiSnt$ an \ADIF sentence in \pnf.
  Then, the model-checking problem $\AStr \models \varphiSnt$ can be decided in
  \PSpace \wrt $\card{\AStr}$.
\end{theorem}

As is the case of \ADIF, we do not know whether \Meta enjoys a \emph{prenex
normal form}, even when we only take into consideration the two meta quantifiers
$\LEExs[][_{\pm\WSet}]$ and $\LAAll[][_{\pm\WSet}]$.

\begin{openproblem}[name = \Meta Prenex Normal Form]
  \label{opn:metapnf}
  Is every \Meta formula equivalent to a \Meta formula in \pnf?
\end{openproblem}




\subsection{Second-Order \& Team Logics}
\label{sec:metthr;sub:stl}

We have previously shown that \ADIF is a conservative extension of \DIF.
However, its game-theoretic determinacy gives us a considerably more expressive
logic than \DIF, with a full-fledged second-order flavour, even in the absence
of a contradictory negation.
Indeed, the meta-theory interpretation allows us to show that every \SOL and \TL
formula can be interpreted in the \ADF fragment of \ADIF.
\Viceversa, every \ADF formula, over a restricted class of hyperteams, can be
interpreted by corresponding \SOL sentences and \TL formulae.
This implies that, from a descriptive-complexity viewpoint, \ADF sentences cover
at least the entire polynomial hierarchy \PH~\citep{Imm99}.

Every non-null hyperteam $\HypAsg \in \HypAsgSet(\xVec)$ defined over a sequence
of variables $\xVec \in \VarSet[][*]$, which is \emph{at most equipotent} to the
domain of the underlying structure $\AStr$, \ie, $\card{\HypAsg} \leq
\card{\AStr}$, can be encoded by a $k$-ary relation $\RRel$, with $k \defeq
\card{\xVec} + 1$, whose interpretation $\RRel[][\AStr] \defeq \RelFun{\HypAsg}
\subseteq \ASet[][k]$ is defined (up to isomorphism) as follows: for every team
$\TeamAsg \in \HypAsg$, there is an element $\aElm \in \ASet$ and, \viceversa,
for every element $\aElm \in \ASet$, there is a team $\TeamAsg \in \HypAsg$ such
that
\[
  \asgElm \in \TeamAsg
\text{\ \ \iff\ \ }
  \AStr \uplus \{ \RelFun{\HypAsg} \}, {\asgElm}[\yvarElm \mapsto \aElm]
  \cmodels[\FOL] \RRel(\xVec\yvarElm),
\]
for all assignments $\asgElm \in \AsgSet(\xVec)$.
It is not clear whether there exists other relational encodings of infinite
hyperteams with greater cardinality than the domain of the structure.
Now, by Theorems~\ref{thm:hst}, every \ADIF formula in \pnf can be translated
into an equivalent \Meta formula, where the semantics of the meta quantifier can
be easily modelled via second-order quantifications.
This leads to the result below, stating that every \ADF-definable hyperteam
(under the above restriction) is \SOL-definable too.

\begin{theorem}[name = \ADF-\SOL Interpretation, restate = thmadifsolint]
  \label{thm:adifsolint}
  For every \ADF formula $\varphiFrm$ in \pnf with quantifier prefix $\qntElm
  \in \QntSet$ over a signature $\LSig$, set of variables $\free{\varphiFrm}
  \subseteq \VSet \subseteq \VarSet$ with $\VSet \cap \var{\qntElm} =
  \emptyset$, and relation symbol $\RRel \not\in \LSig$ with $\art{\RRel} =
  \card{\VSet} + 1$, there exist two \SOL sentences $\PhiSnt[\QEA]$ and
  $\PhiSnt[\QAE]$ over signature $\LSig \uplus \{ \RRel \}$ such that, for all
  $\LSig$-structures $\AStr$ and non-null hyperteams $\HypAsg \in
  \HypAsgSet(\VSet)$ with $\card{\HypAsg} \leq \card{\AStr}$, the following
  equivalence hold true: $\AStr, \HypAsg \cmodels[][\alpha] \varphiFrm$ \iff
  $\AStr \uplus \{ \RelFun{\HypAsg} \} \cmodels[\SOL] \PhiSnt[\alpha]$.
\end{theorem}

Using a similar approach, every non-empty non-null hyperteam $\HypAsg \in
\HypAsgSet(\VSet)$ defined over a set of variables $\VSet \subseteq \VarSet$,
with $\card{\HypAsg} \!\leq\! \card{\AStr}$, can be encoded in a team
$\TeamFun{\HypAsg, \yvarElm} \in \TeamAsgSet(\VSet \cup \yvarElm)$, with
$\yvarElm \not\in \VSet$, as follows: for every team $\TeamAsg \in \HypAsg$,
there is an element $\aElm \in \ASet$ and, \viceversa, for every element $\aElm
\in \ASet$, there is a team $\TeamAsg \in \HypAsg$ such that
\[
  \asgElm \in \TeamAsg
\text{\ \ \iff\ \ }
  {\asgElm}[\yvarElm \mapsto \aElm] \in \TeamFun{\HypAsg, \yvarElm},
\]
for all assignments $\asgElm \in \AsgSet(\VSet)$.
Since every \SOL-definable relation can be encoded in a \TL-definable
team~\citep{KV09,KN09}, the next result easily follows from the previous one.

\begin{corollary}[name = \ADF-\TL Interpretation, restate = coradiftlint]
  \label{cor:adiftlint}
  For every \ADF formula $\varphiFrm$ in \pnf with quantifier prefix $\qntElm
  \in \QntSet$, set of variables $\free{\varphiFrm} \subseteq \VSet \subseteq
  \VarSet$ with $\VSet \cap \var{\qntElm} = \emptyset$, and variable $\yvarElm
  \not\in \VSet$, there exist two \TL formulae $\PhiFrm[\QEA]$ and
  $\PhiFrm[\QAE]$ with $\free{\PhiFrm[\QEA]} = \free{\PhiFrm[\QEA]} = \VSet \cup
  \yvarElm$ such that, for all structures $\AStr$ and non-empty non-null
  hyperteams $\HypAsg \in \HypAsgSet(\VSet)$ with $\card{\HypAsg} \leq
  \card{\AStr}$, the following equivalence hold true: $\AStr, \HypAsg
  \cmodels[][\alpha] \varphiFrm$ \iff $\AStr, \TeamFun{\HypAsg, \yvarElm}
  \cmodels[\TL]\! \PhiFrm[\alpha]$.
\end{corollary}

It is unknown whether the above two interpretation results still hold when the
constraint $\card{\HypAsg} \leq \card{\AStr}$ on the size of the hyperteam and
the domain of the structure is violated.

\begin{openproblem}[name = \ADF-\SOL/\TL Interpretations]
  \label{opn:adifint}
  Is it possible to obtain interpretation results in a similar vein to
  Theorem~\ref{thm:adifsolint} and Corollary~\ref{cor:adiftlint}, when
  $\card{\HypAsg} > \card{\AStr}$?
\end{openproblem}

In addition, it is not clear what the distinguishability power of \ADIF is \wrt
the cardinality of the hyperteams, especially in the infinite case.

\begin{openproblem}[name = Hyperteam Cardinality]
  \label{opn:hypcar}
  Is there an \ADIF satisfiable formula $\varphiFrm$ such that, if $\AStr,
  \HypAsg \cmodels[][\alpha] \varphiFrm$, then $\card{\HypAsg} >
  \card{\AStr} \geq \omega$, for some $\alpha \in \{ \QEA, \QAE \}$?
\end{openproblem}

For the converse direction of the interpretation results, given an
$\LSig$-structure $\AStr$, a relation symbol $\RRel \in \LSig$, and a sequence
of variables $\xVec \in \VarSet[][\art{\RRel}]$, we denote by
$\TeamFun{\RRel[][\AStr], \xVec} \in \TeamAsgSet(\xVec)$ the standard encoding
in a team (up to isomorphism) of the interpretation $\RRel[][\AStr]$ of $\RRel$
defined as follows:
\[
  \asgElm \in \TeamFun{\RRel[][\AStr], \xVec}
\text{\ \ \iff\ \ \ }
  \AStr, \asgElm \cmodels[\FOL] \RRel(\xVec),
\]
for all assignments $\asgElm \in \AsgSet(\xVec)$.
Every \SOL sentence can be put in a canonical form, where every quantification
over functions can be simulated by a meta quantifier that only depends on the
variables to which the function is applied.
Thus, by exploiting Theorem~\ref{thm:hst}, the result below can be proved.

\begin{theorem}[name = \SOL-\ADF Interpretation, restate = thmsoladfint]
  \label{thm:soladfint}
  For every \SOL sentence $\PhiSnt$ over a signature $\LSig$, relation symbol
  $\RRel \in \LSig$, and sequence of variables $\xVec \in
  \VarSet^{\art{\RRel}}$, there exists an \ADF formula $\varphiFrm$ in \pnf over
  signature $\LSig \setminus \RRel$ with $\sup{\varphiFrm} = \free{\varphiFrm} =
  \xVec$ such that, for all $\LSig$-structures $\AStr$, the following
  equivalence hold true: $\AStr \cmodels[\SOL]\! \PhiSnt$ \iff $\AStr
  \!\setminus\! \RRel, \left\{ \TeamFun{\RRel[][\AStr], \xVec} \right\}
  \cmodels[][\QEA] \varphiFrm$.
\end{theorem}

By using the translation from \TL to \SOL (see~\citep{Vaa07,Har79}, for the
sentences, and~\citep{KV09,KN09}, for the formulae), we can show the following.

\begin{corollary}[name = \TL-\ADF Interpretation, restate = cortladfint]
  \label{cor:tladfint}
  For every \TL formula $\PhiFrm$, there exists an \ADF formula $\varphiFrm$ in
  \pnf with $\sup{\varphiFrm} = \free{\varphiFrm} = \free{\PhiFrm}$ such that,
  for all structures $\AStr$ and teams $\TeamAsg \in
  \TeamAsgSet[\subseteq](\free{\PhiFrm})$, the following equivalence holds
  true: $\AStr, \TeamAsg \cmodels[\TL]\! \PhiFrm$ \iff $\AStr, \{ \TeamAsg \}
  \cmodels[][\QEA] \varphiFrm$.
\end{corollary}







\section{Game-Theoretic Semantics}
\label{sec:gamthrsem}

As discussed in Section~\ref{sec:adif}, the alternating Hodges semantic relation
$\AStr \models \varphiSnt$ implicitly ensures the existence of a \emph{semantic
game} $\GameName[\varphiSnt][\AStr]$, played by Eloise and Abelard, with the the
property that Eloise wins the game \iff the \ADIF sentence $\varphiSnt$ is
indeed satisfied on the structure $\AStr$.
We conclude the work by formalising such a game, thus providing a
\emph{game-theoretic semantics} for \ADIF and a proof of its adequacy \wrt both
the compositional semantics of Definition~\ref{def:sem(adif)} and the
Herbrand-Skolem semantics of Theorem~\ref{thm:hst}.
Thanks to Corollary~\ref{cor:tladfint}, this result also provides an indirect
game-theoretic semantics for \TL, a result that, as far as we know, was still
missing~\citep{Vaa07}.
Note that, unlike for \DIF~\citep{HS97,MSS11}, $\GameName[\varphiSnt][\AStr]$
needs to be a \emph{zero-sum perfect-information} game in order to comply with
the game-theoretic determinacy of the logic (see Corollary~\ref{cor:det}), that
for sentences is reflected in the law of excluded middle (see
Corollary~\ref{cor:lawexcmid}).

A two-player turn-based \emph{arena} $\ArenaName = \tuple {\PosSet} {\PPosSet}
{\OPosSet} {\iposElm} {\MovRel}$ is a tuple where
\begin{inparaenum}[(i)]
  \item
    $\PosSet$ is the set of all \emph{positions},
  \item
    $\PPosSet, \OPosSet \subseteq \PosSet$ are the sets of positions owned by
    \emph{Eloise} and \emph{Abelard} with $\PPosSet \cap \OPosSet = \emptyset$,
  \item
    $\iposElm \in \PosSet$ is the \emph{initial position}, and
  \item
    $\MovRel \subseteq (\PPosSet \cup \OPosSet) \times \PosSet$ is the binary
    left-total relation describing all possible \emph{moves}.
\end{inparaenum}
A \emph{path} $\pthElm \in \PthSet \subseteq \PosSet[][\infty]$ is a finite or
infinite sequence of positions compatible with the move relation, \ie,
$((\pthElm)_{i}, (\pthElm)_{i + 1}) \in \MovRel$, for all $i \in
\numco{0}{\card{\pthElm} - 1}$; it is \emph{initial} if $\card{\pthElm} > 0$ and
$\fst{\pthElm} = \iposElm$.
A \emph{history} for player $\alpha \in \{ \PlrSym, \OppSym \}$ is a finite
initial path $\hstElm \in \HstSet[\alpha] \subseteq \PthSet \cap (\PosSet[][*]
\cdot \PosSet[\alpha])$ terminating in an $\alpha$-position.
A \emph{play} $\playElm \in \PlaySet \subseteq \PthSet$ is a maximal (\ie,
non-extendable) initial path.
A \emph{strategy} for player $\alpha \!\in\! \{ \PlrSym, \OppSym \}$ is a
function $\strElm[\alpha] \!\in\! \StrSet[\alpha] \subseteq {\HstSet[\alpha] \to
\PosSet}$ mapping each $\alpha$-history $\hstElm \in \HstSet[\alpha]$ to a
position ${\strElm[\alpha](\hstElm) \in \PosSet}$ compatible with the move
relation, \ie, $(\lst{\hstElm}, \strElm[\alpha](\hstElm)) \in \MovRel$.
The \emph{induced play} of a pair of strategies $(\pstrElm, \ostrElm) \in
\PStrSet \times \OStrSet$ is the unique play $\pthElm \in \PlaySet$ such that
$(\pthElm)_{i + 1} \!=\! \pstrElm((\pthElm)_{\leq i})$, if $(\pthElm)_{i} \in
\PPosSet$, and $(\pthElm)_{i + 1} \!=\! \ostrElm((\pthElm)_{\leq i})$,
otherwise, for all $i \in \numco{0}{\card{\pthElm} - 1}$.
A \emph{game} $\GameName = \tuple {\ArenaName} {\WinSet}$ is a tuple, where
$\ArenaName$ is an arena and $\WinSet \subseteq \PlaySet$ is the set of
\emph{winning} plays for Eloise; the complement $\PlaySet \setminus \WinSet$ is
\emph{winning} for Abelard.
Eloise (\resp, Abelard) \emph{wins} the game if she (\resp, he) has a strategy
$\pstrElm \in \PStrSet$ (resp, $\ostrElm \in \OStrSet$) such that, for all
opponent strategies $\ostrElm \in \OStrSet$ (\resp, $\pstrElm \in \PStrSet$),
the corresponding induced play does (\resp, does not) belong to $\WinSet$.
A game is \emph{determined} if one of the two players wins.

With the notation put in place, we can now describe the semantic game, called
\emph{independence game}, where not only the players perform the choices
corresponding to the operators in the formula, but also check that the choices
of the opponent conform to the associated independence constraints.
Although a specific move for each \ADIF syntactic construct can be given, for
the sake of a simpler presentation, we only define the moves for the
quantifiers.
Any quantifier-free \FOL formula $\psiFrm$, indeed, can be interpreted as a
monolithic atomic relation, whose truth can be immediately evaluated once an
assignment on all its free variables is given.
For this reason, we assume $\varphiSnt = \qntElm \psiFrm$ to be in \pnf, for
some quantifier prefix $\qntElm \in \QntSet$, where no variable is quantified
twice.
Finally, as a standard assumption from a descriptive-complexity
viewpoint~\citep{Imm99,GKLMSVVW05}, we restrict to finite structures only.
The general case, as well as the lift of the approach to formulae, will be the
focus of the future works.

The game for $\varphiSnt = \qntElm \psiFrm$ consists of two recurrent
stages/phases, called \emph{decision} and \emph{challenge}.
The decision phase is almost identical to a classic Hintikka's \FOL
game~\citep{HS97}, where the player associated with the current subformula
$\phiFrm = \Qnt[][_{\pm\WSet}] \varElm \ldotp \phiFrm'$ of $\varphiSnt$ chooses
a value for the bound variable $\varElm$ to be stored in the current assignment
$\asgElm$.
Once all quantifiers are eliminated, however, instead of declaring the winner by
simply evaluating the truth of $\AStr, \asgElm \cmodels[\FOL] \psiFrm$, the game
enters the challenge phase.
Here the players, following again the order of quantification, are asked to
confirm or change their choices.
Making a change here is intended to allow for verifying that the independence
constraints declared in $\qntElm$ are satisfied; after all, if the opponent's
choice is indeed independent of the player's one, such a change should not make
any difference in the satisfaction of the formula.
In more detail, the player associated with $\phiFrm$ can either
\begin{inparaenum}[(i)]
  \item
    confirm her/his own choice maintaining both the assignment $\asgElm$ and
    phase unchanged or
  \item
    or challenge the adversary, by modifying the value assigned to the variable
    $\varElm$ in $\asgElm$, deleting all values for the variables quantified in
    $\qntElm$ after $\varElm$, and reverting to the decision phase.
\end{inparaenum}
In both cases, the control is passed on to the player of the formula $\phiFrm'$
in the scope of the quantifier $\Qnt[][_{\pm\WSet}] \varElm$, so as to allow
her/him to reply to the challenge.
As it should be evident from the alternation of phases, unlike the semantic game
for \FOL, $\GameName[\varphiSnt][\AStr]$ is an \emph{infinite-duration} game
that allows for both finite and infinite plays.
The finite ones necessarily terminates in a position of the challenge phase with
current subformula $\psi$, where the winner can be determined by evaluating the
truth of $\AStr, \asgElm \cmodels[\FOL] \psiFrm$.
The infinite plays, instead, are won by the player able to force the adversary
to change infinitely often the values of one of her/his own variables $\varElm$
in a way that violates the independence constraints, without being able, at the
same time, to force the challenger to do the same on a variable subsequent to
$\varElm$ in $\qntElm$.
We clarify this point later on.

The formalisation of the arena $\ArenaName[\varphiSnt][\AStr]$ underlying the
independence game $\GameName[\varphiSnt][\AStr]$ is reported below, where
$\psf{\varphiSnt}$ denotes the smallest set of subformulae of $\varphiSnt$,
called \emph{prefix subformulae}, such that
\begin{inparaenum}[(i)]
  \item
    $\varphiSnt \in \psf{\varphiSnt}$ and
  \item
    if $\phiFrm = \Qnt[][_{\pm\WSet}] \varElm \ldotp \phiFrm' \in
    \psf{\varphiSnt}$ then $\phiFrm' \in \psf{\varphiSnt}$.
\end{inparaenum}
As an example, for the sentence $\varphiSnt[4] = \LExs \varElm \ldotp
\LAll[][+\emptyset] \yvarElm \ldotp \LNeg (\varElm = \yvarElm)$ reported in
Example~\ref{exm:snt}, we have $\psf{\varphiSnt[4]} = \{ \varphiSnt[4],
\LAll[][+\emptyset] \yvarElm \ldotp \LNeg (\varElm = \yvarElm), \LNeg (\varElm =
\yvarElm) \}$.

\begin{construction}[Independence Arena]
  \label{cns:semarn}
  For a finite structure $\AStr$ and a \pnf \ADIF sentence $\varphiSnt \!=\!
  \qntElm \psiFrm$, with $\psi \!\in\! \FOL$, the \emph{independence arena}
  $\ArenaName[\varphiSnt][\AStr] =\! \tuplel {\PosSet} {\PPosSet}
  {\OPosSet,}\!\!\! \allowbreak \tupler {\iposElm} {\MovRel}$ is defined as
  prescribed in the following:
  \begin{enumerate}[1)]
    \item\label{cns:semarn(pos)}
      the set of positions $\PosSet \subset \psf{\varphiSnt} \times \AsgSet
      \times \{ \PhsISym, \PhsIISym \}$ contains those triples $(\phiFrm,
      \asgElm, \blacklozenge)$ of a prefix subformula $\phiFrm \in
      \psf{\varphiSnt}$ of $\varphiSnt$, an assignment $\asgElm \in \AsgSet$,
      and a phase flag $\blacklozenge \in \{ \PhsISym, \PhsIISym \}$ such that
      $\asgElm \in \AsgSet(\free{\phiFrm})$, if $\blacklozenge = \PhsISym$, and
      $\asgElm \in \AsgSet(\free{\psiFrm})$, otherwise;
    \item\label{cns:semarn(eap)}
      the set $\PPosSet$ of Eloise's (\resp, $\OPosSet$ of Abelard's) positions
      contains the triples of the form $(\LExs[][_{\pm\WSet}] \!\varElm\!
      \ldotp\! \phiFrm'\!\!, \asgElm, \blacklozenge)$ or $(\psiFrm, \asgElm,
      \PhsISym)$ (\resp, $(\LAll[][_{\pm\WSet}] \!\varElm\! \ldotp\!
      \phiFrm'\!\!, \asgElm, \blacklozenge)$);
    \item\label{cns:semarn(ini)}
      the initial position $\iposElm \defeq (\varphiSnt, \emptyfun, \PhsISym)$
      contains the original sentence $\varphiSnt$ associated with the empty
      assignment $\emptyfun$ and the phase flag $\PhsISym$;
    \item\label{cns:semarn(mov)}
      the move relation $\MovRel \subseteq \PosSet \times \PosSet$ contains
      exactly those pairs of positions $(\posElm[1], \posElm[2]) \in \PosSet
      \times \PosSet$ satisfying one of the conditions below:
      \begin{enumerate}[a)]
        \item\label{cns:semarn(mov:noc)}
          $\posElm[1] = (\Qnt[][_{\pm\WSet}] \varElm \ldotp \phiFrm', \asgElm,
          \blacklozenge)$ and $\posElm[2] = (\phiFrm', \asgElm, \blacklozenge)$,
          with $\varElm \not\in \free{\phiFrm'}$;
        \item\label{cns:semarn(mov:chc)}
          $\posElm[1] = (\Qnt[][_{\pm\WSet}] \varElm \ldotp \phiFrm'\!, \asgElm,
          \PhsISym)$ and $\posElm[2] = (\phiFrm'\!, {\asgElm}[\varElm
          \!\mapsto\! \aElm], \PhsISym)$, for some $\aElm \!\in\! \ASet$;
        \item\label{cns:semarn(mov:swp)}
          $\posElm[1] = (\psiFrm, \asgElm, \PhsISym)$ and $\posElm[2] =
          (\varphiSnt, \asgElm, \PhsIISym)$;
        \item\label{cns:semarn(mov:cnf)}
          $\posElm[1] = (\Qnt[][_{\pm\WSet}] \varElm \ldotp \phiFrm', \asgElm,
          \PhsIISym)$ and $\posElm[2] = (\phiFrm', \asgElm, \PhsIISym)$;
        \item\label{cns:semarn(mov:chl)}
          $\posElm[1] = (\Qnt[][_{\pm\WSet}] \varElm \ldotp \phiFrm'\!, \asgElm,
          \PhsIISym)$ and $\posElm[2] \!=\! (\phiFrm'\!, \asgElm'[\varElm
          \!\mapsto\! \aElm], \PhsISym)$, for some $\aElm \!\in\! \ASet$ with
          $\aElm \neq \asgElm(\varElm)$, where $\asgElm' \defeq \asgElm \!\rst
          \free{\Qnt[][_{\pm\WSet}] \varElm \ldotp \phiFrm'}$.
      \end{enumerate}
  \end{enumerate}
\end{construction}
Intuitively, a position $(\phiFrm, \asgElm, \blacklozenge)$ maintains the
information about the formula $\phiFrm$ that still has to be played against, the
assignment $\asgElm$ containing the variables whose values have already been
chosen, and a flag $\blacklozenge$ identifying the phase, either $\PhsISym$ or
$\PhsIISym$.
Item~\ref{cns:semarn(mov:noc)} forces the trivial move for the vacuous
quantifications, Item~\ref{cns:semarn(mov:chc)} defines the moves for the
decision phase, Item~\ref{cns:semarn(mov:swp)} switches from the decision to the
challenge phase, Item~\ref{cns:semarn(mov:cnf)} defines the confirmation of the
choice already made, and, finally, Item~\ref{cns:semarn(mov:chl)} describes the
challenge to the adversary, where the phase is reverted to the decision one, the
value for the variable involved in the challenge is changed, and all values for
the subsequent variables are deleted.

The winning condition for the game is defined as follows.
Since the winner of finite plays is easy to determine, as it only depends on
whether the assignment in the last position satisfies $\psiFrm$, we shall focus
on the infinite ones.
Let us consider an arbitrary prefix subformula $\phiFrm = \Qnt[][_{\pm\WSet}]
\varElm \ldotp \phiFrm' \in \psf{\varphiSnt}$ with $\varElm \in \VSet \defeq
\free{\phiFrm'}$.
By $\FCls[\phiFrm] \colon \AsgSet(\VSet) \to \pow{\FncSet[\pm\WSet]{}}$ we
denote the map associating each assignment $\asgElm \in \AsgSet(\VSet)$ defined
over the variables in $\VSet$ with the set $\FCls[\phiFrm](\asgElm) \defeq \set{
\FFun \in \FncSet[\pm\WSet]{} }{ \FFun(\asgElm) = \asgElm(\varElm) }$ of all the
$\pm\WSet$-functions compatible with the value assigned to $\varElm$ in
$\asgElm$.
In addition, by $\BCls[\phiFrm] \colon \HstSet \to \pow{\FncSet[\pm\WSet]{}}$,
with $\HstSet \defeq \HstSet[\PlrSym] \cup \HstSet[\OppSym]$, we denote the map
assigning to each history $\hstElm \in \HstSet$ the set
$\BCls[\phiFrm](\hstElm)$ of all the $\pm\WSet$-functions compatible with the
most recent assignments along $\hstElm$.
Formally:
\begin{itemize}[$\bullet$]
  \item
    $\BCls[\phiFrm](\iposElm) \defeq \FncSet[\pm\WSet]{}$;
  \item
    $\BCls[\phiFrm](\hstElm \cdot (\phiFrm', \asgElm, \PhsISym)) \defeq
    \begin{cases}
      {\FCls[\phiFrm](\asgElm)},
    & \text{if } {\BCls[\phiFrm](\hstElm) \cap \FCls[\phiFrm](\asgElm)} =
      \emptyset; \\
      {\BCls[\phiFrm](\hstElm) \cap \FCls[\phiFrm](\asgElm)},
    & \text{otherwise};
    \end{cases}$
  \item
    $\BCls[\phiFrm](\hstElm \cdot \posElm) \defeq \BCls[\phiFrm](\hstElm)$, in
    all other cases, \ie, $\posElm \neq (\phiFrm', \_, \PhsISym)$.
\end{itemize}
Essentially, the bucket $\BCls[\phiFrm](\hstElm)$ maintains the most updated set
of Herbrand/Skolem functions for the variable $\varElm$ that the associated
player can use to reply to all the variables which $\varElm$ depends upon.
When a play starts, no choice has been made yet, so $\BCls[\phiFrm](\iposElm)$
is full.
Once a position $(\phiFrm', \asgElm, \PhsISym)$ is reached after a history
$\hstElm$, a fresh value $\asgElm(\varElm)$ for $\varElm$ has just been chosen
to resolve the quantifier $\Qnt$, so the bucket is updated by removing from
$\BCls[\phiFrm](\hstElm)$ all the functions that are not compatible with this
new value.
If such resulting set becomes empty, the player is caught cheating and the
bucket is replenished taking into account only the choice just made.

In general there are two reasons for a player to cheat.
Either she/he is changing the value of the variable to challenge the adversary
to prove that he/she is complying with the independence constraints
(Item~\ref{cns:semarn(mov:chl)}), or she/he chooses a new value because is
unable to both satisfy her/his goal and comply with the constraints on her/his
variables (Item~\ref{cns:semarn(mov:chc)}).
Obviously, the second type of cheating, called \emph{defensive cheat}, can, in
turn, induce one of the first type, called \emph{challenge cheat}.
Hence, complex chains of different types of cheating can occur.
In order to identify which player is the last one who was forced to cheat, we
consider an arbitrary map $\prtFun \colon \psf{\varphiSnt} \to \SetN$ assigning
to each prefix subformula $\phiFrm = \Qnt[][_{\pm\WSet}] \varElm \ldotp \phiFrm'
\in \psf{\varphiSnt}$ a priority $\prtFun(\phiFrm)$ such that
\begin{inparaenum}[i)]
  \item
    $\prtFun(\phiFrm)$ is even \iff $\Qnt = \LAll$ and
  \item
    $\prtFun(\phiFrm) < \prtFun(\phiFrm')$.
\end{inparaenum}
To each history $\hstElm \in \HstSet$ we can then assign the sequence of cheats
$\chtFun{\hstElm}$ occurring in it via the map $\chtFun{} \colon \HstSet \to
\SetN[][*]$ as follows:
\begin{itemize}[$\bullet$]
  \item
    $\chtFun{\iposElm} \defeq 0$;
  \item
    $\chtFun{\hstElm \!\cdot\! (\phiFrm', \asgElm, \PhsISym)} \defeq
    \chtFun{\hstElm} \!\cdot\! \prtFun(\phiFrm)$, whenever
    $\BCls[\phiFrm](\hstElm) \cap \FCls[\phiFrm](\asgElm) \!=\! \emptyset$;
  \item
    $\chtFun{\hstElm \cdot \posElm} \defeq \chtFun{\hstElm} \cdot 0$, in all
    other cases.
\end{itemize}
This construction easily lifts to infinite plays $\pthElm \in \PlaySet[][\omega]
\defeq \PlaySet \cap \PosSet[][\omega]$ through the map $\chtFun{} \colon
\PlaySet[][\omega] \to \SetN[][\omega]$ such that $(\chtFun{\pthElm})_{i} =
\chtFun{(\pthElm)_{\leq i}}$, for all $i \in \SetN$.
Finally, $\prtFun(\pthElm)$ denotes the maximal priority seen infinitely often
along $\chtFun{\pthElm}$.
Note that every infinite play necessarily contains at least infinitely many
challenge cheats (Item~\ref{cns:semarn(mov:chl)}).
Thus, $\prtFun(\pthElm)$ uniquely identifies the right-most variable in
$\qntElm$ over which the corresponding player cheated, without being able, at
the same time, to force the adversary to do the same.
If $\prtFun(\pthElm)$ is even, Abelard is cheating infinitely often, so he loses
the play $\pthElm$, which is, therefore, won by Eloise.

\begin{construction}[Independence Game]
  \label{cns:semgam}
  For a finite structure $\AStr$ and a \pnf \ADIF sentence $\varphiSnt = \qntElm
  \psiFrm$, with $\psi \in \FOL$, the \emph{independence game}
  $\GameName[\varphiSnt][\AStr] = \tuple {\ArenaName} {\WinSet}$ is defined as
  prescribed in the following:
  \begin{enumerate}[1)]
    \item\label{cns:semgam(arn)}
      $\ArenaName$ is the independence arena $\ArenaName[\varphiSnt][\AStr]$
      defined in Construction~\ref{cns:semarn};
    \item\label{cns:semgam(win)}
      $\WinSet \subseteq \PlaySet$ is the set of all the plays $\pthElm$
      satisfying the following conditions:
      \begin{enumerate}[a)]
        \item\label{cns:semgam(win:inf)}
          if $\pthElm$ is infinite then $\prtFun(\pthElm)$ is even;
        \item\label{cns:semgam(win:fin)}
          if $\pthElm$ is finite then $\lst{\pthElm} = (\psiFrm, \asgElm,
          \PhsIISym)$ and $\AStr, \asgElm \cmodels[\FOL] \psiFrm$, for some
          assignment $\asgElm \in \AsgSet(\free{\psiFrm})$.
      \end{enumerate}
  \end{enumerate}
\end{construction}

\begin{example}
  \label{exm:runexmgam}
  Let us consider the sentence $\varphiSnt[7]$ of Example~\ref{exm:runexm} from
  Section~\ref{sec:adif;sub:syn}, which is true in the binary structure $\AStr$
  of that example.
  Therefore, Eloise, who controls the values of the variables $\varElm$ and
  $\zvarElm$, must have a strategy to win the independence game
  $\GameName[{\varphiSnt[7]}][\AStr]$.
  One possibility is to choose, during the decision phase, the constant function
  $\fFun[\varElm] = 0$ for $\varElm$ and the identity function
  $\fFun[\zvarElm](\varElm) = \varElm$ for $\zvarElm$.
  Clearly, she wins any finite play where Abelard chooses the constant function
  $\fFun[\yvarElm] = 0$ for $\yvarElm$, since the resulting assignment satisfies
  both $(\varElm = \yvarElm)$ and $(\yvarElm = \zvarElm)$.
  Let us assume, then, that he chooses $\fFun[\yvarElm] = 1$, instead, in the
  decision phase.
  Since at the end of this phase Eloise knows she is losing, she will challenge
  Abelard by changing her function $\fFun[\varElm]$ for $\varElm$ to the
  constant $1$.
  This raises the priority of the current play fragment to $1$.
  Now, if Abelard sticks to function $\fFun[\yvarElm] = 1$ for $\yvarElm$, he
  loses, since $\fFun[\zvarElm](\varElm) = \varElm$ would now give $\zvarElm$
  value $1$ as well, leading to a finite play.
  So he needs to modify his choice to $\fFun[\yvarElm] = 0$, this time raising
  the priority of the play fragment to $2$ and generating a challenge for
  Eloise on $\zvarElm$.
  Eloise, however, can stick to the identity function and make way to a new
  challenge phase.
  Now, since Eloise is losing with the current assignment, she will challenge
  once again, choosing $\fFun[\varElm] = 0$ and raising priority $1$.
  Abelard is then forced to change function and raise priority $2$ and we are
  back to where we started.
  This cyclic process ends up forming an infinite play whose maximal priority is
  $2$, since Eloise can force Abelard to defensively change bucket infinitely
  often, thus satisfying her winning condition.
\end{example}

It is worth noting that the game devised above bears some similarities with the
\emph{team-building game} proposed by~\cite{Bra13} for \DL~\citep{Vaa07}. Both
ours and his are complete-information games extending Hintikka's game for \FOL.
In addition, Bradfield's game also checks the uniformity of the choices made by
Eloise by means of a challenge mechanism, where the sentence is played over
repeatedly by the players.
The similarities, however, end here as the two games differ significantly in
nature.
First, the repeated evaluations of a sentence $\varphiSnt$ in Bradfield's game
allow him to build teams during a play, one for each dependence atom occurring
in $\varphiSnt$.
Each team is then used to check whether Eloise's choices have been made in
accordance to the dependency constraint encoded by the corresponding atom.
All these teams are explicitly recorded in each state of his game, together with
the partial assignment recording the choices made by the players so far in the
current repetition.
In this sense, then, Bradfield's arena is intrinsically second order, as it
records sets of assignments in each state and contains moves that update such
sets.
Second, Bradfield's game on finite structures only admits finite plays and its
winning condition, then, boils down to a simple reachability.
On the contrary, our game is played in a purely first-order arena, whose states
only keep track of players choices collected in the partial assignment.
Moreover, it always admits infinite plays, where players can repeatedly
challenge each other forever.
The second-order power of our game, then, resides entirely in the winning
condition, where the priority-based mechanism accounts for the alternation of
the quantifiers along the, possibly infinite, repeated evaluations of the
sentence.

To conclude, by exploiting Theorem~\ref{thm:hst}, it is possible to prove the
adequacy of the game-theoretic semantics \wrt the model-theoretic one of
Definition~\ref{def:sem(met)} and, in turn, \wrt the compositional one of
Definition~\ref{def:sem(adif)}, where the Herbrand/Skolem functions obtained by
the evaluation of the existential (\resp, universal) meta quantifiers of the
\Meta sentence $\hspFun{\qntElm} \psiFrm$ (resp, $\LNeg \hspFun{\qntElm}
\psiFrm$) induce a winning strategy for Eloise (\resp, Abelard) in
$\GameName[\varphiSnt][\AStr]$.
This also implies the determinacy of the independence game, without the need to
rely on topological determinacy theorems, as those of~\cite{Mar75,Mar85}.

\begin{theorem}[name = Game-Theoretic Semantics, restate = thmgamthrsem]
  \label{thm:gamthrsem}
  For a finite structure $\AStr$ and an \ADIF sentence $\varphiFrm$ in prenex
  form, there exists an independence game $\GameName[\varphiFrm][\AStr]$ such
  that $\AStr \models \varphiFrm$ (\resp, $\AStr \not\models \varphiFrm$) \iff
  $\GameName[\varphiFrm][\AStr]$ is won by Eloise (\resp, Abelard).
\end{theorem}









\section{Discussion}

We have introduced Alternating Dependence/Independence-Friendly Logic (\ADIF), a
conservative extension of Independence-Friendly Logic (\IF), that incorporates
negation in a very natural way and avoids the indeterminacy of the logic.
This is achieved by means of a generalisation of team semantics, where the
choices of both players are represented in a two-level structure, called
hyperteam.
This allows us to treat the two players symmetrically and force both of them to
make their choices according to the (in)dependence constraints specified in the
corresponding quantifiers.
Interestingly enough, the new semantics grants the proposed logic the full
expressive power of Second-Order Logic (\SOL) and, as a consequence, also of
Team Logic (\TL), without the need of including additional connectives in the
language.
It also allows for restoring the law of excluded middle for sentences and enjoys
the property of game-theoretic determinacy.
For the prenex fragment, a Herbrand-Skolem semantics is also provided that
directly connects \ADIF with \SOL, as well as a game-theoretic semantics on
finite structures, given in terms of a determined turn-based infinite-duration
perfect-information game played on a first-order arena.

Interesting questions that remain open concern whether a prenex normal form
theorem holds for the language.
Equally unsettled is the actual expressive power of \ADIF.
We do show that it is at least as expressive as \SOL and, thus, covers the full
polynomial hierarchy \PH.
The proof for the other direction, however, relies upon the assumption of
equipotency between the hyperteam $\HypAsg$ and the domain $\ASet$ of the
underlying structure $\AName$, which allows us to encode hyperteams by means of
a suitable relation $\RelFun{\HypAsg}$.
There seems to be no straightforward way to do the same for ``big'' hyperteams.
Yet again, it is not clear whether such ``big'' hyperteam actually matter, in
the sense of there being a formula that can distinguish between ``big'' and
``small'' hyperteams.
Usually, similar questions have been addressed by defining suitable
Ehrenfeucht-Fra\"iss\'e games (\EFG) to precisely characterise the expressive
power of the logic.
For this reason, one may think to do the same for \ADIF as well.
The main difficulty we foresee here is, however, the treatment of
quantifications, for which no explicit commitment to a specific valued is made
in the semantics (all choices are evenly encoded in the hyperteam).
In a classic \EFG game, instead, the moves corresponding to the choices of a
value by a quantifier make explicit commitments.
Currently, it is not clear to us how to circumvent this discrepancy.






\begin{section}*{Acknowledgments}

Partially supported by the GNCS 2020 project ``Ragionamento Strategico e Sintesi
Automatica di Sistemi Multi-Agente''.

\end{section}



  \newpage
  \appendix



\section{Proofs of Section~\ref{sec:adif}}
\label{app:adif}

Before each proof of a theorem, we display its dependency graph: the vertices are the results used to prove the theorem (they can be lemmata, propositions, other theorems, etc). There is an edge from Result~1 to Result~2 \iff Result~1 is explicitly used in Result~2's proof.

\newcommand{\asgExt}[2]{#2 \!\rst[][\WSet]\xspace}
\newcommand{\asgExtW}[1]{\asgExt{\WSet}{#1}}
\newcommand{\asgExtWX}{\asgExtW{\TeamAsg}}

Let $\WSet \subseteq \VarSet$ and $\HypAsg \in \HypAsgSet$.
For a team $\TeamAsg \in \HypAsg \!\rst[\WSet]$, we denote by $\asgExtWX$
one (arbitrarily chosen) of the teams $\YSet \in \HypAsg$ such that $\YSet
\!\rst[\WSet] = \TeamAsg$.

\lmmdlti*
\begin{proof}
  First, observe that, by Proposition~\ref{prp:empnultrv},
  $\dual{\dual{\HypAsg}} \eqv \HypAsg$ holds for every non-proper hyperteam
  $\HypAsg$.

  Next, we show that $\HypAsg \subseteq \dual{\dual{\HypAsg}}$, for a proper
  hyperteam $\HypAsg$.
  Let $\TeamAsg \in \HypAsg$.
  Observe that, since $X$ is proper, $\TeamAsg' \cap \TeamAsg \neq \EmpTeamAsg$
  for all $\TeamAsg' \in \dual{\HypAsg}$.
  For every $\asgElm \in \TeamAsg$, fix a choice function $\chcFun_{\asgElm} \in
  \ChcSet{\HypAsg}$ such that $\chcFun_{\asgElm}(\TeamAsg) = \asgElm \in
  \TeamAsg$.
  Now, consider $\dual{\chcFun} \in \ChcSet{\dual{\HypAsg}}$ such that
  $\dual{\chcFun}(\img{\chcFun_{\asgElm}}) = \asgElm$ for all $\asgElm \in
  \TeamAsg$, and $\dual{\chcFun}(\TeamAsg') \in \TeamAsg \cap
  \TeamAsg'$ for all the other teams $\TeamAsg' \in \dual{\HypAsg}
  \setminus \{ \img{\chcFun_{\asgElm}} \mid \asgElm \in \TeamAsg \}$.
  Clearly, $\TeamAsg = \img{\dual{\chcFun}} \in \dual{\dual{\HypAsg}}$, hence
  $\HypAsg \subseteq \dual{\dual{\HypAsg}}$.

  Since $\HypAsg \subseteq \dual{\dual{\HypAsg}}$ implies $\HypAsg \inc
  \dual{\dual{\HypAsg}}$, it suffices to prove that $\dual{\dual{\HypAsg}} \inc
  \HypAsg$ holds to obtain $\HypAsg \eqv \dual{\dual{\HypAsg}}$.
  To this end, let $\dual{\TeamAsg'} = \img{\dual{\chcFun}} \in
  \dual{\dual{\HypAsg}}$, for some $\dual{\chcFun} \in \ChcSet{\dual{\HypAsg}}$.
  We show that there is $\TeamAsg \in \HypAsg$ such that $\TeamAsg \subseteq
  \dual{\TeamAsg'}$.
  Assume, towards a contradiction, that this is not the case, \ie, for all
  $\TeamAsg \in \HypAsg$ there is $\asgElm_{\TeamAsg} \in \TeamAsg \setminus
  \dual{\TeamAsg'}$.
  Then, define $\chcFun \in \ChcSet{\HypAsg}$ as: $\chcFun(\TeamAsg) =
  \asgElm_{\TeamAsg}$ for all $\TeamAsg \in \HypAsg$.
  Clearly, $\dual{\chcFun}(\img{\chcFun}) \notin \dual{\TeamAsg'}$, thus raising
  a contradiction.
  Now, the thesis follows from the observation that $\HypAsg \eqv
  \dual{\dual{\HypAsg}}$ is equivalent to $\HypAsg \eqv[\VarSet]
  \dual{\dual{\HypAsg}}$, which implies $\HypAsg \eqv[\WSet]
  \dual{\dual{\HypAsg}}$, due to $\WSet \subseteq \VarSet$.
\end{proof}

\lmmdltii*
\begin{proof}
  We consider the three equivalences separately.
  \begin{enumerate}[1)]
  \item First, we show that there exists a team $\TeamAsg \in \HypAsg$ such
    that $\TeamAsg \subseteq \PsiSet$ if and only if for all teams
    $\TeamAsg' \in \dual{\HypAsg}$, it holds that $\TeamAsg' \cap
    \PsiSet \neq \EmpTeamAsg$.

    (\emph{only-if} direction)
    Let $\TeamAsg'$ be a generic element of $\dual{\HypAsg}$.
    Thus, $\TeamAsg' = \img{\chcFun}$ for some $\chcFun \in
    \ChcSet{\HypAsg}$.
    Thus, $\chcFun(\TeamAsg) \in \TeamAsg \cap \TeamAsg'$.
    The thesis follows from $\TeamAsg \subseteq \PsiSet$.

    (\emph{if} direction) By Proposition~\ref{prp:empnultrv}, if $\dual{\HypAsg}
    = \EmpHypAsg$, then $\EmpTeamAsg \in \HypAsg$, and the thesis immediately
    follows since $\EmpTeamAsg \subseteq \PsiSet$.
    If, instead $\dual{\HypAsg} \neq \EmpHypAsg$, then assume, towards a
    contradiction, that for all $\TeamAsg \in \HypAsg$, there is
    $\asgElm_{\TeamAsg} \in \TeamAsg \setminus \PsiSet$.
    Define $\chcFun \in \ChcSet{\HypAsg}$ as: $\chcFun(\TeamAsg) =
    \asgElm_{\TeamAsg}$, for all $\TeamAsg \in \HypAsg$.
    Since $\img{\chcFun} \in \dual{\HypAsg}$ and $\img{\chcFun} \cap \PsiSet =
    \emptyset$, we get a contradiction.

    The rest of the claim, \ie, there exists a team $\TeamAsg' \in
    \dual{\HypAsg}$ such that $\TeamAsg' \subseteq \PsiSet$ if and only if
    for all teams $\TeamAsg \in \HypAsg$, it holds that $\TeamAsg \cap \PsiSet
    \neq \EmpTeamAsg$, follows from above and the fact that $\HypAsg \equiv
    \dual{\dual{\HypAsg}}$ (Lemma~\ref{lmm:dlti}).

  \item (\emph{only-if} direction) Consider $\chcFun \in \ChcSet{\HypAsg}$ such
    that $\chcFun(\TeamAsg) \in \TeamAsg \cap \PsiSet$.
    The thesis follows from $\chcFun(\TeamAsg) \in \img{\chcFun} \in
    \dual{\HypAsg}$.

    (\emph{if} direction) Let $\TeamAsg' = \img{\chcFun} \in
    \dual{\HypAsg}$, for some $\chcFun \in \ChcSet{\HypAsg}$, be such that
    $\TeamAsg' \cap \PsiSet \neq \EmpTeamAsg$ and let $\dual{\asgElm} \in
    \TeamAsg' \cap \PsiSet$.
    Thus, there is $\TeamAsg \in \HypAsg$ such that $\chcFun(\TeamAsg) =
    \dual\asgElm \in \TeamAsg$, which means that $\TeamAsg \cap \PsiSet \neq
    \EmpTeamAsg$, hence the thesis.

  \item The claim follows by instantiating $\PsiSet$ with $\AsgSet \setminus
    {\PsiSet}$ in the previous claim, and observing that \ref{lmm:dltii(aa:org)}
    and \ref{lmm:dltii(aa:dlt)} correspond to the negations of
    \ref{lmm:dltii(ee:org)} and \ref{lmm:dltii(ee:dlt)}, respectively.
    \qedhere
  \end{enumerate}
\end{proof}

%


\lmmempnulhyp*
\begin{proof}
  The claim follows from the more general Lemma~\ref{lmm:empnulhypMeta},
  reported in~\ref{app:metthr}, by instantiating $\FunAsg$ with the empty
  function $\emptyfun$.
\end{proof}

\figGraphThmHypRef

\thmhypref*
\begin{proof}
  The claim follows from the more general Theorem~\ref{thm:hyprefMeta}, reported
  in~\ref{app:metthr}, by instantiating both $\FunAsg$ and $\iotaFun$ with
  the empty function $\emptyfun$.
\end{proof}

\figGraphThmDblDlt

\thmdbldlt*
\begin{proof}
  The claim follows from the more general Theorem~\ref{thm:dbldltMeta}, reported
  in~\ref{app:metthr}, by instantiating $\FunAsg$ with the empty function
  $\emptyfun$.
\end{proof}

\figGraphThmBnlLawI

\thmblnlawi*
\begin{proof}
  Proving that an equivalence (resp., implication) $\varphiFrm[1] \equiv
  \varphiFrm[2]$ (resp., $\varphiFrm[1] \implies \varphiFrm[2]$) holds true
  amounts to showing that both $\varphiFrm[1] \cequiv[][\QEA] \varphiFrm[2]$ and
  $\varphiFrm[1] \cequiv[][\QAE] \varphiFrm[2]$ (\resp, $\varphiFrm[1]
  \implies[][\QEA] \varphiFrm[2]$ and $\varphiFrm[1] \implies[][\QAE]
  \varphiFrm[2]$) hold true.
  However, as a consequence of Theorem~\ref{thm:dbldlt}, we have that
  $\varphiFrm[1] \cequiv[][\alpha] \varphiFrm[2]$ \iff $\varphiFrm[1]
  \cequiv[][\dual{\alpha}] \varphiFrm[2]$ (\resp, $\varphiFrm[1]
  \implies[][\alpha] \varphiFrm[2]$ \iff $\varphiFrm[1]
  \implies[][\dual{\alpha}] \varphiFrm[2]$) for all $\alpha \in \{ \QEA,
  \QAE\}$.
  Therefore, for every claim in the statement of
  the theorem, it is enough to focus on one of the two alternation flags $\QEA$
  and $\QAE$ only.\footnote{Observe that a proof that does not make use of
    Theorem~\ref{thm:dbldlt} is possible.
    Therefore, Theorem~\ref{thm:dbldlt} does not occur in the dependency graph
    of Theorem~\ref{thm:blnlawi}}
  In the following, when proving an equivalence $\varphiFrm[1] \equiv
  \varphiFrm[2]$ (resp., implication $\varphiFrm[1] \implies \varphiFrm[2]$), we
  assume $\HypAsg \in \HypAsgSet[\subseteq](\sup{\varphiFrm[1]} \cup
  \sup{\varphiFrm[2]})$.

  \begin{enumerate}[1)]
  \item
    \begin{enumerate}[a)]
    \item $\AStr, \HypAsg \cmodels[][\QEA] \neg \Ff \iffExpl{sem.}
      \AStr, \HypAsg \not\cmodels[][\QAE] \Ff \iffExpl{sem.} \HypAsg \neq
      \EmpHypAsg \iffExpl{sem.} \AStr, \HypAsg \cmodels[][\QEA] \Tt$.

    \item $\AStr, \HypAsg \cmodels[][\QAE] \neg \Tt \iffExpl{sem.}
      \AStr, \HypAsg \not\cmodels[][\QEA] \Tt \iffExpl{sem.} \HypAsg =
      \EmpHypAsg \iffExpl{sem.} \AStr, \HypAsg \cmodels[][\QAE] \Ff$.

    \item $\HypAsg \cmodels[][\alpha] \neg\neg\varphiFrm \iffExpl{sem.}
      \HypAsg \not\cmodels[][\dual{\alpha}] \neg\varphiFrm \iffExpl{sem.}
      \HypAsg \cmodels[][\alpha] \varphiFrm$.
    \end{enumerate}
  \item
    \begin{enumerate}[a)]
    \item
      First, we prove that $\varphiFrm \land \Ff \equiv \Ff$ holds.
      To this end, we show that if $\AStr, \HypAsg \EAmodels \varphiFrm \land
      \Ff$, then $\AStr, \HypAsg \EAmodels \Ff$, and vice versa.
      By semantics, $\AStr, \HypAsg \EAmodels \varphiFrm \land \Ff$ implies that
      for all $(\HypAsg[1], \HypAsg[2]) \in \parFun{\HypAsg}$, it holds that
      $\AStr, \HypAsg[1] \EAmodels \varphiFrm$ or $\AStr, \HypAsg[2] \EAmodels
      \Ff$. In particular, since $(\EmpHypAsg, \HypAsg) \in \parFun{\HypAsg}$
      and, by Item~\ref{lmm:empnulhyp(ea:emp)} of Lemma~\ref{lmm:empnulhyp},
      $\AStr, \EmpHypAsg \not\cmodels[][\QEA] \varphiFrm$, we have that $\AStr,
      \HypAsg \EAmodels \Ff$.
      Conversely, $\AStr, \HypAsg \EAmodels \Ff$ means that $\emptyset \in
      \HypAsg$.
      Thus, for every $(\HypAsg[1], \HypAsg[2]) \in \parFun{\HypAsg}$, it holds
      that $\emptyset \in \HypAsg[1]$ or $\emptyset \in \HypAsg[2]$. Thanks to
      Item~\ref{lmm:empnulhyp(ea:nul)} of Lemma~\ref{lmm:empnulhyp}, we have
      $\AStr, \HypAsg[1] \EAmodels \varphiFrm$ or $\AStr, \HypAsg[2] \EAmodels
      \Ff$, which, by semantics of $\land$, implies $\AStr, \HypAsg \EAmodels
      \varphiFrm \land \Ff$.
      To conclude, observe that $\varphiFrm \land \Ff \equiv \Ff \land
      \varphiFrm$ holds, due to commutativity of $\land$, formally proved below
      (Item~\ref{thm:blnlawi(com:con)}).

    \item
      First, we prove that $\varphiFrm \land \Tt \equiv \varphiFrm$ holds.
      To this end, we show that if $\AStr, \HypAsg \EAmodels \varphiFrm \land
      \Tt$, then $\AStr, \HypAsg \EAmodels \varphiFrm$, and vice versa.
      By semantics, $\AStr, \HypAsg \EAmodels \varphiFrm \land \Tt$ implies that
      for all $(\HypAsg[1], \HypAsg[2]) \in \parFun{\HypAsg}$, it holds that
      $\AStr, \HypAsg[1] \EAmodels \varphiFrm$ or $\AStr, \HypAsg[2] \EAmodels
      \Tt$. In particular, since $(\HypAsg, \EmpHypAsg) \in \parFun{\HypAsg}$
      and, by Item~\ref{lmm:empnulhyp(ea:emp)} of Lemma~\ref{lmm:empnulhyp},
      $\AStr, \EmpHypAsg \not\cmodels[][\QEA] \Tt$, we have that $\AStr,
      \HypAsg \EAmodels \varphiFrm$.
      Conversely, assume $\AStr, \HypAsg \EAmodels \varphiFrm$ and let
      $(\HypAsg[1], \HypAsg[2]) \in \parFun{\HypAsg}$.
      If $\HypAsg[1] = \HypAsg$, then $\AStr, \HypAsg[1] \EAmodels \varphiFrm$;
      if $\HypAsg[1] \neq \HypAsg$, then $\HypAsg[2] \neq \emptyset$, and thus,
      by semantics of $\Tt$, it holds that $\AStr, \HypAsg[2] \EAmodels \Tt$.
      Therefore, for every $(\HypAsg[1], \HypAsg[2]) \in \parFun{\HypAsg}$, it
      holds that $\AStr, \HypAsg[1] \EAmodels \varphiFrm$ or $\AStr, \HypAsg[2]
      \cmodels[][\QEA] \Tt$, which, by semantics of $\land$, implies $\AStr,
      \HypAsg \EAmodels \varphiFrm \land \Tt$.
      To conclude, observe that $\varphiFrm \land \Tt \equiv \Tt \land
      \varphiFrm$ holds, due to commutativity of $\land$, formally proved below
      (Item~\ref{thm:blnlawi(com:con)}).
    \end{enumerate}

  \item
    \begin{enumerate}[a)]
    \item
      First, we prove that $\varphiFrm \vee \Tt \equiv \Tt$ holds.
      To this end, we show that if $\AStr, \HypAsg \AEmodels \varphiFrm \vee
      \Ff$, then $\AStr, \HypAsg \AEmodels \Tt$, and vice versa.
      By semantics, $\AStr, \HypAsg \AEmodels \varphiFrm \vee \Tt$ implies that
      there is $(\HypAsg[1], \HypAsg[2]) \in \parFun{\HypAsg}$ such  that
      $\AStr, \HypAsg[1] \AEmodels \varphiFrm$ and $\AStr, \HypAsg[2] \AEmodels
      \Tt$.
      By Item~\ref{lmm:empnulhyp(ae:nul)} of Lemma~\ref{lmm:empnulhyp}, it must
      be $\emptyset \notin \HypAsg[i]$, for each $i\in\{1,2\}$, and thus
      $\emptyset \notin \HypAsg$, which, by semantics of $\Tt$, implies $\AStr,
      \HypAsg \AEmodels \Tt$.
      Conversely, assume $\AStr, \HypAsg \AEmodels \Tt$.
      The claim follows from the fact that $(\EmpHypAsg,\HypAsg) \in
      \parFun{\HypAsg}$ is such that $\AStr, \EmpHypAsg \AEmodels \varphiFrm$
      (by Item~\ref{lmm:empnulhyp(ae:emp)} of Lemma~\ref{lmm:empnulhyp}) and
      $\AStr, \HypAsg \AEmodels \Tt$, which implies that $\AStr, \HypAsg
      \AEmodels \varphiFrm \vee \Tt$.
      To conclude, observe that $\varphiFrm \vee \Tt \equiv \Tt \vee
      \varphiFrm$ holds, due to commutativity of $\vee$, formally proved below
      (Item~\ref{thm:blnlawi(com:dis)}).

    \item
      First, we prove that $\varphiFrm \vee \Ff \equiv \varphiFrm$ holds.
      To this end, we show that if $\AStr, \HypAsg \AEmodels \varphiFrm \vee
      \Ff$, then $\AStr, \HypAsg \AEmodels \varphiFrm$, and vice versa.
      By semantics, $\AStr, \HypAsg \AEmodels \varphiFrm \vee \Ff$ implies that
      there is $(\HypAsg[1], \HypAsg[2]) \in \parFun{\HypAsg}$ such  that
      $\AStr, \HypAsg[1] \AEmodels \varphiFrm$ and $\AStr, \HypAsg[2] \AEmodels
      \Ff$.
      From the latter, it follows $\HypAsg[2] = \emptyset$, meaning that
      $\HypAsg[1] = \HypAsg$.
      Therefore, we have $\AStr, \HypAsg \AEmodels \varphiFrm$.
      Conversely, assume $\AStr, \HypAsg \AEmodels \varphiFrm$.
      The claim follows from the fact that $(\HypAsg,\EmpHypAsg) \in
      \parFun{\HypAsg}$ is such that $\AStr, \HypAsg \AEmodels \varphiFrm$ and
      $\AStr, \EmpHypAsg \AEmodels \Ff$ (by semantics of $\Ff$), which implies
      that $\AStr, \HypAsg \AEmodels \varphiFrm \vee \Ff$.
      To conclude, observe that $\varphiFrm \vee \Ff \equiv \Ff \vee
      \varphiFrm$ holds, due to commutativity of $\vee$, formally proved below
      (Item~\ref{thm:blnlawi(com:dis)}).
    \end{enumerate}

  \item Both Items~\ref{thm:blnlawi(com:con)} and~\ref{thm:blnlawi(com:dis)}
    follow from the observation that $(\HypAsg[1], \HypAsg[2]) \in
    \parFun{\HypAsg}$ \iff $(\HypAsg[2], \HypAsg[1]) \in \parFun{\HypAsg}$.

  \item
    \begin{enumerate}[a)]
    \item If $\AStr, \HypAsg \EAmodels \varphiFrm[1] \LCon \varphiFrm[2]$,
      then for all $(\HypAsg[1], \HypAsg[2]) \in \parFun{\HypAsg}$, it holds
      that $\AStr, \HypAsg[1] \EAmodels \varphiFrm[1]$ or $\AStr, \HypAsg[2]
      \EAmodels \varphiFrm[2]$.
      In particular, since $(\HypAsg, \EmpHypAsg) \in \parFun{\HypAsg}$, we have
      that $\AStr, \HypAsg \EAmodels \varphiFrm[1]$.

    \item The claim follows from the observation that partitioning is
      associative.
    \end{enumerate}

  \item
    \begin{enumerate}[a)]
    \item Assume $\AStr, \HypAsg \AEmodels \varphiFrm[1]$.
      The claim follows from the fact that $(\HypAsg,\EmpHypAsg) \in
      \parFun{\HypAsg}$ is such that $\AStr, \HypAsg \AEmodels \varphiFrm[1]$
      and $\AStr, \EmpHypAsg \AEmodels \varphiFrm[2]$ (by
      Item~\ref{lmm:empnulhyp(ae:emp)} of Lemma~\ref{lmm:empnulhyp}), which
      implies that $\AStr, \HypAsg \AEmodels \varphiFrm[1] \vee \varphiFrm[2]$.

    \item The claim follows from the observation that partitioning is
      associative.
    \end{enumerate}

  \item
    \begin{enumerate}[a)]
    \item
      $\AStr, \HypAsg \EAmodels \neg (\neg \varphiFrm[1] \lor \neg
      \varphiFrm[2]) \iffExpl{sem.} \AStr, \HypAsg \not\cmodels[][\QAE] \neg
      \varphiFrm[1] \lor \neg \varphiFrm[2] \iffExpl{}$ it does not holds that
      $\AStr, \HypAsg \AEmodels \neg \varphiFrm[1] \lor \neg \varphiFrm[2]
      \iffExpl{sem.}$ there is no $(\HypAsg[1], \HypAsg[2]) \in
      \parFun{\HypAsg}$ such that $\AStr, \HypAsg[1] \AEmodels
      \neg\varphiFrm[1]$ and $\AStr, \HypAsg[2] \AEmodels \neg\varphiFrm[2]
      \iffExpl{}$ for all $(\HypAsg[1], \HypAsg[2]) \in \parFun{\HypAsg}$ it
      holds that $\AStr, \HypAsg[1] \not\cmodels[][\QAE] \neg\varphiFrm[1]$ or
      $\AStr, \HypAsg[2] \not\cmodels[][\QAE] \neg\varphiFrm[2] \iffExpl{sem.}$
      for all $(\HypAsg[1], \HypAsg[2]) \in \parFun{\HypAsg}$ it holds that
      $\AStr, \HypAsg[1] \EAmodels \varphiFrm[1]$ or $\AStr, \HypAsg[2]
      \EAmodels \varphiFrm[2] \iffExpl{sem.} \AStr, \HypAsg \EAmodels
      \varphiFrm[1] \land \varphiFrm[2]$.

    \item $\AStr, \HypAsg \AEmodels \neg (\neg \varphiFrm[1] \land \neg
      \varphiFrm[2]) \iffExpl{sem.} \AStr, \HypAsg \not\cmodels[][\QEA] \neg
      \varphiFrm[1] \land \neg \varphiFrm[2] \iffExpl{}$ it does not holds that
      $\AStr, \HypAsg \EAmodels \neg \varphiFrm[1] \land \neg \varphiFrm[2]
      \iffExpl{sem.}$ there is $(\HypAsg[1], \HypAsg[2]) \in \parFun{\HypAsg}$
      such that $\AStr, \HypAsg[1] \not\cmodels[][\QEA] \neg\varphiFrm[1]$ and
      $\AStr, \HypAsg[2] \not\cmodels[][\QEA] \neg\varphiFrm[2] \iffExpl{sem.}$
      there is $(\HypAsg[1], \HypAsg[2]) \in \parFun{\HypAsg}$ such that $\AStr,
      \HypAsg[1] \AEmodels \varphiFrm[1]$ and $\AStr, \HypAsg[2] \AEmodels
      \varphiFrm[2] \iffExpl{sem.} \AStr, \HypAsg \AEmodels \varphiFrm[1] \vee
      \varphiFrm[2]$.
    \end{enumerate}

  \item
    \begin{enumerate}[a)]
    \item
      $\AStr, \HypAsg \EAmodels \neg (\LAll[][_{\pm\WSet}] \varElm \ldotp \neg
      \varphiFrm) \iffExpl{sem.}$ $\AStr, \HypAsg \notcmodels[][\QAE]
      \LAll[][_{\pm\WSet}] \varElm \ldotp \neg \varphiFrm \iffExpl{sem.}
      \AStr, \extFun[\denot{\pm\WSet}]{\HypAsg, \varElm} \not\cmodels[][\QAE]
      \neg \varphiFrm$ $\iffExpl{sem.} \AStr, \extFun[\denot{\pm\WSet}]{\HypAsg,
        \varElm} \EAmodels \varphiFrm $ $ \iffExpl{sem.} \AStr, \HypAsg
      \EAmodels \LExs[][_{\pm\WSet}] \varElm \ldotp \varphiFrm$.

    \item $\AStr, \HypAsg \AEmodels \neg (\LExs[][_{\pm\WSet}] \varElm \ldotp
      \neg \varphiFrm) \iffExpl{sem.} \AStr, \HypAsg \not\cmodels[][\QEA]
      \LExs[][_{\pm\WSet}] \varElm \ldotp \neg \varphiFrm \iffExpl{sem.}
      \AStr, \extFun[\denot{\pm\WSet}]{\HypAsg, \varElm} \not\cmodels[][\QEA]
      \neg \varphiFrm$ $\iffExpl{sem.} \AStr, \extFun[\denot{\pm\WSet}]{\HypAsg,
        \varElm} \AEmodels \varphiFrm \iffExpl{sem.} \AStr, \HypAsg \AEmodels
      \LAll[][_{\pm\WSet}] \varElm \ldotp \varphiFrm$.
      \qedhere
    \end{enumerate}
  \end{enumerate}
\end{proof}


\figGraphThmPrfExt

\thmprfext*
\begin{proof}
  The claim follows from the more general Theorem~\ref{thm:prfextMeta}, reported
  in~\ref{app:metthr}, by instantiating $\FunAsg$ with the empty function
  $\emptyfun$.
\end{proof}

The following result states \emph{monotonicity} of the dualization, extension and partition operators \wrt the preorder $\sqsubseteq$.

\begin{lemma}[name = Monotonicity I]
	\label{lmm:mon}
	Let $\HypAsg, \HypAsg' \!\in\! \HypAsgSet$ be two hyperteams with $\HypAsg
	\inc[\WSet] \HypAsg'$, for some $\WSet \subseteq \VarSet$.
	Then, the following hold true:
	\begin{enumerate}[1)]
		\item\label{lmm:mon(dlt)}
		$\dual{\HypAsg'} \inc[\WSet] \dual{\HypAsg}$;
		\item\label{lmm:mon(ext)}
                  \begin{enumerate}[a)]
                  \item\label{lmm:mon(ext:eql)}
                    $\HypAsg \eql[\WSet] \extFun[\USet]{\HypAsg, \varElm}$, if
                    $\varElm \not\in \WSet$, with $\USet \subseteq \VarSet$;
                  \item\label{lmm:mon(ext:dep)}
                    $\extFun[\USet]{\HypAsg, \varElm} \inc[\WSet \cup \{ \varElm
                    \}] \extFun[\USet']{\HypAsg', \varElm}$, with $\varElm \in
                    \VarSet$, $\USet \subseteq \USet' \subseteq \VarSet$, and
                    $\USet \subseteq \WSet$;
		\end{enumerate}
%
%
		\item\label{lmm:mon(par-unrestricted)} 
		for every $(\HypAsg[1]', \HypAsg[2]') \in \parFun[]{\HypAsg'}$,
		there is $(\HypAsg[1], \HypAsg[2]) \in \parFun[]{\HypAsg}$ such
		that $\HypAsg[1] \inc[\WSet] \HypAsg[1]'$ and $\HypAsg[2] \inc[\WSet]
		\HypAsg[2]'$.
	\end{enumerate}
\end{lemma}
\begin{proof}
%
%
  \begin{itemize}
  \item[\ref{lmm:mon(dlt)})] By $\HypAsg \inc[\WSet] \HypAsg'$, there is a
    function $f : \HypAsg \!\rst[\WSet] \rightarrow \HypAsg' \!\rst[\WSet]$ such
    that $f(\TeamAsg \!\rst[\WSet]) \subseteq \TeamAsg \!\rst[\WSet]$ for all
    $\TeamAsg \in \HypAsg$.
    Moreover, for all $\TeamAsg \in \HypAsg$, since $f(\TeamAsg \!\rst[\WSet])
    \subseteq \TeamAsg \!\rst[\WSet]$, there is a function $g_{\TeamAsg} :
    \bigcup \{ \TeamAsg' \in \HypAsg' \mid \TeamAsg' \!\rst[\WSet] = f(\TeamAsg
    \!\rst[\WSet]) \} \rightarrow \TeamAsg$ such that $\asgElm \!\rst[\WSet] =
    (g_{\TeamAsg}(\asgElm)) \!\rst[\WSet]$ for all $\asgElm$ in $\bigcup \{ \TeamAsg' \in \HypAsg' \mid \TeamAsg' \!\rst[\WSet] = f(\TeamAsg
    \!\rst[\WSet]) \}$.
    In order to prove the claim, consider a generic team $\TeamAsg' \in
    \dual{\HypAsg'} \!\rst[\WSet]$.
    We have to show that there is $\TeamAsg \in \dual{\HypAsg}$
    such that $\TeamAsg \!\rst[\WSet] \subseteq \TeamAsg'$.
    By the definition of $\dual{\HypAsg'} \!\rst[\WSet]$, we have that
    $\TeamAsg' = (\img{\chcFun'}) \!\rst[\WSet]$, for some $\chcFun' \in
    \ChcSet{\HypAsg'}$.
    We define $\chcFun \in \ChcSet{\HypAsg}$ as: $\chcFun(\TeamAsg) =
    g_{\TeamAsg}(\chcFun'(\asgExtW{(f(\TeamAsg \!\rst[\WSet]))}))$ for all
    $\TeamAsg \in \HypAsg$.
    Clearly, $(\img{\chcFun}) \!\rst[\WSet] \subseteq (\img{\chcFun'})
    \!\rst[\WSet] = \TeamAsg'$.
    Since $(\img{\chcFun})\in \dual{\HypAsg}$, the thesis holds.

  \item[\ref{lmm:mon(ext:eql)})] The claim follows from the fact that for every
    $\FFun \in \FncSet{}$, $\asgElm \in \AsgSet$, and $\varElm \not\in \WSet$,
    it holds that $\extFun{\asgElm, \FFun, \varElm} \!\rst[\WSet] = \asgElm
    \!\rst[\WSet]$, which implies $\extFun{\TeamAsg,
      \FFun, \varElm} \!\rst[\WSet] = \TeamAsg \!\rst[\WSet]$ for every
    $\TeamAsg \in \HypAsg$ and $\FFun \in \FncSet[\USet]{}$, and the claim
    follows.

  \item[\ref{lmm:mon(ext:dep)})]
    By $\HypAsg \inc[\WSet] \HypAsg'$, there is a function $f : \HypAsg
    \!\rst[\WSet] \rightarrow \HypAsg' \!\rst[\WSet]$ such that $f(\TeamAsg
    \!\rst[\WSet]) \subseteq \TeamAsg \!\rst[\WSet]$ for all $\TeamAsg \in
    \HypAsg$.
    In order to prove the claim, take a generic team $\hat{\TeamAsg} \in
    \extFun[\USet]{\HypAsg, \varElm}$.
    Thus, $\hat{\TeamAsg} = \extFun{\TeamAsg, \FFun, \varElm} =
    \set{\extFun{\asgElm, \FFun, \varElm}}{{ \asgElm \in \TeamAsg }}$, for some
    $\TeamAsg \in \HypAsg$ and $\FFun \in \FncSet[\USet]{}$.
    Let $\TeamAsg' = \asgExtW{(f(\TeamAsg \!\rst[\WSet]))} \in \HypAsg'$.
    Clearly, $\TeamAsg' \!\rst[\WSet] = f(\TeamAsg \!\rst[\WSet]) \subseteq
    \TeamAsg \!\rst[\WSet]$.
    Moreover, $\extFun{\TeamAsg', \FFun, \varElm} \in \extFun[\USet']{\HypAsg',
      \varElm}$, since $\FFun \in \FncSet[\USet]{} \subseteq \FncSet[\USet']{}$
    (as $\USet \subseteq \USet'$).
    To complete the proof, it is enough to show that $\extFun{\TeamAsg', \FFun,
      \varElm} \!\rst[\WSet \cup \{ \varElm \}] \subseteq \hat{\TeamAsg}
    \!\rst[\WSet \cup \{ \varElm \}]$.
    To this purpose, take $\extFun{\asgElm', \FFun, \varElm} \!\rst[\WSet \cup
    \{ \varElm \}]$ for some $\asgElm' \in \TeamAsg'$.
    Observe that $\asgElm' \!\rst[\WSet] \in \TeamAsg' \!\rst[\WSet] =
    f(\TeamAsg \!\rst[\WSet]) \subseteq \TeamAsg \!\rst[\WSet]$, which means
    that there is $\asgElm \in \TeamAsg$ such that $\asgElm \!\rst[\WSet] =
    \asgElm' \!\rst[\WSet]$.
    Since $\USet \subseteq \WSet$, it holds that $\asgElm \!\rst[\USet] =
    \asgElm' \!\rst[\USet]$, which implies $\FFun(\asgElm) = \FFun(\asgElm')$,
    as $\FFun \in \FncSet[\USet]{}$.
    Therefore, $\extFun{\asgElm', \FFun, \varElm} \!\rst[\WSet \cup \{ \varElm
    \}] = \extFun{\asgElm, \FFun, \varElm} \!\rst[\WSet \cup \{ \varElm \}] \in
    \hat{\TeamAsg} \!\rst[\WSet \cup \{ \varElm \}]$.

%

  \item[\ref{lmm:mon(par-unrestricted)})]
    By $\HypAsg \inc[\WSet] \HypAsg'$, there is a function $f : \HypAsg
    \!\rst[\WSet] \rightarrow \HypAsg' \!\rst[\WSet]$ such that $f(\TeamAsg
    \!\rst[\WSet]) \subseteq \TeamAsg \!\rst[\WSet]$ for all $\TeamAsg \in
    \HypAsg$.
    Let $(\HypAsg[1]', \HypAsg[2]') \in \parFun{\HypAsg'}$ and define
    $\HypAsg[i] = \{ \TeamAsg \in \HypAsg \mid \asgExtW{(f(\TeamAsg
      \!\rst[\WSet]))} \in \HypAsg[i]' \}$ for $i \in \{ 1,2 \}$.
    We have to show that $\HypAsg[i] \inc[\WSet] \HypAsg[i]'$ ($i \in \{ 1,2
    \}$).
    To this end, let $\TeamAsg \in \HypAsg[i]$ and consider team
    $\asgExtW{(f(\TeamAsg \!\rst[\WSet]))} \in \HypAsg[i]'$.
    Clearly, $(\asgExtW{(f(\TeamAsg \!\rst[\WSet]))}) \!\rst[\WSet] = f(\TeamAsg
    \!\rst[\WSet]) \subseteq \TeamAsg \!\rst[\WSet]$.
    The thesis follows as $(\asgExtW{(f(\TeamAsg \!\rst[\WSet]))}) \!\rst[\WSet]
    \in \HypAsg[i]' \!\rst[\WSet]$.
    \qedhere
  \end{itemize}

\end{proof}





\section{Proofs of Section~\ref{sec:adq}}
\label{app:adq}

\lmmfoldlt*

\begin{proof}
  The first equivalence follows from Lemma~\ref{lmm:dltii},
  Item~\ref{lmm:dltii(ea)}, by letting $\PsiSet = \set{\asgElm \in
    \AsgSet[\subseteq](\sup{\varphiFrm})}{ \AStr, \asgElm \cmodels[\FOL] \varphiFrm }$.
  The second equivalence follows from the first one and from $\HypAsg \equiv
  \dual{\dual{\HypAsg}}$ (Lemma~\ref{lmm:dlti}).
\end{proof}

\lmmfolqnt*
\begin{proof}

  \begin{itemize}
  \item[($\ref{lmm:folqnt(ea:org)} \Rightarrow \ref{lmm:folqnt(ea:ext)}$)] Let
    $\XSet \in \HypAsg$ be such that $\AStr, \asgElm \cmodels[\FOL] \LExs
    \varElm \ldotp \varphiFrm$ holds for every $\asgElm \in \XSet$. By the
    standard \FOL semantics, for every $\asgElm \in \XSet$, there is an element
    $\aElm[\asgElm] \in \ASet$ such that $\AStr, {\asgElm}[\varElm \mapsto
    \aElm[\asgElm]] \cmodels[\FOL] \varphi$.
    We safely assume that $\aElm[\asgElm_1] = \aElm[\asgElm_2]$ whenever
    $\asgElm_1 \!\rst[\VSet] = \asgElm_2 \!\rst[\VSet]$, for all $\asgElm_1,
    \asgElm_2 \in \XSet$.
    Let $\FFun \in \FncSet[\VSet]{}$ be such that $\FFun(\asgElm) =
    \aElm[\asgElm]$ for every $\asgElm \in \XSet$ and let $\XSet_{\FFun} = \{
    {\asgElm}[\varElm \mapsto \FFun(\asgElm)] : \asgElm \in \XSet \}$.
    Since $\XSet_{\FFun} \in \extFun[\VSet]{\HypAsg, \varElm}$ and $\AStr,
    \asgElm \cmodels[\FOL] \varphi$ holds for every $\asgElm \in \XSet_{\FFun}$,
    the thesis holds.

  \item[($\ref{lmm:folqnt(ea:ext)} \Rightarrow \ref{lmm:folqnt(ea:org)}$)] Let
    $\XSet_{\FFun} = \{ {\asgElm}[\varElm \mapsto \FFun(\asgElm)] : \asgElm \in
    \XSet \} \in \extFun[\VSet]{\HypAsg, \varElm}$, for some $\XSet \in \HypAsg$
    and $\FFun \in \FncSet[\VSet]{}$, be such that $\AStr, \asgElm
    \cmodels[\FOL] \varphiFrm$ holds for every $\asgElm \in \XSet_{\FFun}$.
%
%
    Clearly, by the standard \FOL semantics, this implies that $\AStr, \asgElm
    \cmodels[\FOL] \LExs \varElm \ldotp \varphiFrm$ holds for every $\asgElm \in
    \XSet$, hence the thesis.

  \item[($\ref{lmm:folqnt(ae:org)} \Leftrightarrow \ref{lmm:folqnt(ae:ext)}$)]
    By statement~\ref{lmm:folqnt(ea)} of this lemma, we have that
    $\ref{lmm:folqnt(ea:org)}$ is false if and only if
    $\ref{lmm:folqnt(ea:ext)}$ is false ($\mathit{not} \
    \ref{lmm:folqnt(ea:org)} \Leftrightarrow \mathit{not} \
    \ref{lmm:folqnt(ea:ext)}$, for short).
    By instantiating, in this last equivalence, $\varphi$ with $\neg\varphi$, we
    have $\ref{lmm:folqnt(ea:org)}' \Leftrightarrow \ref{lmm:folqnt(ea:ext)}'$,
    where $\ref{lmm:folqnt(ea:org)}'$ and $\ref{lmm:folqnt(ea:ext)}'$ are
    abbreviations for, respectively:

    \begin{itemize}
    \item for all teams $\XSet \in \HypAsg$, there exists an assignment $\asgElm
      \in \XSet$ such that $\AStr, \asgElm \not\cmodels[\FOL] \exists \varElm
      . \neg \varphi$;
    \item for all teams $\XSet \in \extFun[\VSet]{\HypAsg, \varElm}$, there
      exists an assignment $\asgElm \in \XSet$ such that $\AStr, \asgElm
      \not\cmodels[\FOL] \neg \varphi$.
    \end{itemize}

    By applying standard \FOL semantics for negation and the duality of
    $\exists$ and $\forall$ in standard \FOL, it is straightforward to see that
    $\ref{lmm:folqnt(ea:org)}'$ and $\ref{lmm:folqnt(ea:ext)}'$ correspond to
    $\ref{lmm:folqnt(ae:org)}$ and $\ref{lmm:folqnt(ae:ext)}$, respectively,
    hence the thesis. \qedhere
  \end{itemize}
\end{proof}

\lmmfolblncon*
\begin{proof}

\begin{itemize}
\item[($\ref{lmm:folblncon(ea:org)} \Rightarrow \ref{lmm:folblncon(ea:par)}$)]
%
%
  Let $\XSet \in \HypAsg$ be such that $\AStr, \asgElm \cmodels[\FOL]
  \varphi_{1} \wedge \varphi_{2}$ holds for every $\asgElm \in \XSet$ and
  consider an arbitrary pair $(\HypAsg[1], \HypAsg[2]) \in \parFun[]{\HypAsg}$.
  Since $(\HypAsg[1], \HypAsg[2])$ is a partition of $\HypAsg$, either $\XSet
  \in \HypAsg[1]$ or $\XSet \in \HypAsg[2]$: in the former case, let $i=1$; in
  the latter, let $i = 2$.
  Since $\XSet \in \HypAsg[i]$ and $\AStr, \asgElm \cmodels[\FOL] \varphi_{i} $
  holds for every $\asgElm \in \XSet$, the thesis holds.

\item[($\ref{lmm:folblncon(ea:par)} \Rightarrow \ref{lmm:folblncon(ea:org)}$)]
  Consider the hyperteam $\HypAsg[1]' = \{ \XSet \in \HypAsg : \forall
  \asgElm \in \XSet \ . \ \AStr, \asgElm \cmodels[\FOL] \varphi_1 \}$ and the
  pair $(\HypAsg[1] \defeq \HypAsg \setminus \HypAsg[1]', \HypAsg[2] \defeq
  \HypAsg[1]') \in \parFun[]{\HypAsg}$.
  Observe that, by definition of $\HypAsg[1]$, there is no $\XSet \in
  \HypAsg[1]$ such that $\AStr, \asgElm \cmodels[\FOL] \varphi_{1}$ holds for
  every $\asgElm \in \XSet$.
  Thus, by~$\ref{lmm:folblncon(ea:par)}$, there must exist $\XSet \in
  \HypAsg[2]$ such that $\AStr, \asgElm \cmodels[\FOL] \varphi_{2}$ holds for
  every $\asgElm \in \XSet$.
  By definition of $\HypAsg[2]$, it also holds that $\AStr, \asgElm
  \cmodels[\FOL] \varphi_{1}$ for every $\asgElm \in \XSet$, hence the thesis.

\item[($\ref{lmm:folblncon(ae:org)} \Leftrightarrow
  \ref{lmm:folblncon(ae:org)}$)] By statement~\ref{lmm:folblncon(ea)} of this
  lemma, we have that $\ref{lmm:folblncon(ea:org)}$ is false if and only if
  $\ref{lmm:folblncon(ea:par)}$ is false ($\mathit{not} \
  \ref{lmm:folblncon(ea:org)} \Leftrightarrow \mathit{not} \
  \ref{lmm:folblncon(ea:par)}$, for short).
  By instantiating, in this last equivalence, $\varphi_1$ with $\neg\varphi_1$
  and $\varphi_2$ with $\neg\varphi_2$, we have $\ref{lmm:folblncon(ea:org)}'
  \Leftrightarrow \ref{lmm:folblncon(ea:par)}'$, where
  $\ref{lmm:folblncon(ea:org)}'$ and $\ref{lmm:folblncon(ea:par)}'$ are
  abbreviations for, respectively:

  \begin{itemize}
  \item for all teams $\XSet \in \HypAsg$, there exists an assignment $\asgElm
    \in \XSet$ such that $\AStr, \asgElm \not\cmodels[\FOL] \neg\varphi_{1}
    \wedge \neg\varphi_{2}$;
  \item there exists a pair of hyperteams $(\HypAsg[1],
    \HypAsg[2]) \in \parFun[]{\HypAsg}$ such that, for all indexes $i \in \{ 1,
    2 \}$ and teams $\XSet \in \HypAsg[i]$, there exists an assignment $\asgElm
    \in \XSet$ for which it holds that $\AStr, \asgElm \not\cmodels[\FOL]
    \neg\varphi_{i}$.
  \end{itemize}

  By applying semantics of negation and De Morgan's laws, it is straightforward
  to see that $\ref{lmm:folblncon(ea:org)}'$ and $\ref{lmm:folblncon(ea:par)}'$
  correspond to $\ref{lmm:folblncon(ae:org)}$ and $\ref{lmm:folblncon(ae:org)}$,
  respectively, hence the thesis.      \qedhere

\end{itemize}

\end{proof}

\figGraphThmFolSemAdq

\thmfolsemadq*
\begin{proof}
  Both Items 1 and 2 are proved together, by induction on the structure of the
  formula.

  \begin{itemize}
  \item If $\varphi$ is an atomic formula, \ie, it is $\bot$ or $\top$, or it
    has the form $\RRel(\xVec)$, then the claims immediately follow from the
    semantics (Definition~\ref{def:sem(adif)},
    Items~\ref{def:sem(adif:fbv)}--\ref{def:sem(adif:rel)}).

  \item If $\varphi = \neg \phi$, then we have, by semantics,
    $\AStr, \HypAsg \cmodels[][\alpha]\: \varphi$ if and only if $\AStr, \HypAsg
    \notcmodels[][\dual{\alpha}]\: \phi$.
    If $\alpha = \exists\forall$, then, by inductive hypothesis, it is not the
    case that for every $\XSet \in \HypAsg$ there is $\asgElm \in \XSet$ such
    that $\AStr, \asgElm \cmodels[\FOL] \phi$, which amounts to say that there
    is $\XSet \in \HypAsg$ such that for every $\asgElm \in \XSet$ it holds
    $\AStr, \asgElm \notcmodels[\FOL] \phi$, from which the thesis follows.
    If, instead, $\alpha = \forall\exists$, then, by inductive hypothesis, there
    is no $\XSet \in \HypAsg$ such that for every $\asgElm \in \XSet$ it holds
    $\AStr, \asgElm \cmodels[\FOL] \phi$, which amounts to say that for every
    $\XSet \in \HypAsg$ there is $\asgElm \in \XSet$ such that $\AStr, \asgElm
    \notcmodels[\FOL] \phi$, from which the thesis follows.

  \item If $\varphi = \varphi_1 \wedge\varphi_2$ and $\alpha = \exists\forall$,
    then we have, by semantics, $\AStr, \HypAsg \cmodels[][\alpha]\: \varphi$ if
    and only if for every $(\HypAsg[1], \HypAsg[2]) \in \parFun[]{\HypAsg}$
%
%
    it holds that $\AStr, \HypAsg[1] \cmodels[][\alpha] \varphi_{1}$ or $\AStr,
    \HypAsg[2] \cmodels[][\alpha] \varphi_{2}$.
    By inductive hypothesis, this amounts to say that for every $(\HypAsg[1],
    \HypAsg[2]) \in \parFun[]{\HypAsg}$ there is $i \in \{ 1,2 \}$ and $\XSet
    \in \HypAsg[i]$ such that for every $\asgElm \in \XSet$ it holds $\AStr,
    \asgElm \cmodels[\FOL]\varphi_{i}$.
    The thesis follows from Lemma~\ref{lmm:folblncon},
    Item~\ref{lmm:folblncon(ea)}.

    If $\varphi = \varphi_1 \wedge\varphi_2$ and $\alpha = \forall\exists$, then
    we have, by semantics, $\AStr, \HypAsg \cmodels[][\alpha]\: \varphi$ if and
    only if $\AStr, \dual{\HypAsg} \cmodels[][\dual{\alpha}]\: \varphi$.
    By proceeding as before, i.e., by applying semantics, inductive hypothesis,
    and Lemma~\ref{lmm:folblncon}, Item~\ref{lmm:folblncon(ea)}, we have that
    there is $\XSet' \in \dual{\HypAsg}$ such that for every $\asgElm' \in
    \XSet'$ it holds $\AStr, \asgElm' \cmodels[\FOL]\varphi$.
    The thesis follows from Lemma~\ref{lmm:foldlt}, Item~\ref{lmm:foldlt(ae)}.

  \item If $\varphi = \varphi_1 \vee\varphi_2$ and $\alpha = \forall\exists$,
    then we have, by semantics, $\AStr, \HypAsg \cmodels[][\alpha]\: \varphi$ if
    and only if there is $(\HypAsg[1], \HypAsg[2]) \in \parFun[]{\HypAsg}$ such
    that $\AStr, \HypAsg[1] \cmodels[][\alpha] \varphi_{1}$ and $\AStr,
    \HypAsg[2] \cmodels[][\alpha] \varphi_{2}$.
    By inductive hypothesis, this amounts to say that there is $(\HypAsg[1],
    \HypAsg[2]) \in \parFun[]{\HypAsg}$ such that for every $i \in \{ 1,2 \}$
    and $\XSet \in \HypAsg[i]$ there is $\asgElm \in \XSet$ for which it holds
    $\AStr, \asgElm \cmodels[\FOL]\varphi_{i}$.
    The thesis follows from Lemma~\ref{lmm:folblncon},
    Item~\ref{lmm:folblncon(ae)}.

    If $\varphi = \varphi_1 \vee\varphi_2$ and $\alpha = \exists\forall$, then
    we have, by semantics, $\AStr, \HypAsg \cmodels[][\alpha]\: \varphi$ if and
    only if $\AStr, \dual{\HypAsg} \cmodels[][\dual{\alpha}]\: \varphi$.
    By proceeding as before, i.e., by applying semantics, inductive hypothesis,
    and Lemma~\ref{lmm:folblncon}, Item~\ref{lmm:folblncon(ae)}, we have that
    for every $\XSet' \in \dual{\HypAsg}$ there is $\asgElm' \in \XSet'$ such
    that $\AStr, \asgElm' \cmodels[\FOL]\varphi$.
    The thesis follows from Lemma~\ref{lmm:foldlt}, Item~\ref{lmm:foldlt(ea)}.

  \item If $\varphi = \exists \varElm . \phi$ and $\alpha = \exists\forall$,
    then we have, by semantics, $\AStr, \HypAsg \cmodels[][\alpha]\: \varphi$ if
    and only if $\AStr, \extFun[\sup{\phi} \setminus \{ \varElm \}]{\HypAsg,
      \varElm} \cmodels[][\alpha] \phi$.
    By inductive hypothesis, this amounts to say that there is $\XSet \in
    \extFun[\sup{\phi} \setminus \{ \varElm \}]{\HypAsg, \varElm}$ such that for
    every $\asgElm \in \XSet$ it holds $\AStr, \asgElm \cmodels[\FOL]\phi$.
    The thesis follows from Lemma~\ref{lmm:folqnt}, Item~\ref{lmm:folqnt(ea)}.

    If $\varphi = \exists \varElm . \phi$ and $\alpha = \forall\exists$, then we
    have, by semantics, $\AStr, \HypAsg \cmodels[][\alpha]\: \varphi$ if and
    only if $\AStr, \dual{\HypAsg} \cmodels[][\dual{\alpha}]\: \varphi$.
    By proceeding as before, i.e., by applying semantics, inductive hypothesis,
    and Lemma~\ref{lmm:folqnt}, Item~\ref{lmm:folqnt(ea)}, we have that there is
    $\XSet' \in \dual{\HypAsg}$ such that for every $\asgElm' \in \XSet'$ it
    holds $\AStr, \asgElm' \cmodels[\FOL]\varphi$.
    The thesis follows from Lemma~\ref{lmm:foldlt}, Item~\ref{lmm:foldlt(ae)}.

  \item If $\varphi = \forall \varElm . \phi$ and $\alpha = \forall\exists$,
    then we have, by semantics, $\AStr, \HypAsg \cmodels[][\alpha]\: \varphi$ if
    and only if $\AStr, \extFun[\sup{\phi} \setminus \{ \varElm \}]{\HypAsg,
      \varElm} \cmodels[][\alpha] \phi$.
    By inductive hypothesis, this amounts to say that for every $\XSet \in
    \extFun[\sup{\phi} \setminus \{ \varElm \}]{\HypAsg, \varElm}$ there is
    $\asgElm \in \XSet$ such that $\AStr, \asgElm \cmodels[\FOL]\phi$.
    The thesis follows from Lemma~\ref{lmm:folqnt}, Item~\ref{lmm:folqnt(ae)}.

    If $\varphi = \forall \varElm . \phi$ and $\alpha = \exists\forall$, then we
    have, by semantics, $\AStr, \HypAsg \cmodels[][\alpha]\: \varphi$ if and
    only if $\AStr, \dual{\HypAsg} \cmodels[][\dual{\alpha}] \varphi$.
    By proceeding as before, i.e., by applying semantics, inductive hypothesis,
    and Lemma~\ref{lmm:folqnt}, Item~\ref{lmm:folqnt(ae)}, we have that for
    every $\XSet' \in \dual{\HypAsg}$ there is $\asgElm' \in \XSet'$ such that
    $\AStr, \asgElm' \cmodels[\FOL]\varphi$.
    The thesis follows from Lemma~\ref{lmm:foldlt}, Item~\ref{lmm:foldlt(ea)}.
    \qedhere
  \end{itemize}
\end{proof}

\lmmcylext*
\begin{proof}
	The proof is done by showing the two directions of the equivalence.

	First, we prove the following:
        $$\cylFun{\HypAsg,\varElm}\sqsubseteq \dual{\extFun[\WSet]{\dual{\HypAsg}, \varElm}}.$$
	Let $\Team[u] \in \cylFun{\HypAsg, \varElm}$. There is $\Team\in\HypAsg$
	such that $\Team[u] = \cylFun{\Team, \varElm}$. Remark that for every
	$\TeamInDual\in{\dual{\HypAsg}}$ there is
	${\asgElm[\TeamInDual]}\in\TeamInDual\!\cap\!\Team$. Then, for every
	$\FFun \in \FncSet[\WSet]{}$, it holds that
	${\asgElm[\TeamInDual]}[\varElm \mapsto
	\FFun({\asgElm[\TeamInDual]})]\in\Team[u]$.
        Now, observe that for every $\TeamHat \in \extFun[\WSet]{\dual{
            \HypAsg}, \varElm}$, there is $\TeamInDual \in \dual{\HypAsg}$ and
        $\FFun \in \FncSet[\WSet]{}$ such that $\TeamHat =
        \extFun{\TeamInDual,\FFun,\varElm}$.
        Consider $\chcFun \in \ChcSet{ {\extFun[\WSet]{ \dual{\HypAsg},
              \varElm}}}$ defined as follows.
        For every $\TeamHat \in \extFun[\WSet]{\dual{\HypAsg}, \varElm}$, we
	define $\chcFun(\TeamHat) =
	\asgElm[{\TeamInDual}]{}[\varElm\mapsto\FFun(\asgElm[\TeamInDual])]$.
        We can deduce immediately that $\img{\chcFun}\subseteq\Team[u]$.

	We turn now to showing that
	$$\dual{\extFun[\WSet]{\dual{\HypAsg}, \varElm}}\sqsubseteq\cylFun{\HypAsg,\varElm}.$$
	Let $\TeamInDDual \in \dual{\extFun[\WSet]{\dual{\HypAsg}, \varElm}}$.
        We have $\TeamInDDual = \img{\mathring{\chcFun}}$ for some choice
        function $\mathring{\chcFun} \in \ChcSet{ {\extFun[\WSet]{ \dual{
                \HypAsg}, \varElm}}}$.
        Then,
        \begin{equation}\label{eqn:cyl}
          \forall \FFun \in \FncSet[\WSet]{}, \forall
          \TeamInDual\in{\dual{\HypAsg}}, \exists \asgElm'\in\TeamInDual \text{
            s.t. } \asgElm'[\varElm\mapsto\FFun(\asgElm')]\in\TeamInDDual.
	\end{equation} Toward contradiction, assume that $\cylFun{\Team,
          \varElm} \not\subseteq \TeamInDDual$ for all $\Team \in \HypAsg$. Then
	for all $\Team \in \HypAsg$, there is $\asgElm[\Team]\in\Team$ and
	$\domainValue_{\Team} \in \ASet$ such that
	$\asgElm[\Team]{}[\varElm\mapsto\domainValue_{\Team}]\notin
	\TeamInDDual$. We assume that
	$\domainValue_{\Team[1]}=\domainValue_{\Team[2]}$ if
	$\asgElm[{\Team[1]}]=\asgElm[{\Team[2]}]$ so that each $\asgElm$ is
	associated with only one $\domainValue \in \ASet$. Consider $\chcFun \in
	\ChcSet{\HypAsg}$ such that $\chcFun(\Team) = \asgElm[\Team]$ for all
	$\Team \in \HypAsg$, and $\FFun \in \FncSet[\WSet]{}$ such that
	$\FFun(\asgElm[\Team]) = \domainValue_{\Team}$ for all $\Team \in
	\HypAsg$.
        By construction, for all $\asgElm' \in \img{\chcFun}$, it holds that
	$\asgElm'[\varElm\mapsto\FFun(\asgElm')]\notin\TeamInDDual$ and, since
	$\img{\chcFun} \in \dual{\HypAsg}$, we have a contradiction with
	$\eqref{eqn:cyl}$.
\end{proof}

\lmmteapar*
\begin{proof}
  In the following, we assume index $i$ to range over $\{1,2\}$.
  \begin{enumerate}[1)]
  \item Let $(\HypAsg[1], \HypAsg[2]) \in \parFun{\dual{\HypAsg}}$ and
    $\YSet[i] \in \dual{\HypAsg[i]}$.
    Then, there are $\chcFun[i] \in \ChcSet{\HypAsg[i]}$ such that $\YSet[i] =
    \img{\chcFun[i]}$. Let $\chcFun \in \ChcSet{\dual{\HypAsg}}$ be defined as:
    $\chcFun(\Team) = \chcFun[i](\Team)$ if $\Team \in \HypAsg[i]$, for all
    $\Team \in \dual{\HypAsg}$.
    It clearly holds that $\img{\chcFun}=\img{\chcFun[1]} \cup \img{\chcFun[2]}$
    and $\img{\chcFun} \in \dual{\dual{\HypAsg}}$.
    Finally, thanks to Lemma~\ref{lmm:dlti}, there is $\Team^\star \in \HypAsg$
    such that $\Team^\star \subseteq \img{\chcFun} = \YSet[1] \cup \YSet[2]$.

  \item Let $\Team \in \HypAsg$ and $(\Team[1], \Team[2]) \in \parFun{\Team}$.
    Consider $\HypAsg[1]$ and $\HypAsg[2]$ defined as follows: $\HypAsg[1] =
    \set{\img{\chcFun}}{\chcFun\in\ChcSet\HypAsg \text{ and }
    \chcFun(\Team)\in\Team[1] }$ and $\HypAsg[2] = \HypAsg \setminus
    \HypAsg[1]$.
    Clearly, it holds that $(\HypAsg[1], \HypAsg[2]) \in
    \parFun{\dual{\HypAsg}}$.
    Moreover, for every $\XSet[i]['] \in \HypAsg[i]$, it holds that $\XSet[i][']
    \cap \XSet[i] \neq \emptyset$.
    Let $\chcFun[i] \in \ChcSet{\HypAsg[i]}$ be such that
    $\chcFun[i](\XSet[i][']) \in \YSet[i] \cap \XSet[i]$, for every $\XSet[i][']
    \in \HypAsg[i]$.
    Then, $\img{\chcFun[i]} \in \dual{\HypAsg[i]}$ is such that
    $\img{\chcFun[i]} \subseteq \XSet[i]$.\qedhere
  \end{enumerate}
\end{proof}


\figGraphThmDifSemAdq

\thmdifsemadq*
\begin{proof}
  In the following, we assume index $i$ to range over $\{1,2\}$.

  To begin with, we prove Item~\ref{thm:difsemadq(ea)}.
  The proof is done by structural induction on the formula
  $\varphiFrm[][_\exists]$.
  \begin{itemize}
  \item[(base case)] If $\varphiFrm[][_\exists]=\RRel(\xVec)$ or
    $\varphiFrm[][_\exists]= \neg \RRel(\xVec)$, then the property holds by the
    semantics rules.
  \item[(inductive cases)] Suppose that the property holds for the subformulae
    of $\varphiFrm[][_\exists]$.
    \begin{itemize}
    \item[{($\varphiFrm[][_\exists]=\varphiFrm[1] \land \varphiFrm[2]$)}]
      $\AStr, \HypAsg \cmodels[][\QEA] \varphiFrm[1] \wedge \varphiFrm[2]
      \iffExpl{sem.}$ for all $(\HypAsg[1],\HypAsg[2])\in\parFun{\HypAsg}$ it
      holds that $\AStr, \HypAsg[1] \cmodels[][\QEA] \varphiFrm[1]$ or $\AStr,
      \HypAsg[2] \cmodels[][\QEA] \varphiFrm[2] \iffExpl{ind.hp.}$ for all
      $(\HypAsg[1],\HypAsg[2])\in\parFun{\HypAsg}$ it holds that there is
      $\TeamAsg[1] \in \HypAsg[1]$ for which it holds $\AStr, \TeamAsg[1]
      \cmodels[\DIF][\forall] \varphiFrm[1]$ or there is $\TeamAsg[2] \in
      \HypAsg[2]$ for which it holds $\AStr, \TeamAsg[2] \cmodels[\DIF][\forall]
      \varphiFrm[2] \iffExpl{}$ there is $\TeamAsg \in \HypAsg$ such that
      $\AStr, \TeamAsg \cmodels[\DIF][\forall] \varphiFrm[1]$ and $\AStr,
      \TeamAsg \cmodels[\DIF][\forall] \varphiFrm[2] \iffExpl{\DIF-sem.}$ there
      is $\TeamAsg \in \HypAsg$ such that $\AStr, \TeamAsg
      \cmodels[\DIF][\forall] \varphiFrm[1] \wedge \varphiFrm[2]$.

    \item[{($\varphiFrm[][_\exists]=\varphiFrm[1] \lor \varphiFrm[2]$)}]
      If $\AStr, \HypAsg \cmodels[][\QEA] \varphiFrm[1] \lor \varphiFrm[2]$,
      then $\AStr, \dual{\HypAsg} \cmodels[][\QAE] \varphiFrm[1] \lor
      \varphiFrm[2]$.
      By semantics, there is $(\HypAsg[1],\HypAsg[2])\in\parFun{\dual{\HypAsg}}$
      such that $\AStr, \HypAsg[1] \AEmodels \varphiFrm[1]$ and $\AStr,
      \HypAsg[2] \AEmodels \varphiFrm[2]$, which amounts to say that there is
      $(\HypAsg[1],\HypAsg[2])\in\parFun{\dual{\HypAsg}}$ such that $\AStr,
      \dual{\HypAsg[1]} \EAmodels \varphiFrm[1]$ and $\AStr, \dual{\HypAsg[2]}
      \EAmodels \varphiFrm[2] $.
      By inductive hypothesis, there are $\Team[1] \in \dual{\HypAsg[1]}$ and
      $\Team[2] \in \dual{\HypAsg[2]}$ such that $\AStr, \Team[1] \IFmodels
      \varphiFrm[1]$ and $\AStr, \Team[2] \IFmodels \varphiFrm[2]$.
      By Item~\ref{lmm:teapar(hyp)} of Lemma~\ref{lmm:teapar}, there is $\Team
      \in \HypAsg$ such that $\Team \subseteq \Team[1] \!\cup\! \Team[2]$.
      Then, $\AStr, \Team \IFmodels \varphiFrm[1] \lor \varphiFrm[2]$.

      Conversely, if there is $\Team \in \HypAsg$ such that $\AStr, \Team
      \IFmodels \varphiFrm[1] \lor \varphiFrm[2]$, then there is $(\TeamAsg[1],
      \TeamAsg[2]) \in \parFun[]{\TeamAsg}$ such that $\AStr, \TeamAsg[i]
      \cmodels[\DIF][\forall] \varphiFrm[i]$.
      By Item~\ref{lmm:teapar(tea)} of Lemma~\ref{lmm:teapar}, there are
      $(\HypAsg[1], \HypAsg[2]) \in \parFun{\dual{\HypAsg}}$ and $\YSet[i] \in
      \dual{\HypAsg[i]}$ such that $\YSet[i] \subseteq \Team[i]$.
      Then, it holds that $\AStr, \YSet[i] \IFmodels \varphiFrm[i]$.
      By inductive hypothesis, we have $\AStr, \dual{\HypAsg[i]} \EAmodels
      \varphiFrm[i]$, or, equivalently, $\AStr, \HypAsg[i] \cmodels[][\QAE]
      \varphiFrm[i]$.
      Therefore, there is $(\HypAsg[1],\HypAsg[2])\in\parFun{\dual{\HypAsg}}$
      such that $\AStr, \HypAsg[i] \cmodels[][\QAE] \varphiFrm[i]$, which
      implies $\AStr, \dual{\HypAsg} \cmodels[][\QAE] \varphiFrm[1] \lor
      \varphiFrm[2]$, and we can conclude $\AStr, \HypAsg \EAmodels
      \varphiFrm[1] \lor \varphiFrm[2]$.

    \item[{$(\varphiFrm[][_\exists] = \LExs[][\pm\WSet] x. \varphiFrm)$}]
      $\AStr, \HypAsg \EAmodels\; \LExs[][\pm\WSet] x. \varphiFrm \iffExpl{sem.}
      \AStr, \extFun[\denot{\pm\WSet}]{\HypAsg, x} \EAmodels \varphiFrm
      \iffExpl{ind.hp.}$ there is $\Team \in \extFun[\denot{\pm\WSet}]{\HypAsg,
        x}$ such that $\AStr, \Team\IFmodels \varphiFrm$ $\iffExpl{def.}$ there
      are $\Team \in \HypAsg$ and $\FFun \in \FncSet[\denot{\pm\WSet}]{}$ such
      that $\AStr,
      \extFun[]{\Team,\FFun,x}\IFmodels\varphiFrm\iffExpl{\DIF-sem.}$ there is
      $\Team \in \HypAsg$ such that $\AStr, \Team \IFmodels \LExs[][\pm\WSet]
      x.\ \varphiFrm$.

    \item[{$(\varphiFrm[][_\exists] = \LAll[][-\emptyset] x. \varphiFrm)$}]
      $\AStr, \HypAsg \EAmodels\; \LAll[][-\emptyset] x. \varphiFrm
\iffExpl{sem.}
      \AStr, \dual{\HypAsg} \AEmodels\; \LAll[][-\emptyset] x. \varphiFrm
      \iffExpl{sem.}  \AStr, \extFun[\VarSet]{\dual{\HypAsg},
        x} \AEmodels \varphiFrm$ $\iffExpl{Thm.\ref{thm:dbldlt}} \AStr,
      \dual{\extFun[\VarSet]{\dual{\HypAsg}, x}} \EAmodels \varphiFrm
      \iffExpl{Lemma~\ref{lmm:cylext}} \AStr, \cylFun{\HypAsg, x} \EAmodels
      \varphiFrm \iffExpl{ind.hp.}$ there is $ \Team \in \cylFun{\HypAsg, x}$
      such that $\AStr, \Team\IFmodels \varphiFrm \iffExpl{def.}$ there is
      $\Team \in \HypAsg$ such that $\AStr,
      \cylFun{\Team,x}\IFmodels\varphiFrm\iffExpl{\DIF-sem.}$ there is $\Team
      \in \HypAsg$ such that $\AStr, \Team \IFmodels \LAll[][-\emptyset] x.\
      \varphiFrm$.
    \end{itemize}
  \end{itemize}

  We turn now to proving Item~\ref{thm:difsemadq(ae)}.
  We proceed by structural induction on the formula $\varphiFrm[][_\forall]$.
  \begin{itemize}
  \item[(base case)] If $\varphiFrm[][_\forall]=\RRel(\xVec)$ or
    $\varphiFrm[][_\forall]= \neg \RRel(\xVec)$, then the property holds by the
    semantics rules.
  \item[(inductive cases)] Suppose that the property holds for the subformulae
    of $\varphiFrm[][_\forall]$.
    \begin{itemize}
    \item[{($\varphiFrm[][_\forall]=\varphiFrm[1] \wedge \varphiFrm[2]$)}]
      We assume that $\AStr, \HypAsg \cmodels[][\QAE] \varphiFrm[1] \wedge
      \varphiFrm[2]$ and we show that for all teams $\TeamAsg \in \HypAsg$, it
      holds that $\AStr, \TeamAsg \cmodels[_{\DIF}][\exists] \varphiFrm[1]
      \wedge \varphiFrm[2]$, which amount to showing that for all teams
      $\TeamAsg \in \HypAsg$ and $(\TeamAsg[1], \TeamAsg[2]) \in
      \parFun[]{\TeamAsg}$, it holds that $\AStr, \TeamAsg[1]
      \cmodels[_{\DIF}][\exists] \varphiFrm[1]$ or $\AStr, \TeamAsg[2]
      \cmodels[_{\DIF}][\exists] \varphiFrm[2]$.
      To this end, we let $\TeamAsg \in \HypAsg$ and $(\TeamAsg[1], \TeamAsg[2])
      \in \parFun[]{\TeamAsg}$.
      By Item~\ref{lmm:teapar(tea)} of Lemma~\ref{lmm:teapar}, there are
      $(\HypAsg[1], \HypAsg[2]) \in \parFun[]{\dual{\HypAsg}}$, $\YSet[1] \in
      \dual{\HypAsg[1]}$, and $\YSet[2] \in \dual{\HypAsg[2]}$, such that
      $\YSet[1] \subseteq \TeamAsg[1]$ and $\YSet[2] \subseteq \TeamAsg[2]$.
      From $\AStr, \HypAsg \cmodels[][\QAE] \varphiFrm[1] \wedge \varphiFrm[2]$,
      it follows that $\AStr, \dual{\HypAsg} \cmodels[][\QEA] \varphiFrm[1]
      \wedge \varphiFrm[2]$.
      By semantics, for all $(\HypAsg[1],\HypAsg[2])\in\parFun{\dual{\HypAsg}}$
      it holds that $\AStr, \HypAsg[1] \EAmodels \varphiFrm[1]$ or $\AStr,
      \HypAsg[2] \EAmodels \varphiFrm[2]$, which, by Theorem~\ref{thm:dbldlt},
      amounts to saying that for all
      $(\HypAsg[1],\HypAsg[2])\in\parFun{\dual{\HypAsg}}$ it holds that $\AStr,
      \dual{\HypAsg[1]} \AEmodels \varphiFrm[1]$ or $\AStr, \dual{\HypAsg[2]}
      \AEmodels \varphiFrm[2]$.
      By inductive hypothesis, for all
      $(\HypAsg[1],\HypAsg[2])\in\parFun{\dual{\HypAsg}}$ it holds that $\AStr,
      \Team[1] \cmodels[\DIF][\exists] \varphiFrm[1]$ for all $\Team[1] \in
      \dual{\HypAsg[1]}$ or it holds that $\AStr, \Team[2]
      \cmodels[\DIF][\exists] \varphiFrm[2]$ for all $\Team[2] \in
      \dual{\HypAsg[2]}$.
      Equivalently, for all $(\HypAsg[1],\HypAsg[2])\in\parFun{\dual{\HypAsg}}$,
      $\Team[1] \in \dual{\HypAsg[1]}$, and $\Team[2] \in \dual{\HypAsg[2]}$, it
      holds that $\AStr, \Team[1] \cmodels[\DIF][\exists] \varphiFrm[1]$ or
      $\AStr, \Team[2] \cmodels[\DIF][\exists] \varphiFrm[2]$.
      Therefore, we have that $\AStr, \YSet[1] \cmodels[\DIF][\exists]
      \varphiFrm[1]$ or $\AStr, \YSet[2] \cmodels[\DIF][\exists] \varphiFrm[2]$,
      and, due to $\YSet[1] \subseteq \TeamAsg[1]$ and $\YSet[2] \subseteq
      \TeamAsg[2]$, we conclude $\AStr, \Team[1] \cmodels[\DIF][\exists]
      \varphiFrm[1]$ or $\AStr, \Team[2] \cmodels[\DIF][\exists] \varphiFrm[2]$.

      Conversely, assume that for all teams $\TeamAsg \in \HypAsg$, it holds
      that $\AStr, \TeamAsg \cmodels[_{\DIF}][\exists] \varphiFrm[1] \wedge
      \varphiFrm[2]$, which amounts to saying that for all $\TeamAsg \in
      \HypAsg$ and $(\TeamAsg[1], \TeamAsg[2]) \in \parFun[]{\TeamAsg}$, it
      holds that $\AStr, \TeamAsg[1] \cmodels[_{\DIF}][\exists] \varphiFrm[1]$
      or $\AStr, \TeamAsg[2] \cmodels[_{\DIF}][\exists] \varphiFrm[2]$.
      First, we show that for all
      $(\HypAsg[1],\HypAsg[2])\in\parFun{\dual{\HypAsg}}$, $\Team[1] \in
      \dual{\HypAsg[1]}$, and $\Team[2] \in \dual\HypAsg[2]$, it holds that
      $\AStr, \Team[1] \cmodels[_{\DIF}][\exists] \varphiFrm[1]$ or $\AStr,
      \Team[2] \cmodels[_{\DIF}][\exists] \varphiFrm[2]$.
      To this end, let $(\HypAsg[1],\HypAsg[2])\in\parFun{\dual{\HypAsg}}$,
      $\Team[1] \in \dual{\HypAsg[1]}$, and $\Team[2] \in \dual\HypAsg[2]$.
      By Item~\ref{lmm:teapar(hyp)} of Lemma~\ref{lmm:teapar}, there exists a
      team $\TeamAsg \in \HypAsg$ such that $\TeamAsg \subseteq \Team[1] \cup
      \Team[2]$.
      Let $\TeamAsg[1]['] = \Team[1] \cap \Team$ and $\TeamAsg[2]['] = \Team
      \setminus \TeamAsg[1][']$.
      Clearly, $(\TeamAsg[1]['], \TeamAsg[2][']) \in \parFun[]{\TeamAsg}$,
      $\TeamAsg[1]['] \subseteq \Team[1]$, and $\TeamAsg[2]['] \subseteq
      \Team[2]$.
      By assumption, it holds that $\AStr, \TeamAsg[1][']
      \cmodels[_{\DIF}][\exists] \varphiFrm[1]$ or $\AStr, \TeamAsg[2][']
      \cmodels[_{\DIF}][\exists] \varphiFrm[2]$.
      From $\TeamAsg[1]['] \subseteq \Team[1]$, and $\TeamAsg[2]['] \subseteq
      \Team[2]$, it follows $\AStr, \Team[1] \cmodels[_{\DIF}][\exists]
      \varphiFrm[1]$ or $\AStr, \Team[2] \cmodels[_{\DIF}][\exists]
      \varphiFrm[2]$.
      Therefore, we have showed that for all
      $(\HypAsg[1],\HypAsg[2])\in\parFun{\dual{\HypAsg}}$, $\Team[1] \in
      \dual{\HypAsg[1]}$, and $\Team[2] \in \dual\HypAsg[2]$, it holds that
      $\AStr, \Team[1] \cmodels[_{\DIF}][\exists] \varphiFrm[1]$ or $\AStr,
      \Team[2] \cmodels[_{\DIF}][\exists] \varphiFrm[2]$.
      This amount to saying that for all
      $(\HypAsg[1],\HypAsg[2])\in\parFun{\dual{\HypAsg}}$, it holds that $\AStr,
      \Team[1] \cmodels[_{\DIF}][\exists] \varphiFrm[1]$ for all $\Team[1] \in
      \dual{\HypAsg[1]}$ or it holds that $\AStr, \Team[2]
      \cmodels[_{\DIF}][\exists] \varphiFrm[2]$ for all $\Team[2] \in
      \dual\HypAsg[2]$.
      By inductive hypothesis, we have that for all
      $(\HypAsg[1],\HypAsg[2])\in\parFun{\dual{\HypAsg}}$, it holds that $\AStr,
      \dual{\HypAsg[1]} \EAmodels \varphiFrm[1]$ or $\AStr, \dual{\HypAsg[2]}
      \EAmodels \varphiFrm[2]$, which eventually amounts to saying $\AStr,
      \HypAsg \AEmodels \varphiFrm[1] \wedge \varphiFrm[2]$.

    \item[{($\varphiFrm[][_\forall]=\varphiFrm[1] \vee \varphiFrm[2]$)}]
      $\AStr, \HypAsg \cmodels[][\QAE] \varphiFrm[1] \vee \varphiFrm[2]
      \iffExpl{sem.}$ there is $(\HypAsg[1],\HypAsg[2])\in\parFun{\HypAsg}$ such
      that $\AStr, \HypAsg[1] \cmodels[][\QAE] \varphiFrm[1]$ and $\AStr,
      \HypAsg[2] \cmodels[][\QAE] \varphiFrm[2] \iffExpl{ind.hp.}$ there is
      $(\HypAsg[1],\HypAsg[2])\in\parFun{\HypAsg}$ such that for all
      $\TeamAsg[1] \in \HypAsg[1]$ it holds $\AStr, \TeamAsg[1]
      \cmodels[\DIF][\exists] \varphiFrm[1]$ and for all $\TeamAsg[2] \in
      \HypAsg[2]$ it holds $\AStr, \TeamAsg[2] \cmodels[\DIF][\exists]
      \varphiFrm[2] \iffExpl{}$ for all $\TeamAsg \in \HypAsg$ it holds that
      $\AStr, \TeamAsg \cmodels[\DIF][\exists] \varphiFrm[1]$ or $\AStr,
      \TeamAsg \cmodels[\DIF][\exists] \varphiFrm[2] \iffExpl{\DIF-sem.}$ for
      all $\TeamAsg \in \HypAsg$ it holds that $\AStr, \TeamAsg
      \cmodels[\DIF][\exists] \varphiFrm[1] \vee \varphiFrm[2]$.

    \item[{$(\varphiFrm[][_\forall] = \LExs[][-\emptyset] x. \varphiFrm)$}]
      $\AStr, \HypAsg \AEmodels\; \LExs[][-\emptyset] x. \varphiFrm
\iffExpl{sem.}
      \AStr, \dual{\HypAsg} \EAmodels\; \LExs[][-\emptyset] x. \varphiFrm
      \iffExpl{sem.}  \AStr, \extFun[\VarSet]{\dual{\HypAsg},
        x} \EAmodels \varphiFrm$ $\iffExpl{Thm.\ref{thm:dbldlt}} \AStr,
      \dual{\extFun[\VarSet]{\dual{\HypAsg}, x}} \AEmodels \varphiFrm
      \iffExpl{Lemma~\ref{lmm:cylext}} \AStr, \cylFun{\HypAsg, x} \AEmodels
      \varphiFrm \iffExpl{ind.hp.}$ for all $ \Team \in \cylFun{\HypAsg, x}$ it
      holds that $\AStr, \Team\cmodels[\DIF][\exists] \varphiFrm \iffExpl{def.}$
      for all $\Team \in \HypAsg$ it holds that $\AStr,
      \cylFun{\Team,x}\cmodels[\DIF][\exists]\varphiFrm\iffExpl{\DIF-sem.}$ for
      all $\Team \in \HypAsg$ it holds that $\AStr, \Team
      \cmodels[\DIF][\exists] \LExs[][-\emptyset] x.\ \varphiFrm$.

    \item[{$(\varphiFrm[][_\forall] = \LAll[][\pm\WSet] x. \varphiFrm)$}]
      $\AStr, \HypAsg \AEmodels\; \LAll[][\pm\WSet] x. \varphiFrm \iffExpl{sem.}
      \AStr, \extFun[\denot{\pm\WSet}]{\HypAsg, x} \AEmodels \varphiFrm
      \iffExpl{ind.hp.}$ for all $\Team \in \extFun[\denot{\pm\WSet}]{\HypAsg,
        x}$ it holds that $\AStr, \Team\cmodels[\DIF][\exists] \varphiFrm$
      $\iffExpl{def.}$ for all $\Team \in \HypAsg$ and $\FFun \in
      \FncSet[\denot{\pm\WSet}]{}$ it holds that $\AStr,
      \extFun[]{\Team,\FFun,x}\cmodels[\DIF][\exists]\varphiFrm\iffExpl{\DIF-sem.}$
      for all $\Team \in \HypAsg$ it holds that $\AStr, \Team
      \cmodels[\DIF][\exists] \LAll[][\pm\WSet] x.\ \varphiFrm$.
      \qedhere
    \end{itemize}
  \end{itemize}
\end{proof}





\section{Proofs of Section~\ref{sec:metthr}}
\label{app:metthr}

\begin{lemma}[name = Generalised Empty \& Null Hyperteams]
  \label{lmm:empnulhypMeta}
  The following hold true for every \Meta formula $\varphiFrm$, function
  assignment $\FunAsg \in \FunAsgSet$, and hyperteam $\HypAsg \in
  \HypAsgSet[\subseteq](\sup{\varphiFrm} \setminus \dom{\FunAsg})$.
  \begin{enumerate}[1)]
  \item
    \begin{inparaenum}[a)]
    \item
      $\AStr, \FunAsg, \EmpHypAsg \notcmodels[][\QEA] \varphiFrm$; \hfill
    \item
      $\AStr, \FunAsg, \HypAsg \cmodels[][\QEA] \varphiFrm$, where $\EmpTeamAsg
      \in \HypAsg$;
    \end{inparaenum}
  \item
    \begin{inparaenum}[a)]
    \item
      $\AStr, \FunAsg, \EmpHypAsg \cmodels[][\QAE] \varphiFrm$; \hfill
    \item
      $\AStr, \FunAsg, \HypAsg \notcmodels[][\QAE] \varphiFrm$, where
      $\EmpTeamAsg \in \HypAsg$.
    \end{inparaenum}
  \end{enumerate}
\end{lemma}
\begin{proof}
  We proceed by structural induction on the size of $\varphiFrm$.
  For the four inductive cases concerning the two binary Boolean connectives and
  the two standard quantifiers, it is useful to recall that, thanks to
  Proposition~\ref{prp:empnultrv}, $\dual{\EmpHypAsg} = \NulHypAsg$ and
  $\dual{\HypAsg} = \EmpHypAsg$ \iff $\EmpTeamAsg \in \HypAsg$.
  \begin{itemize}[$\bullet$]
    \item\textbf{[Base case $\varphiFrm = \Ff$]}
      Both subitems of Item~\ref{lmm:empnulhyp(ea)} directly follow from the
      meta-variant -- $\AStr, \FunAsg, \HypAsg \cmodels[][\QEA] \Ff$ \iff
      $\EmpTeamAsg \in \HypAsg$ -- of Item~\ref{def:sem(adif:fbv:ea)} of
      Definition~\ref{def:sem(adif)}.
      Similarly, Item~\ref{lmm:empnulhyp(ae)} follows from the variant --
      $\AStr, \FunAsg, \HypAsg \cmodels[][\QAE] \Ff$ \iff $\HypAsg = \EmpHypAsg$
      -- of Item~\ref{def:sem(adif:fbv:ae)} of the same definition.
    \item\textbf{[Base case $\varphiFrm = \Tt$]}
      Both subitems of Item~\ref{lmm:empnulhyp(ae)} directly follow from the
      meta-variant -- $\AStr, \FunAsg, \HypAsg \cmodels[][\QAE] \Tt$ \iff
      $\EmpTeamAsg \not\in \HypAsg$ -- of Item~\ref{def:sem(adif:tbv:ae)} of
      Definition~\ref{def:sem(adif)}.
      Similarly, Item~\ref{lmm:empnulhyp(ea)} follows from the variant --
      $\AStr, \FunAsg, \HypAsg \cmodels[][\QEA] \Tt$ \iff $\HypAsg \neq
      \EmpHypAsg$ -- of Item~\ref{def:sem(adif:tbv:ea)} of the same definition.
    \item\textbf{[Base case $\varphiFrm = \RRel(\xVec)$]}
      By observing that $\extFun{\EmpHypAsg, \FunAsg} = \EmpHypAsg$ and
      $\EmpTeamAsg \in \HypAsg$ \iff $\EmpTeamAsg \in \extFun{\HypAsg,
        \FunAsg}$, it is easy to see that Items~\ref{lmm:empnulhyp(ea)}
      and~\ref{lmm:empnulhyp(ae)} immediately follows from
      Items~\ref{def:sem(met:rel:ea)} and~\ref{def:sem(met:rel:ae)} of
      Definition~\ref{def:sem(met)}, respectively.
    \item\textbf{[Inductive case $\varphiFrm = \neg \phiFrm$]}
      Item~\ref{lmm:empnulhyp(ea:emp)} (\resp, Item~\ref{lmm:empnulhyp(ea:nul)},
      Item~\ref{lmm:empnulhyp(ae:emp)}, and Item~\ref{lmm:empnulhyp(ae:nul)})
      follows from the meta-variant -- $\AStr, \FunAsg, \HypAsg
      \cmodels[][\alpha] \neg \phiFrm$ \iff $\AStr, \FunAsg, \allowbreak \HypAsg
      \notcmodels[][\dual{\alpha}] \phiFrm$ -- of Item~\ref{def:sem(adif:neg)}
      of Definition~\ref{def:sem(adif)} and Item~\ref{lmm:empnulhyp(ae:emp)}
      (\resp, Item~\ref{lmm:empnulhyp(ae:nul)},
      Item~\ref{lmm:empnulhyp(ea:emp)}, and Item~\ref{lmm:empnulhyp(ea:nul)}) of
      the inductive hypothesis applied to $\phiFrm$.
    \item\textbf{[Inductive case $\varphiFrm = \phiFrm[1] \LCon[][]
        \phiFrm[2]$]} Items~\ref{lmm:empnulhyp(ae:emp)}
      and~\ref{lmm:empnulhyp(ae:nul)} directly follow from
      Items~\ref{lmm:empnulhyp(ea:nul)} and~\ref{lmm:empnulhyp(ea:emp)},
      respectively, via the meta-variant -- $\AStr, \FunAsg, \HypAsg
      \cmodels[][\QAE] \phiFrm[1] \LCon[][] \phiFrm[2]$ \iff $\AStr,
      \FunAsg, \dual{\HypAsg} \cmodels[][\QEA] \phiFrm[1] \LCon[][]
      \phiFrm[2]$~-- of Item~\ref{def:sem(adif:con:ae)} of
      Definition~\ref{def:sem(adif)}.
      We can therefore focus on the latter two.
      \begin{itemize}
        \item\textbf{[Item~\ref{lmm:empnulhyp(ea:emp)}]}
          By the meta-variant of Item~\ref{def:sem(adif:con:ea)} of
          Definition~\ref{def:sem(adif)}, it holds that $\AStr, \FunAsg,
          \EmpHypAsg \notcmodels[][\QEA] \varphiFrm$ \iff there exists a
          partitioning $(\HypAsg[1], \HypAsg[2]) \in \parFun{\EmpHypAsg}$
          such that $\AStr, \FunAsg, \HypAsg[1] \notcmodels[][\QEA] \phiFrm[1]$
          and $\AStr, \FunAsg, \HypAsg[2] \notcmodels[][\QEA] \phiFrm[2]$.
          Now, from the inductive hypothesis applied to $\phiFrm[1]$ and
          $\phiFrm[2]$, it follows that $\AStr, \FunAsg, \EmpHypAsg
          \notcmodels[][\QEA] \phiFrm[1]$ and $\AStr, \FunAsg, \EmpHypAsg
          \notcmodels[][\QEA] \phiFrm[2]$.
          Moreover, $(\EmpHypAsg, \EmpHypAsg) \in
          \parFun{\EmpHypAsg}$.
          Thus, the thesis clearly holds.
        \item\textbf{[Item~\ref{lmm:empnulhyp(ea:nul)}]}
          By the meta-variant of Item~\ref{def:sem(adif:con:ea)} of
          Definition~\ref{def:sem(adif)}, it holds that $\AStr, \FunAsg, \HypAsg
          \cmodels[][\QEA] \varphiFrm$ \iff, for all partitioning $(\HypAsg[1],
          \HypAsg[2]) \in \parFun{\HypAsg}$, it holds that $\AStr,
          \FunAsg, \HypAsg[1] \cmodels[][\QEA] \phiFrm[1]$ or ${\AStr, \FunAsg,
          \HypAsg[2] \cmodels[][\QEA] \phiFrm[2]}$, where $\EmpTeamAsg \in
          \HypAsg$.
          Now, from the inductive hypothesis applied to $\phiFrm[1]$ and
          $\phiFrm[2]$, it follows that $\AStr, \FunAsg, \HypAsg'
          \cmodels[][\QEA] \phiFrm[1]$ and $\AStr, \FunAsg, \HypAsg'
          \cmodels[][\QEA] \phiFrm[2]$, for every hyperteam $\HypAsg'$
          such that $\EmpTeamAsg \in \HypAsg'$.
          Moreover, for every partitioning $(\HypAsg[1], \HypAsg[2]) \in
          \parFun{\HypAsg}$, one can observe that $\EmpTeamAsg \in
          \HypAsg[1]$ or $\EmpTeamAsg \in \HypAsg[2]$.
          Thus, the thesis clearly holds.
      \end{itemize}
    \item\textbf{[Inductive case $\varphiFrm = \phiFrm[1] \LDis[][]
        \phiFrm[2]$]} Items~\ref{lmm:empnulhyp(ea:emp)}
      and~\ref{lmm:empnulhyp(ea:nul)} directly follow from
      Items~\ref{lmm:empnulhyp(ae:nul)} and~\ref{lmm:empnulhyp(ae:emp)},
      respectively, via the meta-variant -- $\AStr, \FunAsg, \HypAsg
      \cmodels[][\QEA] \phiFrm[1] \LDis[][] \phiFrm[2]$ \iff $\AStr, \FunAsg,
      \dual{\HypAsg} \cmodels[][\QAE] \phiFrm[1] \LDis[][] \phiFrm[2]$ -- of
      Item~\ref{def:sem(adif:dis:ea)} of Definition~\ref{def:sem(adif)}.
      We can therefore focus on the latter two.
      \begin{itemize}
        \item\textbf{[Item~\ref{lmm:empnulhyp(ae:emp)}]}
          By the meta-variant of Item~\ref{def:sem(adif:dis:ae)} of
          Definition~\ref{def:sem(adif)}, it holds that $\AStr, \FunAsg,
          \EmpHypAsg \cmodels[][\QAE] \varphiFrm$ \iff there exists a partitioning
          $(\HypAsg[1], \HypAsg[2]) \in \parFun{\EmpHypAsg}$ such that
          $\AStr, \FunAsg, \HypAsg[1] \cmodels[][\QAE] \phiFrm[1]$ and $\AStr,
          \FunAsg, \HypAsg[2] \cmodels[][\QAE] \phiFrm[2]$.
          Now, by the inductive hypothesis applied to $\phiFrm[1]$ and
          $\phiFrm[2]$, it follows that $\AStr, \FunAsg, \EmpHypAsg
          \cmodels[][\QAE] \phiFrm[1]$ and $\AStr, \FunAsg, \EmpHypAsg
          \cmodels[][\QAE] \phiFrm[2]$.
          Moreover, $(\EmpHypAsg, \EmpHypAsg) \in
          \parFun{\EmpHypAsg}$.
          Thus, the thesis clearly holds.
        \item\textbf{[Item~\ref{lmm:empnulhyp(ae:nul)}]}
          By the meta-variant of Item~\ref{def:sem(adif:dis:ae)} of
          Definition~\ref{def:sem(adif)}, it holds that $\AStr, \FunAsg, \HypAsg
          \notcmodels[][\QAE] \varphiFrm$ \iff, for all partitioning $(\HypAsg[1],
          \HypAsg[2]) \in \parFun{\HypAsg}$, it holds that $\AStr,
          \FunAsg, \HypAsg[1] \notcmodels[][\QAE] \phiFrm[1]$ or ${\AStr,
          \FunAsg, \HypAsg[2] \notcmodels[][\QAE] \phiFrm[2]}$, where
          $\EmpTeamAsg \in \HypAsg$.
          Now, by the inductive hypothesis applied to $\phiFrm[1]$ and
          $\phiFrm[2]$, it follows that $\AStr, \FunAsg, \HypAsg'
          \notcmodels[][\QAE] \phiFrm[1]$ and $\AStr, \FunAsg, \HypAsg'
          \notcmodels[][\QAE] \phiFrm[2]$, for every hyperteam $\HypAsg'$
          such that $\EmpTeamAsg \in \HypAsg'$.
          Moreover, for every partitioning $(\HypAsg[1], \HypAsg[2]) \in
          \parFun{\HypAsg}$, one can observe that $\EmpTeamAsg \in
          \HypAsg[1]$ or $\EmpTeamAsg \in \HypAsg[2]$.
          Thus, the thesis clearly holds.
      \end{itemize}
    \item\textbf{[Inductive case $\varphiFrm = \LExs[][_{\pm\WSet}] \varElm
        \ldotp \phiFrm$]} Items~\ref{lmm:empnulhyp(ae:emp)}
      and~\ref{lmm:empnulhyp(ae:nul)} directly follow from
      Items~\ref{lmm:empnulhyp(ea:nul)} and~\ref{lmm:empnulhyp(ea:emp)},
      respectively, via the meta-variant -- $\AStr, \FunAsg, \HypAsg
      \cmodels[][\QAE] \LExs[][_{\pm\WSet}] \varElm \ldotp \phiFrm$ \iff $\AStr,
      \FunAsg, \dual{\HypAsg} \cmodels[][\QEA] \LExs[][_{\pm\WSet}] \varElm
      \ldotp \phiFrm$ -- of Item~\ref{def:sem(adif:exs:ae)} of
      Definition~\ref{def:sem(adif)}.
      We can therefore focus on the latter two.
      \begin{itemize}
      \item\textbf{[Item~\ref{lmm:empnulhyp(ea:emp)}]}
        By the meta-variant of Item~\ref{def:sem(adif:exs:ea)} of
        Definition~\ref{def:sem(adif)}, it holds that $\AStr, \FunAsg,
        \EmpHypAsg \notcmodels[][\QEA] \varphiFrm$ \iff $\AStr, \FunAsg,
        \extFun[\denot{\pm\WSet}]{\EmpHypAsg, \varElm} \notcmodels[][\QEA]
        \phiFrm$.
          Now, by the inductive hypothesis on $\phiFrm$, it follows that $\AStr,
          \FunAsg, \EmpHypAsg \notcmodels[][\QEA] \phiFrm$.
          Moreover, $\extFun[\denot{\pm\WSet}]{\EmpHypAsg, \varElm} =
          \EmpHypAsg$.
          Thus, the thesis clearly holds.
        \item\textbf{[Item~\ref{lmm:empnulhyp(ea:nul)}]}
          By the meta-variant of Item~\ref{def:sem(adif:exs:ea)} of
          Definition~\ref{def:sem(adif)}, it holds that $\AStr, \FunAsg, \HypAsg
          \cmodels[][\QEA] \varphiFrm$ \iff $\AStr, \FunAsg,
          \extFun[\denot{\pm\WSet}]{\HypAsg, \varElm} \cmodels[][\QEA] \phiFrm$,
          where $\EmpTeamAsg \in \HypAsg$.
          Now, by the inductive hypothesis on $\phiFrm$, it follows that $\AStr,
          \FunAsg, \HypAsg' \cmodels[][\QEA] \phiFrm$, for each hyperteam
          $\HypAsg'$ with $\EmpTeamAsg \in \HypAsg'$.
          Moreover, $\EmpTeamAsg \in \extFun[\denot{\pm\WSet}]{\HypAsg,
            \varElm}$.
          Thus, the thesis clearly holds.
      \end{itemize}
    \item\textbf{[Inductive case $\varphiFrm = \LAll[][_{\pm\WSet}] \varElm
        \ldotp \phiFrm$]} Items~\ref{lmm:empnulhyp(ea:emp)}
      and~\ref{lmm:empnulhyp(ea:nul)} directly follow from
      Items~\ref{lmm:empnulhyp(ae:nul)} and~\ref{lmm:empnulhyp(ae:emp)},
      respectively, via the meta-variant -- $\AStr, \FunAsg, \HypAsg
      \cmodels[][\QEA] \LAll[][_{\pm\WSet}] \varElm \ldotp \phiFrm$ \iff $\AStr,
      \FunAsg, \dual{\HypAsg} \cmodels[][\QAE] \LAll[][_{\pm\WSet}] \varElm
      \ldotp \phiFrm$ -- of Item~\ref{def:sem(adif:all:ea)} of
      Definition~\ref{def:sem(adif)}.
      We can therefore focus on the latter two.
      \begin{itemize}
      \item\textbf{[Item~\ref{lmm:empnulhyp(ae:emp)}]}
        By the meta-variant of Item~\ref{def:sem(adif:all:ae)} of
        Definition~\ref{def:sem(adif)}, it holds that $\AStr, \FunAsg,
        \EmpHypAsg \cmodels[][\QAE] \varphiFrm$ \iff $\AStr, \FunAsg,
        \extFun[\denot{\pm\WSet}]{\EmpHypAsg, \varElm} \cmodels[][\QAE]
        \phiFrm$.
          Now, by the inductive hypothesis on $\phiFrm$, it follows that $\AStr,
          \FunAsg, \EmpHypAsg \cmodels[][\QAE] \phiFrm$.
          Moreover, $\extFun[\denot{\pm\WSet}]{\EmpHypAsg, \varElm} =
          \EmpHypAsg$.
          Thus, the thesis clearly holds.
        \item\textbf{[Item~\ref{lmm:empnulhyp(ae:nul)}]}
          By the meta-variant of Item~\ref{def:sem(adif:all:ae)} of
          Definition~\ref{def:sem(adif)}, it holds that $\AStr, \FunAsg, \HypAsg
          \notcmodels[][\QAE] \varphiFrm$ \iff $\AStr, \FunAsg,
          \extFun[\denot{\pm\WSet}]{\HypAsg, \varElm} \notcmodels[][\QAE]
          \phiFrm$, where $\EmpTeamAsg \in \HypAsg$.
          Now, by the inductive hypothesis on $\phiFrm$, it follows that $\AStr,
          \FunAsg, \HypAsg' \notcmodels[][\QAE] \phiFrm$, for each
          hyperteam $\HypAsg'$ with $\EmpTeamAsg \in \HypAsg'$.
          Moreover, $\EmpTeamAsg \in \extFun[\denot{\pm\WSet}]{\HypAsg,
            \varElm}$.
          Thus, the thesis clearly holds.
      \end{itemize}
    \item\textbf{[Inductive case $\varphiFrm = \LEExs[][_{\pm\WSet}] \varElm
      \ldotp \phiFrm$]}
      Since the semantics of the existential meta quantifier does not depend on
      the alternation flag $\alpha$, we consider the two satisfaction (\resp,
      non-satisfaction) cases altogether.
      \begin{itemize}
        \item\textbf{[Items~\ref{lmm:empnulhyp(ea:emp)}
          and~\ref{lmm:empnulhyp(ae:nul)}]}
          By Item~\ref{def:sem(met:eexs)} of Definition~\ref{def:sem(met)}, it
          holds that $\AStr, \FunAsg, \HypAsg \notcmodels[][\alpha]
          \LEExs[][_{\pm\WSet}] \varElm \ldotp \phiFrm$ \iff, for all functions
          $\FFun \in \FncSet[\denot{\pm\WSet}]{}$, it holds that $\AStr,
          {\FunAsg}[\varElm \mapsto \FFun], \HypAsg \notcmodels[][\alpha]
          \phiFrm$.
          Now, by the inductive hypothesis on $\phiFrm$, it follows that $\AStr,
          \FunAsg', \HypAsg \notcmodels[][\alpha] \phiFrm$, for every function
          assignment $\FunAsg'$, where either $\alpha = \QEA$ and $\HypAsg =
          \EmpHypAsg$ or $\alpha = \QAE$ and $\EmpTeamAsg \in \HypAsg$.
          Thus, the thesis clearly holds.
        \item\textbf{[Items~\ref{lmm:empnulhyp(ea:nul)}
          and~\ref{lmm:empnulhyp(ae:emp)}]}
          By Item~\ref{def:sem(met:eexs)} of Definition~\ref{def:sem(met)}, it
          holds that $\AStr, \FunAsg, \HypAsg \cmodels[][\alpha]
          \LEExs[][_{\pm\WSet}] \varElm \ldotp \phiFrm$ \iff there exists a
          function $\FFun \in \FncSet[\denot{\pm\WSet}]{}$ such that $\AStr,
          {\FunAsg}[\varElm \mapsto \FFun], \HypAsg \cmodels[][\alpha] \phiFrm$.
          Now, by the inductive hypothesis on $\phiFrm$, it follows that $\AStr,
          \FunAsg', \HypAsg \cmodels[][\alpha] \phiFrm$, for every function
          assignment $\FunAsg'$, where either $\alpha = \QAE$ and $\HypAsg =
          \EmpHypAsg$ or $\alpha = \QEA$ and $\EmpTeamAsg \in \HypAsg$.
          Thus, the thesis clearly holds.
      \end{itemize}
    \item\textbf{[Inductive case $\varphiFrm = \LAAll[][_{\pm\WSet}] \varElm
      \ldotp \phiFrm$]}
      Since the semantics of the universal meta quantifier does not depend on
      the alternation flag $\alpha$, we consider the two satisfaction (\resp,
      non-satisfaction) cases altogether.
      \begin{itemize}
        \item\textbf{[Items~\ref{lmm:empnulhyp(ea:emp)}
          and~\ref{lmm:empnulhyp(ae:nul)}]}
          By Item~\ref{def:sem(met:aall)} of Definition~\ref{def:sem(met)}, it
          holds that $\AStr, \FunAsg, \HypAsg \notcmodels[][\alpha]
          \LAAll[][_{\pm\WSet}] \varElm \ldotp \phiFrm$ \iff there exists a
          function $\FFun \in \FncSet[\denot{\pm\WSet}]{}$ such that $\AStr,
          {\FunAsg}[\varElm \mapsto \FFun], \HypAsg \notcmodels[][\alpha]
          \phiFrm$.
          Now, by the inductive hypothesis on $\phiFrm$, it follows that $\AStr,
          \FunAsg', \HypAsg \notcmodels[][\alpha] \phiFrm$, for every function
          assignment $\FunAsg'$, where either $\alpha = \QEA$ and $\HypAsg =
          \EmpHypAsg$ or $\alpha = \QAE$ and $\EmpTeamAsg \in \HypAsg$.
          Thus, the thesis clearly holds.
        \item\textbf{[Items~\ref{lmm:empnulhyp(ea:nul)}
          and~\ref{lmm:empnulhyp(ae:emp)}]}
          By Item~\ref{def:sem(met:aall)} of Definition~\ref{def:sem(met)}, it
          holds that $\AStr, \FunAsg, \HypAsg \cmodels[][\alpha]
          \LAAll[][_{\pm\WSet}] \varElm \ldotp \phiFrm$ \iff, for all functions
          $\FFun \in \FncSet[\denot{\pm\WSet}]{}$, it holds that $\AStr,
          {\FunAsg}[\varElm \mapsto \FFun], \HypAsg \cmodels[][\alpha]
          \phiFrm$.
          Now, by the inductive hypothesis on $\phiFrm$, it follows that $\AStr,
          \FunAsg', \HypAsg \cmodels[][\alpha] \phiFrm$, for every function
          assignment $\FunAsg'$, where either $\alpha = \QAE$ and $\HypAsg =
          \EmpHypAsg$ or $\alpha = \QEA$ and $\EmpTeamAsg \in \HypAsg$.
          Thus, the thesis clearly holds.
          \qedhere
      \end{itemize}
  \end{itemize}
\end{proof}

\begin{lemma}[name = Extension Monotonicity]
  \label{lmm:monMeta}
  For all sets of variables $\WSet \subseteq \VarSet$, function assignments
  $\FunAsg \in \FunAsgSet$, and hyperteams $\HypAsg[1], \HypAsg[2] \in
  \HypAsgSet$, where ${\HypAsg[1] \inc[\WSet] \HypAsg[2]}$ and $\FunAsg(\varElm)
  \in \FncSet[\WSet]{}$, for all $\varElm \in \dom{\FunAsg} \cap \WSet$, it
  holds that $\extFun{\HypAsg[1], \FunAsg} \inc[\WSet] \extFun{\HypAsg[2],
    \FunAsg}$.
\end{lemma}
\begin{proof}
  Let $\TeamAsg[1] \in \extFun{\HypAsg[1], \FunAsg} \! \rst[\WSet]$.
  We show that there is $\TeamAsg[2] \in \extFun{\HypAsg[2], \FunAsg}
  \!\rst[\WSet]$ such that $\TeamAsg[2] \subseteq \TeamAsg[1]$.
  By $\TeamAsg[1] \in \extFun{\HypAsg[1], \FunAsg} \!\rst[\WSet]$, it holds that
  $\TeamAsg[1] = \extFun{\TeamAsg[1]['], \FunAsg} \!\rst[\WSet]$ for some
  $\TeamAsg[1]['] \in \HypAsg[1]$.
  By $\HypAsg[1] \inc[\WSet] \HypAsg[2]$, there is $\TeamAsg[2]['] \in
  \HypAsg[2]$ such that $\TeamAsg[2]['] \rst[\WSet] \subseteq \TeamAsg[1][']
  \rst[\WSet]$.
  Thus, $\extFun{\TeamAsg[2]['], \FunAsg} \!\rst[\WSet] \in \extFun{\HypAsg[2],
    \FunAsg} \!\rst[\WSet]$.
  From $\TeamAsg[2]['] \rst[\WSet] \subseteq \TeamAsg[1]['] \rst[\WSet]$ and the
  fact that $\FunAsg(\varElm) \in \FncSet[\WSet]{}$ holds for all $\varElm \in
  \dom{\FunAsg} \cap \WSet$, it follows that $\extFun{\TeamAsg[2]['], \FunAsg}
  \!\rst[\WSet] \inc \extFun{\TeamAsg[1]['], \FunAsg} \!\rst[\WSet] =
  \TeamAsg[1]$.
  Hence the thesis.
\end{proof}

\figGraphThmHypRefMeta

\begin{theorem}[name = Generalised Hyperteam Refinement]
  \label{thm:hyprefMeta}
  The following hold true for every \Meta formula $\varphiFrm$, function
  assignment $\FunAsg \in \FunAsgSet$, function $\iotaFun : \dom{\iotaFun} \to
  \pow{\VarSet}$, with $\dom{\FunAsg} \subseteq \dom{\iotaFun}$,
  and hyperteams $\HypAsg, \HypAsg' \in \HypAsgSet[\subseteq](\sup{\varphiFrm}
  \setminus \dom{\FunAsg})$, with $\FunAsg(\varElm) \in
  \FncSet[\iotaFun(\varElm)]{}$, for all $\varElm \in \dom{\FunAsg}$, and
  ${\HypAsg \inc[\free{\varphiFrm,\iotaFun}] \HypAsg'}$:
  \begin{enumerate}[1)]
  \item
    if $\AStr, \FunAsg, \HypAsg \cmodels[][\QEA] \varphiFrm$ then $\AStr,
    \FunAsg, \HypAsg' \cmodels[][\QEA] \varphiFrm$;
  \item
    if $\AStr, \FunAsg, \HypAsg' \cmodels[][\QAE] \varphiFrm$ then $\AStr,
    \FunAsg, \HypAsg \cmodels[][\QAE] \varphiFrm$.
  \end{enumerate}
\end{theorem}
\begin{proof}
  Due to $\HypAsg \inc[\free{\varphiFrm,\iotaFun}] \HypAsg'$, there is a
  function $f : \HypAsg \!\rst[\free{\varphiFrm,\iotaFun}] \rightarrow \HypAsg'
  \!\rst[\free{\varphiFrm,\iotaFun}]$, such that $f(\TeamAsg) \subseteq
  \TeamAsg$ for every $\TeamAsg\in\HypAsg \!\rst[\free{\varphiFrm,\iotaFun}]$.
  The claim is proved by induction on the structure of the formula and the
  alternation flag $\alpha$.
  \begin{itemize}
  \item If $\varphi = \bot$, then $\AStr, \FunAsg, \HypAsg
    \cmodels[][\QEA]\! \varphiFrm$ implies $\EmpTeamAsg \in \HypAsg$, which
    means that $\EmpTeamAsg \in \HypAsg \!\rst[\free{\varphiFrm,\iotaFun}]$.
    By $\HypAsg \inc[\free{\varphiFrm,\iotaFun}] \HypAsg'$, we have $\EmpTeamAsg
    \in \HypAsg' \!\rst[\free{\varphiFrm,\iotaFun}]$.
    Thus, $\EmpTeamAsg \in \HypAsg'$, which amounts to $\AStr, \FunAsg, \HypAsg'
    \cmodels[][\QEA]\! \varphiFrm$.

    On the other hand, we also have that $\AStr, \FunAsg, \HypAsg'
    \cmodels[][\QAE]\!
    \varphiFrm$ implies $\HypAsg' = \EmpHypAsg$, which means that $\HypAsg'
    \!\rst[\free{\varphiFrm,\iotaFun}] = \EmpHypAsg$.
    By $\HypAsg \inc[\free{\varphiFrm,\iotaFun}] \HypAsg'$, we have $\HypAsg
    \!\rst[\free{\varphiFrm,\iotaFun}] = \EmpHypAsg$.
    Thus, $\HypAsg = \EmpHypAsg$, which amounts to $\AStr, \FunAsg, \HypAsg
    \cmodels[][\QAE]\! \varphiFrm$.

  \item If $\varphi = \top$, then $\AStr, \FunAsg, \HypAsg
    \cmodels[][\QEA]\! \varphiFrm$ implies $\HypAsg \neq \EmpHypAsg$, which
    means that $\HypAsg \!\rst[\free{\varphiFrm,\iotaFun}] \neq \EmpHypAsg$.
    By $\HypAsg \inc[\free{\varphiFrm,\iotaFun}] \HypAsg'$, we have $\HypAsg'
    \!\rst[\free{\varphiFrm,\iotaFun}] \neq \EmpHypAsg$.
    Thus, $\HypAsg' \neq \EmpHypAsg$, which amounts to $\AStr, \FunAsg, \HypAsg'
    \cmodels[][\QEA]\! \varphiFrm$.

    On the other hand, we also have that $\AStr, \FunAsg, \HypAsg'
    \cmodels[][\QAE]\!
    \varphiFrm$ implies $\EmpTeamAsg \notin \HypAsg'$, which means that
    $\EmpTeamAsg \notin \HypAsg' \!\rst[\free{\varphiFrm,\iotaFun}]$.
    By $\HypAsg \inc[\free{\varphiFrm,\iotaFun}] \HypAsg'$, we have $\EmpTeamAsg \notin
    \HypAsg \!\rst[\free{\varphiFrm,\iotaFun}]$.
    Thus, $\EmpTeamAsg \notin \HypAsg$, which amounts to $\AStr, \FunAsg,
    \HypAsg \cmodels[][\QAE]\! \varphiFrm$.

  \item If $\varphi = \RRel(\xVec)$, then $\AStr, \FunAsg, \HypAsg
    \cmodels[][\QEA] \varphi$ implies the existence of a team $\TeamAsg \in
    \extFun{\HypAsg, \FunAsg}$ such that, for all assignments $\asgElm \in
    \TeamAsg$, it holds that $\xVec[][\asgElm] \in \RRel[][\AStr]$.
    By $\HypAsg \inc[\free{\varphiFrm,\iotaFun}] \HypAsg'$ and
    Lemma~\ref{lmm:monMeta} (notice that $\iotaFun(\varElm) \subseteq
    \free{\varphiFrm,\iotaFun}$, for all $\varElm \in \dom{\FunAsg} \cap
    \free{\varphiFrm,\iotaFun}$), we have that $\extFun{\HypAsg, \FunAsg}
    \inc[\free{\varphiFrm,\iotaFun}] \extFun{\HypAsg', \FunAsg}$, and thus there
    is a team $\TeamAsg' \in \extFun{\HypAsg', \FunAsg}$ such that $\TeamAsg'
    \!\rst[\free{\varphiFrm,\iotaFun}] \subseteq \TeamAsg
    \!\rst[\free{\varphiFrm,\iotaFun}]$, which implies $\TeamAsg' \!\rst[\xVec]
    \subseteq \TeamAsg \!\rst[\xVec]$, since $\xVec \subseteq
    \free{\varphiFrm,\iotaFun}$.
    The thesis follows from the fact that $\xVec[][\asgElm] \in \RRel[][\AStr]$
    if and only if $\xVec[][{\asgElm \!\rst[\xVec]}] \in \RRel[][\AStr]$ holds,
    for every $\asgElm \in \AsgSet$.

    On the other hand, we also have that $\AStr, \FunAsg, \HypAsg'
    \cmodels[][\QAE] \varphi$ implies that for all teams $\TeamAsg' \in
    \extFun{\HypAsg', \FunAsg}$, there exists an assignment $\asgElm' \in
    \TeamAsg'$ such that $\xVec[][\asgElm'] \in \RRel[][\AStr]$.
    By $\HypAsg \inc[\free{\varphiFrm,\iotaFun}] \HypAsg'$ and
    Lemma~\ref{lmm:monMeta}, we have that $\extFun{\HypAsg, \FunAsg}
    \inc[\free{\varphiFrm,\iotaFun}] \extFun{\HypAsg', \FunAsg}$, and thus for
    every team $\TeamAsg \in \extFun{\HypAsg, \FunAsg}$ there is a team
    $\TeamAsg' \in \extFun{\HypAsg',
      \FunAsg}$ such that $\TeamAsg' \!\rst[\free{\varphiFrm,\iotaFun}]
    \subseteq \TeamAsg \!\rst[\free{\varphiFrm,\iotaFun}]$.
    The thesis follows from the same argument used above.

  \item If $\varphi = \neg \phiFrm$, then $\AStr, \FunAsg, \HypAsg
    \cmodels[][\QEA]\! \varphiFrm$ implies $\AStr, \FunAsg, \HypAsg
    \notcmodels[][\QAE] \phiFrm$.
    By inductive hypothesis, this implies $\AStr, \FunAsg, \HypAsg'
    \notcmodels[][\QAE]\: \phiFrm$, which amounts to $\AStr, \FunAsg, \HypAsg'
    \cmodels[][\QEA]\: \varphi$.

    On the other hand, we also have that $\AStr, \FunAsg, \HypAsg'
    \cmodels[][\QAE]\: \varphi$ implies $\AStr, \FunAsg, \HypAsg'
    \notcmodels[][\QEA]\: \phiFrm$.
    By inductive hypothesis, this implies $\AStr, \FunAsg, \HypAsg
    \notcmodels[][\QEA] \phiFrm$, which amounts to $\AStr, \FunAsg, \HypAsg
    \cmodels[][\QAE]\: \varphi$.

  \item Let $\varphi = \phiFrm[1] \LCon[][] \phiFrm[2]$.
    We assume $\AStr, \FunAsg, \HypAsg \cmodels[][\QEA] \varphi$ and we show
    that $\AStr, \FunAsg, \HypAsg[1]' \cmodels[][\QEA] \phiFrm[1]$ or $\AStr,
    \FunAsg, \HypAsg[2]' \cmodels[][\QEA] \phiFrm[2]$ holds for all
    $(\HypAsg[1]', \HypAsg[2]') \in \parFun[]{\HypAsg'}$.
    To this end, let $(\HypAsg[1]', \HypAsg[2]') \in
    \parFun[]{\HypAsg'}$.
    By Lemma~\ref{lmm:mon}, item~\ref{lmm:mon(par-unrestricted)}, there is
    $(\HypAsg[1], \HypAsg[2]) \in \parFun[]{\HypAsg}$ such that $\HypAsg[1]
    \inc[\free{\varphiFrm,\iotaFun}] \HypAsg[1]'$ and $\HypAsg[2] \inc[\free{\varphiFrm,\iotaFun}]
    \HypAsg[2]'$, and, by the semantics of $\LCon[][]$, we have that
    $(\HypAsg[1], \HypAsg[2]) \in \parFun[]{\HypAsg}$ implies that $\AStr,
    \FunAsg, \HypAsg[1] \cmodels[][\QEA] \phiFrm[1]$ or $\AStr, \FunAsg,
    \HypAsg[2] \cmodels[][\QEA] \phiFrm[2]$.
    Moreover, since $\free{\phiFrm[1],\iota} \subseteq \free{\varphiFrm,\iota}$
    and $\free{\phiFrm[2],\iota} \subseteq \free{\varphiFrm,\iota}$, we have
    that $\HypAsg[1] \inc[\free{\phiFrm[1],\iota}] \HypAsg[1]'$ and $\HypAsg[2]
    \inc[\free{\phiFrm[2],\iota}] \HypAsg[2]'$.
    Finally, by inductive hypothesis it holds that $\AStr, \FunAsg, \HypAsg[1]'
    \cmodels[][\QEA] \phiFrm[1]$ or $\AStr, \FunAsg, \HypAsg[2]'
    \cmodels[][\QEA] \phiFrm[2]$.

    On the other hand, we also have that $\AStr, \FunAsg, \HypAsg'
    \cmodels[][\QAE]\: \varphi$ if and only if $\AStr, \FunAsg, \dual{\HypAsg'}
    \cmodels[][\QEA]\: \varphi$.
    By inductive hypothesis and Lemma~\ref{lmm:mon}, Item~\ref{lmm:mon(dlt)},
    this implies $\AStr, \FunAsg, \dual{\HypAsg} \cmodels[][\QEA]\: \varphi$,
    which amounts to $\AStr, \FunAsg, \HypAsg \cmodels[][\QAE]\: \varphi$.

  \item Let $\varphi = \phiFrm[1] \LDis[][] \phiFrm[2]$.
    In this case, we first prove the second item of the claim.
    We assume $\AStr, \FunAsg, \HypAsg' \cmodels[][\QAE] \varphi$ and we show
    that there is $(\HypAsg[1], \HypAsg[2]) \in \parFun[]{\HypAsg}$ such that
    $\AStr, \FunAsg, \HypAsg[1] \!\cmodels[][\QAE]\! \phiFrm[1]$ and $\AStr,
    \FunAsg, \HypAsg[2] \!\cmodels[][\QAE]\! \phiFrm[2]$.
    By the semantics of $\LDis[][]$, we have that there is $(\HypAsg[1]',
    \HypAsg[2]') \in \parFun[]{\HypAsg'}$ such that $\AStr, \FunAsg, \HypAsg[1]'
    \!\cmodels[][\QAE]\! \phiFrm[1]$ and $\AStr, \FunAsg, \HypAsg[2]'
    \!\cmodels[][\QAE]\! \phiFrm[2]$.
    By Lemma~\ref{lmm:mon}, item~\ref{lmm:mon(par-unrestricted)}, there is
    $(\HypAsg[1], \HypAsg[2]) \in \parFun[]{\HypAsg}$ such that $\HypAsg[1]
    \inc[\free{\varphiFrm,\iota}] \HypAsg[1]'$ and $\HypAsg[2]
    \inc[\free{\varphiFrm,\iota}] \HypAsg[2]'$.
    Moreover, since $\free{\phiFrm[1],\iota} \subseteq \free{\varphiFrm,\iota}$
    and $\free{\phiFrm[2],\iota} \subseteq \free{\varphiFrm,\iota}$, we have
    that $\HypAsg[1] \inc[\free{\phiFrm[1],\iota}] \HypAsg[1]'$ and $\HypAsg[2]
    \inc[\free{\phiFrm[2],\iota}] \HypAsg[2]'$.
    Finally, by inductive hypothesis it holds that $\AStr, \FunAsg, \HypAsg[1]
    \cmodels[][\QAE] \phiFrm[1]$ and $\AStr, \FunAsg, \HypAsg[2]
    \cmodels[][\QAE] \phiFrm[2]$.

    On the other hand, we also have that $\AStr, \FunAsg, \HypAsg
    \cmodels[][\QEA]\: \varphi$ if and only if $\AStr, \FunAsg, \dual{\HypAsg}
    \cmodels[][\QAE]\: \varphi$.
    By inductive hypothesis and Lemma~\ref{lmm:mon}, Item~\ref{lmm:mon(dlt)},
    this implies $\AStr, \FunAsg, \dual{\HypAsg'} \cmodels[][\QAE]\: \varphi$,
    which amounts to $\AStr, \FunAsg, \HypAsg' \cmodels[][\QEA]\: \varphi$.

  \item If $\varphi = \LExs[][_{\pm\WSet}] \varElm \ldotp
    \phiFrm$, then $\AStr, \FunAsg, \HypAsg \cmodels[][\QEA]\: \varphi$ implies
    $\AStr, \FunAsg, \extFun[\denot{\pm\WSet}]{\HypAsg, \varElm}
    \cmodels[][\QEA] \phiFrm$.
    If $\varElm \in \free{\phiFrm,{\iotaFun}[\varElm \mapsto
      \emptyset]}$, then $\denot{\pm\WSet} \subseteq
    \free{\varphiFrm,\iotaFun}$, and thus, by Lemma~\ref{lmm:mon},
    item~\ref{lmm:mon(ext:dep)}, we have $\extFun[\denot{\pm\WSet}]{\HypAsg,
      \varElm} \inc[\free{\varphiFrm,\iotaFun} \cup \{ \varElm \}]
    \extFun[\denot{\pm\WSet}]{\HypAsg', \varElm}$.
    Since $\free{\phiFrm, {\iotaFun}[\varElm \mapsto \emptyset]} \subseteq
    \free{\varphiFrm, \iotaFun} \cup \{ \varElm \}$, we have
    $\extFun[\denot{\pm\WSet}]{\HypAsg, \varElm} \inc[\free{\phiFrm,
      {\iotaFun}[\varElm \mapsto \emptyset]}]
    \extFun[\denot{\pm\WSet}]{\HypAsg', \varElm}$.
    From the inductive hypothesis, it follows $\AStr, \FunAsg,
    \extFun[\denot{\pm\WSet}]{\HypAsg', \varElm} \cmodels[][\QEA] \phiFrm$,
    which amounts to $\AStr, \FunAsg, \HypAsg' \cmodels[][\QEA] \varphi$.
    If, instead, $\varElm \notin \free{\phiFrm,{\iotaFun}[\varElm \mapsto
      \emptyset]}$, then $\free{\varphiFrm,\iotaFun} =
    \free{\phiFrm,{\iotaFun}[\varElm \mapsto \emptyset]}$, which means that
    $\varElm \notin \free{\varphiFrm,\iotaFun}$.
    By Lemma~\ref{lmm:mon}, Item~\ref{lmm:mon(ext:eql)}, we have that $\HypAsg
    \eql[\free{\varphiFrm,\iotaFun}] \extFun[\denot{\pm\WSet}]{\HypAsg,
      \varElm}$ and $\HypAsg' \eql[\free{\varphiFrm,\iotaFun}]
    \extFun[\denot{\pm\WSet}]{\HypAsg', \varElm}$, which means that
    $\extFun[\denot{\pm\WSet}]{\HypAsg, \varElm} \inc[\free{\phiFrm,
      {\iotaFun}[\varElm \mapsto \emptyset]}]
    \extFun[\denot{\pm\WSet}]{\HypAsg', \varElm}$, as
    $\free{\varphiFrm,\iotaFun} = \free{\phiFrm,{\iotaFun}[\varElm \mapsto
      \emptyset]}$.
    By inductive hypothesis, it holds that $\AStr, \FunAsg,
    \extFun[\denot{\pm\WSet}]{\HypAsg', \varElm} \cmodels[][\QEA] \phiFrm$,
    which amounts to $\AStr, \FunAsg, \HypAsg' \cmodels[][\QEA]\: \varphi$.

    On the other hand, we also have, by semantics, $\AStr, \FunAsg, \HypAsg'
    \cmodels[][\QAE]\: \varphi$ if and only if $\AStr, \FunAsg, \dual{\HypAsg'}
    \cmodels[][\QEA]\: \varphi$.
    By inductive hypothesis and Lemma~\ref{lmm:mon}, Item~\ref{lmm:mon(dlt)},
    this implies $\AStr, \FunAsg, \dual{\HypAsg} \cmodels[][\QEA]\: \varphi$,
    which amounts to $\AStr, \FunAsg, \HypAsg \cmodels[][\QAE]\: \varphi$.

  \item If $\varphi = \LAll[][_{\pm\WSet}] \varElm \ldotp \phiFrm$, then
    $\AStr, \FunAsg, \HypAsg' \cmodels[][\QAE] \varphi$ implies $\AStr, \FunAsg,
    \extFun[\denot{\pm\WSet}]{\HypAsg', \varElm} \cmodels[][\QAE] \phiFrm$.

    If $\varElm \in \free{\phiFrm,{\iotaFun}[\varElm \mapsto
      \emptyset]}$, then $\denot{\pm\WSet} \subseteq
    \free{\varphiFrm,\iotaFun}$, and thus, by Lemma~\ref{lmm:mon},
    item~\ref{lmm:mon(ext:dep)}, we have $\extFun[\denot{\pm\WSet}]{\HypAsg,
      \varElm} \inc[\free{\varphiFrm,\iotaFun} \cup \{ \varElm \}]
    \extFun[\denot{\pm\WSet}]{\HypAsg', \varElm}$.
    Since $\free{\phiFrm, {\iotaFun}[\varElm \mapsto \emptyset]} \subseteq
    \free{\varphiFrm, \iotaFun} \cup \{ \varElm \}$, we have
    $\extFun[\denot{\pm\WSet}]{\HypAsg, \varElm} \inc[\free{\phiFrm,
      {\iotaFun}[\varElm \mapsto \emptyset]}]
    \extFun[\denot{\pm\WSet}]{\HypAsg', \varElm}$.
    From the inductive hypothesis, it follows $\AStr, \FunAsg,
    \extFun[\denot{\pm\WSet}]{\HypAsg, \varElm} \cmodels[][\QAE] \phiFrm$,
    which amounts to $\AStr, \FunAsg, \HypAsg \cmodels[][\QAE] \varphi$.
    If, instead, $\varElm \notin \free{\phiFrm,{\iotaFun}[\varElm \mapsto
      \emptyset]}$, then $\free{\varphiFrm,\iotaFun} =
    \free{\phiFrm,{\iotaFun}[\varElm \mapsto \emptyset]}$, which means that
    $\varElm \notin \free{\varphiFrm,\iotaFun}$.
    By Lemma~\ref{lmm:mon}, Item~\ref{lmm:mon(ext:eql)}, we have that $\HypAsg
    \eql[\free{\varphiFrm,\iotaFun}] \extFun[\denot{\pm\WSet}]{\HypAsg,
      \varElm}$ and $\HypAsg' \eql[\free{\varphiFrm,\iotaFun}]
    \extFun[\denot{\pm\WSet}]{\HypAsg', \varElm}$, which means that
    $\extFun[\denot{\pm\WSet}]{\HypAsg, \varElm} \inc[\free{\phiFrm,
      {\iotaFun}[\varElm \mapsto \emptyset]}]
    \extFun[\denot{\pm\WSet}]{\HypAsg', \varElm}$, as
    $\free{\varphiFrm,\iotaFun} = \free{\phiFrm,{\iotaFun}[\varElm \mapsto
      \emptyset]}$.
    By inductive hypothesis, it holds that $\AStr, \FunAsg,
    \extFun[\denot{\pm\WSet}]{\HypAsg, \varElm} \cmodels[][\QAE] \phiFrm$,
    which amounts to $\AStr, \FunAsg, \HypAsg \cmodels[][\QAE]\: \varphi$.

    On the other hand, we also have, by semantics, $\AStr, \FunAsg, \HypAsg
    \cmodels[][\QEA] \varphi$ if and only if $\AStr, \FunAsg, \dual{\HypAsg}
    \cmodels[][\QAE] \varphi$.
    By inductive hypothesis and Lemma~\ref{lmm:mon}, Item~\ref{lmm:mon(dlt)},
    this implies $\AStr, \FunAsg, \dual{\HypAsg'} \cmodels[][\QAE]\: \varphi$,
    which amounts to $\AStr, \FunAsg, \HypAsg' \cmodels[][\QEA]\:
    \varphi$.

  \item If $\varphiFrm = \LEExs[][_{\pm\WSet}] \varElm \ldotp \phiFrm$, then
    $\AStr, \FunAsg, \HypAsg \cmodels[][\QEA]\: \varphi$ implies $\AStr,
    {\FunAsg}[\varElm \mapsto \FFun], \HypAsg \cmodels[][\QEA] \phiFrm$, for
    some function $\FFun \in \FncSet[\denot{\pm\WSet}]{}$.
    By inductive hypothesis, we have $\AStr, {\FunAsg}[\varElm \mapsto \FFun],
    \HypAsg' \cmodels[][\QEA] \phiFrm$, from which $\AStr, \FunAsg, \HypAsg'
    \cmodels[][\QEA]\: \varphi$ follows.

    On the other hand, we also have that $\AStr, \FunAsg, \HypAsg'
    \cmodels[][\QAE]\: \varphi$ implies $\AStr, {\FunAsg}[\varElm \mapsto
    \FFun], \HypAsg' \cmodels[][\QAE] \phiFrm$, for some function $\FFun \in
    \FncSet[\denot{\pm\WSet}]{}$.
    By inductive hypothesis, we have $\AStr, {\FunAsg}[\varElm \mapsto \FFun],
    \HypAsg \cmodels[][\QAE] \phiFrm$, from which $\AStr, \FunAsg, \HypAsg
    \cmodels[][\QAE]\: \varphi$ follows.

  \item Finally, let $\varphiFrm = \LAAll[][_{\pm\WSet}] \varElm \ldotp
    \phiFrm$.
    Then, $\AStr, \FunAsg, \HypAsg \cmodels[][\QEA]\: \varphi$ implies $\AStr,
    {\FunAsg}[\varElm \mapsto \FFun], \HypAsg \cmodels[][\QEA] \phiFrm$, for
    all functions $\FFun \in \FncSet[\denot{\pm\WSet}]{}$.
    By inductive hypothesis, we have that $\AStr, {\FunAsg}[\varElm \mapsto
    \FFun], \HypAsg' \cmodels[][\QEA] \phiFrm$ holds for all functions $\FFun
    \in \FncSet[\denot{\pm\WSet}]{}$, which amounts to $\AStr, \FunAsg, \HypAsg'
    \cmodels[][\QEA]\: \varphi$.

    On the other hand, we also have that $\AStr, \FunAsg, \HypAsg'
    \cmodels[][\QAE]\: \varphi$ implies $\AStr, {\FunAsg}[\varElm \mapsto
    \FFun], \HypAsg' \cmodels[][\QAE] \phiFrm$, for all functions $\FFun \in
    \FncSet[\denot{\pm\WSet}]{}$.
    By inductive hypothesis, we have that $\AStr, {\FunAsg}[\varElm \mapsto
    \FFun], \HypAsg \cmodels[][\QAE] \phiFrm$ holds for all functions $\FFun \in
    \FncSet[\denot{\pm\WSet}]{}$, which amounts to $\AStr, \FunAsg, \HypAsg
    \cmodels[][\QAE]\: \varphi$.\qedhere
  \end{itemize}

\end{proof}

\figGraphThmDblDltMeta

\begin{theorem}[name = Generalized Double Dualisation]
  \label{thm:dbldltMeta}
  For every \ADIF formula $\varphiFrm$, function assignment $\FunAsg \in
  \FunAsgSet$, and hyperteam $\HypAsg \in \HypAsgSet[\subseteq](\sup{\varphiFrm}
  \setminus \dom{\FunAsg})$, it holds that $\AStr, \FunAsg, \HypAsg
  \cmodels[][\alpha] \varphiFrm$ \iff $\AStr, \FunAsg, \dual{\dual{\HypAsg}}
  \cmodels[][\alpha] \varphiFrm$.
  Moreover, if $\FunAsg$ is acyclic, then it also holds that $\AStr, \FunAsg,
  \HypAsg \cmodels[][\alpha] \varphiFrm$ \iff $\AStr, \FunAsg, \dual{\HypAsg}
  \cmodels[][\dual{\alpha}] \varphiFrm$.
\end{theorem}
\begin{proof}
  The fact that $\AStr, \FunAsg, \HypAsg \cmodels[][\alpha] \varphi$ \iff
  $\AStr, \FunAsg, \dual{\dual{\HypAsg}} \cmodels[][\alpha] \varphi$ immediately
  follows from $\HypAsg \eqv[\free{\varphiFrm,\iota}] \dual{\dual{\HypAsg}}$,
  for every function $\iota \in \VarSet \pto \pow{\VarSet}$
  (Lemma~\ref{lmm:dlti}), and Theorem~\ref{thm:hyprefMeta}.

  We turn now to proving that $\AStr, \FunAsg, \HypAsg \cmodels[][\alpha]
  \varphi$ \iff $\AStr, \FunAsg, \dual{\HypAsg} \cmodels[][\dual{\alpha}]
  \varphi$.
  As a preliminary result, notice that if $\FunAsg$ is acyclic, then for every
  $\TeamAsg \in \AsgSet(\USet)$, for some $\USet \subseteq \VarSet$, there is a
  bijection $\tau$ from $\TeamAsg$ and $\extFun{\TeamAsg, \FunAsg}$, with
  $\tau(\asgElm) \rst[\USet] = \asgElm$.
  Consequently, it holds that $\extFun{\dual{\HypAsg}, \FunAsg} =
  \dual{\extFun{\HypAsg, \FunAsg}}$.
  The proof is done by case analysis of the syntax of the formula.

  \begin{itemize}
  \item If $\varphi = \bot$, then we have:
    \begin{itemize}
    \item $\AStr, \FunAsg, \HypAsg \cmodels[][\QEA] \varphi
      \iffExpl{sem.}  \EmpTeamAsg \in \HypAsg
      \iffExpl{Prop.~\ref{prp:empnultrv}} \dual{\HypAsg} = \EmpHypAsg
      \iffExpl{sem.} \AStr, \FunAsg, \dual{\HypAsg} \cmodels[][\QAE] \varphi$,
      and

    \item $\AStr, \FunAsg, \HypAsg \cmodels[][\QAE] \varphi \iffExpl{sem.}
      \HypAsg = \EmpHypAsg \iffExpl{Prop.~\ref{prp:empnultrv}} \HypAsg \equiv
      \EmpHypAsg \iffExpl{Lemma~\ref{lmm:dlti}} \dual{\dual{\HypAsg}} \equiv
      \EmpHypAsg \iffExpl{Prop.~\ref{prp:empnultrv}} \dual{\dual{\HypAsg}} =
      \EmpHypAsg \iffExpl{Prop.~\ref{prp:empnultrv}} \EmpTeamAsg \in
      \dual{\HypAsg} \iffExpl{sem.} \AStr, \FunAsg, \dual{\HypAsg}
      \cmodels[][\QEA] \varphi$.

    \end{itemize}

  \item If $\varphi = \top$, then we have:
    \begin{itemize}
    \item $\AStr, \FunAsg, \HypAsg \cmodels[][\QEA] \varphi
      \iffExpl{sem.} \HypAsg \neq \EmpHypAsg \iffExpl{Prop.~\ref{prp:empnultrv}}
      \HypAsg \not \equiv \EmpHypAsg \iffExpl{Lemma~\ref{lmm:dlti}}
      \dual{\dual{\HypAsg}} \not \equiv \EmpHypAsg
      \iffExpl{Prop.~\ref{prp:empnultrv}} \dual{\dual{\HypAsg}} \neq \EmpHypAsg
      \iffExpl{Prop.~\ref{prp:empnultrv}} \EmpTeamAsg \not\in \dual{\HypAsg}
      \iffExpl{sem.} \AStr, \FunAsg, \dual{\HypAsg} \cmodels[][\QAE] \varphi$,
      and

    \item $\AStr, \FunAsg, \HypAsg \cmodels[][\QAE] \varphi
      \iffExpl{sem.}  \EmpTeamAsg \not\in \HypAsg
      \iffExpl{Prop.~\ref{prp:empnultrv}} \dual{\HypAsg} \neq \EmpHypAsg
      \iffExpl{sem.} \AStr, \FunAsg, \dual{\HypAsg} \cmodels[][\QEA] \varphi$.

    \end{itemize}

  \item If $\varphi = \RRel(\xVec)$, then the claim follows from the
    semantics, Lemma~\ref{lmm:dltii}, Item~\ref{lmm:dltii(ea)}, and the fact
    that $\extFun{\dual{\HypAsg}, \FunAsg} = \dual{\extFun{\HypAsg, \FunAsg}}$.

  \item If $\varphi = \neg \psi$, then we have: $\AStr, \FunAsg, \HypAsg
    \cmodels[][\alpha]\: \varphi \iffExpl{sem.}\AStr, \FunAsg, \HypAsg
    {\notcmodels[][\dual{\alpha}]}\: \psi \iffExpl{ind.hp.}\AStr, \FunAsg,
    \dual{\HypAsg} \notcmodels[][\alpha]\: \psi \iffExpl{sem.}\AStr, \FunAsg,
    \dual{\HypAsg} \cmodels[][\dual{\alpha}]\: \varphi$.

  \item If $\varphi = \varphi_1 \wedge \varphi_2$, then we have:
    \begin{itemize}
    \item $\AStr, \FunAsg, \dual{\HypAsg} \cmodels[][\QAE]\: \varphi
      \iffExpl{sem.}  \AStr, \FunAsg, \dual{\dual{\HypAsg}} \cmodels[][\QEA]\:
      \varphi \iffExpl{Thm.~\ref{thm:dbldltMeta} (part 1)} \AStr, \FunAsg,
      \HypAsg \cmodels[][\QEA]\: \varphi$, and

    \item $\AStr, \FunAsg, \HypAsg \cmodels[][\QAE]\: \varphi \iffExpl{sem.}
      \AStr, \FunAsg, \dual{\HypAsg} \cmodels[][\QEA]\: \varphi$.

    \end{itemize}

  \item If $\varphi = \varphi_1 \vee \varphi_2$, then we have:
    \begin{itemize}
    \item $\AStr, \FunAsg, \HypAsg \cmodels[][\QEA]\: \varphi \iffExpl{sem.}
      \AStr, \FunAsg, \dual{\HypAsg} \cmodels[][\QAE]\: \varphi$, and

    \item $\AStr, \FunAsg, \dual{\HypAsg} \cmodels[][\QEA]\: \varphi
      \iffExpl{sem.}  \AStr, \FunAsg, \dual{\dual{\HypAsg}} \cmodels[][\QAE]\:
      \varphi \iffExpl{Thm.~\ref{thm:dbldltMeta} (part 1)} \AStr, \FunAsg,
      \HypAsg \cmodels[][\QAE]\: \varphi$.

    \end{itemize}

  \item If $\varphi = \LExs[][_{\pm\WSet}] \varElm \ldotp \phiFrm$, then we
    have:
    \begin{itemize}
    \item $\AStr, \FunAsg, \dual{\HypAsg} \cmodels[][\QAE]\: \varphi
      \iffExpl{sem.}\AStr, \FunAsg, \dual{\dual{\HypAsg}} \cmodels[][\QEA]\:
      \varphi \iffExpl{Thm.~\ref{thm:dbldltMeta} (part 1)} \AStr, \FunAsg,
      \HypAsg \cmodels[][\QEA]\: \varphi$, and

    \item $\AStr, \FunAsg, \HypAsg \cmodels[][\QAE]\: \varphi \iffExpl{sem.}
      \AStr, \FunAsg, \dual{\HypAsg} \cmodels[][\QEA]\: \varphi$.
    \end{itemize}

  \item If $\varphi = \LAll[][_{\pm\WSet}] \varElm \ldotp \phiFrm$, then we
    have:
    \begin{itemize}
    \item $\AStr, \FunAsg, \HypAsg \cmodels[][\QEA]\: \varphi \iffExpl{sem.}
      \AStr, \FunAsg, \dual{\HypAsg} \cmodels[][\QAE]\: \varphi$;

    \item $\AStr, \FunAsg, \dual{\HypAsg} \cmodels[][\QEA]\: \varphi
      \iffExpl{sem.}  \AStr, \FunAsg, \dual{\dual{\HypAsg}} \cmodels[][\QAE]\:
      \varphi \iffExpl{Thm.~\ref{thm:dbldltMeta} (part 1)} \AStr, \FunAsg,
      \HypAsg \cmodels[][\QAE]\: \varphi$.
      \qedhere
    \end{itemize}
  \end{itemize}
\end{proof}

\figGraphThmPrfExtMeta

\begin{theorem}[name = Generalized Prefix Extension]
  \label{thm:prfextMeta}
  Let $\qntElm \phi$ be a \ADIF formula, where $\qntElm \in \QntSet$ is a
  quantifier prefix and $\phi$ is an arbitrary \ADIF formula.
  Then, $\AStr, \FunAsg, \HypAsg \cmodels[][\alpha] \qntElm \phi$ \iff $\AStr,
  \FunAsg, \extFun[\alpha]{\HypAsg, \qntElm} \cmodels[][\alpha] \phi$, for all
  acyclic function assignments $\FunAsg \in \FunAsgSet$ and hyperteams $\HypAsg
  \in \HypAsgSet[\subseteq](\sup{\qntElm \phi} \setminus \dom{\FunAsg})$.
\end{theorem}
\begin{proof}
  We proceed by induction on the
  structure of the quantification prefix $\qntElm \in \QntSet[][]$.
  \begin{itemize}[$\bullet$]
    \item\textbf{[Base case $\qntElm = \varepsilon$]}
      Since $\extFun[\alpha]{\HypAsg, \qntElm} = \HypAsg$, there is really
      nothing to prove as the statement is trivially true.
    \item\textbf{[Inductive case $\qntElm = \Qnt[][_{\pm\WSet}] \varElm \ldotp
      \qntElm'$]}
        We proceed by a case analysis on the coherence of the quantifier
        $\Qnt$ with the alternation flag $\alpha$.
        \begin{itemize}[$-$]
          \item\textbf{[$\Qnt$ is $\alpha$-coherent]}
            By the meta-variants of Items~\ref{def:sem(adif:exs:ea)}
            and~\ref{def:sem(adif:all:ae)} of Definition~\ref{def:sem(adif)},
            it holds that $\AStr, \FunAsg, \HypAsg \!\cmodels[][\alpha]\!
            \qntElm \phi$ \iff\ \ $\AStr, \FunAsg,
            \extFun[\denot{\pm\WSet}]{\HypAsg,\!
            \varElm} \!\cmodels[][\alpha]\!\!\! \qntElm' \phi$ \iff\ \ $\AStr,
            \FunAsg, \extFun[\alpha]{\HypAsg, \Qnt[][_{\pm\WSet}] \varElm}
            \cmodels[][\alpha] \qntElm' \phi$.
            Now, by the inductive hypothesis, it follows that $\AStr, \FunAsg,
            \extFun[\alpha]{\HypAsg, \Qnt[][_{\pm\WSet}] \varElm}
            \cmodels[][\alpha] \qntElm' \phi$ \iff $\AStr, \FunAsg, \allowbreak
            \extFun[\alpha]{\extFun[\alpha]{\HypAsg, \Qnt[][_{\pm\WSet}]
            \varElm}, \qntElm'} \cmodels[][\alpha] \phi$ \iff $\AStr, \FunAsg,
            \extFun[\alpha]{\HypAsg, \qntElm} \cmodels[][\alpha] \phi$, which
            concludes the proof of this case.
          \item\textbf{[$\Qnt$ is $\dual{\alpha}$-coherent]}
            By the meta-variants of Items~\ref{def:sem(adif:exs:ae)}
            and~\ref{def:sem(adif:all:ea)} of Definition~\ref{def:sem(adif)},
            it holds that $\AStr, \FunAsg, \HypAsg \cmodels[][\alpha] \qntElm
            \phi$ \iff $\AStr, \FunAsg, \dual{\HypAsg} \cmodels[][\dual{\alpha}]
            \qntElm \phi$.
            Now, by the meta-variants of Items~\ref{def:sem(adif:exs:ea)}
            and~\ref{def:sem(adif:all:ae)} of the same definition, $\AStr,
            \FunAsg, \dual{\HypAsg} \cmodels[][\dual{\alpha}] \qntElm \phi$ \iff
            $\AStr, \FunAsg, \extFun[\denot{\pm\WSet}]{\dual{\HypAsg}, \varElm}
            \cmodels[][\dual{\alpha}] \qntElm' \phi$.
            Thanks to Theorem~\ref{thm:dbldltMeta}, $\AStr, \FunAsg,
            \extFun[\denot{\pm\WSet}]{\dual{\HypAsg}, \varElm}
            \!\cmodels[][\dual{\alpha}]\! \qntElm' \phi$ \iff $\AStr, \FunAsg,
            \dual{\extFun[\denot{\pm\WSet}]{\dual{\HypAsg}, \varElm}}
            \cmodels[][\alpha] \qntElm' \phi$ \iff $\AStr, \FunAsg,
            \extFun[\alpha]{\HypAsg,
              \Qnt[][_{\pm\WSet}] \varElm} \cmodels[][\alpha] \qntElm' \phi$.
            Summing up, $\AStr, \FunAsg, \HypAsg \cmodels[][\alpha] \qntElm
            \phi$ \iff $\AStr, \FunAsg, \extFun[\alpha]{\HypAsg,
            \Qnt[][_{\pm\WSet}] \varElm} \cmodels[][\alpha] \qntElm' \phi$.
            At this point, by the inductive hypothesis, it follows that $\AStr,
            \FunAsg, \extFun[\alpha]{\HypAsg, \Qnt[][_{\pm\WSet}] \varElm}
            \cmodels[][\alpha] \qntElm' \phi$ \iff $\AStr, \FunAsg,
            \extFun[\alpha]{\extFun[\alpha]{\HypAsg, \Qnt[][_{\pm\WSet}]
            \varElm}, \qntElm'} \cmodels[][\alpha] \phi$ \iff $\AStr, \FunAsg,
            \extFun[\alpha]{\HypAsg, \qntElm} \cmodels[][\alpha] \phi$, which
            concludes the proof of this case as well.
            \qedhere
        \end{itemize}
  \end{itemize}
\end{proof}

\lmmextint*
\begin{proof}
  We first prove Items~\ref{lmm:extint(inccoh)} and~\ref{lmm:extint(intcoh)}
  altogether, where $\Qnt$ is $\alpha$-coherent, and then we proceed with the
  remaining ones separately.
  In particular, for these last two, we make use, given an arbitrary function
  $\FFun \in \FncSet[\pm\WSet]{}$, of the auxiliary notation $\prj{\PsiSet,
  \FFun, \varElm} \defeq \set{ \asgElm \in \AsgSet(\VSet) }{ \extFun{\asgElm,
  \FFun, \varElm} \in \PsiSet }$ satisfying the following two properties, for
  every team $\TeamAsg \!\in\! \TeamAsgSet(\VSet)$:
  \begin{inparaenum}[(i)]
    \item
      $\extFun{\TeamAsg, \FFun, \varElm} \!\subseteq \PsiSet$ \iff $\TeamAsg
      \subseteq \prj{\PsiSet, \FFun, \varElm}$;
    \item
      $\extFun{\TeamAsg, \FFun, \varElm} \cap \PsiSet \neq \EmpTeamAsg$ \iff
      $\TeamAsg \cap \prj{\PsiSet, \FFun, \varElm} \neq \EmpTeamAsg$.
  \end{inparaenum}
  \begin{itemize}[$\bullet$]
    \item\textbf{[Items~\ref{lmm:extint(inccoh)}
      and~\ref{lmm:extint(intcoh)}] }
      By definition of the extension function, when $\Qnt$ is $\alpha$-coherent,
      we have that
      \[
        \extFun[\alpha]{\HypAsg, \Qnt[][_{\pm\WSet}] \varElm}
      =
        \extFun[\denot{\pm\WSet}]{\HypAsg, \varElm}
      =
        \set{ \extFun{\XSet, \FFun, \varElm} }{ \TeamAsg \in \HypAsg, \FFun \in
        \FncSet[\denot{\pm\WSet}]{} }.
      \]
      Thus, for every possible team $\TeamAsg' \in \TeamAsgSet(\VSet \cup
      \{ \varElm \})$, it holds that $\TeamAsg' \in \extFun[\alpha]{\HypAsg,
      \Qnt[][_{\pm\WSet}] \varElm}$ \iff there exists a function $\FFun \in
      \FncSet[\denot{\pm\WSet}]{}$ and a team $\TeamAsg \in \HypAsg$ such that
      $\TeamAsg' = \extFun{\TeamAsg, \FFun, \varElm}$.
      Hence, both equivalences immediately follows.
    \item\textbf{[Item~\ref{lmm:extint(incnch)}]}
      Since $\Qnt$ is $\dual{\alpha}$-coherent, $\extFun[\alpha]{\HypAsg,
        \Qnt[][_{\pm\WSet}] \varElm} =
      {\dual{\extFun[\denot{\pm\WSet}]{\dual{\HypAsg}, \varElm}}}$, and thus
      Condition~\ref{lmm:extint(incnch:org)} holds \iff there is a team $\Team'
      \in \dual{\extFun[\denot{\pm\WSet}]{\dual{\HypAsg}, \varElm}}$ such that
      $\TeamAsg' \subseteq \PsiSet$.
      By Item~\ref{lmm:dltii(ea)} of Lemma~\ref{lmm:dltii}, this holds
       \iff for all teams $\TeamAsg' \in
      \extFun[\denot{\pm\WSet}]{\dual{\HypAsg}, \varElm} =
      \extFun[\alpha]{\dual{\HypAsg}, \dual{\Qnt}[][_{\pm\WSet}] \varElm}$, it
      holds that $\TeamAsg' \cap \PsiSet \allowbreak \neq \EmpTeamAsg$.
      Thanks to Item~\ref{lmm:extint(intcoh)}, the latter is true \iff for all
      functions $\FFun \in \FncSet[\denot{\pm\WSet}]{}$ and teams $\TeamAsg \in
      \dual{\HypAsg}$, it holds that $\extFun{\TeamAsg, \FFun, \varElm} \cap
      \PsiSet \neq \EmpTeamAsg$, and thus $\TeamAsg \cap \prj{\PsiSet, \FFun,
        \varElm} \neq \EmpTeamAsg$.
      At this point, again by Item~\ref{lmm:dltii(ea)} of Lemma~\ref{lmm:dltii},
      for all $\TeamAsg \in \dual{\HypAsg}$, it holds that $\TeamAsg \cap
      \prj{\PsiSet, \FFun, \varElm} \neq \EmpTeamAsg$ \iff there exists a team
      $\TeamAsg \in \HypAsg$ such that $\TeamAsg \subseteq \prj{\PsiSet, \FFun,
        \varElm}$, and thus $\extFun{\TeamAsg, \FFun, \varElm} \subseteq
      \PsiSet$.
      Therefore, the following equivalence concludes the proof: for all
      functions $\FFun \in \FncSet[\denot{\pm\WSet}]{}$ and teams $\TeamAsg \in
      \dual{\HypAsg}$, it holds that $\TeamAsg \cap \prj{\PsiSet, \FFun,
        \varElm} \neq \EmpTeamAsg$ \iff for all functions $\FFun \in
      \FncSet[\denot{\pm\WSet}]{}$ there exists a team $\TeamAsg \in \HypAsg$
      such that $\extFun{\TeamAsg, \FFun, \varElm} \subseteq \PsiSet$, which
      coincides with Condition~\ref{lmm:extint(incnch:int)}.

    \item\textbf{[Item~\ref{lmm:extint(intnch)}]}
      Since $\Qnt$ is $\dual{\alpha}$-coherent, $\extFun[\alpha]{\HypAsg,
        \Qnt[][_{\pm\WSet}] \varElm} =
      {\dual{\extFun[\denot{\pm\WSet}]{\dual{\HypAsg}, \varElm}}}$, and thus
      Condition~\ref{lmm:extint(intnch:org)} holds \iff for all teams $\Team'
      \in \dual{\extFun[\denot{\pm\WSet}]{\dual{\HypAsg}, \varElm}}$, it holds
      that $\TeamAsg' \cap \PsiSet \neq \emptyset$.
      By Item~\ref{lmm:dltii(ea)} of Lemma~\ref{lmm:dltii}, this holds \iff
      there exists a team $\TeamAsg \in
      \extFun[\denot{\pm\WSet}]{\dual{\HypAsg}, \varElm} =
      \extFun[\alpha]{\dual{\HypAsg}, \dual{\Qnt}[][_{\pm\WSet}]
        \varElm}$ such that $\TeamAsg \subseteq \PsiSet$.
      Thanks to Item~\ref{lmm:extint(inccoh)}, the latter is true \iff there
      exist a function $\FFun \in \FncSet[\denot{\pm\WSet}]{}$ and a team
      $\TeamAsg \in \dual{\HypAsg}$ such that $\extFun{\TeamAsg, \FFun, \varElm}
      \subseteq \PsiSet$, and thus $\TeamAsg \subseteq \prj{\PsiSet, \FFun,
        \varElm}$.
      At this point, again by Item~\ref{lmm:dltii(ea)} of Lemma~\ref{lmm:dltii},
      there exists a team $\TeamAsg \in \dual{\HypAsg}$ such that $\TeamAsg
      \subseteq \prj{\PsiSet, \FFun, \varElm}$ \iff for all teams $\TeamAsg' \in
      \HypAsg$, it holds that $\TeamAsg' \cap \prj{\PsiSet, \FFun, \varElm} \neq
      \EmpTeamAsg$, and thus $\extFun{\TeamAsg', \FFun, \varElm} \cap \PsiSet
      \neq \EmpTeamAsg$.
      Therefore, the following equivalence concludes the proof: there exist a
      function $\FFun \in \FncSet[\denot{\pm\WSet}]{}$ and a team $\TeamAsg \in
      \dual{\HypAsg}$ such that $\TeamAsg \subseteq \prj{\PsiSet, \FFun,
        \varElm}$ \iff there exists a function $\FFun \in
      \FncSet[\denot{\pm\WSet}]{}$ such that for all teams $\TeamAsg' \in
      \HypAsg$, it holds that $\extFun{\TeamAsg', \FFun, \varElm} \cap \PsiSet
      \neq \EmpTeamAsg$, which coincides with
      Condition~\ref{lmm:extint(intnch:int)}.
      \qedhere
  \end{itemize}
\end{proof}

\figGraphThmQntInt

\thmqntint*
\begin{proof}
  First, observe that, by a generalisation of Theorem~\ref{thm:folsemadq} to
  \Meta, the following two equivalences hold true, where we define
  $\denot{\phiFrm} \defeq \set{ \asgElm \in \AsgSet[\subseteq](\sup{\phiFrm}) }{
    \AStr, \asgElm \cmodels[\FOL] \phiFrm }$ for every \FOL formula $\phiFrm$
  and acyclic function assignments $\FunAsg \in \FunAsgSet$:
  \begin{enumerate}[a)]
    \item\label{thm:qntint(ea)}
      $\AStr, \FunAsg, \HypAsg \cmodels[][\QEA] \phiFrm$ \iff $\TeamAsg
      \subseteq \denot{\phiFrm}$, for some team $\TeamAsg \in \extFun{\HypAsg,
      \FunAsg}$;
    \item\label{thm:qntint(ae)}
      $\AStr, \FunAsg, \HypAsg \cmodels[][\QAE] \phiFrm$ \iff $\TeamAsg \cap
      \denot{\phiFrm} \neq \EmpTeamAsg$, for all teams $\TeamAsg \in
      \extFun{\HypAsg, \FunAsg}$.
  \end{enumerate}
  which are equivalent to the following, respectively:
  \begin{itemize}[$\bullet$]
  \item
    $\AStr, \FunAsg, \HypAsg \cmodels[][\QEA] \phiFrm$ \iff $\extFun{\TeamAsg,
      \FunAsg} \subseteq \denot{\phiFrm}$, for some team $\TeamAsg \in \HypAsg$;
  \item
    $\AStr, \FunAsg, \HypAsg \cmodels[][\QAE] \phiFrm$ \iff $\extFun{\TeamAsg,
      \FunAsg} \cap \denot{\phiFrm} \neq \EmpTeamAsg$, for all teams $\TeamAsg
    \in \HypAsg$.
  \end{itemize}
  For technical convenience, given $\PsiSet \subseteq \AsgSet$ and $\USet
  \subseteq \VarSet$, let us introduce the notation $\prj{\PsiSet, \USet,
    \FunAsg} \defeq \set{ \asgElm \in \PsiSet }{ \forall \varElm \in
    \dom{\FunAsg} \setminus \USet \ldotp \asgElm(\varElm) =
    \FunAsg(\varElm)(\asgElm) } \!\rst[\USet]$ satisfying the following two
  properties, for every team $\TeamAsg \in \TeamAsgSet(\USet)$:
  \begin{inparaenum}[(i)]
    \item
      $\extFun{\TeamAsg, \FunAsg} \subseteq \PsiSet$ \iff $\TeamAsg \subseteq
      \prj{\PsiSet, \USet, \FunAsg}$;
    \item
      $\extFun{\TeamAsg, \FunAsg} \cap \PsiSet \neq \EmpTeamAsg$ \iff $\TeamAsg
      \cap \prj{\PsiSet, \USet, \FunAsg} \neq \EmpTeamAsg$.
  \end{inparaenum}
  In the light of this notation, we can rewrite the last two equivalences above
  as follows:
  \begin{itemize}[$\bullet$]
  \item
    $\AStr, \FunAsg, \HypAsg \cmodels[][\QEA] \phiFrm$ \iff $\TeamAsg \subseteq
    \prj{\denot{\phiFrm}, \var{\TeamAsg}, \FunAsg}$, for
    some team $\TeamAsg \in \HypAsg$;
  \item
    $\AStr, \FunAsg, \HypAsg \cmodels[][\QAE] \phiFrm$ \iff $\TeamAsg \cap
    \prj{\denot{\phiFrm}, \var{\TeamAsg}, \FunAsg} \neq
    \EmpTeamAsg$, for all teams $\TeamAsg \in \HypAsg$.
  \end{itemize}
    By applying to a formula $\Qnt[][_{\pm\WSet}] \varElm \ldotp \phiFrm$, where
    $\Qnt \in \{ \exists, \forall \}$, a combination of
    Theorem~\ref{thm:prfextMeta} with what we have just derived, we obtain the
    two equivalences below:
  \begin{enumerate}[i)]
  \item\label{thm:qntint(qnt:ea)}
    $\AStr, \FunAsg, \HypAsg \cmodels[][\QEA] \Qnt[][_{\pm\WSet}] \varElm \ldotp
    \phiFrm$ \iff there exists a team $\TeamAsg \in \extFun[\QEA]{\HypAsg,
      \Qnt[][_{\pm\WSet}] \varElm}$ such that $\TeamAsg \subseteq
    \prj{\denot{\phiFrm}, \var{\TeamAsg}, \FunAsg}$;
  \item\label{thm:qntint(qnt:ae)}
    $\AStr, \FunAsg, \HypAsg \cmodels[][\QAE] \Qnt[][_{\pm\WSet}] \varElm \ldotp
    \phiFrm$ \iff, for all teams $\TeamAsg \in \extFun[\QAE]{\HypAsg,
      \Qnt[][_{\pm\WSet}] \varElm}$, it holds that $\TeamAsg \cap
    \prj{\denot{\phiFrm}, \var{\TeamAsg}, \FunAsg} \neq \EmpTeamAsg$.
  \end{enumerate}
  At this point, we proceed by a case analysis on the type of quantifier $\Qnt$
  and the alternation flag $\alpha$, where we exploit the fact that, given
  $\TeamAsg \in \TeamAsgSet$,
%
%
  if
  $\varElm \notin \var{\TeamAsg}$, then for every function $\FFun \in
  \FncSet[\denot{\pm\WSet}]{}$, there exists a function $\FFun[][\star] \in
  \FncSet[\denot{\pm\WSet}]{}$ and, \viceversa, for every function
  $\FFun[][\star] \in \FncSet[\denot{\pm\WSet}]{}$, there exists a function
  $\FFun \in \FncSet[\denot{\pm\WSet}]{}$ such that
  \[
    \extFun{\extFun{\TeamAsg, \FFun, \varElm}, \FunAsg} = \extFun{\TeamAsg,
    {\FunAsg}[\varElm \mapsto \FFun[][\star]]}.
  \]
  Notice that, since $\FunAsg$ is acyclic, $\varElm \not\in \denot{\pm\WSet}$,
  and $\dom{\FunAsg} \cap \denot{\pm\WSet} = \emptyset$, it holds that
  ${\FunAsg}[\varElm \mapsto \FFun[][\star]]$ is acyclic as well.

  \begin{itemize}[$\bullet$]
  \item\textbf{[$\Qnt = \LExs$ \& $\alpha = \QEA$]}
    By Equivalence~\ref{thm:qntint(qnt:ea)}) and Item~\ref{lmm:extint(inccoh)}
    of Lemma~\ref{lmm:extint}, $\AStr, \FunAsg, \HypAsg \cmodels[][\QEA]
    \LExs[][_{\pm\WSet}] \varElm \ldotp \phiFrm$ \iff there exist a function
    $\FFun \in \FncSet[\denot{\pm\WSet}]{}$ and a team $\TeamAsg \in \HypAsg$
    such that $\extFun{\TeamAsg, \FFun, \varElm} \subseteq \prj{\denot{\phiFrm},
      \var{\TeamAsg} \cup \{ \varElm \}, \FunAsg}$, and thus
    $\extFun{\extFun{\TeamAsg, \FFun, \varElm}, \FunAsg} \subseteq
    \denot{\phiFrm}$.
      This means that $\AStr, \FunAsg, \HypAsg \cmodels[][\QEA]
      \LExs[][_{\pm\WSet}] \varElm \ldotp \phiFrm$ \iff there exist a function
      $\FFun[][\star] \in \FncSet[\denot{\pm\WSet}]{}$ and a team $\TeamAsg \in
      \HypAsg$ such that $\extFun{\TeamAsg, {\FunAsg}[\varElm \mapsto
      \FFun[][\star]]} \subseteq \denot{\phiFrm}$ \iff there exists a function
      $\FFun[][\star] \in \FncSet[\denot{\pm\WSet}]{}$ such that $\TeamAsg
      \subseteq \denot{\phiFrm}$, for some team $\TeamAsg \in \extFun{\HypAsg,
      {\FunAsg}[\varElm \mapsto \FFun[][\star]]}$.
    By Equivalence~\ref{thm:qntint(ea)}), the latter statement can be rewritten
    as: there exists a function $\FFun[][\star] \in \FncSet[\denot{\pm\WSet}]{}$
    such that $\AStr, {\FunAsg}[\varElm \mapsto \FFun[][\star]], \HypAsg
    \cmodels[][\QEA] \phiFrm$; this in turn is equivalent to $\AStr, \FunAsg,
    \HypAsg \cmodels[][\QEA] \LEExs[][_{\pm\WSet}] \varElm \ldotp \phiFrm$,
    due to Item~\ref{def:sem(met:eexs)} of Definition~\ref{def:sem(met)}.
      This concludes the proof of Item~\ref{thm:qntint(exs)} for $\alpha
      = \QEA$.
    \item\textbf{[$\Qnt = \LExs$ \& $\alpha = \QAE$]}
      By Equivalence~\ref{thm:qntint(qnt:ae)}) and Item~\ref{lmm:extint(intnch)}
      of Lemma~\ref{lmm:extint}, $\AStr, \FunAsg, \HypAsg \cmodels[][\QAE]
      \LExs[][_{\pm\WSet}] \varElm \ldotp \phiFrm$ \iff there exists a function
      $\FFun \in \FncSet[\denot{\pm\WSet}]{}$ such that, for all teams $\TeamAsg
      \in \HypAsg$, it holds true that $\extFun{\TeamAsg, \FFun,
        \varElm} \cap \prj{\denot{\phiFrm}, \var{\TeamAsg} \cup \{ \varElm \},
        \FunAsg} \neq
      \EmpTeamAsg$, and thus $\extFun{\extFun{\TeamAsg, \FFun, \varElm},
        \FunAsg} \cap \denot{\phiFrm} \neq \EmpTeamAsg$.
      This means that $\AStr, \FunAsg, \HypAsg \cmodels[][\QAE]
      \LExs[][_{\pm\WSet}] \varElm \ldotp \phiFrm$ \iff there exists a function
      $\FFun[][\star] \in \FncSet[\denot{\pm\WSet}]{}$ such that, for all teams
      $\TeamAsg \in \HypAsg$, it holds that $\extFun{\TeamAsg, {\FunAsg}[\varElm
      \mapsto \FFun[][\star]]} \cap \denot{\phiFrm} \neq \EmpTeamAsg$ \iff
      there exists a function $\FFun[][\star] \in \FncSet[\denot{\pm\WSet}]{}$
      such that $\TeamAsg \cap \denot{\phiFrm} \neq \EmpTeamAsg$, for all teams
      $\TeamAsg \in \extFun{\HypAsg, {\FunAsg}[\varElm \mapsto
      \FFun[][\star]]}$.
    By Equivalence~\ref{thm:qntint(ae)}), the latter statement can be rewritten
    as: there exists a function $\FFun[][\star] \in \FncSet[\denot{\pm\WSet}]{}$
    such that $\AStr, {\FunAsg}[\varElm \mapsto \FFun[][\star]], \HypAsg
    \cmodels[][\QAE] \phiFrm$; this in turn is equivalent to $\AStr, \FunAsg,
    \HypAsg \cmodels[][\QAE] \LEExs[][_{\pm\WSet}] \varElm \ldotp \phiFrm$,
    due to Item~\ref{def:sem(met:eexs)} of Definition~\ref{def:sem(met)}.
      This concludes the proof of Item~\ref{thm:qntint(exs)} for $\alpha
      = \QAE$.
    \item\textbf{[$\Qnt = \LAll$ \& $\alpha = \QEA$]}
      By Equivalence~\ref{thm:qntint(qnt:ea)}) and Item~\ref{lmm:extint(incnch)}
      of Lemma~\ref{lmm:extint}, $\AStr, \FunAsg, \HypAsg \cmodels[][\QEA]
      \LAll[][_{\pm\WSet}] \varElm \ldotp \phiFrm$ \iff, for all functions
      $\FFun \in \FncSet[\denot{\pm\WSet}]{}$, there exists a team $\TeamAsg
      \!\in\! \HypAsg$ such that $\extFun{\TeamAsg, \FFun, \varElm}
      \!\subseteq\! \prj{\denot{\phiFrm}, \var{\TeamAsg} \cup \{ \varElm \},
        \FunAsg}$, and thus $\extFun{\extFun{\TeamAsg, \FFun, \varElm}, \FunAsg}
      \subseteq \denot{\phiFrm}$.
      This means that $\AStr, \FunAsg, \HypAsg \cmodels[][\QEA]
      \LAll[][_{\pm\WSet}] \varElm \ldotp \phiFrm$ \iff, for all functions
      $\FFun[][\star] \in \FncSet[\denot{\pm\WSet}]{}$, there exists a team
      $\TeamAsg \in \HypAsg$ such that $\extFun{\TeamAsg, {\FunAsg}[\varElm
      \mapsto \FFun[][\star]]} \subseteq \denot{\phiFrm}$ \iff, for all
      functions $\FFun[][\star] \in \FncSet[\denot{\pm\WSet}]{}$, it holds that
      $\TeamAsg \subseteq \denot{\phiFrm}$, for some team $\TeamAsg \in
      \extFun{\HypAsg, {\FunAsg}[\varElm \mapsto \FFun[][\star]]}$.
      By Equivalence~\ref{thm:qntint(ea)}), the latter statement can be
      rewritten as: for all functions $\FFun[][\star] \in
      \FncSet[\denot{\pm\WSet}]{}$, it holds that $\AStr, {\FunAsg}[\varElm
      \mapsto \FFun[][\star]], \HypAsg \cmodels[][\QEA] \phiFrm$; this in
      turn is equivalent to $\AStr, \FunAsg, \HypAsg \cmodels[][\QEA]
      \LAAll[][_{\pm\WSet}] \varElm \ldotp \phiFrm$, due to
      Item~\ref{def:sem(met:aall)} of Definition~\ref{def:sem(met)}.
      This concludes the proof of Item~\ref{thm:qntint(all)} for $\alpha
      = \QEA$.
    \item\textbf{[$\Qnt = \LAll$ \& $\alpha = \QAE$]}
      By Equivalence~\ref{thm:qntint(qnt:ae)}) and Item~\ref{lmm:extint(intcoh)}
      of Lemma~\ref{lmm:extint}, $\AStr, \FunAsg, \HypAsg \cmodels[][\QAE]
      \LAll[][_{\pm\WSet}] \varElm \ldotp \phiFrm$ \iff, for all functions
      $\FFun \in \FncSet[\denot{\pm\WSet}]{}$ and teams $\TeamAsg \in \HypAsg$,
      it holds that $\extFun{\TeamAsg, \FFun, \varElm} \cap
      \prj{\denot{\phiFrm}, \var{\HypAsg} \cup \{ \varElm \}, \FunAsg} \neq
      \EmpTeamAsg$, and thus $\extFun{\extFun{\TeamAsg, \FFun, \varElm},
        \FunAsg} \cap \denot{\phiFrm} \neq \EmpTeamAsg$.
      This means that $\AStr, \FunAsg, \HypAsg \cmodels[][\QAE]
      \LAll[][_{\pm\WSet}] \varElm \ldotp \phiFrm$ \iff, for all functions
      $\FFun[][\star] \in \FncSet[\denot{\pm\WSet}]{}$ and teams $\TeamAsg \in
      \HypAsg$, it holds that $\extFun{\TeamAsg, {\FunAsg}[\varElm \mapsto
      \FFun[][\star]]} \cap \denot{\phiFrm} \neq \EmpTeamAsg$ \iff, for all
      functions $\FFun[][\star] \in \FncSet[\denot{\pm\WSet}]{}$, it holds that
      $\TeamAsg \cap \denot{\phiFrm} \neq \EmpTeamAsg$, for all teams $\TeamAsg
      \in \extFun{\HypAsg, {\FunAsg}[\varElm \mapsto \FFun[][\star]]}$.
      By Equivalence~\ref{thm:qntint(ae)}), the latter statement can be
      rewritten as: for all functions $\FFun[][\star] \in
      \FncSet[\denot{\pm\WSet}]{}$, it holds that $\AStr, {\FunAsg}[\varElm
      \mapsto \FFun[][\star]], \HypAsg \cmodels[][\QAE] \phiFrm$; this in
      turn is equivalent to $\AStr, \FunAsg, \HypAsg \cmodels[][\QAE]
      \LAAll[][_{\pm\WSet}] \varElm \ldotp \phiFrm$, due to
      Item~\ref{def:sem(met:aall)} of Definition~\ref{def:sem(met)}.
      This concludes the proof of Item~\ref{thm:qntint(all)} for $\alpha
      = \QAE$.
      \qedhere
  \end{itemize}
\end{proof}

\figGraphThmHST

\thmhst*
\begin{proof}
  The proof proceeds by structural induction on the quantifier prefix $\qntElm
  \!\in\! \QntSet$.
  \begin{itemize}[$\bullet$]
    \item\textbf{[Base case $\qntElm = \varepsilon$]}
      Since $\hspFun{\qntElm} = \varepsilon$, there is really nothing to prove
      as the statement is trivially true.
    \item\textbf{[Inductive case $\qntElm = \qntElm' \ldotp \Qnt[][_{\pm\WSet}]
      \varElm$]}
      By Theorem~\ref{thm:prfextMeta}, it holds that
      $\AStr, \FunAsg, \HypAsg \cmodels[][\alpha] \qntElm \phi$ \iff $\AStr,
      \FunAsg, \extFun[\alpha]{\HypAsg, \qntElm'} \cmodels[][\alpha]
      \Qnt[][_{\pm\WSet}] \varElm \ldotp \phi$.
      A case analysis on the type of quantifier is now required.
      \begin{itemize}[$\bullet$]
        \item\textbf{[$\Qnt = \exists$]}
          By Item~\ref{thm:qntint(exs)} of Theorem~\ref{thm:qntint}, $\AStr,
          \FunAsg, \extFun[\alpha]{\HypAsg, \qntElm'} \cmodels[][\alpha]
          \LExs[][_{\pm\WSet}] \varElm \ldotp \phi$ \iff $\AStr, \FunAsg,
          \allowbreak \extFun[\alpha]{\HypAsg, \qntElm'} \cmodels[][\alpha]
          \LEExs[][_{\pm\WSet}] \varElm \ldotp \phiFrm$, since $\phiFrm$ is
          a \FOL formula, being $\qntElm \phi$ in \pnf.
          Thus, by Item~\ref{def:sem(met:eexs)} of
          Definition~\ref{def:sem(met)}, we have that $\AStr, \FunAsg, \HypAsg
          \cmodels[][\alpha] \qntElm \phi$ \iff there exists a function $\FFun
          \in \FncSet[\denot{\pm\WSet}]{}$ such that $\AStr, {\FunAsg}[\varElm
          \mapsto \FFun], \extFun[\alpha]{\HypAsg, \qntElm'} \cmodels[][\alpha]
          \phiFrm$.
          Observe that ${\FunAsg}[\varElm \mapsto \FFun]$ is still acyclic, due
          to assumptions made at page~\pageref{pag:Qassumption} on $\QntSet$
          and the fact that $\dom{\FunAsg} \cap \dep{\qntElm} = \emptyset$.
          Again by Theorem~\ref{thm:prfextMeta}, $\AStr, {\FunAsg}[\varElm
          \mapsto \FFun], \extFun[\alpha]{\HypAsg, \qntElm'} \cmodels[][\alpha]
          \phiFrm$ is equivalent to $\AStr, {\FunAsg}[\varElm \mapsto \FFun],
          \HypAsg \cmodels[][\alpha] \qntElm' \phiFrm$, which in turn, by the
          inductive hypothesis applied to $\qntElm' \phiFrm$, is equivalent to
          $\AStr, {\FunAsg}[\varElm \mapsto \FFun], \HypAsg \cmodels[][\alpha]
          \hspFun{\qntElm'} \phiFrm$.
          Summing up, we have $\AStr, \FunAsg, \HypAsg \cmodels[][\alpha]
          \qntElm \phi$ \iff there exists a function $\FFun \in
          \FncSet[\denot{\pm\WSet}]{}$ such that $\AStr, {\FunAsg}[\varElm
          \mapsto \FFun], \HypAsg \cmodels[][\alpha] \hspFun{\qntElm'} \phiFrm$.
          At this point, again by Item~\ref{def:sem(met:eexs)} of
          Definition~\ref{def:sem(met)}, we obtain $\AStr, \FunAsg, \HypAsg
          \cmodels[][\alpha] \qntElm \phi$ \iff $\AStr, \FunAsg, \HypAsg
          \cmodels[][\alpha] \LEExs[][_{\pm\WSet}] \varElm \ldotp
          \hspFun{\qntElm'} \phiFrm$ \iff $\AStr, \FunAsg, \HypAsg
          \cmodels[][\alpha] \hspFun{\qntElm} \phiFrm$, where the latter
          equivalence is due to the definition of the $\hspFun{}$ function
          satisfying the equality $\hspFun{\qntElm} = \hspFun{\qntElm' \ldotp
          \LExs[][_{\pm\WSet}] \varElm} = \LEExs[][_{\pm\WSet}] \varElm \ldotp
          \hspFun{\qntElm'}$.
          This concludes the proof of the existential case.
        \item\textbf{[$\Qnt = \forall$]}
          By Item~\ref{thm:qntint(all)} of Theorem~\ref{thm:qntint}, $\AStr,
          \FunAsg, \extFun[\alpha]{\HypAsg, \qntElm'} \cmodels[][\alpha]
          \LAll[][_{\pm\WSet}] \varElm \ldotp \phi$ \iff $\AStr, \FunAsg,
          \allowbreak \extFun[\alpha]{\HypAsg, \qntElm'} \cmodels[][\alpha]
          \LAAll[][_{\pm\WSet}] \varElm \ldotp \phiFrm$, since $\phiFrm$ is
          a \FOL formula, being $\qntElm \phi$ in \pnf.
          Thus, by Item~\ref{def:sem(met:aall)} of
          Definition~\ref{def:sem(met)}, we have that $\AStr, \FunAsg, \HypAsg
          \cmodels[][\alpha] \qntElm \phi$ \iff, for all functions $\FFun \in
          \FncSet[\denot{\pm\WSet}]{}$, it holds that $\AStr, {\FunAsg}[\varElm
          \mapsto \FFun], \extFun[\alpha]{\HypAsg, \qntElm'} \cmodels[][\alpha]
          \phiFrm$.
          Observe again that ${\FunAsg}[\varElm \mapsto \FFun]$ is acyclic.
          By Theorem~\ref{thm:prfextMeta}, $\AStr, {\FunAsg}[\varElm \mapsto
          \FFun], \extFun[\alpha]{\HypAsg, \qntElm'} \cmodels[][\alpha] \phiFrm$
          is equivalent to $\AStr, {\FunAsg}[\varElm \mapsto \FFun], \HypAsg
          \cmodels[][\alpha] \qntElm' \phiFrm$, which in turn, by the inductive
          hypothesis applied to $\qntElm' \phiFrm$, is equivalent to $\AStr,
          {\FunAsg}[\varElm \mapsto \FFun], \HypAsg \cmodels[][\alpha]
          \hspFun{\qntElm'} \phiFrm$.
          Summing up, we have $\AStr, \FunAsg, \HypAsg \cmodels[][\alpha]
          \qntElm \phi$ \iff, for all functions $\FFun \in
          \FncSet[\denot{\pm\WSet}]{}$, it holds that $\AStr, {\FunAsg}[\varElm
          \mapsto \FFun], \HypAsg \cmodels[][\alpha] \hspFun{\qntElm'} \phiFrm$.
          At this point, again by Item~\ref{def:sem(met:aall)} of
          Definition~\ref{def:sem(met)}, we obtain $\AStr, \FunAsg, \HypAsg
          \cmodels[][\alpha] \qntElm \phi$ \iff $\AStr, \FunAsg, \HypAsg
          \cmodels[][\alpha] \LAAll[][_{\pm\WSet}] \varElm \ldotp
          \hspFun{\qntElm'} \phiFrm$ \iff $\AStr, \FunAsg, \HypAsg
          \cmodels[][\alpha] \hspFun{\qntElm} \phiFrm$, where the latter
          equivalence is due to the definition of the $\hspFun{}$ function
          satisfying the equality $\hspFun{\qntElm} = \hspFun{\qntElm' \ldotp
          \LAll[][_{\pm\WSet}] \varElm} = \LAAll[][_{\pm\WSet}] \varElm \ldotp
          \hspFun{\qntElm'}$.
          This concludes the proof of the universal case too.
          \qedhere
      \end{itemize}
  \end{itemize}
\end{proof}

\figGraphThmAdifSolInt

\thmadifsolint*
\begin{proof}
  For simplicity, we just consider the case of \ADF formulae; given the set of
  variables $\VSet$, a conversion from \AIF formulae to \ADF formulae is
  immediate.
  As first step, consider a formula $\varphiFrm = \qntElm \phiFrm$ in prenex
  form, where $\qntElm$ is a quantifier prefix and $\phiFrm$ a quantifier-free
  matrix.
  Then, by Theorem~\ref{thm:hst}, we further transform the latter into the
  equivalent Meta-\ADF formula $\hspFun{\qntElm} \phiFrm$.
  Obviously, $\hspFun{\qntElm} = (\Qnt[i][_{+\WSet[i]}]\, \varElm[i])_{i =
  1}^{k}$, for some $k \in \SetN$, where $\WSet[i] \subseteq \VarSet$ and
  $\Qnt[i] \in \{ \LEExs, \LAAll \}$.
  Now, let $\der{\qntElm} \defeq (\der{\Qnt}[i] \funElm[i])_{i = 1}^{k}$ be the
  second-order function-quantifier prefix, where
  \begin{inparaenum}[(i)]
    \item
      the arity of each function symbol $\funElm[i]$ equals the number of
      variables $\varElm[i]$ depend on, \ie, $\art{\funElm[i]} =
      \card{\WSet[i]}$, and
    \item
      each second-order quantifier symbol $\der{\Qnt}[i] \in \{ \LExs, \LAll \}$
      is existential \iff the first-order quantifier symbol $\Qnt[i]$ is
      existential too.
  \end{inparaenum}
  At this point, the \SOL sentences $\PhiSnt[\QEA]$ and $\PhiSnt[\QAE]$ can be
  defined as follows, where $\der{\phiFrm}$ is obtained from the matrix
  $\phiFrm$ by replacing each occurrence of a variable $\varElm[i]$ with the
  corresponding term $\funElm[i](\wVec[i])$, where $\wVec[i]$ is a vector of all
  the variables in $\WSet[i]$:
  \begin{enumerate}[1)]
    \item
      $\PhiSnt[\QEA] \defeq \der{\qntElm} \ldotp \LExs \yvarElm \ldotp \LAll
      \xVec \ldotp \neg \RRel(\xVec\yvarElm) \LDis \der{\phiFrm}$;
    \item
      $\PhiSnt[\QAE] \defeq \der{\qntElm} \ldotp \LAll \yvarElm \ldotp \LExs
      \xVec \ldotp \RRel(\xVec\yvarElm) \LCon \der{\phiFrm}$.
  \end{enumerate}
  To conclude, the correctness of the translation can be proved by a simple
  induction on the length of the quantifier prefix $\qntElm$, where, as base
  case, we exploit the extension of Theorem~\ref{thm:folsemadq} to Meta-\ADIF.
\end{proof}

\figGraphThmSolAdfInt

\thmsoladfint*
\begin{proof}
  To begin with, let us assume \wlogx (see~\cite{KN09} for a proof) that the
  \SOL sentence $\PhiSnt$ is of the form
  \[
    (\Qnt[i] \funElm[i])_{i = 1}^{k} \ldotp \LAll \xVec \ldotp (\RRel(\yVec)
    \leftrightarrow \terElm[1] = \terElm[2]) \wedge \psiFrm,
  \]
  which in addition complies with the following constraints:
  \begin{enumerate}[a)]
    \item
      $\var{\PhiSnt} \cap \VSet = \emptyset$, \ie, no variable in $\VSet$ occurs
      in $\PhiSnt$;
    \item
      $\yVec \subseteq \xVec$, \ie, the vector of variables $\yVec$ used in the
      atom $\RRel(\yVec)$ is included in the vector of universally-quantified
      variables $\xVec$;
    \item
      every function $\funElm[i]$ only appears in a single term
      $\terElm[{\funElm[i]}] = \funElm[i](\wVec[i])$;
    \item
      every term $\terElm$ (including $\terElm[1]$ and $\terElm[2]$) is of the
      form $\funElm[i](\wVec)$, for some index $i \in \numcc{1}{k}$ and vector
      of variables $\wVec \subseteq \xVec$;
    \item
      the relation $\RRel$ does not occur in the \FOL formula $\psiFrm$.
  \end{enumerate}
  Now, let $\qntElm \defeq (\der{\Qnt}[i][_{+\WSet[i]}] \zvarElm[i])_{i =
  k}^{1}$ be the first-order quantifier prefix, where
  \begin{inparaenum}[(i)]
    \item
      the set of dependence variables $\WSet[i]$ coincides with the vector of
      variables $\wVec[i]$ used in the term $\terElm[{\funElm[i]}]$
      corresponding to the function $\funElm[i]$, and
    \item
      each first-order quantifier symbol $\der{\Qnt}[i] \in \{ \LExs, \LAll \}$
      is existential \iff the second-order quantifier symbol $\Qnt[i]$ is
      existential too.
  \end{inparaenum}
  Notice that the order of quantification is reversed \wrt the one in $(\Qnt[i]
  \funElm[i])_{i = 1}^{k}$.
  At this point, the \ADF formula $\varphiFrm$ can be defined as follows, where
  \begin{inparaenum}[(1)]
    \item
      $(\yVec = \VSet)$ denotes a shortcut for a conjunction of equalities
      between every variable in $\yVec$ and the associated variable in $\VSet$
      (the association is the one used in the team encoding
      $\TeamFun{\RRel[][\AStr], \VSet}$),
    \item
      $\zvarElm[1]'$ and $\zvarElm[2]'$ are the variables corresponding to the
      functions used in the terms $\terElm[1]$ and $\terElm[2]$, and
    \item
      $\psiFrm'$ is the \FOL formula obtained from $\psiFrm$ by replacing each
      occurrence of a term $\terElm[{\funElm[i]}]$ with the corresponding
      variable $\zvarElm[i]$:
  \end{inparaenum}
  \[
    \varphiFrm \defeq \LAll \xVec \ldotp \qntElm \ldotp ((\yVec = \VSet)
    \leftrightarrow \zvarElm[1]' = \zvarElm[2]') \wedge \psiFrm'.
  \]
  To conclude, the correctness of the translation can be shown by first
  applying Theorem~\ref{thm:hst} to $\varphiFrm$, obtaining the Meta-\ADF
  formula
  \[
    \hspFun{\qntElm} \ldotp \LAll \xVec \ldotp ((\yVec = \VSet)
    \leftrightarrow \zvarElm[1]' = \zvarElm[2]') \wedge \psiFrm',
  \]
  and then proceeding with a standard induction on the length of the quantifier
  prefix $(\Qnt[i] \funElm[i])_{i = 1}^{k}$.
\end{proof}





\section{Proofs of Section~\ref{sec:gamthrsem}}
\label{app:gamthrsem}

In order to prove Theorem~\ref{thm:gamthrsem}, we shall first prove two
additional lemmata.
The first one states a Skolemisation property for \Meta. A sentence of \Meta in
prenex form that only has meta quantifiers $\LEExs$ or $\LAAll$ can be viewed as
an \SOL formula.
Therefore, we can use classic Skolem results to define a function
$\SkolemizedSOF_{\varElm}$ for the first existentially quantified variable
$\varElm$ such that if $\FunAsg$ is a function assignment of variables
(universally) quantified before $\varElm$, then $\FunAsg{}[\varElm \mapsto
  \SkolemizedSOF_{\varElm}(\FunAsg)]$ satisfies the subformula that follows the
quantification of $\varElm$.
We need some notation first.  For a quantifier prefix $\qntElm =
\Qnt[0][\WSet_0]\varElm[0] \ldots \Qnt[n][\WSet_n]\varElm[n]$ and a quantifier
symbol $\Qnt \in \{\LEExs, \LAAll\}$, the set $\var[{\Qnt}]{\qntElm} =
\{\varElm[i] | \Qnt[i] = \Qnt\}$ collects all the variables quantified in
$\qntElm$ using the specific symbol $\Qnt$.
A \emph{Skolemisation} for $\qntElm$ is a sequence $(\SkolemizedSOF_{\varElm[i]}
:(\prod_{j<i} \FncSet[{\WSet_j}]{}) \rightarrow
\FncSet[{\WSet_i}]{})_{\varElm[i]\in \var[{\Sigma}]{\qntElm}}$ of functions, one
for each existentially quantifier variable $\varElm[i]$ of $\qntElm$ and each
one intuitively mapping the interpretations of the variables preceding
$\varElm[i]$ in $\qntElm$ to some interpretation for $\varElm[i]$.
A \emph{Skolem extension} of $\FunAsg$ \wrt a Skolemisation
$(\SkolemizedSOF_{\varElm[i]})_{\varElm[i]\in \var[{\Sigma}]{\qntElm}}$ for $\qntElm$ is a
function assignment $\FunAsg'$ such that: \emph{(i)} $\dom{\FunAsg'} =
\dom{\FunAsg} \cup \var{\qntElm}$; \emph{(ii)} $\FunAsg'(\varElm) = \FunAsg(\varElm)$, for
$\varElm \in \dom{\FunAsg}\setminus \var{\qntElm}$; and \emph{(iii)}
$\FunAsg'(\varElm[i]) = \SkolemizedSOF_{\varElm[i]}((\FunAsg'(\varElm[j]))_{j<i})$,
if $\varElm[i] \in \var[{\Sigma}]{\qntElm}$.
%
Observe that $\FunAsg$ assigns a function to each variable in $\qntElm$, using
the Skolemisation for the existentially quantified variables and arbitrary
functions for the universally quantified ones. We can now state the following
lemma.

\begin{lemma}[\Meta Skolemisation]
	\label{lmm:MetaSkol}
	Let $\HypAsg$ be a hyperteam, $\FunAsg$ a function assignment and $\varphiFrm
  = \qntElm \psiFrm$ a \Meta formula in prenex form, where $\qntElm =
  \Qnt[0][\WSet_0]\varElm[0] \ldots \Qnt[n][\WSet_n]\varElm[n]$ with $\Qnt[i]
  \in \{\LEExs, \LAAll\}$ for $i\leq n$. The following holds: $\AStr, \FunAsg,
  \HypAsg \cmodels[][\alpha] \varphiFrm$ \iff there exists a Skolemisation
  $(\SkolemizedSOF_{\varElm[i]})_{\varElm[i]\in \var[{\Sigma}]{\qntElm}}$ for $\qntElm$ such
  that $\AStr,\FunAsg', \HypAsg \cmodels[][\alpha] \psiFrm$, for all Skolem
  extensions $\FunAsg'$ of $\FunAsg$ \wrt $(\SkolemizedSOF_{\varElm[i]})_{\varElm[i]\in
    \var[{\Sigma}]{\qntElm}}$.
\end{lemma}



\begin{proof}
  We prove the result by induction on the size of $\var[{\Sigma}]{\qntElm}$.
	\begin{description}
	\item[Base case ${\var[{\Sigma}]{\qntElm}} = \emptyset$.] The only
    Skolemisation for $\qntElm$ is the empty sequence of functions. A simple
    application of the semantic rules for the universal quantifiers, applied to
    $\LAAll \varElm[i]$ for each $i\leq n$, gives the result.
  \item[Inductive case.] Suppose the property holds for all formulae with
    $|\var[{\Sigma}]{\qntElm}| < n$.
    We construct $\SkolemizedSOF_{\varElm}$ for each $\varElm \in \var[{\Sigma}]{\qntElm}$
    with the desired properties.
    Let $i_0$ be the smallest integer such that $\varElm[{i_0}] \in
    \var[{\Sigma}]{\qntElm}$, so that we can set $\varphiFrm = \LAAll[][\WSet_{0}]
    \varElm[0]\ldots\LAAll[][\WSet_{i_0-1}] \varElm[{i_{0}-1}]
    \LEExs[][\WSet_{i_0}]\varElm[i_0] \varphiFrm'$ and $\varphiFrm' =
    \Qnt[i_0+1][\WSet_{i_0+1}]\varElm[i_0+1]\ldots \Qnt[n][\WSet_{n}]\varElm[n]
    \psi = \qntElm'\psi$.
    By application of the semantic rules for the $i_0-1$ universal quantifiers
    and the last existential one, we obtain that $\AStr, \FunAsg, \HypAsg
    \cmodels[][\alpha] \varphiFrm$ \iff for every sequence of functions
    $(\FFun[{\varElm[j]}])_{j < i_0}$, with $\FFun[{\varElm[j]}] \in
    \FncSet[{\WSet_j}]{}$, there is function $\FFun[{\varElm[i_0]}] \in
    \FncSet[{\WSet_{i_0}}]{}$ such that $\AStr, \FunAsg', \HypAsg
    \cmodels[][\alpha] \varphiFrm'$, with $\FunAsg'= \FunAsg{}[{\varElm[0]}
      \mapsto {\FFun[{\varElm_0}]},\ldots, {\varElm[i_0]} \mapsto
              {\FFun[{\varElm[i_0]}]}]$.
%
    Now, since $\var[\Sigma]{\qntElm'} < n$, by inductive hypothesis $\AStr,
    \FunAsg', \HypAsg \cmodels[][\alpha] \varphiFrm'$ \iff there is
    Skolemisation $(\SkolemizedSOF'_{\varElm[i]})_{\varElm[i]\in\var[{\Sigma}]{\qntElm'}}$
    for $\qntElm'$ such that
    $\AStr, \FunAsg'', \HypAsg \cmodels[][\alpha] \psi$, for every Skolem
    extension $\FunAsg''$ of $\FunAsg'$.
    We then have that $\AStr, \FunAsg, \HypAsg \cmodels[][\alpha] \varphiFrm$
    \iff for all sequences of functions $(\FFun[{\varElm[j]}])_{j < i_0}$, there
    exist a function $\FFun[{\varElm[i_{0}]}]$ and a Skolemisation
    $(\SkolemizedSOF'_{\varElm[i]})_{\varElm[i]\in\var[{\Sigma}]{\qntElm'}}$ for $\qntElm'$
    such that $\AStr, \FunAsg'', \HypAsg \cmodels[][\alpha] \psi$, for every
    Skolem extension $\FunAsg''$ of $\FunAsg{}[{\varElm[0]} \mapsto
      {\FFun[{\varElm_0}]},\ldots, {\varElm[i_0]} \mapsto
      {\FFun[{\varElm[i_0]}]}]$ \wrt
    $(\SkolemizedSOF'_{\varElm[i]})_{\varElm[i]\in\var[{\Sigma}]{\qntElm'}}$.
    Since the choices of $\FFun[{\varElm[i_{0}]}]$ and of the Skolemisation
    $(\SkolemizedSOF'_{\varElm[i]})_{\varElm[i]\in\var[{\Sigma}]{\qntElm'}}$ depend on the
    sequence $(\FFun[{\varElm[j]}])_{j < i_0}$, obviously there exists a
    Skolemisation $(\SkolemizedSOF_{\varElm[i]})_{\varElm[i]\in\var[{\Sigma}]{\qntElm}}$
    for $\qntElm$ such that $\AStr, \FunAsg'', \HypAsg \cmodels[][\alpha] \psi$,
    for all Skolem extension $\FunAsg''$ of $\FunAsg$ \wrt
    $(\SkolemizedSOF_{\varElm[i]})_{\varElm[i]\in\var[{\Sigma}]{\qntElm}}$. Indeed, for all
    sequences $(\FFun[{\varElm[j]}])_{j < i_0}$, the function
    $\FFun[{\varElm[i_{0}]}]$ and the Skolemisation
    $(\SkolemizedSOF'_{\varElm[i_k]})_{\varElm[i]\in\var[{\Sigma}]{\qntElm'}}$ defined as
    follow satisfy the properties shown above:
    \begin{itemize}
  \item $\FFun[{\varElm[i_{0}]}] = \SkolemizedSOF_{\varElm[i_0]}((\FFun[{\varElm[j]}])_{j <
    i_0})$;
  \item $\SkolemizedSOF'_{\varElm[i_k]}((\FFun[{\varElm[j]}])_{i_{0} < j < i_k})
    = \SkolemizedSOF_{\varElm[i_k]}((\FFun[{\varElm[j]}])_{j < i_k})$, for all
    $\varElm[i_k] \in \var[\Sigma]{\qntElm'}$ and sequence of functions
    $(\FFun[{\varElm[j]}])_{i_{0} < j < i_k}$.
    \end{itemize}
  \end{description}
  \vspace{-1em}
\end{proof}

The second lemma proves a property of the independence game
$\GameName[\varphiFrm][\AStr]$ defined in Construction~\ref{cns:semgam} for an
\ADIF sentence $\varphiFrm$ and a structure $\AStr$.  It states that, after a
history $\hstElm$, no matter how the functions in each bucket are chosen, the
only assignment that is coherent with the functions in the bucket is the one
associated with the last position of $\hstElm$.
In the following, we consider an \ADIF sentence $\varphiFrm = \qntElm \psiFrm$
in prenex form, with $\qntElm = \Qnt[0][\WSet_0]\varElm[0]
\ldots \Qnt[n][\WSet_n]\varElm[n]$ for $\Qnt[i]\in \set{\forall\exists}{}$, and
$\psiFrm$ quantifier free.
For every subformula $\phiFrm= \Qnt[i][\WSet_i]\varElm[i] \phiFrm'$, we rename
the buckets $\BCls[\phiFrm](\pthElm)$ by $\BCls[{\varElm[i]}](\pthElm)$ and
associate priorities with variables by setting $\prtFun(\varElm[i]) =
\prtFun(\phiFrm)$. Let $\Buckets \defeq 2^{\FncSet{}}$ denote the set of all
buckets. For convenience, we set $\VarX = \{\varElm[0], \ldots, \varElm[n]\}$
and $\VarX[i] = \{\varElm[0], \ldots, \varElm[i]\}$, $\VarX[\exists] =
\{\varElm[i] \in \VarX| \Qnt[i] = \exists\}$ and $\VarX[\forall] = \VarX
\setminus \VarX[\exists]$.  We also introduce choice functions over buckets.
Basically, a choice function over buckets chooses, for each variable $\varElm$,
a function $\FFun$ in the bucket of $\varElm$. It takes both
$\BCls[\varElm](\pthElm)$ and $\varElm$ in input because there might be multiple
variables with the same bucket (for instance, when they all depend exactly on
the same variables and the same value have been played for all of them during
the play).
\[
\ChoiceFunsBuckets= \set{ \choiceFunBucket \colon (\Buckets \times \VarX)
\rightarrow \FncSet{} }{ \forall \BSet \in \Buckets, \forall \varElm \in \VarX,
\choiceFunBucket(\BSet, \varElm) \in \BSet }
\]
Given a function $\FFun[j] \in \FncSet[+\WSet_{j}]{} $ for each variable
$\varElm[j] \in \VarX[i]$ with $i\leq n$, we define $\asgChc
\in \AsgSet(\VarX[i])$ as the unique assignment $\asgElm$ such that
$\asgElm(\varElm_j) = \FFun[{j}](\asgElm\!\rst\!  \free{\Qnt[j][_{\WSet[j]}]
  \varElm[j] \ldotp \phiFrm'})$ for every $j \leq i$. We say that $\asgElm$ is
\emph{coherent} with $(\FFun_j)_{j\leq i}$.

\begin{lemma}[Buckets soundness]
  \label{lmm:BucketSoundness}
  For every choice function $\choiceFunBucket \in \ChoiceFunsBuckets$ over
  buckets and every play $\pthElm = \rho\posElm$ with $\posElm = (\phiFrm,
  \asgElm, \clubsuit)$ where $\clubsuit \in \{\PhsISym,\PhsIISym\}$, the
  following holds:
  \begin{itemize}
    \item if $\phiFrm = \Qnt[i][\WSet_i] \varElm_i \ldotp \phiFrm'$ and
      $\clubsuit = \PhsISym$, it holds $\asgElm =
      \asgElm((\choiceFunBucket(\BCls[{\varElm[j]}](\pthElm),\varElm[j]))_{j\leq
        i})$.
    \item if $\phiFrm = \psiFrm$ or $\clubsuit = \PhsIISym$ it holds $\asgElm =
      \asgElm((\choiceFunBucket(\BCls[{\varElm[j]}](\pthElm),\varElm[j]))_{j\leq
        n})$.
  \end{itemize}

\end{lemma}
\begin{proof}
  We prove this lemma by induction on the history $\hstElm$.

  For the base case history, the property is trivial.

  For the induction case, suppose the lemma holds for a history $\hstElm =
  \hstElm' \posElm'$ with $\posElm' = (\phiFrm', \asgElm', \clubsuit)$. Consider
  a history of the form $\hstElm \posElm$. There are two cases to consider:
  either $\clubsuit= \PhsISym$, or $\clubsuit = \PhsIISym$.
\begin{description}
  \item[($\clubsuit= \PhsISym$)] There are again two cases to look at:
   \begin{enumerate}
   \item if $\phiFrm' = \psiFrm$ then the only
     possible successor position $\posElm$ in the game is $(\varphiFrm,
     \asgElm', \PhsIISym)$. So, by definition of the bucket and direct
     application of the inductive hypothesis, the property holds for $\hstElm
     \posElm$.
    \item if $\phiFrm' = \Qnt[i][\WSet_i] \varElm_i \ldotp \phiFrm$, then
      $\posElm$ is of the form $(\phiFrm, \asgElm, \PhsISym)$.  The only bucket
      that might change is $\BCls[{\varElm[{i}]}]$.  By definition, any function
      $\FFun \in \BCls[{\varElm[{i}]}](\hstElm \posElm)$ satisfies
      $\FFun(\asgElm') = \asgElm'(\varElm[i])$ and the property holds for
      $\hstElm \posElm$. Indeed, by inductive hypothesis, $\asgElm(\varElm[j])
      = \asgElm'(\varElm[j])$, for every $\varElm[j]$ with $j<i$.
    \end{enumerate}
 \item[($\clubsuit= \PhsIISym$)] If $\phiFrm' = \psiFrm$, there is no reachable
   position. So, the only possibility is $\phiFrm' = \Qnt[i][\WSet_i]
   \varElm_i \ldotp \phiFrm$. There are again two possibilities:
\begin{enumerate}
	\item $\posElm$ is of the form $(\phiFrm, \asgElm', \PhsIISym)$. In this case,
    by the definition of bucket and a direct application of the inductive
    hypothesis, the property immediately follows for $\hstElm \posElm$.
	\item $\posElm$ is of the form $(\phiFrm,
    \mathring{\asgElm}{}[{\varElm[i]\mapsto a}], \PhsISym)$, for some $\aElm
    \!\in\! \ASet$ with $\aElm \neq \asgElm'(\varElm[i])$, with
    $\mathring{\asgElm} \defeq \asgElm' \!\rst \free{\Qnt[i][_{\WSet[i]}]
      \varElm[i] \ldotp \phiFrm'}$.  By inductive hypothesis, for every $\FFun
    \in \BCls[{\varElm[i]}](\hstElm)$, we have $\FFun(\mathring{\asgElm}) =
    \asgElm'(\varElm[i])$. Then, $\FCls[\phiFrm](\mathring{\asgElm}) \cap
    \BCls[{\varElm[i]}](\hstElm) = \emptyset$.  Thus,
    $\BCls[{\varElm[i]}](\hstElm \posElm) = \FCls[\phiFrm](\mathring{\asgElm})$
    and, by definition, we have $\FFun (\mathring{\asgElm}\rst\WSet[i]) =
    \mathring{\asgElm}(\varElm[i])$, for every $\FFun \in
    \BCls[{\varElm[i]}](\hstElm \posElm)$. Since the other buckets have not
    changed, the property holds for $\hstElm \posElm$.
\end{enumerate}
  \end{description}
\vspace{-1em}
\end{proof}

\figGraphThmGamThrSem

\thmgamthrsem*
\begin{proof}
  We prove that if the sentence is true in $\AStr$, then Eloise wins the game and
  if the sentence is false, then Abelard wins the game.

  First, suppose that the sentence $\varphiFrm$ is true in $\AStr$. By
  Theorem~\ref{thm:hst}, $\varphiFrm$ is equivalent to the \Meta sentence
  $\hspFun{\qntElm}\psiFrm$. So, by Lemma~\ref{lmm:MetaSkol} and recalling that
  in $\hspFun{\qntElm}$ the order of the quantifiers is reversed, we can
  conclude that there is a Skolemisation $(\SkolemizedSOF_{\varElm[i]} :
  \prod_{j>i} \FncSet[{\WSet_j}]{} \rightarrow \FncSet[{\WSet_i}]{})_{\varElm[i]
  \in \VarX[\exists]}$ such that $\AStr, \FunAsg, \TrvHypAsg \models
  \varphiFrm$, for every Skolem extension $\FunAsg$ of the empty assignment \wrt
  $(\SkolemizedSOF_{\varElm[i]})_{\varElm[i] \in \VarX[\exists]}$.
  We now define a strategy for Eloise and then prove that it is
  winning. Intuitively, the strategy consists in looking, by means of the
  buckets, at one possible function assignment of the variables controlled by
  Abelard and, then, applying what is prescribed by the Skolemisation
  $(\SkolemizedSOF_{\varElm[i]})_{\varElm[i] \in \VarX[\exists]}$ to select the
  values for the variables controlled by Eloise. Formally, let us fix a choice
  function $\choiceFunBucket \in \ChoiceFunsBuckets$ on the buckets. Given a
  history $\hstElm$, we define $\FFun[i][\hstElm]$ for $i\in\{0,\ldots,n\}$ as
  follows. If $\varElm[i] \in \VarX[\forall]$ then $\FFun[i][\hstElm] =
  \choiceFunBucket(\BCls[{\varElm[i]}](\hstElm), \varElm[i])$, otherwise,
  $\FFun[i][\hstElm] =
  \SkolemizedSOF_{\varElm[i]}((\FFun[j][\hstElm])_{j>i})$. When Eloise must make
  a move for the variable $\varElm[i]$ at the history $\hstElm = \hstElm'
  \posElm'$, with $\posElm' =(\phiFrm, \asgElm, \_)$, she moves to the position
  $\posElm = (\phiFrm', \asgElm', \_)$ with $\asgElm'(\varElm[i]) =
  \FFun[i][\hstElm](\asgElm)$. Observe that this strategy does not depend on the
  current phase of the game.



  Consider now a finite play $\pthElm = \rho\posElm$, with $\posElm = (\psiFrm,
  \asgElm, \PhsIISym)$, compatible with the strategy.
  We define a choice function $\choiceFunBucketWinning$ as follows: for all
  $\varElm[i]\in \XSet$
  $$\choiceFunBucketWinning(\BCls[{\varElm[i]}](\pthElm), \varElm[i]) =
  \FFun[i][{\pthElm}]$$
  The function $\choiceFunBucketWinning$ is a choice function since, if
  $\varElm[i] \in \VarX[\forall]$, by definition, $\FFun[i][\pthElm] \in
  \BCls[{\varElm[i]}](\pthElm)$ and if $\varElm[i] \in \VarX[\exists]$, then
  because Eloise played according to $\FFun[i][\pthElm]$, this function is in
  the bucket of $\varElm[i]$. The Lemma~\ref{lmm:BucketSoundness} ensures that
  the assignment $\asgElm$ is coherent with
  $(\choiceFunBucketWinning(\BCls[\varElm](\pthElm),\varElm))_{\varElm\in\VarX}$. By
  definition of $(\SkolemizedSOF_{\varElm[i]})_{\varElm[i] \in \VarX[\exists]}$,
  it holds that $\asgElm \models \psiFrm$. Therefore, the play is won by Eloise.

  Let us now consider an infinite play $\pthElm \in \PlaySet[][\omega]$
  compatible with the strategy. Toward a contradiction, suppose that the
  priority $\prtFun{(\pthElm)}$ of the play is odd.  Then, there must be a
  variable $\varElm[i] \in \VarX[\exists]$ such that \emph{(i)}
  $\prtFun(\varElm[i])$ appears infinitely often in $\chtFun{\pthElm}$ and
  \emph{(ii)} for all $j > i$ the priority $\prtFun(\varElm[j])$ appears only a
  finite number of time.
  Recall that if a variable $\varElm$ is not ``caught cheating'' after a finite
  prefix $\pthElm' = \rho \posElm$ of $\pthElm$, then
  $\BCls[{\varElm}](\pthElm') \subseteq \BCls[{\varElm}](\rho)$.
  But then, since we assumed the domain to be finite,
  starting from some index $N$ along the play $\pthElm$, the buckets for each
  $\varElm[j]$, with $j > i$, remain constant forever.
  Let us denote the constant bucket of $\varElm[j]$ by
  $\BCls[{\varElm[j]}]'$. Then, the strategy for $\varElm[i]$ is for Eloise to
  chose the values of $\varElm[i]$ by means of the function $\FFun[i] =
  \SkolemizedSOF_{\varElm[i]}((\choiceFunBucket(\BCls[{\varElm[j]}]'(\pthElm),
  \varElm[j]))_{j>i})$, which is constant since the buckets of variables
  $\varElm[j]$ do not change. By definition, for every prefix of $\pthElm$ of
  size greater than $N$, the bucket of $\varElm[i]$ contains $\FFun[i]$. Then
  the bucket of $\varElm[i]$ is never emptied and $\varElm[i]$ would never get
  ``caught cheating''.
  This is a contradiction.
  We proved that if the sentence is true, then Eloise has a winning strategy in
  $\GameName[\varphiFrm][\AStr]$.

  The second part of the proof proceeds similarly, in that we can apply the same
  exact reasoning, with only the roles of Eloise and Abelard exchanged, to
  obtain a winning strategy for Abelard.


\end{proof}



  \newpage
  \footnotesize
  \bibliographystyle{elsarticle-harv}
  \bibliography{References}

\end{document}
